\documentclass[oneside,12 pt]{book}
\usepackage{psfig}
\usepackage{latexsym}

\font\mybb=msbm9 at 10pt
\def\bb#1{\hbox{\mybb#1}}

\font\myBB=msbm9 at 12pt
\def\BB#1{\hbox{\myBB#1}}

\newcommand{\be}{\begin{equation}}
\newcommand{\ee}{\end{equation}}
\newcommand{\ba}{\begin{eqnarray}}
\newcommand{\ea}{\end{eqnarray}}

\newcommand{\dsl} {\kern.06em \hbox{ \raise.15ex
\hbox{$/$} \kern-.56em\hbox{$\partial$}}}
\newcommand{\Dsl}{\not\!\! D}

\newcommand{\eeq}{\end{equation}}
\newcommand{\eeqarr}{\end{eqnarray}}
\newcommand{\ZZ}{{\rm \kern 0.275em Z \kern -0.92em Z}\;}


\begin{document}
\setlength{\baselineskip}{1.5\baselineskip}

\begin{titlepage}
\begin{center}
{\LARGE\bf Born-Infeld action and Supersymmetry \\ } {(in
spanish)}
\end{center}

\vskip 1.5in

\centerline{\bf A thesis submitted to } %
\centerline{\bf Universidad de Cuyo} \centerline{\bf for the}
\centerline{\bf Ph.D. Degree in Physics}

\vskip 1.5in

 \centerline{\bf by} \centerline{\bf Guillermo A. Silva}
 \centerline{\bf Departamento de F\'{\i}sica}
 \centerline{\bf Facultad de Ciencias Exactas}
 \centerline{\bf Universidad Nacional de La Plata}

\vskip 1.5in

\centerline{\Large\bf  2000}

\end{titlepage}


\chapter*{Abstract}

In the thesis we analize different problems related to the
supersymmetric extension of the Dirac-Born-Infeld action. In
chapter 2 we introduce the DBI action and show how it appears in
string theory, we discuss also it's connection with Dp-branes.
Chapter 3 is a self contained introduction to supersymmetry, with
emphasis on BPS states. In chapter 4  we construct the $N=2$
supersymmetric extension of the Born-Infeld-Higgs in three
space-time dimensions and discuss it's BPS states and Bogomol'nyi
bounds.  In chapter 5 we construct the $N=1$ supersymmetric
extension of the non-abelian Born-Infeld theory in four space-time
dimensions. Chapter 6 deals with the analisis of BPS and non-BPS
solutions of the Dirac-Born-Infeld action and their interpretation
in superstring theory.  Chapter 7 is devoted to the conclusions.
Three appendix complete the work.


\chapter*{Agradecimientos}

En primer lugar quiero agradecer a Fidel, por el infatigable est\'
\i  mulo que me brind\'o en estos tres a\~nos, por su incisivo
car\'acter a la hora de discutir y por todo lo que me ense\~n\'o.
A los chicos de La Plata y Buenos Aires: charles, sonia, mart\' \i
n, daniel, nico, leo, gast\'on y mauricio, todas las horas que
charlamos tomando caf\'e y aprendiendo.

Finalmente nada en esta tesis hubiera sido posible sin el apoyo de
muchos amigos, $(ales,~jaila,~gordi~y~lio)\times 2$, toda la
familia, y en especial de vos, SIL.

\pagestyle{plain}
\pagenumbering{roman}

\tableofcontents



\chapter{Introducci\'{o}n}

\pagenumbering{arabic}

\vspace{.7cm}

La motivaci\'on original de Max Born y Leopold Infeld al formular
la teor\' \i a hoy conocida como de Born-Infeld \cite{B}-\cite{BI}
fue la de atribuir, a nivel cl\'asico, un origen
electromagn\'etico a la masa de las part\' \i culas conocidas
hasta ese momento, mediante una modificaci\'on de las ecuaciones
de Maxwell. La idea b\'{a}sica era considerar a las part\' \i culas de
materia como singularidades del campo, de manera que la noci\'on
de masa pudiera expresarse en t\'erminos de la energ\' \i a de la
configuraci\'on de campo (masa electromagn\'etica). En lenguaje
moderno, Born e Infeld quer\' \i an ver aparecer a las part\' \i
culas  como solitones del modelo. En realidad no lo lograron: la
soluci\'on que presentaron, hoy llamada BI\'{o}n, es una soluci\'on
con fuentes (a diferencia del caso solit\'onico) de las ecuaciones
de movimiento.

La observaci\'on hecha por Born e Infeld, que condujo a postular
una acci\'on, se bas\'{o} en una analog\' \i a con la relatividad
especial, donde el imponer una cota m\'axima para la velocidad,
fuerza a reemplazar la acci\'on de Newton para la part\' \i cula
libre, por la acci\'{o}n relativista. Aplicando una condici\'{o}n an\'{a}loga
en relaci\'{o}n con una nueva constante b\'{a}sica $\beta$, reemplazaron
la acci\'{o}n de Maxwell por una que hoy lleva sus nombres.  La
constante dimensional $\beta$ est\'a relacionada con el valor
m\'aximo (finito) que puede tomar el campo electromagn\'etico.
Esto conduce, al calcular la energ\' \i a de la soluci\'on, a un
resultado finito. En la propuesta de Born e Infeld para
interpretar al electr\'{o}n,  la masa ten\' \i a origen
electromagn\'etico, y se resolv\' \i a el problema de autoenerg\'
\i a infinita.

El descubrimiento del neutr\'on mostr\'o que la masa no estaba
indisolublemente ligada a la carga \cite{diri}. Las dificultades
encontradas en la cuantificaci\'{o}n de la teor\' \i a y el \'{e}xito de
la electrodin\'{a}mica cu\'{a}ntica desarrollada por Dyson, Feynman,
Schwinger y Tomonaga, provocaron el olvido de la teor\' \i a de
Born-Infeld por aproximadamente 50 a\~nos. Reci\'{e}n en los a\~nos
'80 se la vi\'{o} resurgir en el contexto de la teor\' \i a de
cuerdas.

\vspace {.5cm}

La supersimetr\' \i a naci\'o del estudio realizado por Y.
Gol'fand y E. Likhtman de las \'algebras de Lie gradadas
\cite{gl}. Este tipo de  \'algebras de Lie hace aparecer, adem\'as
de los conmutadores usuales, anticonmutadores entre algunos de los
generadores del \'algebra. Independientemente, J. Wess y B. Zumino
\cite{wezu} concibieron la idea de la supersimetr\' \i a al
generalizar a $d=4$ la supersimetr\' \i a del modelo de cuerdas
bidimensional de Ramond-Neveu-Schwarz.

Desde un punto de vista algebraico la supersimetr\' \i a consiste
en extender el \'algebra de Poincar\'e mediante el agregado de
generadores espinoriales-fermi\'onicos $Q$ (cargas
supersim\'{e}tricas) \cite{wess}-\cite{lykken}. La presencia de estas
cargas conduce a que toda representaci\'on irreducible cuente con
igual n\'umero de estados bos\'onicos y fermi\'onicos. El teorema
Haag-{\L}opusza\'nski-Sohnius  \cite{haag} muestra que la m\'axima
simetr\' \i a unitaria que puede tener una teor\' \i a en
in\-te\-rac\-ci\'on, es el producto directo de una simetr\' \i a
de gauge por el grupo generado por el \'algebra de super-Poincar\'{e},
pudiendo contener este, en general, $N$ cargas espinoriales
(supersimetr\' \i a extendida).

En principio, la motivaci\'on para el estudio de las teor\' \i as
supersim\'{e}tricas es puramente te\'orica. Sin embargo, han sido
intensamente estudiadas pues presentan interesantes
pro\-pie\-da\-des formales. En particular, tienen un
comportamiento UV mejorado debido a la compensaci\'on de
divergencias de las contribuciones bos\'onicas y fermi\'onicas (en
ciertos modelos $N>1$ es posible mostrar que las contribuciones
perturbativas se anulan a partir de cierto orden). La invarianza
supersim\'etrica impone fuertes v\' \i nculos en la construcci\'on
de posibles modelos, pudi\'endose en ciertos casos hallar
completamente la acci\'on efectiva Wilsoniana de la teor\' \i a
\cite{sw}. Desde un punto de vista fenomenol\'ogico, la
incorporaci\'on de la supersimetr\' \i a al modelo est\'andar
mejora el comportamiento de las constantes de acoplamiento con la
energ\' \i a permitiendo la formulaci\'on de teor\' \i as de gran
unificaci\'on.

Es sabido que en ciertos modelos bos\'onicos
\cite{inst}-\cite{Bogo},\cite{NS1}-\cite{NS2} es posible encontrar
soluciones a las ecuaciones de segundo orden de Euler-Lagrange
considerando ecuaciones de primer orden, conocidas como ecuaciones
BPS (por sus descubridores  E. Bogomol'nyi, M. Prasad y C.
Sommerfield). El m\'etodo est\'{a}ndar para obtener tales ecuaciones
consiste en reescribir la expresi\'on para la energ\' \i a de una
configuraci\'on de campo de manera que quede acotada inferiormente
por una cantidad que tiene car\'{a}cter topol\'ogico. Las
configuraciones de campos que saturan tal cota satisfacen las
ecuaciones de Euler-Lagrange como asi tambi\'en las ecuaciones de
BPS. A nivel cl\'asico las soluciones de estas ecuaciones
corresponden a multi-solitones o instantones \cite{raja}.

Esta estructura se reinterpret\'o al aparecer en la extensi\'on
supersim\'etrica de dichos modelos, donde permiti\'o adem\'as
hacer predicciones exactas acerca del espectro cu\'antico de la
teor\' \i a. En efecto, estudiando teor\' \i as supersim\'etricas
donde el vac\' \i o es degenerado y en las cuales existen cargas
topol\'ogicas no nulas (solitones) se observ\'o
\cite{OW},\cite{LLW}-\cite{cbpf} que las extensiones centrales
admitidas por el \'{a}lgebra supersim\'{e}trica tienen lugar en los
modelos, en t\'erminos de las cargas topol\'ogicas. La presencia
de extensiones centrales en el \'algebra super-Poincar\'e se
indentific\'o entonces con la existencia de cargas topol\'ogicas
en la teor\' \i a.

Esta observaci\'on permiti\'o demostrar que la cota de
Bogomol'nyi, mencionada mas arriba a nivel cl\'asico, es v\'alida
a nivel cu\'antico, tomada como valor de expectaci\'on sobre
cualquier estado f\' \i sico, como consecuencia de la unitariedad
de la teor\' \i a. Estudiando las representaciones irreducibles
del \'algebra supersim\'etrica se mostr\'o que los estados que
saturan la cota, conocidos como estados BPS, o sea estados cuya
masa es igual a su carga topol\'ogica, presentan la propiedad de
ser invariante frente a algunas de las cargas supersim\'etricas y
de mantener a nivel cu\'antico la igualdad entre masa y carga. Las
ecuaciones BPS aparecen entonces de imponer que dichas cargas sean
nulas sobre el estado BPS (se llaman estados
$\frac12,\frac14,\ldots$-BPS a los estados que son invariantes
frente a $\frac12,\frac14,\ldots$ del n\'umero total de
supersimetr\' \i as). Las cargas no nulas actuando sobre el estado
generan el resto de los estados del multiplete supersim\'{e}trico.

\vspace{.5cm}

La teor\' \i a de cuerdas \cite{tring} tuvo su origen en un
intento de describir las propiedades de las interacciones fuertes
mediante la construcci\'on del modelo resonante dual.
Experimentalmente se observaba una enorme proliferaci\'on de
hadrones (resonancias hadr\'onicas) que parec\' \i an tener
espines indefinidamente altos. Una ley muy simple que relacionaba
la masa con el esp\' \i n (trayectorias de Regge). R\'apidamente,
sin embargo, qued\'o claro que la consistencia del modelo
resonante dual, luego reconocido como proveniente de la
cuantificaci\'on de una cuerda relativista, predec\' \i a un
conjunto de part\' \i culas no masivas que no inclu\' \i a a la
part\' \i cula pseudo-escalar necesaria en el l\' \i mite quiral
de las interacciones fuertes (pi\'on). Este hecho, sumado a un
decaimiento exponencial (en lugar de la observada ley de
potencias) para las secciones eficaces de alto momento
transferido, dej\'o en claro que las teor\' \i as de cuerdas no
pod\' \i an dar una descripci\'on acabada de las interacciones
fuertes. Entre 1973 y 1974 una teor\' \i a de campos, la
cromodin\'amica cu\'antica, emergi\'o como alternativa para la
descripci\'on de las interacciones fuertes explicando, entre otras
cosas, el comportamiento de ley de potencias para las secciones
eficaces.

Abandonada transitoriamente para describir las interacciones
fuertes, aparecieron nuevas motivaciones en otros \'{a}mbitos para
estudiar la rica estructura de las teor\' \i as de cuerdas. Su
excelente comportamiento UV, la presencia de un estado no masivo
de esp\' \i n $s=2$ (que podr\' \i a asociarse con el gravit\'{o}n) y
la consistencia de la teor\' \i a para $d>4$, sugirieron estudiar
a la teor\' \i a de cuerdas como una posible teor\' \i a cu\'{a}ntica
donde se encontrar\' \i an unificadas todas las interacciones
conocidas incluyendo la gravedad \cite{ss}. El an\'alisis de
consistencia de la teor\' \i a requiri\'o la introducci\'on de la
supersimetr\' \i a. Con esta estructura adicional, la teor\' \i a
nacida de los modelos duales pas\'o a llamarse teor\' \i a de
supercuerdas.

Paralelamente a los estudios anteriormente citados Alexander
Polyakov \cite{poly} formul\'o la teor\' \i a de cuerdas en
t\'erminos de integrales funcionales, lo que permiti\'o
interpretar los antiguos resultados de consistencia de los modelos
duales. Fue a partir de esta \'ultima formulaci\'on que la
acci\'on de Born-Infeld reapareci\'{o} en nuestros d\' \i as. En
efecto, usando el formalismo de Polyakov, E. Fradkin y A.
Tsey\-tlin obtuvieron la acci\'on de Born-Infeld como acci\'on de
baja energ\' \i a para los modos vectoriales $A_{\mu}$ de la
teor\' \i a de cuerdas abiertas \cite{Tse}-\cite{tse}. Con
condiciones de contorno de Neumann para los extremos de la cuerda
acoplada a un campo de gauge abeliano, la acci\'on de Born-Infeld
se obtuvo en la aproximaci\'on a orden \'{a}rbol (de la teor\' \i a de
cuerdas) y en el l\' \i mite de campos electromagn\'eticos
cuasi-constantes (despreciando derivadas de $F_{\mu\nu}$).

Posteriormente, la acci\'{o}n de Born-Infeld fue obtenida mediante el
m\'etodo de campos de fondo, ``background fields method"
\cite{AN}. En este \'ultimo trabajo se la obtuvo acoplando una
cuerda bos\'onica abierta, con condiciones de contorno de Neumann,
a un campo abeliano de gauge y exigiendo invarianza conforme de la
teor\' \i a (nuevamente en la aproximaci\'on a orden \'arbol para
la teor\' \i a de cuerdas y en el l\' \i mite de campos
electromagn\'{e}ticos cuasi-constantes). La invarianza conforme
implica la anulaci\'on de las funciones beta (Callan-Zymanzik),
las cuales dan ecuaciones de movimiento para $F_{\mu\nu}$. A
partir de \'{e}stas es posible derivar una acci\'on efectiva, que
coincide con la de Born-Infeld obtenida en \cite{Tse}. Lo notable
de ambos c\'alculos es que el resultado obtenido es exacto en
$\alpha'$, al orden calculado.

En 1989, estudiando la teor\' \i a de cuerdas bos\'onicas
compactificada toroidalmente, se concluy\'{o} \cite{Lei} que la teor\'
\i a de cuerdas deber\' \i a contener objetos macrosc\'{o}picos
extendidos en p-dimensiones espaciales. Se llam\'{o} a estos objetos
Dp-branas y se determin\'{o} que interactuaban en el l\' \i mite de
bajas energ\' \i as con cuerdas abiertas y cerradas.
Inmediatamente, usando la t\'{e}cnica de campos de fondo, se dedujo en
\cite{lei2} la acci\'{o}n efectiva para los modos no masivos
(provenientes del sector de cuerdas abiertas) de estos objetos
``solit\'{o}nicos". La misma se obtuvo pidiendo invarianza conforme a
una teor\' \i a de cuerdas abiertas con condiciones de contorno
mixtas Neumann-Dirichlet para los extremos de las
cuerdas\footnote{El formalismo tradicional fijaba condiciones de
Neumann en los extremos de la cuerda, debido a que son invariantes
de Lorentz. Las condiciones de Dirichlet para los extremos de la
cuerda no son invariantes Lorentz, lo cual es natural si estamos
en presencia de un solit\'{o}n.}. El resultado que se obtuvo para
(p+1)-coordenadas $X^\mu$ con condiciones de Neumann y
(26-(p+1))-coordenadas $X^i$ con condiciones Dirichlet fue la
acci\'{o}n de Dirac-Born-Infeld en (p+1)-dimensiones. En el caso
supersim\'{e}trico (teor\' \i a de supercuerdas), la din\'{a}mica de bajas
energ\' \i as para la Dp-brana contiene entonces en su parte
bos\'onica 10-(p+1) campos escalares (modos de Dirichlet) que se
interpretan como las oscilaciones de la Dp-brana sumergida en el
espacio-tiempo 10-dimensional, junto con un campo de gauge
contenido en el volumen de mundo de la Dp-brana (modos de
Neumann).

La importancia de estos estados para la teor\' \i a de
supercuerdas no se advirti\'o hasta que Joseph Polchinski
descubri\'o que estaban cargados frente a ciertas (p+1)-formas
contenidas en la teor\' \i a \cite{Pol2}-\cite{Pol}.  El c\'{a}lculo
mostr\'{o} que las Dp-branas son objetos no perturbativos que
satisfacen una generalizaci\'on de la igualdad entre masa y carga.
El trabajo de Polchinski mostr\'o tambi\'en que las Dp-branas son
un nuevo tipo de solitones de la teor\' \i a de cuerdas. En
particular, su tensi\'on (``masa") es inversamente proporcional a
la constante de acoplamiento de las cuerdas $T\sim1/g$ (a
diferencia de los solitones usuales cuya masa $M\sim1/g^2$), y que
admiten una descripci\'{o}n muy simple y exacta como hipersuperficies
donde pueden t\'{e}rminar cuerdas abiertas. Las Dp-branas son los
estados de la teor\' \i a cargados respecto de (p+1)-formas y
debido al trabajo de R. Leigh \cite{lei2}, la din\'amica de bajas
energ\' \i as est\'a dada por la acci\'on de Dirac-Born-Infeld. Es
posible ver que ciertas soluciones cl\'asicas solit\'onicas
halladas en el contexto de SUGRA+SYM corresponden a la
descripci\'on macrosc\'opica de una superposici\'on de Dp-branas
\cite{ads}.

En 1995 Edward Witten \cite{Wi} mostr\'{o} la necesidad de extender,
en el contexto de D-branas superpuestas, la acci\'on de
Born-Infeld al caso no abeliano. En la citada referencia se mostr\'{o}
que al superponer $\cal N$ Dp-branas el espectro no masivo se
modifica de manera tal que la acci\'{o}n efectiva para la
configuraci\'{o}n debe ser descripta por una teor\' \i a de gauge con
grupo $U(\cal N)$. Dado que la acci\'{o}n de bajas energ\' \i as para
una Dp-brana es la acci\'{o}n de Dirac-Born-Infeld, es natural
conjeturar que en el caso de $\cal N$ Dp-branas, la teor\' \i a
efectiva ser\'{a} la generalizaci\'{o}n no abeliana de la teor\' \i a de
Dirac-Born-Infeld con grupo de gauge $U(\cal N)$. En el contexto
de supercuerdas es entonces importante tener una definici\'on para
la acci\'on no abeliana de Born-Infeld ya que brindar\' \i a la
posibilidad de estudiar el l\' \i mite de bajas energ\' \i as de
un conjunto de D-branas superpuestas. Esta definici\'{o}n debe ser
compatible con supersimetr\'{\i}a.

Con el trabajo de Polchinski \cite{Pol2} en mente y basados en
argumentos debidos a Andrew Strominger \cite{stro}, Curtis Callan
y Juan Maldacena estudiaron soluciones cl\'asicas de la acci\'on
de Dirac-Born-Infeld \cite{CM}. Las  soluciones las hallaron
usando argumentos BPS (ver cap. \ref{herra}) y correspond\' \i an
para la teor\' \i a (p+1)-dimensional, al potencial coulombiano de
un electr\'on puntual junto con la excitaci\'on de un campo
escalar. Computando la energ\' \i a \cite{CM} interpretaron desde
la perspectiva de la teor\' \i a de cuerdas al electr\'on en el
volumen de mundo de la Dp-brana como la intersecci\'on de una
cuerda fundamental con la Dp-brana. La autoenerg\' \i a infinita
de la soluci\'on se debe a la extensi\'on semi-infinita de la
cuerda que intersecta la Dp-brana. Simult\'aneamente al trabajo
anterior, Gary Gibbons \cite{G} encontr\'o las mismas soluciones,
mostrando c\'{o}mo las soluciones BPS separan naturalmente dos
familias mas generales de soluciones.

Con el objeto de validar la imagen de Polchinski respecto de las
Dp-branas, en \cite{CM} se investig\'{o} la propagaci\'on de una
perurbaci\'on normal tanto a la cuerda como a la D3-brana para una
soluci\'on de fondo de tipo BPS. El resultado obtenido fue el
esperado, esto es, la condici\'{o}n efectiva para la cuerda sujeta a
la brana es del tipo Dirichlet. Posteriormente, en \cite{Hashi} se
construyeron soluciones no-BPS el\'ectricas para la acci\'on de
DBI y se discutieron  tambi\'en  soluciones BPS magn\'eticas. Un
estudio detallado de las soluciones BPS di\'onicas fue presentado
en \cite{BLM}.

\vspace{.5cm}

Los trabajos descriptos en esta tesis tratan de dar respuesta a
algunos de los problemas discutidos mas arriba. As\'{\i}, en el
cap\'{\i}tulo \ref{susya} construiremos la extensi\'{o}n supersim\'{e}trica
$N=2$ del modelo abeliano de Born-Infeld-Higgs en tres dimensiones
espacio-temporales. Discutiremos la cota de Bogomol'nyi y las
ecuaciones BPS del modelo, compar\'andolas con las conocidas
soluciones de v\'ortice del modelo supersim\'etrico de
Maxwell-Higgs \cite{dVS},\cite{ed}. Analizaremos tambi\'{e}n la
sensibilidad de las ecuaciones BPS en modelos no polin\'{o}micos
generales para el campo de gauge. En el cap\' \i tulo \ref{nstr}
discutiremos la extensi\'on no abeliana de la acci\'on de
Born-Infeld en conexi\'{o}n con supersimetr\' \i a. El primer paso en
la construcci\'{o}n de la teor\' \i a de Born-Infeld no abeliana
consiste en reemplazar el tensor de campo $F_{\mu\nu}$ por su
extensi\'{o}n no abeliana $F_{\mu\nu}^{a}t^{a}$ y reemplazar la
m\'{e}trica $g_{\mu\nu}$ por $g_{\mu\nu}\cal I$ donde $\cal I$ es la
matriz identidad en el espacio del grupo. Luego, dado que el
lagrangiano debe ser un escalar tanto en el espacio-tiempo como en
el grupo, es necesario incluir una operaci\'{o}n de ``traza" (o
determinante) sobre los \' \i ndices de grupo. Existen en la
literatura trabajos que sugieren distintas prescripciones para la
traza\cite{H}-\cite{Tse2}; discutiremos en particular la
sugerencia de A. Tseytlin \cite{Tse2} de simetrizar los
generadores del grupo, debido a la observaci\'{o}n \cite{Bre} de que
es posible establecer una cota de Bogomol'nyi para dicha acci\'{o}n,
lo que sugiere que la misma ser\' \i a supersimetrizable. En el
cap\' \i tulo \ref{nobps} nos concentraremos en el caso de
D3-branas y construiremos ex\-pl\' \i\-ci\-ta\-men\-te soluciones
di\'onicas no-BPS con el campo de gauge U(1) acoplado a un campo
escalar. Analizaremos luego las soluciones en conexi\'on con la
geometr\' \i a de la deformaci\'on de la brana por efecto de la
tensi\'on de la cuerda-$(n,m)$  que soporta cargas el\'ectricas y
magn\'eticas \cite{Wi},\cite{Sch}. Estudiando la energ\' \i a de
estas configuraciones no-BPS, compararemos los resultados con los
obtenidos en los casos BPS y no-BPS puramente el\'ectricos
\cite{CM}-\cite{BLM}. Estudiaremos tambi\'en peque\~nas
excitaciones, transversales tanto a la brana como a la cuerda, de
manera de examinar si la respuesta de las soluciones no-BPS es
consistente con la interpretaci\'on en la cual el sistema
D3-brana+cuerda descripto corresponde a las condiciones de borde
adecuadas (Dirichlet).



\chapter{Ideas b\'asicas\label{herra}}
\begin{center}
 {\begin{minipage}{6truein}
 { \sl Discutiremos varias ideas que es necesario precisar
 para el desarrollo de la tesis.}
 \end{minipage}}
\end{center}

\section{La Teor\' \i  a de Born-Infeld}

El lagrangiano propuesto por Born e Infeld tiene la forma
\cite{BI}\footnote{Ver ap\'endice \ref{a1} para una descripci\'on
mas detallada de la propuesta original.}
\ba
  {\cal L}_{{BI}}\!\!\!&=\!\!\!&{\beta ^2}\left( \sqrt{-\det (
 g_{\mu \nu })\phantom{\frac 1\beta}} - \sqrt{-\det \left(
 g_{\mu \nu }+\frac 1\beta F_{\mu \nu }\right) } \right)\\
  \!\!\!&=\!\!\!&{\beta ^2}\left(1 - \sqrt{1-\frac
 1{\beta ^2} (\vec E^2-\vec B^2) -\frac 1{\beta ^4}
 (\vec E\cdot\vec B)^2} \right)
 \label{lbii}
\ea
La segunda l\' \i nea se obtiene en el caso de espacio de
Minkowski, $g_{\mu\nu}=diag(+,-,-,-)$.

El punto esencial en la teor\' \i a de BI es la diferencia entre
el campo el\'{e}ctrico (ponderomotriz) $\vec E$ y el vector de
desplazamiento (inducci\'{o}n) $\vec D$.  El electr\'on de BI que
denominaremos BI\'on satisface
\be
 {\rm div}\vec D=q_e \delta^{(3)}(\vec r)
\ee
Obviamente $\vec D$ diverge en el origen, pero dado que en
ausencia de campos magn\'eticos, el campo el\'ectrico $\vec E$ y
el vector de desplazamiento $\vec D$ est\'an relacionados por
\be
 \vec D=\left.{\frac {\delta {\cal L}_{{BI}}}
 {\delta \vec E}}\right\rfloor_{\vec B=0}=
  \frac{\vec E} {\sqrt{1-\frac 1{\beta^2} \vec E^2}}
  ~~\Longrightarrow~~\vec E=\frac {\vec D}
  {\sqrt{1+\frac 1{\beta^2}\vec D^2}}
\ee
vemos que el valor del campo el\'ectrico en el origen, donde esta
situada la fuente, es finito $\vec E(\vec r=0)=\beta$. Esta
propiedad conduce, al calcular la energ\' \i a de la soluci\'on, a
un resultado finito (ver cap. \ref{nobps}).

Sintetizando, para soluciones puramente el\'ectricas tenemos:

\noindent
 {\underline{Teor{\'{\i}}a  de Maxwell}}
\ba
 &{\cal L}=\frac12 \vec E^2
 &\Longrightarrow~~ \vec\nabla\cdot\vec E=q\,\delta^{(3)}\longrightarrow
 \vec E\sim\frac q{r^2}\nonumber\\
 &\vec D=\vec E&\Longrightarrow~~
 E=\int d^3x\; \frac 12\vec E^2\sim
 {\rm lim_{\Lambda\to\infty}} \frac {q^2}{\Lambda}~~
 ({\rm divergente})\nonumber
\ea
\noindent\underline{Teor{\'{\i}}a de Born-Infeld}
\ba
 & {\cal L}=\beta^2 (1-\sqrt{1-\frac1{\beta^2}\vec E^2})
 &\Longrightarrow~~ \vec\nabla\cdot\vec D=q\delta\longrightarrow
 \vec E\sim\frac q{\sqrt{r^4+q^2/\beta^2}}\nonumber\\
 &\vec D=\frac {\vec E}{\sqrt{1-\vec E^2/\beta^2}}&\Longrightarrow~~
 E=\int d^3x \;(\vec D\cdot\vec E-{\cal L})\sim q^{3/2} \beta^{1/2}
 ~~({\rm finita})
 \nonumber
\ea

\section{Acci\'on de Born-Infeld en teor\' \i a de cuerdas  }

Veamos sint\'eticamente la derivaci\'{o}n de la acci\'{o}n de BI en el
contexto de la teor\' \i a de cuerdas siguiendo  \cite{Tse}. La
acci\'{o}n efectiva off-shell para los modos no masivos de una teor\'
\i a  bos\'onica de cuerdas abiertas y cerradas se define, en
espacio eucl\' \i deo, como
\be
 \Gamma[\phi,A_\mu,g_{\mu\nu},b_{\mu\nu}]=\sum_{\chi=2,1,0,-1,..}
 g_{op}^{-2\chi}\int {\cal D}h_{ab}{\cal D} X^\mu~ e^{-S_1-S_2}
 \label{efect}
\ee

\be
 S_1=\int_{{\cal M}}d^2z\left(\frac 1 {4\pi\acute{\alpha}}\sqrt h h^{ab}
 \partial_a X^\mu \partial_b X^\nu g_{\mu\nu}
 +i\varepsilon^{ab}\partial_a X^\mu \partial_b X^\nu b_{\mu\nu}
 +\frac 1 {4\pi} \sqrt h R\, \phi\right)
 \label{s1}
\ee

\be
 S_2=i\int_{\partial{\cal M}}d\tau \dot X^\mu A_\mu
 \label{s2}
\ee

Aqu\' \i~ los $X^\mu$ describen el embedding de la hoja de mundo
${\cal M}$ de la cuerda  en el espacio-tiempo ($\mu,\nu=1,...,D$),
$z^a$ (a=1,2) son las coordenadas de ${\cal M}$, $h_{ab}$ es la
m\'{e}trica en ${\cal M}$, $R$ es la curvatura escalar de $h_{ab}$,
$\partial {\cal M}$ es el borde de ${\cal M}$ parametrizado por
$\tau$ y $g_{op}$ es la constante de acoplamiento respecto de cual
est\'a definida  la teor\' \i a perturbativa. En (\ref{efect}) se
suma sobre todas las posibles hojas de mundo virtuales ${\cal M}$
compactas, orientables y con borde $\partial {\cal M}$, las cuales
topol\'{o}gicamente corresponden a discos con un n\'{u}mero arbitrario de
bordes $n$ ($\partial{\cal M}=\cup_{i=1}^n \gamma_i$) y manijas
$k$ ($\chi=2-2k-n$) ( ${\cal M }$ tiene la topolog\' \i a de un
disco en la aproximaci\'on a orden \'{a}rbol, un anillo a un loop,..,
luego, la expansi\'{o}n en (\ref{efect}) corresponde a una expansi\'{o}n
en las clases topol\'{o}gicas). En cada clase topol\'{o}gica fijamos el
borde, integramos sobre todas las posibles superficies con
condiciones de contorno fijas $X^\mu(z)\rfloor_{\partial {\cal
M}}=c^\mu(\tau)$ y luego sumamos sobre todas las posibles curvas
de borde $c^\mu(\tau)$. Las cuerdas (abiertas y cerradas) se mueve
sobre campos de fondo correspondientes a sus modos no masivos: el
dilat\'on $\phi$, el campo vectorial $A_\mu$, el gravit\'on
$g_{\mu\nu}$ y el tensor antisim\'etrico $b_{\mu\nu}$.

La acci\'{o}n efectiva definida por (\ref{efect}) se puede interpretar
de distintas formas. Una primera interpretaci\'on es que
(\ref{efect}) es la generatriz funcional de las amplitudes de
scattering off-shell sobre fondos arbitrarios: derivando respecto
de las fuentes $\phi,A_\mu,..$ obtenemos el valor de expectaci\'{o}n
de los correspondientes operadores de v\'{e}rtice para una cuerda
propag\'andose en un fondo de campos $\phi, A_\mu,...$ (poniendo a
cero las fuentes luego de derivar y fijando las condiciones de
capa de masa obtenemos las funciones de correlaci\'on para los
modos no masivos de la cuerda on-shell). Una segunda
interpretaci\'on es que $\Gamma$ en (\ref{efect}) es el an\'alogo
directo de la acci\'on efectiva est\'andar en teor\' \i a de
campos (generatriz funcional de las funciones de Green
irreducibles), esto significa que contiene toda la din\'amica
cu\'antica y no debe ser, consecuentemente, cuantizada. El punto
es que los argumentos de $\Gamma$ son los valores de fondo para
los campos (correspondientes a modos virtuales de cuerdas) que se
propagan en loops (manijas y agujeros en la hoja de mundo
corresponden al an\'alogo de los lazos en teor\' \i a de campos).
Dado que esta formulaci\'on es off-shell, el vac\' \i o de la
teor\' \i a se halla minimizando $\Gamma$
\be
 \delta \Gamma/\delta \phi=0,~~~~~\delta \Gamma/\delta A_\mu=0,
 ~~~~~\delta \Gamma/\delta g_{\mu\nu}=0,...
 \label{invconf}
\ee
Confirmar que el vac\' \i o es estable significa coorroborar la
ausencia de fantasmas y taquiones en el espectro de peque\~nas
fluctuaciones y la consistencia de las amplitudes de n-puntos al
ser calculadas sobre el fondo correspondiente. Un posible criterio
necesario de consistencia es la invarianza conforme de $S_1+S_2$
definidas por (\ref{s1})-(\ref{s2}) cuando tomamos como campos de
fondo de la cuerda las soluciones de (\ref{invconf}) (este fue el
m\'etodo utilizado en \cite{AN} para hallar la acci\'on de BI).

Para obtener la acci\'on de BI en el contexto de la cuerda
bos\'onica \footnote{Es posible realizar un c\'alculo similar en
la teor\' \i a de supercuerdas, obteni\'endose el mismo resultado
\cite{tse}.} nos restringiremos por simplicidad a fondos triviales
para los campos que provienen del sector de cuerdas cerradas
($g_{\mu\nu}=\delta_{\mu\nu},\phi=b_{\mu\nu}=0$). Nos limitaremos
tambi\'{e}n al c\'{a}lculo a orden \'{a}rbol, $\chi=1$ (disco), para la cuerda
bos\'onica en $D=26$. Elegimos condiciones de contorno de Neumann
en todas las direcciones, para los extremos de la cuerda. Siendo
la integral sobre las m\'{e}tricas $h_{ab}$ trivial, separamos $X^\mu$
en parte constante y no constante, $X^\mu=x^\mu+\xi^\mu$, quedando
\be
 \int {\cal D}X^\mu e^{-S_1-S_2}=\int d^{D}\!x\int{\cal D}\xi^\mu
 \exp(-\frac 1 {4\pi\acute{\alpha}}
 \int d^2z\,\partial_a \xi^\mu\partial_a \xi^\mu
 -i\int d\tau \,\dot\xi^\mu A_\mu(x+\xi))
 \label{inta}
\ee
donde las funciones $\xi^\mu(z)$ representan fluctuaciones de la
hoja de mundo ${\cal M}$ de la cuerda  que satisfacen condiciones
de contorno de Neumann. La factorizaci\'on que permite reescribir
a $\Gamma$ como una integral sobre el espacio tiempo ${\BB R}^{D}$
se basa en que la acci\'on para la cuerda libre (\ref{s1}) con
$g_{\mu\nu}=\delta_{\mu\nu}$ es invariante frente a traslaciones
$X^\mu \to X^\mu+a^\mu$ con $a^\mu=cte$. Luego la funci\'on de
partici\'on $\Gamma[0,0,\delta_{\mu\nu},0]$ contiene la
contribuci\'on de estos modos cero (el volumen de ${\BB R}^{D}$)
como un factor. Esta invarianza se rompe en presencia de un fondo
no trivial de manera que la integral sobre ${\BB R}^{D}$ no se
factoriza. Procedamos a evaluar la integral funcional en todos los
puntos interiores de ${\cal M}$, reduciendo (\ref{inta}) a una
integral sobre la frontera. A tal efecto introducimos las
variables $\eta^\mu(\tau)$ definidas como la restricci\'on de
$\xi^\mu$ a la frontera (el borde del disco) $\partial{\cal
M}=S^1$. Insertando en la integral funcional (\ref{inta}) el
siguiente ``1"
\be
 1=\int{\cal D}\eta^\mu\delta\left(\xi^\mu\rfloor_{\partial{\cal M}}
 -\eta^\mu\right)
\ee
y representando la funci\'on $\delta$ mediante una integral
funcional sobre $\nu_\mu(\tau)$, llegamos a una expresi\'on
gaussiana para los campos de integraci\'on $\xi^\mu$ y $\nu_\mu$
\ba
 \int{\cal D}\xi^\mu e^{-S_1-S_2[\,\xi|_{\partial{\cal M}}]}=
 \int{\cal D}\xi^\mu{\cal D}\eta^\mu{\cal D}\nu_\mu
 &&\!\!\!\!\!\!\!\!\exp\left(
 -\frac 1 {4\pi\acute{\alpha}}
 \int d^2z\,\partial_a \xi^\mu\partial_a \xi^\mu
 -i\int d\tau \,\dot\eta^\mu A_\mu(x+\eta)\right.\nonumber\\
 &&\left.-\int d\tau\, \nu_\mu
 (\xi^\mu\rfloor_{\partial{\cal M}}
 -\eta^\mu)\right)
\ea
Integrando los campos gaussianos obtenemos
\be
 \int{\cal D}\eta^\mu \exp\left(-\frac1 {4\pi\acute\alpha}\;\eta\,
 G^{-1}\eta-i\int d\tau \,\dot\eta^\mu A_\mu(x+\eta)
 \right)
 \label{prima}
\ee
donde
\be
 \eta\,G^{-1}\eta=\int d\tau d\tau'\, \eta^\mu(\tau)\, G^{-1}
 (\tau,\tau')\eta^\mu(\tau')
\ee
y $G^{-1}$ esta definida de la siguiente manera. Encontrada la
funci\'on de Neumann $N(z,z')$ para el operador de Laplace
$\Box=\partial_a\partial_a$, $\Box N=-\delta^{(2)}(z-z')$,
calculamos la restricci\'on de la misma a la frontera del disco
$\partial \cal M$, $G(\tau,\tau')=N(z(\tau),z'(\tau'))$. $G^{-1}$
es la inversa de este \'ultimo operador,
$G^{-1}G=\delta^{(1)}(\tau-\tau')$. La integral (\ref{prima})
contiene, en general, en el exponente, infinitas potencias de
$\eta$. Expandiendo $A_\mu$ en potencias de $\eta$ tenemos
\be
 \int d\tau \,\dot\eta^\mu A^\mu(x+\eta)=\frac12F_{\nu\mu}(x)\int d\tau
 \,\dot\eta^\mu\eta^\nu+\frac13\partial_\rho F_{\nu\mu}(x)\int d\tau
 \,\dot\eta^\mu\eta^\nu\eta^\rho+{\cal O}(\partial^2F,\eta^4)
\ee
En general, la acci\'on efectiva $\Gamma[A_\mu]=\int d^{D}\!x\,
{\cal L}(x)$, depender\'a de $F_{\mu\nu}$ y todas sus derivadas.
En la aproximaci\'on $F_{\mu\nu}(x)=\,$cte. ($\partial^{(n)}
F\approx0$), la integral que nos queda en $\eta^\mu$ es gaussiana
\be
 \int{\cal D}\eta^\mu \exp\left(-\frac1 {4\pi\acute\alpha}\;\eta\,
 G^{-1}\eta+\frac i2F_{\mu\nu}\int d\tau \,\dot\eta^\mu
 \eta^\nu\right)
\ee
Llevando $F_{\mu\nu}$ a la forma est\'andar $F=\epsilon\otimes f$,
donde $\epsilon$ son matrices antim\'etricas $2\times2$ y $f$ es
una matriz diagonal con autovalores $f_i$ ($i=1,...,\frac D2$), el
calculo se reduce al producto
\be
 \prod_{i=1}^{D/2}\int {\cal D}\eta_i\exp\left(
 -\frac 12 \eta_i G^{-1}\eta_i-\frac 12 \tilde f_i^2\,
 \dot\eta_i G \dot\eta_i
 \right)=\prod_{i=1}^{D/2}\int {\cal D}\eta_i\exp\left(
 -\frac 12 \eta_i G^{-1}\eta_i-\frac 12 \tilde f_i^2\,
 \eta_i {\ddot G}  \eta_i \right)
\ee
donde hemos reescaleado $\eta\to\sqrt{2\pi\acute\alpha}\,\eta$ y
$\tilde f_k=2\pi\acute\alpha\, f_k$. En la \'{u}ltima expresi\'{o}n debe
entenderse ${\ddot G}=(d^2/d\tau d\tau')G(\tau,\tau')$. La acci\'{o}n
efectiva para los modos vectoriales no masivos $\Gamma(A_\mu)$
queda \cite{Tse} \footnote{Expresamos la acci\'on efectiva, como
es convencional, en t\'erminos de la constante de acoplamiento
$g_{cl}$ para cuerdas cerradas, recordando que $g_{op}^2=g_{cl}$.}
\be
 \Gamma[A_\mu]=\frac {Z(0)}{g_{cl}}\int d^D x \prod_{i=1}^{D/2}
 \frac 1 {\sqrt{\det \Delta_i}}
 \label{nom}
\ee
\be
 Z(0)= \int{\cal D}\xi^\mu e^{-S_1},\quad\quad \Delta_i=1+\tilde f^2_i \ddot G\cdot G
\ee
Dado que para el caso del disco unitario
\be
 N(z,z')=-\frac 1 {2\pi} \ln\, |z-z'||z-\bar z'^{-1}|\Longrightarrow
 G(\theta,\theta')=N(e^{i\theta},e^{i\theta'})=-\frac 1 {2\pi}
 \ln(2-2\cos\zeta)
\ee
donde $\zeta=\theta-\theta'$ y $0\leq\theta\leq2\pi$, entonces
\be
 \ddot G\cdot G=\frac 1 \pi \sum_{m=1}^\infty\cos m\zeta\equiv
 \delta(\zeta)\Longrightarrow \prod_{i=1}^{D/2}
 \frac 1 {\sqrt{\det \Delta_i}}=\prod_{i=1}^{D/2}
 \prod_{m=1}^{\infty}\frac 1 {\sqrt{1+\tilde f^2_i}}=\prod_{i=1}^{D/2}
 \sqrt{1+\tilde f^2_i}
\ee
donde en el \'{u}ltimo paso regularizamos el producto infinito usando
la funci\'{o}n $\zeta$ de Riemann. La acci\'on efectiva (\ref{nom})
para los modos vectoriales no masivos a orden arbol en teor\' \i a
de cuerdas, resulta ser la acci\'{o}n de Born-Infeld \cite{Tse}
\ba
 \Gamma[A_\mu]&=&\frac {Z_0}{\acute\alpha^{D/2}g_{cl}}\int d^D x
 \sqrt{\det (\delta_{\mu\nu}+2\pi\acute\alpha F_{\mu\nu})}
 \label{opalala}
\ea
Aqu\' \i~$Z_0=\,$cte. es la funci\'{o}n de partici\'{o}n libre para el
disco y hemos factorizado la dependencia en $\acute\alpha$. Al ser
deducido de la teor\' \i  a de cuerdas, el par\'{a}metro de
Born-Infeld $\beta$ que determina el valor m\'{a}ximo de campo resulta
ser igual a la tensi\'{o}n de las cuerdas fundamentales $\beta=T$.

\section{La teor\' \i a de Dirac-Born-Infeld y Dp-branas}

La acci\'{o}n de Dirac-Born-Infeld  en $p+1$ dimensiones,
correspondiente a la acci\'on efectiva (bajas energ\' \i as) de
una Dp-brana, se define como
\be
 S_{DBI}=-T_p\int d^{p+1}x\,\sqrt{-\det(g_{\mu\nu}+2\pi\acute\alpha
 F_{\mu\nu})}
 \label{dbii}
\ee
donde
\be
 g_{\mu\nu}=(\partial X^M/\partial x^\mu)(\partial X^N/\partial
 x^\nu)G_{MN}
\ee
($\mu,\nu=0,\ldots,p\,; M,N=0,\ldots,D-1$). $G_{MN}$ es la
m\'etrica del espacio-tiempo D-dimensional donde se encuentra
embebida la brana, $X^M=X^M(x^\mu)$ representan la posici\'on de
la p-brana en el espacio-tiempo D-dimensional y $g_{\mu\nu}$ se
interpreta como la m\'etrica inducida sobre la superficie de la
brana por la m\'etrica del espacio ambiente $G_{MN}$.
$F_{\mu\nu}=\partial_\mu A_\nu-\partial_\nu A_\mu$ es el tensor
electromagn\'etico correspondiente al campo de gauge contenido en
el volumen de mundo de la brana y $T_p$ es la tensi\'{o}n de la
Dp-brana.

La acci\'on (\ref{dbii}) es invariante frente a
reparametrizaciones en su volumen de mundo. Al fijar esta
invarianza, por ejemplo en el gauge est\'atico (ver cap.
\ref{nstr}) aparecen los campos escalares que corresponden a las
oscilaciones transversales de la brana. La acci\'on (\ref{dbii})
con $F_{\mu\nu}=0$ coincide con el modelo para el electr\'on
propuesto por Dirac \cite{dirac} donde se lo representaba como una
c\'ascara (2-brana) cargada. La tensi\'on que tiende a contraer la
superficie contraresta la repulsi\'on coulombiana que experimenta
la misma. La idea de Dirac de re\-pre\-sen\-tar distintas part\'
\i culas (electr\'on y mu\'on)  como estados excitados de un mismo
objeto fundamental es anterior a la teor\' \i a de cuerdas.
Parad\'ojicamente las membranas (2-branas) reaparecieron
recientemente en el contexto de teor\' \i a M como los objetos
fundamentales de la acci\'on de bajas energ\' \i as (SUGRA
$d=11$).

Una $(p+1)$-forma $A_{p+1}=A_{M_1 M_2\ldots M_{p+1}}
dX^{M_1}\wedge dX^{M_2}\ldots \wedge dX^{M_{p+1}}$ se acopla
naturalmente a una $p$-brana, esto es un objeto extendido con $p$
dimensiones espaciales. El acoplamiento es la integral de la
$(p+1)$-forma sobre el volumen de mundo $(p+1)$-dimensional de la
$p$-brana
\be
 S_{WZ}=\rho_p\int_{v.m.}A_{p+1}\equiv\rho_p \int d^{p+1}x
 \,A_{M_1M_2\ldots M_{p+1}}(X)
 \,\partial_1 X^{M_1}\partial_2 X^{M_2}\ldots\partial_{p+1} X^{M_{p+1}}
 \label{wz}
\ee
$\rho_p$ es la densidad de carga de la p-brana y denotamos
$\partial_\mu\equiv\partial/\partial x^\mu$. Por ejemplo: una
part\' \i cula puntual o $0$-brana describe una trayectoria
unidimensional en el espacio tiempo y se acopla a un campo
vectorial $A_{M}$, el objeto cargado frente a un tensor de dos \'
\i ndices antisim\'etrico es una cuerda o 1-brana ya que describe
en el espacio tiempo una superficie bidimensional. En la teor\' \i
a perturbativa  de supercuerdas no existen estados (p-branas) con
$p\ne1$, solo contamos con cuerdas fundamentales (1-branas) que de
hecho son las cargas elementales para el tensor antisim\'{e}trico
$b_{\mu\nu}$. El hecho de que no existan estados perturbativos
cargados respecto de los campos bos\'onicos de Ramond se sigue de
que el v\'ertice c\'ubico $\langle string|A_{p+1}| string
\rangle$, que representa la emisi\'on de un estado R-R a partir de
un estado de cuerda fundamental, involucra un n\'{u}mero impar de
v\'{e}rtices fermi\'{o}nicos izquierdos (y derechos), luego la amplitud de
emisi\'{o}n se cancela autom\'{a}ticamente a todo orden en la teor\' \i a
perturbativa. Este argumento muestra que las cuerdas no se acoplan
el\'ectricamente a la 2-forma R-R, presente en la teor\' \i a IIA.
En presencia de fronteras (Dp-branas) este argumento falla. La
propiedad $\frac12$-BPS (multiplete corto, ver cap. \ref{susy})
para una Dp-brana est\'{a}tica, se sigue de
\be
 T_{p}=\rho_{p}
 \label{inva}
\ee
y del an\'{a}lisis del \'{a}lgebra supersim\'{e}trica al ser extendida para
contemplar cargas centrales tensoriales \cite{tow}. La acci\'{o}n
completa para la Dp-brana (\ref{dbii}) debe entonces ser corregida
sum\'andosele el t\'{e}rmino de Wess-Zumino (\ref{wz}),
correspondiente al acoplamiento de la Dp-brana con la (p+1)-forma,
este t\'ermino sin embargo no afecta las ecuaciones de movimiento
para los campos.

\section{El electr\'on \`a la Strominger-Callan-Maldacena}

Los argumentos de C. Callan y J. Maldacena (CM) para  hallar
soluciones a la acci\'on de DBI fueron los siguientes \cite{CM}: a
partir de la variaci\'on supersim\'etrica del gaugino (obtenida
por reducci\'on dimensional)
\be
 \delta \lambda=\Gamma^{MN}F_{MN}\epsilon
 =(\Gamma^{\mu\nu}F_{\mu\nu}+2\Gamma^{\mu i}\partial_\mu X_i)
 \,\epsilon~~~~,
 \label {trbps}
\ee
un background BPS corresponde a $\delta \lambda=0$ para alg\'un
$\epsilon\ne0$. Si estamos interesados en soluciones de Coulomb
$A_0=q_e/4\pi r^{p-2}$, vemos de (\ref{trbps}) que si proponemos
$X_9=q_e/4\pi r^{p-2}\Longrightarrow F_{9r}\equiv\partial_r
X^9=F_{0r}$, tenemos
\be
 (\Gamma^{0r}+\Gamma^{9r})\,\epsilon=0
 \Rightarrow(\Gamma^{0}+\Gamma^{9})\,\epsilon=0
\ee
Esta ecuaci\'on tiene soluciones no triviales para la mitad de las
componentes de $\epsilon$, de lo que concluimos que nuestra
soluci\'on es $1/2$-BPS (dichos $\epsilon\ne0$ no alteran la
soluci\'on o como se dice usualmente nuestra soluci\'on preserva
la mitad de las supersimetr\' \i as). Las soluciones propuestas
por CM corresponden a soluciones del modelo DBI linealizado o sea
son soluciones del modelo Maxwell-Higgs (MH) (ver discusi\'on mas
abajo).

Computando la energ\' \i a (ver cap. \ref{nobps}) se llega a la
conclusi\'on de que el electr\'on BPS con potencial coulombiano
(lo que implica excitar un campo escalar) en el volumen de mundo
de la brana, debe ser interpretado al embeber la brana en $D=10$
como la intersecci\'on de una cuerda fundamental con la Dp-brana.
La autoenerg\' \i a infinta de la soluci\'on se debe a la
extensi\'on semi-infinita de la cuerda que intersecta la Dp-brana.
Notemos que el estado BPS no manifiesta energ\' \i a de
interacci\'on, la energ\' \i a del estado es la suma de las
energ\' \i as de sus constituyentes.  Esto se observa tambi\'en al
tratar soluciones BPS mas generales centradas en varios puntos.

Recalquemos una propiedad muy importante y general de todas las
soluciones BPS: la energ\' \i a de las mismas es simplemente su
carga topol\'ogica. Esto indica que para soluciones BPS, la
energ\' \i a es independientes de la din\'amica asociada con
dichas soluciones (ver cap. \ref{susya}). La presente soluci\'on
fue calculada para la acci\'on de MH y sin embargo resulta ser
tambi\'en soluci\'on de la acci\'on DBI. En el cap\'{\i}tulo
\ref{susya} discutiremos en un contexto mas sencillo este mismo
fen\'omeno: la insensibilidad de las ecuaciones BPS respecto a la
din\'amica asociada al campo de gauge.



\chapter{La supersimetr\'{\i}a y sus representaciones\label{susy}}
\begin{center}
 {\begin{minipage}{6truein}
 { \sl Discutiremos la supersimetr\' \i a  y sus representaciones }
 \end{minipage}}
\end{center}

\section {Introducci\'{o}n}

A mediados de los '60, la creciente importancia de los grupos de
simetr\' \i as internas como $SU(2),SU(3),...$ llev\'o a los f\'
\i sicos a investigar sobre la posibilidad de combinar, de manera
no trivial, el grupo de Poincar\'e del espacio-tiempo con un grupo
de simetr\' \i a interna. Luego de varios intentos frustrados, el
interrogante fue resuelto por S. Coleman y J. Mandula de manera
negativa en el llamado ``no-go theorem" \cite{cole}: en $d>2$, una
teor\' \i a cu\'{a}ntica relativista, con un espectro discreto de
estados masivos de una part\' \i cula y con amplitudes de
dispersi\'{o}n no nulas, tiene como \'{u}nicas cargas tensoriales
conservadas, el cuadrivector energ\' \i a-impulso $P_\mu$ y el
tensor de momento angular $M_{\mu\nu}$; o sea los generadores del
grupo de Poincar\'e\footnote{En presencia de part\'{\i}culas no masivas
es posible extender el grupo de Poincar\'e al grupo conforme
\cite{mack}.}. {\it El resto de las cargas conservadas deben ser
escalares de Lorentz}. Como regla, estas cargas escalares de
Lorentz deben satisfacer el \'{a}lgebra de un grupo de Lie semi-simple
que puede contener factores $U(1)$ adicionales.

Es conveniente hacer algunos comentarios acerca del teorema
``no-go". Si consideramos la dispersi\'{o}n de dos cuerpos, una vez
tenida en cuenta la conservaci\'{o}n de impulso angular y de energ\'
\i a, la \'{u}nica variable libre es el \'{a}ngulo de dispersi\'{o}n $\theta$.
Si existiera un grupo de Lie que transformara no trivialmente
frente al grupo de Poincar\'e, tendr\' \i amos generadores
adicionales (de caracter tensorial) asociados al espacio-tiempo.
Las leyes de conservaci\'{o}n resultantes limitar\' \i an la amplitud
de dispersi\'{o}n de dos cuerpos, de manera que el \'{a}ngulo $\theta$
tomar\' \i a solo valores discretos. Sin embargo, el proceso de
dispersi\'{o}n se supone anal\' \i tico en el \'{a}ngulo $\theta$, de
manera que debemos concluir que, de existir estas cargas
adicionales, el proceso no depende del \'{a}ngulo. En s\' \i ntesis,
si existe un grupo de Lie combinado no trivialmente con el grupo
de Poincar\'e, la din\'{a}mica resultante para los procesos de
dispersi\'{o}n es trivial.

Los argumentos anteriores muestran la imposibi\-li\-dad de
si\-me\-tr\' \i\-as no tri\-via\-les que co\-necten part\' \i
culas de distintos espines, si {\it todas} las part\' \i culas
tienen esp\' \i n entero o {\it todas} tienen esp\' \i n
semientero. Si tuvi\'{e}ramos una simetr\' \i a que conectase part\'
\i culas de esp\' \i n entero con part\' \i culas de esp\' \i n
semientero, tendr\' \i amos cargas de caracter espinorial y el
teorema ``no-go" no ser\' \i a aplicable. El cambio es dr\'{a}stico.
Estas cargas espinoriales estar\' \i an dadas por integrales
espaciales de campos espinoriales locales. La conexi\'{o}n esp\' \i
n-estad\' \i stica en la teor\' \i a cu\'{a}ntica de campos dice que
para separaciones tipo espacio los campos espinoriales
anticonmutan. Luego, la conexi\'{o}n esp\' \i n-estad\' \i stica dicta
que es el anticonmutador de las cargas el que est\'{a} determinado, de
manera que estamos frozados a inventar un nuevo tipo de \'{a}lgebra de
Lie donde la operaci\'{o}n $[,]$ no siempre es antisim\'{e}trica (como en
un conmutador), sino que ocasionalmente puede ser sim\'{e}trica (como
el anticonmutador). Este tipo de \'{a}lgebras se denominan
super-\'{a}lgebras de Lie o \'{a}lgebras gradadas.

En este esp\' \i ritu, Y. Golfand y E. Likhtman \cite{gl}
demostraron que, generalizando el concepto de grupo de Lie, era
posible encontrar un grupo de simetr\' \i a que incluyera al grupo
de Poincar\'e, y a un grupo de simetr\' \i a interna de manera no
trivial. Nuestra discusi\'{o}n de la supersimetr\' \i a seguir\'a este
punto de vista.

Dado que la estructura de un grupo de Lie, al menos en un entorno
de la identidad, est\'{a} determinada por su \'{a}lgebra de Lie,
introducimos generadores $Q_\alpha$ y sus adjuntos $\bar
Q_{\dot\beta}$ (en notaci\'on de Weyl, ver ap\'endice para las
convenciones), transformando en las representaciones
(espinoriales) del grupo de Lorentz $(\frac 12,0)$ y $(0,\frac
12)$, que satisfacen relaciones de anti-conmutaci\'{o}n\footnote{Por
convenci\'on, los \' \i ndices $A,B$ se colocan abajo al tratar
espinores pertenecientes a la representaci\'on $(0,\frac 12)$.}
\ba
 \{Q_\alpha^A,\bar Q_{\dot\beta B}\}&=&Q_\alpha^A\bar Q_{\dot\beta
 B}+\bar Q_{\dot\beta B} Q_\alpha^A \\
 &=&2 \sigma^\mu_{\alpha\dot\beta}P_\mu\, \delta^{A}_B
 \label{lico}
\ea
(los \' \i ndices adicionales $A,B=1,...,N$ caracterizan
extensiones con un n\'{u}mero $N$ de supersimetr\' \i as). El lado
derecho de la ec.(\ref{lico}) se obtiene bas\'{a}ndose en
consideraciones de simetr\' \i a frente al grupo de Lorentz, ya
que debe transformar como $(\frac 12,\frac12)$, y hermiticidad.
Asumamos entonces que el grupo de supersimetr\' \i a contiene
generadores ``bos\'{o}nicos" $P_\mu,\, M_{\mu\nu},\, T_s$, que
satisfacen reglas de conmutaci\'{o}n, asi como generadores fermi\'{o}nicos
$Q_\alpha^A\, (A=1,2,\dots,N)$ que satisfacen reglas de
anti-conmutaci\'{o}n. La generalizaci\'{o}n del teorema de Coleman-Mandula
incluyendo generadores espinoriales efectuada por
Haag-{\L}opusza\'nski-Sohnius muestra que el \'{a}lgebra de simetr\'
\i a m\'as general que puede tener la matriz S es \cite{haag},
\ba
 \left[ P_\mu, P_\nu\right]&=&0
 \label{puli}\\
 \left[M_{\mu\nu}, P_\rho\right]
 &=&i(\eta _{\nu\rho}P_\mu - \eta_{\mu\rho}P_\nu)\\
 \left[M_{\mu\nu},M_{\rho\sigma}\right]
 &=& -i(\eta_{\mu\rho}M_{\nu\sigma}-\eta_{\mu\sigma}M_{\nu\rho}\\
 && -\eta_{\nu\rho}M_{\mu\sigma} + \eta_{\nu\sigma}M_{\mu\rho}) \\
 \left[T_{\ell}, T_k\right] &=& i{C_{\ell k}}^j T_j \\
 \left[P_\mu, T_{\ell}\right] &=&\left[ M_{\mu\nu}, T_{\ell}\right] = 0
 \label{ocvio}\\
 \{ Q^A_{\alpha},\bar Q _{\dot\beta B} \}
 &=& 2\sigma ^{\mu}_{\alpha\dot\beta} P_{\mu} \,\delta^A_B
 \label{bepi}\\
 \{ Q^A_{\alpha},Q^B_{\beta}\}
 &=&2\epsilon_{\alpha\beta} Z^{AB}
 \label{esi}\\
 \{ \bar Q _{\dot\alpha A},\bar Q _{\dot\beta B}\}
 &=& -2\epsilon_{\dot\alpha\dot\beta}Z_{AB}^*
 \label{esi1}\\
 \left[ Q^A_{\alpha},P_\mu\right]
 &=& \left[ \bar Q_{\dot\alpha A},P_\mu\right] = 0
 \label{q1}\\
 \left[ Q^A_{\alpha},M_{\mu\nu}\right]
 &=&(\sigma _{\mu\nu})_\alpha^{~\beta}Q^A_{\beta} \\
 \left[\bar Q ^{\dot\alpha}_A, M_{\mu\nu}\right]
 &=&(\bar\sigma _{\mu\nu})_{~\dot\beta}^{\dot\alpha}\bar Q^{\dot\beta}_A
 \label{q2}\\
 \left[Q^A_{\alpha},T_{\ell}\right]&=&S_{l~B}^A Q^B_{\alpha}
 \label{ca1}\\
 \left[\bar Q _{\dot\alpha A}, T_{\ell}\right]
 &=& -S^{*~B}_{\ell A} \bar Q _{\dot\alpha B}
 \label{ca2}
\ea
donde los escalares complejos de Lorentz $Z^{AB}=U^{AB}+iV^{AB}$
se denominan cargas centrales y son antisim\'etricos en $A,B$
\footnote{Tomamos la convenci\'on \cite{wess} $Z^{AB}=-Z_{AB}$}.
Mediante el uso de las identidades de Jacobi se muestra que
conmutan con todos los elementos del \'algebra super-Poincar\'e y
que generan una sub\'algebra abeliana invariante, contenida en el
\'algebra de Lie generada por los $T_\ell\,$\footnote{ Recordemos
que la generalizaci\'on del teorema de Coleman-Mandula permit\' \i
a grupos semi-simples y factores $U(1)$ adicionales.}. Las podemos
expresar entonces como
\be
  Z^{AB}=a^{\ell AB}T_{\ell}
\ee
donde $a^{\ell}$ son matrices antisim\'etricas. $S_{\ell}$,
$a_{\ell}$ deben satisfacer
\be
 S_{\ell~C}^{A} a^{kCB}= -a^{kAC} S^{*~B}_{\ell\, C}
\ee
Las relaciones (\ref{ocvio}) son consecuencia del teorema de
Coleman-Mandula. Las ecs.(\ref{q1})-(\ref{q2}) establecen que las
supercargas $Q_\alpha^A,\,\bar Q_{\dot\alpha A}$ transforman como
espinores frente al grupo de Poincar\'e, y (\ref{ca1})-(\ref{ca2})
que est\'an cargadas frente al grupo de simetr\' \i a interna.
Como veremos mas adelante, en presencia de cargas centrales
$Z^{AB}\neq0$ ($N>1$), las ec.(\ref{bepi})-(\ref{esi1}) son las
responsables de que pueden aparecer estados BPS en una teor\' \i a
supersim\'{e}trica.

Es \'util expresar los anticonmutadores del \'algebra
supersim\'{e}trica (\ref{bepi})-(\ref{esi1}) mediante espinores de
Majorana $Q_a$ ($a=1,...,4$). Siguiendo las convenciones del
ap\'endice se obtiene
\be
 \{Q^A_a,Q^B_b\}=2(\Gamma^\mu C)_{ab}P_\mu\delta^{AB}+2C_{ab}U^{AB}
 +2i(\Gamma^5C)_{ab}V^{AB}
 \label{amsa}
\ee
que tambi\'en se suele expresar, dado que $Q^A$ es Majorana, como
\be
 \{Q^A,\bar Q^B\}=-2\left(\Gamma^\mu P_\mu\delta^{AB}+U^{AB}
 +i\Gamma^5 V^{AB}\right)
 \label{techito}
\ee

En la pr\'oxima secci\'on discutiremos las distintas
representaciones irreducibles del \'algebra super-Poincar\'e
(\ref{puli})-(\ref{ca2}).

\section{Representaciones de la supersimetr\' \i a}

Dado que las supercargas $Q$ conmutan con el operador 4-impulso
$P_\mu$, $P^2$ es un casimir para el \'algebra super-Poincar\'e.
Luego, todos los estados en una representaci\'on dada  tienen la
misma masa. El nuevo casimir que reemplaza al cuadrado del vector
de Pauli-Ljubansk\' \i~ es $C^2$, donde \cite{lykken}\footnote{La
convenci\'on para el vector de Pauli-Ljubansk\' \i~ es
\be
 W_\mu = \frac 12 \varepsilon _{\mu\nu\rho\sigma}P^\nu M^{\rho\sigma}
 \quad .
\ee
Cuyo autovalor es $w^2=-m^2s(s+1)$, $s=0,\frac 12 ,1,\ldots$, para
representaciones masivas, y en el caso no masivo, con $w^2=0$,
queda $W_\mu = \lambda P_\mu$, donde $\lambda$ es la helicidad.}
\ba
 C^2 &=& C_{\mu\nu}C^{\mu\nu}\\
 C_{\mu\nu}&=& B_\mu P_\nu - B_\nu P_\mu\\
 B_\mu &=& W_\mu - \frac14 \bar Q\bar\sigma_\mu Q
 \label{estasi}
\ea
Para que una teor\' \i a sea supersim\'etrica, el contenido de
particulas de la misma debe formar una representaci\'on del
\'algebra escrita en la secci\'on anterior. Como es conocido las
representaciones del grupo de Poincar\'e se encuentran mediante el
m\'etodo de representaciones inducidas de Wigner\cite{wein}. El
m\'etodo consiste en encontrar representaciones de un subgrupo del
grupo de Poincar\'e y luego boostearlas para encontrar la
representaci\'on de todo el grupo. El procedimiento es elegir un
valor para el momento $p^\mu$ que satisfaga $p^2=0$ o $p^2=m^2$,
seg\'un el caso a considerar, y luego, hallado el subgrupo $H$ que
deja invariante el $p^\mu$ elegido representarlo sobre los estados
$|\,p^\mu \rangle$. En lo que sigue discutiremos las
representaciones del \'algebra (\ref{bepi})-(\ref{esi1}), en el
sistema de reposo de la part\' \i cula. La imagen f\' \i sica es
simple: las propiedades de una part\' \i cula est\'an determinadas
por su comportamiento en un dado sistema de referencia. Embebiendo
el \'algebra supersim\'{e}trica en el \'algebra super-Poincar\'e,
encontramos las representaciones de esta \'ultima mediante el
formalismo de Wigner. Como veremos al estudiar las posibles
representaciones del \'algebra supersim\'etrica, la supersimetr\'
\i a relaciona bosones con fermiones y viceversa

\subsection{Ausencia de cargas centrales}

En ausencia de cargas centrales, el \'algebra para las supercargas
$Q,\, \bar Q$ es
\ba
 \{ Q^A_{\alpha},\bar Q _{\dot\beta B} \}
 &=& 2\sigma ^{\mu}_{\alpha\dot\beta} P_{\mu} \,\delta^A_B
 \label{bepis}\\
 \{ Q^A_{\alpha},Q^B_{\beta}\}
 &=&
 \{ \bar Q _{\dot\alpha A},\bar Q _{\dot\beta B}\}
 =0
 \label{esi1s}
\ea
\underline{Representaciones irreducibles no masivas}: En el caso
no masivo elegimos el sistema de re\-fe\-ren\-cia, donde
$p^\mu=m(1,0,0,1)$, el \'algebra supersim\'{e}trica queda
\be
 \{ Q^{A}_{\alpha}, \bar{Q}_{\dot{\beta}B} \} =
 \left(
 \begin{array}{lr}
    0 & 0 \\
    0 & 4m
 \end{array} \right )
 \delta^{A}_{B}
\ee
Esta relaci\'on nos dice, dado que $\bar Q_A=(Q^A)^\dagger$, que
en una teor\' \i a unitaria, todo estado f\' \i sico $|fis
\rangle$ satisface
\be
 Q_1|fis \rangle= \bar Q_{\dot{1}}|fis \rangle =0~~~~~~
 \Longrightarrow~~~~~~Q_1 = \bar Q_{\dot{1}}=0~~
 {\rm sobre~estados~f\acute{\i}sicos}
 \label{fisura}
\ee
Solo la mitad de los generadores $Q$ est\'an activos\footnote
{Dado que la mitad de las cargas supersim\'{e}tricas dejan invariantes
a los estados f\' \i sicos, es l\' \i cito decir que las
representaciones no masivas son representaciones $\frac12$-BPS.}.
Reescaleando los generadores no nulos
\be
 a^A=\frac{1}{2\sqrt{m}}Q^A_2\,,\qquad
 (a^A)^\dagger=\frac{1}{2\sqrt{m}}\bar Q_{\dot 2A}\,,
\ee
el \'algebra supersim\'{e}trica toma la forma
\be
 \{ a^A, (a^B)^\dagger \} = \delta^{AB}\,,\quad  \{ a^A, a^B \}=0\,,
 \quad \{ (a^A)^\dagger, (a^B)^\dagger \} = 0\,.
 \label{nomas}
\ee
Recordando que $A,B=1,...,N$ reconocemos (\ref{nomas}) como un
\'algebra de $N$ osciladores fermi\'onicos, que tiene una
representaci\'on $2^N$-dimensional. La dimensi\'on de las
representaciones irreducibles no masivas es entonces
$dim(rep)|_{m=0}=2^N$.

Analicemos el efecto de los $Q$'s sobre la helicidad de un estado.
Calculando el conmutador del vector de Pauli--Ljubansk\' \i~ con
las supercargas se obtiene
\be
 {[}W_\mu,\bar Q^{\dot\alpha}_A{]}=-iP^\nu
 (\bar\sigma_{\mu\nu})^{\dot\alpha}_{~\dot\beta}\bar
 Q^{\dot\beta}_A
\ee
donde hemos usado la identidad (\ref{dudu1}). Aplicando este
\'ultimo operador a un estado no masivo con helicidad $\lambda$ e
impulso $p$, en el sistema que elegimos, vemos que
\be
 W_0\,\bar Q^{\dot 1}_A|p,\lambda\rangle=p^0
 (\lambda-\frac12) \bar Q^{\dot1}_A|p,\lambda\rangle
\ee
o en forma equivalente
\be
 W_0\,\bar Q_{\dot2A}|p,\lambda\rangle=p^0
 (\lambda-\frac12) \bar Q_{\dot2A}|p,\lambda\rangle
\ee
de manera que la acci\'on de los operadores $\bar Q_{\dot2A}$
sobre estados con helicidad definida es disminuir $\lambda$ a
$\lambda-\frac 12$.

La construcci\'on del supermultiplete es simplemente representar
el \'algebra fermi\'onica (\ref{nomas}). Definimos el vac\'{\i}o
$|\Omega_\lambda \rangle$ como el estado anulado por todos los
operadores de des\-truc\-ci\'on $a^A|\Omega_\lambda \rangle=0$ y
con helicidad definida $\lambda$ \footnote{El vac\'{\i}o en el caso no
masivo es no degenerado.}. Los estados del multiplete los
generamos por acci\'on de los operadores de creaci\'on
$(a^A)^\dagger$ actuando sobre el vac\'{\i}o
\be
  (a^{A_1})^\dagger...(a^{A_k})^\dagger |\Omega_\lambda \rangle
  ~,~~~~~~~~~~~A_i=1,...,N
  \label{estados}
\ee
Debido a antisimetr\' \i a el estado con $k~(a^A)^\dagger$'s
tendr\'a degeneraci\'on ${\bf C}^N_k$ y helicidad $\lambda- k/2$.
El multiplete contar\'a con estados desde $\lambda$ a
$\lambda-N/2$. Esquem\'aticamente
\be
 \begin{array}{rrrrrr}
 \underline{N=1}\,:& \quad |\lambda \rangle,
 & |\lambda-1/2\rangle\hphantom{,}& & &\\
 \underline{N=2}\,:& \quad |\lambda \rangle,& 2\,|\lambda-1/2\rangle,
 &|\lambda-1\rangle\hphantom{,}&&\\
 \underline{N=4}\,:& \quad |\lambda \rangle,&4\,|\lambda-1/2\rangle,
 & 6\,|\lambda-1 \rangle,&4\,|\lambda-3/2\rangle,&|\lambda-2 \rangle
 \end{array}
\ee
Las representaciones no son necesariamente invariantes CPT.
Multipletes invariantes CPT se obtienen cuando $\lambda=N/4$. En
el caso particular $N=1$, al pedir invarianza CPT debemos
considerar la uni\'on de dos representaciones irreducibles no
masivas, obteni\'endose cuatro estados
$\lambda,\,\lambda-\frac12,\,-\lambda+\frac12$ y $-\lambda$.

Si pretendemos realizar la supersimetr\'{\i}a en una teor\'{\i}a de campos,
los estados de part\'{\i}cula necesariamente vendr\'an
a\-com\-pa\-\~na\-das por su conjugado CPT. Escribamos las
posibles representaciones CPT invariantes y su interpretaci\'on en
t\'erminos de campos
\be
 \begin{array}{ccccc}
 \underline{N=1} &\!\!(|\Omega_{-\frac12}\rangle):&
 \!\!\!\{|-1/2\rangle\,,|-1\rangle\}
 &\!\!\!\oplus~ CPT:&\!\!\!\!\!\!\!\!\{|1/2\rangle\,,|1\rangle\} \\
 &\!\!\!\!(|\Omega_{0}\rangle)~\,:&
 \!\!\!\{|0\rangle\,,|-1/2\rangle\}
 &\!\!\!\oplus~ CPT:&\!\!\!\!\!\!\!\!\{|0\rangle\,,|1/2\rangle\} \\
 &\!\!\!\!(|\Omega_{\frac12}\rangle):&
 \!\!\!\{|1/2\rangle\,,|0\rangle\}
 &\!\!\!\oplus~ CPT:&\!\!\!\!\!\!\!\!\{|-1/2\rangle\,,|0\rangle\} \\
 &\!\!\!\!(|\Omega_{1}\rangle)\,:&\!\!\!\{|1\rangle \,,|1/2\rangle\}
 &\!\!\!\oplus~ CPT:&\!\!\!\!\!\!\!\!\!\!\!\!\!
 \{|-1\rangle\,, |-1/2\rangle\}\\
 \underline{N=2} &\!\!\!(|\Omega_{0}\rangle)\,:&\!\!\!
 \{|0\rangle \,,2|-1/2\rangle\,,|-1\rangle\} &\!\!\!\oplus~ CPT:&
 \!\!\!\!\!\!\!\!\!\!\!\!\!\!\!\!\!\!\!\!\!
 \{|0\rangle\,,2|1/2\rangle\,,|1\rangle\}\\
 &\!\!\!(|\Omega_{\frac12}\rangle):&\!\!\!
 \{|1/2\rangle,2|0\rangle,|-1/2\rangle\}
 &\!\!\!(\rm autoconjugada~CPT)&\\
 &\!\!\!(|\Omega_{1}\rangle):&\!\!\!
 \{|1\rangle \,,2|1/2\rangle\,,|0\rangle\} &\!\!\!\oplus~ CPT:&
 \!\!\!\!\!\!\!\!\!\!\!\!\!\!\!\!\!\!\!\!\!
 \{|-1\rangle\,,2|-1/2\rangle\,,|0\rangle\}\\
 \underline{N=4}&\!\!\!(|\Omega_{1}\rangle):&\!\!\!
 \{|1\rangle\,,4|1/2\rangle\,,6|0\rangle\,,4|-1/2\rangle\,
 ,|-1\rangle\}&\!\!\!(\rm autoconjugada~CPT)&
 \end{array}
 \label{tabla1}
\ee
Para $N=1$ la representaci\'on contiene un campo escalar complejo
y un espinor de Majorana (multiplete quiral $N=1$ $\Phi=(\phi
,\psi )$) cuando partimos de un vac\'{\i}o $|\Omega_\lambda\rangle$ con
helicidad $\lambda=0,1/2$ y sumamos la representaci\'on conjugada
CPT. De hecho, la representaci\'on correspondiente al multiplete
quiral, es la uni\'on de las representaciones obtenidas a partir
de $|\Omega_0\rangle$ y $|\Omega_{1/2}\rangle$. La
representaci\'on correspondiente al multiplete vectorial $N=1$
($V=(A_\mu ,\lambda )$) contiene un campo vectorial no masivo y un
espinor de Majorana. Esta representaci\'on se obtiene partiendo de
un vac\'{\i}o con helicidad $\lambda=1,-1/2$ y sumando la
correspondiente conjugada CPT. La representaci\'on se obtiene
uniendo los multipletes correspondientes a $|\Omega_{1}\rangle$ y
$|\Omega_{-1/2}\rangle$.

Para $N=2$ y $\lambda =1/2$ tenemos una representaci\'on
autoconjugada CPT que contiene un espinor de Majorana y un escalar
complejo. El hipermultiplete $N=2$ corresponde a esta
representaci\'on pero duplicada. La representaci\'on tiene el
mismo contenido de campos que dos multipletes quirales $N=1$. Para
$N=2$ y $\lambda=-1$, tenemos un campo vectorial no masivo, dos
espinores de Majorana y un escalar complejo (multiplete vectorial
$N=2$). Este \'ultimo multiplete $N=2$ es equivalente en contenido
de campos a la suma de un multiplete vectorial y un multiplete
quiral $N=1$ . La representaci\'{o}n $N=4$ es autoconjugada CPT y
acomoda un campo vectorial no masivo, dos fermiones de Dirac y
tres escalares complejos (multiplete vectorial $N=4$). Nos
detuvimos en $N=4$ ya que un n\'umero mayor de supersimetr\' \i as
necesariamente involucra part\' \i culas con helicidad $\lambda>1$

De acuerdo con el m\'etodo de representaciones inducidas de
Wigner, la representaci\'on del super-grupo de Poincar\'e se
obtiene boosteando los estados descriptos m\'as arriba.

\vspace{.3cm}

\noindent \underline{Representaciones irreducibles masivas}: Para
estados masivos siempre podemos ir al sistema de reposo de la
part\' \i cula, donde $p^\mu=m(1,0,0,0)$. Definiendo
\be
 a^A_\alpha=Q^A_\alpha/\sqrt{2m}\,,\quad (a^A_\alpha)^\dagger=
 {\bar Q}_{\dot\alpha A}/\sqrt{2m}
\ee
El \'algebra supersim\'{e}trica se reduce a
\be
 \{a^A_1,(a_1^B)^\dagger\}=\delta^{AB}\,,\qquad
 \{a^A_2,(a_2^B)^\dagger\}=\delta^{AB}\,
\ee
donde el resto de los conmutadores son nulos. Nuevamente nos
encontramos ante un \'algebra de Clifford ahora, con $4N$
operadores, el doble que en el caso no masivo. Analizando el
operador $C$ definido en (\ref{estasi}) vemos que en el sistema en
reposo podemos escribir
\ba
 C^2 &=& 2m^4J_iJ^i \\
 J_i &\equiv& S_i-{1\over 4m}\bar Q\bar\sigma_iQ
\ea
donde $S_i$ es el operador de esp\' \i n
$S_i=\frac12\varepsilon_{ijk}M^{jk}$ ($i=1,2,3$). Dado que $S_i$ y
$\bar\sigma_i$ obedecen el \'algebra de $SU(2)$, tenemos
\be
 [ J_i,J_j ] = i\epsilon _{ijk} J_k
\ee
luego, $J^2$ tiene autovalores $j(j+1)$, con $j$ entero o
semientero. Definimos el vac\'{\i}o como en la secci\'on anterior y
construimos los posibles estados por aplicaci\'on de los
operadores $(a^A)^\dagger$ actuando sobre el mismo (ver
ec.(\ref{estados})). Sin embargo, en el caso masivo el vac\'{\i}o puede
estar degenerado. Debido a que $J_i$ actuando sobre el vac\'{\i}o se
reduce al operador de esp\' \i n, $|\Omega_s \rangle$ es un
autoestado del operador de esp\' \i n
\be
 |\Omega_s \rangle= |{m,s,s_3} \rangle
\ee
A nivel $k$ tenemos una degeneraci\'on ${\bf C}^{2N}_k$ quedando
la dimensi\'on de la representaci\'on masiva
$dim(rep)|_{m\ne0}=2^{2N}(2j+1)$\footnote{El factor $(2j+1)$
proviene, en el caso masivo, de la degeneraci\'on del vac\'{\i}o.}; la
mitad de los estados son fermi\'onicos y la otra mitad
bos\'onicos. El m\'aximo esp\' \i n que podemos alcanzar desde el
vac\'{\i}o $|\Omega_s \rangle$ es $s+N/2$, en discrepancia con el
c\'alculo naive $s+(2N)/2=s+N$\,, esto se debe a que ahora tenemos
las dos componentes de las cargas supersim\'{e}tricas $Q_{1,2}^A$,
luego, t\'erminos del tipo
$(a^A_1)^\dagger(a^A_2)^\dagger=\frac{1}{2} \epsilon^{\alpha\beta}
(a^A_\alpha)^\dagger(a^A_\beta)^\dagger$ son escalares y no
disminuyen el esp\' \i n. El estado que obtenemos aplicando $k=2N$
operadores de creaci\'{o}n $(a^A)^\dagger$ tiene el mismo esp\' \i n
que el vac\'{\i}o. En el caso $N=1$ partiendo de un vac\'{\i}o $|\Omega_0
\rangle$, tenemos el siguientes multiplete
\be
 \underline{N=1}~ (|\Omega_0 \rangle): \quad |\Omega_0\rangle,\quad a^\dagger_\alpha |\Omega_0\rangle,
 \quad\frac{1}{\sqrt{2}} \epsilon^{\alpha\beta} a^\dagger_\alpha a^\dagger_\beta
 |\Omega_0\rangle
 \label{massiva}
\ee
El n\'{u}mero de estados es  $2^2=4$. Tenemos
$(0)\oplus(1/2)\oplus(0)$ \footnote{Con $(j)$ denotamos un estado
con esp\' \i n total $j$ y degeneraci\'{o}n $2j+1$.}. Que corresponden
a dos campos escalares reales y un espinor de Weyl $(\phi,
\lambda)$. Dado que la operaci\'{o}n de paridad intercambia
$(a_1)^\dagger$ con $(a_2)^\dagger$, el \'{u}ltimo estado en
(\ref{massiva}) corresponde a un campo pseudoescalar.

Partiendo de un vac\'{\i}o $|\Omega_j\rangle$ de esp\' \i n $j$ y
degeneraci\'{o}n $2j+1$, la representaci\'{o}n contiene $(2j+1)2^{2N}$
estados. El espectro se construye combinando la representaci\'{o}n
$j=0$ del \'{a}lgebra con un esp\' \i n $j$, usando las reglas de
adici\'{o}n de momento angular. El resultado es
\be
 \underline{N=1}~ (|\Omega_j \rangle): \quad (j)\oplus(j+1/2)\oplus(j-1/2)\oplus(j)
\ee
Para $j=1/2$ obtenemos un campo de gauge, un fermi\'{o}n de Dirac (dos
fermiones de Weyl) y un campo escalar real $(\varphi,\psi,A_\mu)$,
conteniendo el multiplete $4+4=8$ estados on-shell. Para el
contenido de campos en el caso $ N,~j$ arbitrario ver \cite{wess}.

\subsection{Cargas centrales no nulas}

En presencia de cargas centrales el \'algebra toma la forma
\ba
 \{ Q^{A}_{\alpha}, \bar{Q}_{\dot{\beta}B} \} &=& 2
 \sigma^\mu_{\alpha \dot{\beta}} P_\mu \delta^{A}_{B}\\
 \{ Q^A_\alpha, Q^B_\beta \}&=&2\epsilon_{\alpha\beta} Z^{AB} \\
 \{\bar{Q}_{\dot{\alpha}A},\bar{Q}_{\dot{\beta}B} \}
 &=&-2\epsilon_{\dot{\alpha}\dot{\beta}} Z^{*}_{AB}
 \label{susyC}
\ea
donde $Z$ y $Z^*$ son las matrices de carga central,
antisim\'etricas en $A,B$, y tomamos por convenci\'on
$Z^{AB}=-Z_{AB}$. En presencia de las extensiones centrales, no es
posible interpretar a $\bar Q$ y $Q$ como operadores de creaci\'on
y destrucci\'on sin rediagonalizar la base. Nos limitaremos al
caso de $N$ par. Mediante una transformaci\'on unitaria actuando
sobre los \' \i ndices internos de las supercargas
$Q_\alpha^A\to\tilde Q^A_\alpha=U^A_BQ^B_\alpha$ (o
$Q_\alpha\to\tilde Q_\alpha=UQ_\alpha$) con $U^\dagger=U$, podemos
llevar a $Z\to\tilde Z=UZU^T$ a una forma antisim\'etrica en
bloques sobre la diagonal $\tilde Z=\epsilon\otimes D$
(descomposici\'on de Zumino)
\be
 \tilde Z=\left(\begin{array}{cccc}
   (Z_1\epsilon^{ab}) & 0 & \ldots & 0 \\
    0&(Z_2\epsilon ^{ab})&\ldots &0 \\
   \vdots &\vdots &\ddots &\vdots \\
    0&0& \ldots & (Z_{{N\over 2}}\epsilon ^{ab}) \
 \end{array}
 \right)
\ee
donde $\epsilon^{ab}=-\epsilon_{ab}=i\sigma^2$. En esta
descomposici\'on los ``autovalores" $Z_1,Z_2,...,Z_{N/2}$ son
reales y no negativos. De manera que los \' \i ndices $A,B$ que
cuentan el n\'umero de supersimetr\' \i as pueden ser
descompuestos en $(a,L)$, $(b,M)$, con $a,b=1,2$ que provienen del
tensor $\epsilon$, y $L,M=1,...,N/2$ provenientes de la matriz
$D$. Escribiendo ahora el \'algebra (\ref{susyC}) en el sistema en
reposo (para estados masivos) y en la base de Zumino, vemos que
solo es necesario considerar un \'algebra $N=2$,
\ba
 \{Q^a_\alpha,{\bar Q}_{\dot\beta b}\}&=&
 2m\sigma^0_{\alpha\dot\beta} \delta^{a}_{b}
 \label{ey}\\
 \{Q^a_\alpha,Q^b_\beta\}&=&2\epsilon_{\alpha\beta}
 \epsilon^{ab}Z\\
 \{{\bar Q}_{\dot\alpha a},{\bar Q}_{\dot\beta b}\}
 &=&-2\epsilon_{\dot\alpha\dot\beta}\epsilon_{ab} Z
 \label{veiv}
\ea
Si definimos las siguientes supercargas
\ba
 a_{\alpha}&=& {1\over{2}}
 \left[ Q_{\alpha}^{1} + \epsilon_{\alpha\beta}
 (Q^2_\beta)^\dagger\right]\\
 b_{\alpha} &=& {1\over{2}}
 \left[ Q_{\alpha}^{1} - \epsilon_{\alpha\beta}
 (Q^2_\beta)^\dagger \right]
\ea
el \'algebra (\ref{ey})-(\ref{veiv}) se reduce a
\ba
 \{a_\alpha,(a_\beta)^\dagger\}&=&
 (m+Z) \delta_{\alpha\beta}\nonumber\\
 \{b_\alpha,(b_\beta)^\dagger\}&=&
 (m- Z) \delta_{\alpha\beta}
 \label{rec}
\ea
donde el resto de los conmutadores son cero. La representaci\'on
la construimos definiendo el vac\'{\i}o $|\Omega \rangle$ como el
estado aniquilado por los operadores de destrucci\'on $a_\alpha,
b_\alpha$ y actuando con los operadores de creaci\'on
$(a_\alpha)^\dagger, (b_\alpha)^\dagger$ sobre el mismo. Habiendo
diagonalizado el \'algebra $N=2$, el \'algebra en el caso de tener
$N$ supersimetr\' \i as consiste simplemente en agregar un \' \i
ndice $L=1,...,N/2$ a los operadores $a$, $b$ y a las cargas
centrales
\ba
 \{a^L_\alpha,(a^M_\beta)^\dagger\}&=&
 (m+Z_L) \delta_{\alpha\beta}\delta^L_M\\
 \{b^L_\alpha,(b^M_\beta)^\dagger\}&=&
 (m- Z_L) \delta_{\alpha\beta}\delta^L_M
 \label{juju}
\ea
De estas expresiones se deduce una desigualdad conocida como
``cota de Bogomol'nyi". Tomando el valor de expectaci\'on de
(\ref{juju}) para un estado f\' \i sico arbitrario tenemos para
$L=M$ y $\alpha=\beta$
\be
 \vert\; (a^L)^\dagger|fis \rangle \vert^2
 +\vert\; (a^L)|fis \rangle \vert^2=
 (m+Z_L)~~~\Longrightarrow~~~(m+Z_L)\geq0
\ee
\be
 \vert\; (b^L)^\dagger|fis \rangle \vert^2
 +\vert\; (b^L)|fis \rangle \vert^2=
 (m-Z_L)~~~\Longrightarrow~~~(m-Z_L)\geq0
\ee
de donde se sigue la ``cota de Bogomol'nyi"
\be
 m\geq Z_L~~~~~~-\!\!\!\!\!\!\vee\; Z_L
 \label{coti}
\ee
Esta ecuaci\'on demuestra que:
\begin{itemize}
\item En el caso no masivo no es posible (dado que las $Z_L\geq 0$)
la presencia de cargas centrales.
\item En el caso masivo la masa de las representaciones est\'a
acotada inferiormente por la m\'axima carga central.
\end{itemize}
En la proxima secci\'on discutiremos un efecto interesante
conocido como acortamiento del supermultiplete, que sucede cuando
la masa de la representaci\'on satisface $m=Z_L$ para alguna
$Z_L$.

\section{Estados saturados o Estados BPS}

Hasta aqu\' \i~ hemos discutido las distintas representaciones
irreducibles del \'algebra supersim\'{e}trica y dedujimos que para
\'algebras con supersimetr\'{\i}a extendida $N>1$, en presencia de
cargas centrales, la masa de las representaciones est\'a acotada
inferiormente por el valor de la m\'axima carga central.

El caso en que ninguna carga central $Z_L$ iguala a la masa de la
representaci\'on, es an\'alogo al caso masivo sin cargas centrales
discutido anteriormente. Tenemos $2N$ operadores de creaci\'on
(ver ec.(\ref{juju})) y la dimensi\'on de la representaci\'on es
$2^{2N}(2j+1)$.

Nos interesa discutir ahora el caso en que alguna de las cargas
centrales $Z_L$, coincide con la masa de la representaci\'on. En
este caso vemos de la ec.(\ref{juju}), que los osciladores $b^L$ y
$(b^L)^\dagger$ proyectan sobre estados nulos, y nos encontramos
ante una situaci\'on similar al caso no masivo (ver
ec.(\ref{fisura})). A este tipo de representaciones se las
denomina representaciones cortas del \'algebra supersim\'{e}trica. En
el caso particular en que todas las cargas centrales coincidan con
la masa de la representaci\'on, tenemos que todos los osciladores
$b^L$ son nulos, reduci\'endose entonces a la mitad el n\'umero de
operadores de creaci\'on (representaciones $\frac12$-BPS). En este
caso la dimensi\'on de la representaci\'on masiva pasa de
$2^{2N}(2j+1)$ a $2^N(2j+1)$ estados. En particular para $j=0$, la
dimensi\'on de la representaci\'on masiva $\frac12$-BPS, coincide
con la de la representaci\'on no masiva. Este fen\'omeno es
importante, ya que nos permite afirmar que el fen\'omeno de Higgs
en teor\' \i as de gauge supersim\'etricas, es posible. Partiendo
de un lagrangiano supersim\'{e}trico para campos no masivos, es
posible obtener un espectro masivo, conservando el n\'umero de
grados de libertad, si el mismo es BPS. \footnote{ Los estados
BPS, o estados saturados, son importantes en el testeo de las
conjeturas de dualidad entre distintas teor\' \i as
supersim\'etricas \cite{dual}-\cite{kir} y en el an\'alisis de
fen\'omenos no perturbativos\cite{sw}. Algunas de las razones son:
\begin{itemize}
\item Debido a su relaci\'on con las cargas centrales, a pesar de
ser masivos, forman multipletes de supersimetr\'{\i}a mas cortos que
los multipletes masivos generales. Su masa est\'a dada en
t\'erminos de su carga y de los valores de expectaci\'on de los
campos de Higgs (m\'odulos).
\item Son los \'unicos estados que pueden hacerse no masivos al
variar las constantes de acoplamiento y los valores de
expectaci\'on de los Higgs.
\item Cuando est\'an en reposo no hay fuerza neta entre ellos.
\item Se supone que la f\'ormula de masa es exacta si se escribe
en t\'erminos de los valores renormalizados de las cargas y los
m\'odulos.
\item En puntos gen\'ericos del espacio de m\'odulos son estables.
\end{itemize}
El an\'alisis del \'algebra supersim\'{e}trica en distintos modelos de
teor\' \i as de campos muestra que,  en ciertos casos, los
solitones de la teor\' \i a corresponden a estados BPS. Dado que
estos estados solit\'onicos son los relevantes en la descripci\'on
no perturbativa, es posible hacer predicciones sobre el r\'egimen
de acoplamiento fuerte, basados en las propiedades BPS de los
mismos.}

Como veremos en el cap. \ref{susya} la cota de Bogomol'nyi
ec.(\ref{coti}) est\'a relacionada con la cota algebraica derivada
por Bogomol'nyi \cite{Bogo} en modelos puramente bos\'onicos
mediante m\'etodos algebraicos, de all\' \i~ su nombre. De hecho,
en la extensi\'on supersim\'etrica de dichos modelos (basada en un
\'algebra sin cargas centrales) la presencia de soluciones
solit\'onicas da origen a extensiones centrales en el \'algebra
supersim\'{e}trica realizando (\ref{susyC}) \cite{OW}. Las cargas
centrales tienen su origen en t\'erminos de superficie
(topol\'ogicos) que aparecen al calcular los anticonmutadores de
las cargas supersim\'{e}tricas $Q$ y $\bar Q$. Dado que estas se
expresan, via el teorema de Noether, por integrales de expresiones
locales de los campos del modelo, dichos t\'erminos de superficie,
que usualmente se descartan, deben ser tenidos en cuenta en
presencia de solitones. El resultado es que las cargas centrales,
para una normalizaci\'on adecuada, son iguales a las cargas
topol\'ogicas.

Otro aspecto del an\'alisis de Bogomol'nyi es que las soluciones
que saturaran la cota satisfacen ecuaciones de primer orden
llamadas ecuaciones BPS, en lugar de las ecuaciones de segundo
orden de Euler-Lagrange. Estas ecuaciones surgen, en la
perspectiva supersim\'{e}trica, de observar que para un estado BPS
existen cargas que lo dejan invariante (las correspondientes a los
osciladores $b$ en la ec.(\ref{juju})). Pidiendo invarianza del
estado frente a las transformaciones supersim\'{e}tricas generadas por
dichas cargas obtenemos las ecuaciones
Bogomol'nyi-Prasad-Sommerfield.

\section{Formalismo de supercampos $N=1$}

El \'algebra supersim\'{e}trica (\ref{amsa}) puede ser interpretada
como un \'algebra de Lie con pa\-r\'a\-me\-tros anticonmutantes.
Dado que el grupo de Poincar\'e act\'ua de manera natural sobre
las coordenadas del espacio-tiempo, las transformaciones
supersim\'etricas act\'uan de manera natural sobre una extensi\'on
del espacio-tiempo llamada {\it superespacio}. Al realizar el
super\'algebra sobre el superespacio obtenemos los {\it
supercampos}. Los campos cu\'anticos usuales corresponden a las
distintas componentes de un {\it supercampo}, al ser expandido en
las coordenadas anticonmutantes del superespacio.

\subsection{Superespacio}

Discutiremos el caso de superespacio $N=1$. Con el fin de
interpretar el super\'algebra de Poincare como un \'algebra de
Lie, el superespacio se introduce agregando cuatro grados de
libertad fermi\'onicos $\theta_\alpha,\,\bar\theta^{\dot\alpha}$~,
a los bos\'onicos $x^\mu$ ya existentes. La propiedad fermi\'onica
se manifiesta en que estas coordenadas anticonmutan\footnote{Los
\' \i ndices $\alpha,\,\dot\alpha$ se suben y bajan siguiendo las
convenciones del ap\'endice.}
\be
 \{\theta_\alpha ,\theta_\beta\}=
 \{\bar\theta^{\dot\alpha},\bar\theta^{\dot\beta}\}
 =\{\theta_\alpha ,\bar\theta^{\dot\beta}\} =0
\ee
Con estas propiedades para las variables $\theta$ se deduce que
\ba
 {[}\, \theta Q, \bar\theta\bar Q \,{]}
 &=&2\theta \sigma^\mu\bar\theta P_\mu\\
 {[}\, \theta Q, \theta Q \,{]}&=& 0\\
 {[}\, \bar\theta\bar Q ,\bar\theta \bar Q \,{]} &=& 0
 \label{ja}
\ea
Esta definici\'on motiva la definici\'on del elemento del grupo de
super-Poincare como
\be
 G(x,\theta ,\bar\theta ,\omega ) =
 e^{i[-x^\mu P_\mu + \theta Q + \bar\theta\bar Q ]}
 e^{-{i\over 2}\omega_{\mu\nu}M^{\mu\nu}}
 \label{gu}
\ee
Es claro entonces que $(x^\mu,\,
\theta_\alpha,\,\bar\theta^{\dot\alpha})$ parametrizan el espacio
cociente $4+4$-dimensional: super-Poincar\'e $N=1$  m\'odulo
Lorentz. Este espacio cociente es conocido como {\it superespacio
r\' \i gido $N=1$}\footnote{ R\' \i gido se refiere a
supersimetr\' \i a global.}.

A partir del elemento del grupo ec.(\ref{gu}), es posible deducir
la acci\'on del grupo sobre las coordenadas del superespacio.
Usando la f\'ormula de Hausdorff
\be
 e^Ae^B=e^{A+B+1/2[A,B]+...}
\ee
obtenemos\footnote{Los t\'erminos de orden mayor se anulan en el
presente caso, debido a (\ref{ja}).}
\be
 G(a,\xi,\bar\xi)G(x,\theta,\bar\theta)=
 G(x+a+i\theta\sigma\bar\xi-i\xi\sigma\bar\theta,\theta+\xi,
 \bar\theta+\bar\xi)
\ee
Frente a una transformaci\'on supersim\'{e}trica con par\'ametros
$\xi$ y $\bar \xi$, las coordenadas del superespacio transforman
como
\ba
 x^\mu &\rightarrow& x'^\mu=x^\mu + i\theta\sigma^\mu\bar\xi -
 i\xi\sigma^\mu\bar\theta\\
 \theta &\rightarrow&\theta'=\theta + \xi\\
 \bar\theta&\rightarrow& \bar\theta'=\bar\theta +\bar\xi
 \label{trafos}
\ea
Dado que estas transformaciones son implementadas por el operador
$i(\xi^\alpha Q_\alpha + \bar{\xi}_{\dot\alpha}
\bar{Q}^{\dot\alpha})$, es facil obtener una representaci\'on en
t\'erminos de operadores diferenciales actuando en el superespacio
\be
 P_\mu=i\partial_\mu
\ee \ba
 &iQ_\alpha=\partial_\alpha-i\sigma^\mu_{\alpha\dot{\alpha}}
 \bar{\theta}^{\dot{\alpha}}\, \partial_\mu~,~~~~~~~
 &iQ^\alpha=-\partial^\alpha+i \bar{\theta}_{\dot{\alpha}}
 \bar\sigma^{\mu~\dot\alpha\alpha}\, \partial_\mu\\
 &i\bar Q^{\dot\alpha}=\bar\partial^{\dot\alpha}-i\bar\sigma^
 {\mu\,\dot\alpha\alpha}\theta_\alpha\, \partial_\mu~,~~~~~~~
 &i\bar{Q}_{\dot\alpha}=-\bar\partial_{\dot\alpha}+i\theta^\alpha
 \sigma^\mu_{\alpha\dot\alpha}\, \partial_\mu
 \label{cu}
\ea
Las cuales satisfacen el \'algebra (\ref{esi1s}) con $N=1$ a menos
de un signo  \cite{wess},\cite{soh}. Es util definir derivadas
covariantes frente a las transformaci\'ones
supersim\'{e}tricas\footnote{Usando la regla de la cadena es facil ver
que $\partial_\mu=\partial/\partial x^\mu$ es invariante frente a
(\ref{trafos}), pero que
$\partial_\alpha=\partial/\partial\theta^\alpha$ y
$\bar\partial_{\dot\alpha}=\partial/\partial\theta^{\dot\alpha}$
no lo son.}
\ba
 &D_\alpha=\partial_\alpha+i\sigma^\mu_{\alpha\dot{\alpha}}
 \bar{\theta}^{\dot{\alpha}}\, \partial_\mu~,~~~~~~~
 &D^\alpha=-\partial^\alpha-i \bar{\theta}_{\dot{\alpha}}
 \bar\sigma^{\mu~\dot\alpha\alpha}\, \partial_\mu\\
 &\bar D^{\dot\alpha}=\bar\partial^{\dot\alpha}+i\bar\sigma^
 {\mu\,\dot\alpha\alpha}\theta_\alpha\, \partial_\mu~,~~~~~~~
 &\bar D_{\dot\alpha}=-\bar\partial_{\dot\alpha}-i\theta^\alpha
 \sigma^\mu_{\alpha\dot\alpha}\, \partial_\mu
 \label{deriv}
\ea
Estas derivadas anticonmutan con todos los generadores $Q,\,\bar
Q$.

\subsection{Supercampos}

Un supercampo es simplemente una funci\'on de las coordenadas del
superespacio $(x^\mu,\, \theta_\alpha, \,\bar\theta^
{\dot\alpha})$. Como las variables $\theta$ son Grassmann, el
desarrollo de Taylor tiene un n\'umero finito de t\'erminos. Los
campos usuales aparecen como las componentes del supercampo al
expandirlo en $\theta$. Un supercampo escalar arbitrario toma la
forma
\ba
 \Phi(x,\theta,\bar\theta)&=&f(x)+ \theta\varphi (x) +
 \bar\theta\bar\chi(x)+\theta^2 m(x) +\bar\theta^2 n(x)
 + \theta\sigma^\mu\bar\theta A_\mu(x) \nonumber\\
 &&+\theta^2\,\bar\theta\bar\lambda (x)+\bar\theta^2\theta\psi(x)
 +\theta^2\bar\theta^2 d(x)
\ea
Bajo una transformaci\'on supersim\'{e}trica infinitesimal, el
supercampo transforma como
\be
 \delta\Phi(x,\theta,\bar\theta)=
 i[\xi Q+\bar\xi \bar Q]\Phi(x,\theta,\bar\theta)
 \label{straf}
\ee
Dado que conocemos la acci\'on de las supercargas $Q$ como
operadores diferenciales en el superespacio (\ref{cu}) podemos,
desarrollando (\ref{straf}) encontrar las transformaciones
supersim\'{e}tricas de los campos usuales. Se obtiene
\ba
 \delta_\xi f &=& \xi\varphi + \bar\xi\bar\chi\\
 \delta_\xi \varphi _{\alpha} &=& 2m\,\xi _{\alpha}
 - (\bar\xi\bar\sigma^\mu)_\alpha {[}A_\mu+i\partial_\mu f{]}\\
 \delta_\xi \bar\chi^{\dot\alpha}&=&2n\,\bar\xi^{\dot\alpha}
 + (\xi\sigma^\mu)^{\dot\alpha}{[}A_\mu-i\partial _\mu f{]}\\
 \delta_\xi m &=& \bar\xi\bar\lambda+ {i\over 2}\,
 \bar\xi/\!\!\!{\bar\partial}\varphi\\
 \delta_\xi n&=&\xi\psi+ {i\over 2}\,\xi/\!\!\!\partial\bar\chi\\
 \delta_\xi A_\mu&=&\xi\sigma_\mu\bar\lambda+
 \psi\sigma_\mu\bar\xi+i(\xi\partial_\mu\varphi
 -\bar\xi\partial_\mu\bar\chi)-\frac i2
 (\xi\sigma_\mu/\!\!\!{\bar\partial}\varphi
 -\bar\xi\bar\sigma_\mu/\!\!\!\partial\bar\chi)\\
 \delta_\xi \bar\lambda^{\dot\alpha}&=&2d\,\bar\xi^{\dot\alpha}
 +{i\over 2}(\bar\xi\bar\sigma^\mu\sigma^\nu)^{\dot\alpha}
 \partial_\mu A_\nu-i(\xi\sigma^\mu)^{\dot\alpha}\partial_\mu m\\
 \delta_\xi \psi_{\alpha}&=&2d\,\xi_{\alpha}-{i\over 2}
 (\xi\sigma^\mu\bar\sigma^\nu)_\alpha \partial_\mu A_\nu
 -i(\bar\xi\bar\sigma^\mu)_\alpha\partial_\mu n\\
 \delta_\xi d &=& {i\over 2}\partial_\mu \left[\,\xi\sigma^\mu\bar\lambda
 +\bar\xi\bar\sigma^\mu\psi\,\right]
\ea
donde hemos denotado $/\!\!\!\partial=\sigma^\mu\partial_\mu$ y
$/\!\!\!{\bar\partial}=\bar\sigma^\mu\partial_\mu$. La utilidad
del formalismo de supercampos se descubre al observar que la
componente mas alta del supercampo escalar $\Phi$, el campo $d$,
transforma como una derivada total frente a transformaciones
supersim\'{e}tricas\footnote{Las dimensiones de las variables
anticonmutantes, medidas en unidades de energ\' \i a, se deducen
de observar que
\be
 \{Q,Q\}\sim P~~~\Longrightarrow~~~{[}Q{]}=
 \frac12~~~\Longrightarrow~~~{[}\theta{]}=-\frac12
\ee
 }. Luego, ignorando t\'erminos de superficie, la
integral en el espacio-tiempo de la componente mas alta de un
supercampo escalar es invariante frente a supersimetr\'{\i}a. Los
lagrangianos supersim\'etricos pueden entonces construirse tomando
la componente mas alta del supercampo apropiado. La forma elegante
en que se denota la componente ``invariante" frente a
supersimetr\'{\i}a dentro del formalismo es mediante una integral en
todo el superespacio (ver ap\'endice \ref{a2}).

Hemos demostrado que los supercampos escalares forman una
representaci\'on lineal (off-shell) del \'algebra supersim\'{e}trica
$N=1$. Sin embargo esta representaci\'on es reducible. La forma en
que se obtienen las distintas representaciones irreducibles es
imponiendo v\' \i nculos al supercampo. Se\~nalemos sin embargo,
que el supercampo escalar gen\'erico no es completamente
reducible, esto significa que la representaci\'on reducible $\Phi$
{\it no es} una suma de representaciones irreducibles.

\subsection{Representaciones irreducibles}

\noindent\underline{Supercampo quiral}: El supercampo quiral
est\'a caracterizado por la condici\'on\footnote{Debido a que
$\{Q,D\}=\{Q,\bar D\}=\{\bar Q,D\}=\{\bar Q,\bar D\}=0$ tenemos
que, frente a una transformaci\'on supersim\'{e}trica, $D\Phi$
transforma covariantemente
\be
 \Phi\to G\Phi\Longrightarrow D(G\Phi)=GD\Phi~~.
\ee
donde $G$ est\'a dada por (\ref{gu}). Luego, la condici\'on de
quiralidad es invariante. }
\be
 \bar D_{\dot\alpha}\Phi=0
 \label{vi}
\ee
El v\' \i nculo anterior se resuelve facilmente haciendo un cambio
de variables en el superespacio de
$(x,\theta,\bar\theta)\to(y,\theta,\bar\theta)$ donde
\be
 y^\mu=x^\mu+i\theta\sigma^\mu\bar\theta
\ee
El cambio de variables est\'a motivado porque
\ba
 \bar D_{\dot\alpha} y^\mu&=&0\\
 \bar D_{\dot\alpha} \theta^\beta&=&0
\ea
Luego, toda funci\'on de $(y,\theta)$ es un supercampo
quiral\footnote{El cambio de variables transforma las derivadas
covariantes (\ref{deriv}) en
\ba
 D_\alpha&=&\partial_\alpha+2i(\sigma^\mu\bar{\theta})_{\alpha}
 \, \partial_\mu\\
 \bar D^{\dot\alpha}&=&\bar\partial^{\dot\alpha}
\ea
donde aqu\' \i~ $\partial_\mu=\partial/\partial y^\mu$.}. La
soluci\'on para el v\' \i nculo es
\be
 \Phi(y,\theta) = \phi(y) + \sqrt{2}\theta\psi (y) +\theta^2 F(y)
 \label{fic}
\ee
donde $\phi(y)$ y $F(y)$ son campos escalares complejos, mientras
que $\psi(y)$ es un espinor de Weyl $(\frac12,0)$. Tenemos $4+4=8$
componentes de campo reales off-shell, estas son el doble que las
que aparecen en las representaciones irreducibles on-shell. La
expansi\'on total en $\theta$'s de (\ref{fic}) es\footnote{El
supercampo quiral corresponde al supercampo escalar general con el
siguiente v\' \i nculo
\ba
 \bar\chi (x) &=& 0 \\
 n(x) &=& 0\\
 A_\mu(x) &=& i\partial_\mu f(x)\\
 \bar\lambda (x) &=& -{i\over 2}\partial_\mu\phi\sigma^\mu\\
 \psi (x) &=& 0 \\
 d(x) &=& -\frac14 \Box f(x)
\ea
es facil verificar que las transformaciones supersim\'{e}tricas $N=1$
respetan este v\' \i nculo.}
\ba
 \Phi(y,\theta)&=& \phi(x) + \sqrt{2}\theta\psi (x) +\theta^2 F(x)\\
 &&+i\theta\sigma^\mu\bar\theta\partial_\mu \phi(x)-
 {i\over\sqrt{2}}\theta^2\partial_\mu\psi (x)\sigma^\mu\bar\theta -
 \frac 14 \theta^2\bar\theta^2\Box \phi(x)
\ea
Las transformaciones supersim\'{e}tricas infinitesimales  para los
campos son
\ba
 \delta \phi &=& \sqrt{2}\xi\psi\\
 \delta \psi &=& \sqrt{2}\xi F + \sqrt{2}i\sigma^\mu\bar\xi
 \partial_\mu \phi\\
 \delta F &=& i\sqrt{2}\,\,\partial_\mu(\bar\xi\bar\sigma^\mu
 \psi)
\ea
En notaci\'on de espinores de cuatro componentes, agrupando
$(\xi_\alpha,\bar\xi^{\dot\alpha})\rightarrow\Xi$ y
$(\psi_\alpha,\bar\psi^{\dot\alpha})\rightarrow\Psi$ (ver
ap\'endice \ref{a1}), toman la forma
\ba
 \delta \phi &=& {1\over\sqrt{2}}\,\bar\Xi(1+\Gamma^5)\Psi\\
 \delta \Psi &=& {1\over\sqrt{2}}\left[(1+\Gamma^5)
 (F+i\Gamma^\mu \partial_\mu\phi)+(1-\Gamma^5)
 (F^*+i\Gamma^\mu \partial_\mu\phi^*)\right]\Xi\\
 \delta F &=& {i\over\sqrt{2}}\,\,\partial_\mu\!\left(\bar\Xi(1-\Gamma^5)
 \Gamma^\mu\Psi\right)
\ea
Notemos que la variaci\'on de $F$ es una derivada
total\footnote{La manera de obtener elegantemente la  componente
mas alta de un supercampo quiral es integrando en las variables
quirales (ver ap\'endice \ref{a2}).}. Tenemos entonces que el
supercampo quiral define una representaci\'on lineal off-shell de
supersimetr\'{\i}a $N=1$. La expresi\'on on-shell de las
transformaciones se obtiene al reemplazar el campo auxiliar $F$
por su ecuaci\'on de movimiento (que resulta ser una ecuaci\'on
algebraica). En general las transformaciones on-shell ser\'an no
lineales.

La representaci\'on del campo quiral $\Phi=(\phi,\psi)$
corresponde on-shell al multiplete supersim\'{e}trico que contiene un
campo escalar complejo $\phi$ y un fermi\'on de Weyl $\psi$
(cf.(\ref{tabla1})).

Los supercampos antiquirales se definen de la manera obvia. Si
$\Phi(y,\theta)$ es un campo quiral,
$\Phi^\dagger(y^\dagger,\bar\theta)$ ser\'a un campo antiquiral,
esto es, satisfar\'a
\ba
 D_{\alpha}\Phi^\dagger&=&0\\
 \Phi^\dagger&=&\Phi^\dagger(y^\dagger,\bar\theta)
 ~,~~~~~~y^\dagger=x^\mu-i\theta\sigma^\mu\bar\theta
\ea
Dado que $D_\alpha$ y $\bar D_{\dot\alpha}$ obedecen la regla de
Leibniz, todo producto de supercampos quirales (antiquirales) es
un supercampo quiral (antiquiral).

\vspace{.3cm}

\noindent\underline{Supercampo vectorial}: Este multiplete se
define como un supercampo real
\be
 V(x,\theta,\bar\theta)=V^\dagger(x,\theta,\bar\theta)
\ee
en componentes
\ba
 V(x,\theta,\bar\theta)\!\!\!&=\!\!\!&c+ \theta\psi+
 \bar\theta\bar\psi+\theta^2 m +\bar\theta^2 m^*
 - \theta\sigma^\mu\bar\theta A_\mu \nonumber\\
 \!\!\!&\!\!\!&+i\theta^2(\bar\theta\bar\lambda +\frac1{2}
 \bar\theta/\!\!\!{\bar\partial}\psi)-i\bar\theta^2
 (\theta\lambda-\frac1{2}
 \theta/\!\!\!\partial\bar\psi)
 +\theta^2\bar\theta^2 (\frac12 D-\frac14\Box c)
\ea
Tenemos entonces $4$ escalares reales, 2 espinores de Weyl y un
campo vectorial real. El n\'{u}mero total de componentes es $8+8=16$
componentes reales y corresponde a una representaci\'{o}n irreducible
off-shell con el doble de grados de libertad que la representaci\'{o}n
on-shell masiva correspondiente a $\Omega_{\frac{1}2}$. La
presencia del vector $A_\mu$ en el supermultiplete sugiere
emplearlo para construir modelos supersim\'{e}tricos invariantes de
gauge. Debemos, entonces, definir la generalizaci\'{o}n de la
invarianza de gauge en el caso supersim\'{e}trico.

Definimos las transformaciones de gauge supersim\'{e}tricas
(supergauge) del multiplete vectorial como
\be
 V\rightarrow V+\Lambda+\Lambda^\dagger
 \label{trgauge}
\ee
donde $\Lambda\,(\Lambda^\dagger)$ es un supercampo quiral
(antiquiral). Desarrollando en componentes vemos que es posible
elegir convenientemente $\Lambda$ de manera de llevar al
supercampo $V$ al gauge de ``Wess-Zumino" donde $c=\psi=m=0$
\be
  V_{WZ}=-\theta\sigma^\mu\bar\theta A_\mu+i\theta^2
  \bar\theta\bar\lambda-i\bar\theta^2\theta\lambda
  +\theta^2\bar\theta^2\frac 12D
\ee
La elecci\'on del gauge WZ fija todas las componentes del campo
quiral $\Lambda=(\varphi,\chi,G)$ excepto la correspondiente a la
parte imaginaria del campo $\varphi$. Es facil ver desarrollando
(\ref{trgauge}) en componentes que fijado el gauge de WZ a\'un
tenemos la libertad de realizar transformaciones de gauge
abelianas sobre $A_\mu$
\be
 A_\mu\rightarrow A_\mu+\partial_\mu \varphi^I
\ee
con $\varphi^I=2\,{\rm Im}(\varphi)$. De esta manera, el campo
vectorial $A_\mu$ transforma frente a (\ref{trgauge}) de la manera
esperada. El gauge de Wess-Zumino no es invariante frente a
supersimetr\'{\i}a\footnote{La situaci\'on es an\'aloga a lo que sucede
en las teor\' \i as de gauge usuales, el fijado del gauge de
Coulomb $A_0$ no es invariante de Lorentz, si que\-re\-mos
permanecer en el gauge de Coulomb luego de una transformaci\'on de
Lorentz, debemos realizar una transformaci\'on de gauge adicional.
En nuestro caso, si queremos permanecer en el gauge de WZ luego de
una transformaci\'on supersim\'{e}trica debemos realizar una
transformaci\'on de supergauge adicional. Por lo tanto en
(\ref{trafv}) denotamos por
$\delta_{WZ}\equiv\delta^{(SUSY)}+\delta^{(SG)}$ donde
$\delta^{(SG)}$ corresponde a la transformaci\'on de supergauge
adicional que es necesario realizar sobre el supercampo
$V'=V+\delta^{(SUSY)}V$ para llevarlo al gauge de WZ.}, sin
embargo aun habi\'endolo fijado tenemos la libertad de realizar
transformaciones de gauge abelianas sobre $A_\mu$. Las
transformaciones supersim\'{e}tricas infinitesimales  para los campos
son
\ba
 \delta_{WZ}A_\mu&=&-i(\xi\sigma_\mu\bar\lambda+
 \bar\xi\bar\sigma_\mu\lambda)\\
 \delta_{WZ}\lambda_\alpha&=&i(D\xi_\alpha-F_{\mu\nu}
 (\sigma^{\mu\nu}\xi)_\alpha)\\
 \delta_{WZ}D&=&\partial_\mu(\bar\xi\bar\sigma_\mu\lambda-
 \xi\sigma_\mu\bar\lambda)
 \label{trafv}
\ea
Que en notaci\'on de espinores de cuatro componentes toman la
forma
\ba
 \delta_{WZ}A_\mu&=&-i\,\bar\Xi\Gamma_\mu\Lambda\\
 \delta_{WZ}\Lambda&=&i(D\Gamma^5-F_{\mu\nu}
 \Sigma^{\mu\nu})\Xi\\
 \delta_{WZ}D&=&\partial_\mu(\bar\Xi\Gamma_\mu\Gamma^5\Lambda)
 \label{trafv4}
\ea
A partir de $V$ definimos el supercampo de curvatura
como\footnote{Como veremos mas abajo, en el caso abeliano
$W_\alpha$ es invariante de gauge por lo tanto puedo calcular
(\ref{curva}) en el gauge de WZ, la particularidad de este gauge
es que $V^3=0$. Luego, la exponencial tiene un n\'umero finito de
t\'erminos.}
\ba
 W_\alpha&=&\frac18\bar D^2 e^{-2V}D_\alpha e^{2V}
 \label{curva}\\
 &=&i\lambda_\alpha-\theta_\alpha D+\frac i2
 (\sigma^\mu\bar\sigma^\nu\theta)_\alpha F_{\mu\nu}-
 \theta^2(/\!\!\!\partial\bar\lambda)_\alpha
\ea
$F_{\mu\nu}$ es el tensor de campo electromagn\'{e}tico usual
$F_{\mu\nu}=\partial_\mu A_\nu-\partial_\nu A_\mu$. $W_\alpha$ es
un supercampo quiral e invariante de gauge\footnote{$W_\alpha$ no
es un campo quiral arbitrario, satisface el v\' \i nculo $D^\alpha
W_\alpha=\bar D_{\dot\alpha} \bar W^{\dot\alpha}$.}
\ba
 W_\alpha&\rightarrow&\frac14
 \bar D^2 D_\alpha(V+\Lambda+\Lambda^\dagger)\nonumber\\
 &&=W_\alpha +\frac 14 \bar D^2D_\alpha \Lambda
 \quad\quad\quad\quad\quad\! ({\rm dado~que~ }D_\alpha\Lambda^{\dagger}=0)
 \nonumber\\
 &&=W_\alpha -\frac14 \bar D^{\dot\beta}
 \{ \bar D_{\dot\beta}, D_\alpha \}\Lambda
 \quad \quad ({\rm dado~que~ }\bar D_{\dot\beta}\Lambda = 0) \nonumber\\
 &&= W_\alpha
 \label{inw}
\ea
donde en el \'{u}ltimo paso usamos que
\ba
 \{ \bar D _{\dot\beta}, D_\alpha \} &=&
 -2\sigma^\mu_{\alpha\dot\beta}P_\mu
 \nonumber\\
 {[}\, \bar D ^{\dot\beta},P_\mu \,{]} &=& 0 \quad .
\ea

Los campos de materia, representados por supermultipletes quirales
$\Phi$, cambian frente a una transformaci\'{o}n de supergauge como
\be
 \Phi\rightarrow e^{-2\Lambda}\Phi~~,~~~~~~~\Phi^\dagger
 \rightarrow \Phi^\dagger e^{-2\Lambda^\dagger}
 \label{agi}
\ee
Combinaciones
\be
 \Phi^\dagger e^{2V}\Phi
\ee
son invariantes de gauge.

En la generalizaci\'{o}n no abeliana $V$ pertenece a la representaci\'{o}n
adjunta del \'{a}lgebra $V=V^{a}t_a$ y las transformaciones de gauge
se implementan como
\be
 e^{2V} \rightarrow e^{-i2 \Lambda^\dagger} e^{2V} e^{i2 \Lambda}
 \qquad {\rm donde}, \; \Lambda = \Lambda^{a}t_a
\ee
con ${[}t_a,t_b{]}=if_{abc}t_c$. El supercampo de curvatura se
define por (\ref{curva}), cuya expresi\'{o}n en campos componentes
queda
\be
 W_\alpha=i\lambda_\alpha-\theta_\alpha D+\frac i2
 (\sigma^\mu\bar\sigma^\nu\theta)_\alpha F_{\mu\nu}-
 \theta^2(/\!\!\!\nabla\bar\lambda)_\alpha
\ee
donde ahora las expresiones para la fuerza de campo $F_{\mu\nu}$ y
la derivada covariante actuando sobre el gaugino son
\ba
 F_{\mu\nu}&\equiv&\partial_\mu A_\nu-\partial_\nu
 A_\mu+i{[}A_\mu,A_\nu{]}\\
 (/\!\!\!\nabla\bar\lambda)_\alpha&\equiv&\sigma^\mu_{\alpha\dot\alpha}
 (\partial_\mu\bar\lambda^{\dot\alpha}
 +i{[}A_\mu,\bar\lambda^{\dot\alpha}{]})
\ea
En el caso no abeliano $W_\alpha$ es covariante frente a
transformaciones  de gauge (cf. ec.(\ref{inw}))
\be
 W_\alpha\rightarrow e^{-i2 \Lambda}W_\alpha e^{i2 \Lambda}
 \label{tcal}
\ee
Las transformaciones supersim\'{e}tricas toman la forma (\ref{trafv})
recordando ahora que los campos toman valores en el \'algebra de
Lie del grupo de gauge elegido. El fijado del gauge de WZ en el
caso no abeliano es mas sutil pero una vez realizado, el \'unico
grado de libertad que queda en $\Lambda$ es el correspondiente a
las transformaciones de gauge usuales, las que se realizan tomando
$\Lambda=a$ donde $a$ es un campo real.



\chapter{Teor{\'{\i}}a de  Born-Infeld-Higgs Abeliana
y cotas BPS\label{susya}}
\begin{center}
 {\begin{minipage}{6truein}
 { \sl En este cap\' \i tulo
 estudiamos la extensi\'on supersim\'{e}trica
 $N=2$ del mo\-de\-lo Born-Infeld-Higgs en tres dimensiones
 espacio-temporales y derivamos las ecuaciones de
 Bogomol'nyi-Prasad-Sommerfield (BPS) para el sector bos\'onico. La
 supersimetr\' \i a impone una forma particular para el potencial
 del campo de Higgs y para el acoplamiento del mismo con el campo de gauge
 (cuya din\'amica esta determinada por la acci\'on de Born-Infeld).
 Derivamos las cotas de Bogomol'nyi para la energ\' \i a del v\'ortice
 a partir del \'algebra supersim\'etrica; las ecuaciones
 BPS resultantes coinciden con las obtenidas
 en el modelo de Maxwell-Higgs.

 Discutimos tambi\'en la  es\-truc\-tu\-ra BPS para mo\-de\-los
 no polin\'omicos ge\-ne\-ra\-les,  mos\-tran\-do la
 ineluctabilidad de las ecuaciones de Bogomol'nyi.

 El inter\'es en $3$ dimensiones se debe a que en este espacio-tiempo
 se conocen soluciones de v\'ortice para las ecuaciones de Bogomol'nyi.
 Estos resultados son parte de las contribuciones originales de
 esta tesis \cite{GNSS}.}
 \end{minipage}}
\end{center}

\section{Introducci\'on}

La extensi\'on supersim\'etrica  de la teor\' \i a de Born-Infeld
\cite{B}-\cite{BI} fue analizada en \cite{DP}-\cite{CF} empleando
t\'ecnicas de superespacio. Mas recientemente, se estudiaron las
extensiones supersim\'etricas de la teor\' \i a de Born-Infeld en
$10$ dimensiones, de inter\'es dada su conexi\'on con la
din\'amica de Dp-branas
\cite{Lei}-\cite{Pol},\cite{Tse2}-\cite{Bre},\cite{BG}-\cite{T}.

El objeto del presente cap\' \i tulo es estudiar las relaciones de
Bogomol'nyi y las soluciones BPS para el modelo de
Born-Infeld-Higgs y para modelos no polin\'omicos generales, temas
\' \i ntimamente relacionados a la extensi\'on supersim\'etrica de
los mismos. Con este fin, centramos el an\'alisis en la
extensi\'on supersim\'{e}trica $N=2$ de la teor\' \i a de Born-Infeld
en tres dimensiones espacio-temporales. La elecci\'on del
espacio-tiempo $d=3$ simplifica los c\'alculos sin perder
generalidad. Es posible mostrar que al ser acoplada de manera
particular a un campo de Higgs la teor\' \i a de Born-Infeld,
admite para su sector bos\'onico una cota de Bogomol'nyi
\cite{NS1}. Esto sugiere que dicha teor\' \i a deber\' \i a poder
ser supersimetrizada. Como consecuencia de  la construcci\'on
realizada en \cite{NS1} las ecuaciones BPS resultantes coinciden
con las de la teor\' \i a de Maxwell-Higgs. Este punto ser\'a
discutido hacia el final del presente cap\' \i tulo en conexi\'on
con la extensi\'on supersim\'{e}trica de modelos no polin\'omicos
generales.

Como es sabido, las relaciones de Bogomol'nyi pueden ser halladas
estableciendo una desigualdad entre la energ\' \i a del sistema y
la carga topol\'ogica  \cite{Bogo} o analizando las
representaciones del \'algebra supersim\'{e}trica en presencia de
extensiones centrales \cite{OW}. Para este \'ultimo punto de vista
es muy instructivo derivar expl\' \i citamente, usando el m\'etodo
de Noether, el \'algebra supersim\'etrica, descubri\'endose as\'
\i~ el origen de las propiedades BPS de los estados y mostrando la
igualdad entre la extensi\'on central y  la carga topol\'ogica.
Este procedimiento fue seguido para el modelo de Maxwell-Higgs en
\cite{ed}. En el presente cap\' \i tulo realizamos un an\'alisis
similar para el modelo supersim\'etrico de Born-Infeld-Higgs.

El plan de este cap\' \i tulo es el siguiente: en la primera
secci\'on construiremos la teor\' \i a de Born-Infeld con $N=1$
supersimetr\' \i as en $d=4$ dimensiones dando una f\'ormula
explicita para el lagrangiano fermi\'onico, necesario para la
construcci\'on de las cargas supersim\'etricas. En la secci\'on
siguiente resumiremos el modelo supersim\'etrico de Higgs con
ruptura espont\'anea de la simetr\' \i a de gauge. En la secci\'on
4 procederemos a realizar una reducci\'on dimensional a $d=3$,
obteniendo la teor\' \i a de Born-Infeld-Higgs con $N=2$
supersimetr\' \i as. Mostraremos el origen supersim\'{e}trico de las
ecuaciones BPS en la secci\'on 5;  el \'algebra supersim\'etrica
$N=2$ ser\'a constru\' \i da en la secci\'on 6 en donde se
discutir\'an las cotas de Bogomol'nyi. La sensibilidad de las
ecuaciones BPS a la din\'amica asociada al campo de gauge ser\'a
discutida para modelos polin\'omicos generales en la secci\'on 7.
En la \'ultima secci\'on discutiremos los resultados obtenidos.

\section{El modelo supersim\'etrico de Born-Infeld}

En esta secci\'on construiremos  el modelo de Born-Infeld (BI) con
$N=1$ supersimetr\' \i as en $d=4$. En la pr\'oxima secci\'on lo
acoplaremos a un campo de Higgs y en la subsiguiente obtendremos
mediante la t\'ecnica de reducci\'on dimensional el lagrangiano
supersim\'{e}trico $N=2$ en $d=3$.

\noindent El lagrangiano de la teor\' \i a de Born-Infeld en
espacio plano $d=4$ se define como \cite{BI}
\begin{equation}
 {\cal L}_{{BI}}=\frac{\beta ^2}{e^2}\left( 1 - \sqrt{-\det \left(
 g_{\mu \nu }+\frac 1\beta F_{\mu \nu }\right) } \right)
 \label{1}
\end{equation}
(para las convenciones ver el ap\'endice \ref{a1}).

\noindent Calculando el determinante obtenemos\footnote{El
determinante en ${\bb M}_4$ (Minkowski $d=4$) puede ser calculado
usando
\ba
 \det (g+F)\!\!\!&=\!\!\!&\exp({\rm
 tr}\ln(g+F))\\
 \!\!\!&=\!\!\!&-1-1/2F^2+1/4(F^4-1/2(F^2)^2) \label{bi4}\\
 \!\!\!&=\!\!\!&-1-1/2F^2+1/16(F\tilde F)^2
\ea
donde $F^2=F^{\mu \nu}F_{\mu \nu}$,
$F^4=F_\mu^{~\nu}F_\nu^{~\rho}F_\rho^{~\sigma}F_\sigma^{~\mu}$ y
$F\tilde F= F^{\mu \nu}\tilde F_{\mu \nu}$. Ver el ap\'endice
\ref{a1} para identidades del tensor electromagn\'etico.}
\be
\det \left( g_{\mu \nu }+\frac 1\beta F_{\mu \nu }\right)=-
1-\frac 1{ 2\beta ^2}F^{\mu \nu}F_{\mu \nu}+ \frac 1{16\beta
^4} \left( F^{\mu \nu}\tilde F_{\mu \nu} \right) ^2%
\label{2}%
\ee
lo que nos permite expresar (\ref{1}) en la forma
\ba
 {\cal L}_{{BI}}\!\!\!&=\!\!\!&\frac{\beta ^2}{e^2}\left(1 - \sqrt{1+\frac
 1{2\beta ^2}F^{\mu \nu}F_{\mu \nu} -\frac 1{16\beta ^4}\left(
 F^{\mu\nu}\tilde F_{\mu \nu}\right) ^2} \right)
 \label{tuti}\\
 \!\!\!&=\!\!\!&-\frac 1{4e^2} F^2+\frac 1 {32e^2\beta^2}\left((F^2)^2+
 (F\tilde F)^2\right)-\frac 1 {128e^2\beta^4} F^2 \left((F^2)^2+
 (F\tilde F)^2\right)+\ldots
 \label{3}
\ea
(Aqu\' \i , $\tilde{F}_{\mu \nu} \equiv
\frac{1}{2}\varepsilon_{\mu \nu \rho \sigma}F^{\rho\sigma}$). De
esta expresi\'on vemos que el l\' \i mite a la teor\' \i a de
Maxwell se obtiene haciendo $\beta \to \infty$. El llamado l\' \i
mite ultra Born-Infeld corresponde a $\beta \to 0$ y da una
acci\'on topol\'ogica (ec.(\ref{topo})).
%
%

La extensi\'on supersim\'etrica del lagrangiano de Born-Infeld que
presentaremos sigue los lineamientos de \cite{DP}-\cite{CF}. Es de
notar, sin embargo, que si bien la derivaci\'on de la parte
bos\'onica del lagrangiano ha sido construida en las referencias
anteriormente citadas (ver tambien \cite{BG}-\cite{APS})
desarrollaremos aqu\' \i~ una descripci\'on detallada pues, para
nuestros fines, es necesario conocer tambi\'en parte de los
t\'erminos fermi\'onicos, que son necesarios para el c\'alculo del
\'algebra supersim\'etrica, y estos no fueron calculados en dichos
trabajos.

Comenzamos escribiendo el lagrangiano de BI (\ref{1}) en la forma
\be
{\cal L}_{{BI}} =\frac{\beta ^2}{e^2} \sum_{n=1}^\infty q_{n-1}\!
\left( \frac 1{2\beta ^2}F^{\mu \nu}F_{\mu \nu}-%
\frac 1{16\beta ^4}\left( F^{\mu \nu}\tilde F_{\mu \nu} \right) ^2
\right)^n
\label{4}%
\ee
donde
\begin{eqnarray}
 q_0 \!\!\!&=\!\!\!& - \frac{1}{2} \nonumber \\
 q_n \!\!\!&=\!\!\!&\frac{\left( -1\right) ^{n+1}\left( 2n-1\right) !}
 {4^n\left( n+1\right) !\left( n-1\right) !} ~ ~ ~ ~ {\rm for} ~ ~ n
 \geq 1
 \label{5}
\end{eqnarray}
($q_0=-\frac 1 2 ,~q_1=\frac 1 8 ,~q_2=-\frac 1 {16} ,~q_3=\frac 5
{128} ,~q_4=-\frac 7 {256} ,...$).

La Ec.(\ref{4}) puede ser reescrita como
\be
 {\cal L}_{{BI}}=\frac {\beta^2}{e^2} \sum_{n=1}^\infty
 q_{n-1}\sum_{j=0}^n {n \choose j} \left(  \frac 1{2\beta ^2}F^{\mu
 \nu}F_{\mu \nu} \right) ^j\left( -\frac 1{16} (\frac 1{\beta
 ^2}F^{\mu \nu}\tilde F_{\mu \nu} )^2\right) ^{n-j}
 \label{8}
\ee
El ingrediente b\'asico en la extensi\'on supersim\'etrica de la
acci\'on de BI es el supermultiplete quiral de curvatura
\begin{equation}
 W_\alpha =\frac 14\bar D_{\dot\beta}\bar D^{\dot\beta} D_\alpha V
 \label{9}
\end{equation}
donde $V$ es el supercampo vectorial real $V=V^\dagger$.  Para la
convenci\'on de las variables quirales y derivadas covariantes ver
el cap. \ref{susy}.

Recordemos que el lagrangiano de Maxwell $N=1$ es constru\' \i do
en t\'erminos de los supercampos quirales $W_\alpha$ y su herm\'
\i tico conjugado $\bar W_{\dot \alpha}$ haciendo
\begin{equation}
 {\cal L}_{Maxwell}^{SUSY} = \frac{1}{4e^2}
 \left[ \int \!d^2\theta~ W^2\!\left( y,\theta \right) +\int\!
 d^2\bar \theta ~\bar W^2( y^{\dagger },\bar \theta \,) \right]
 \label{12'}
\end{equation}
cuya expresi\'on en t\'erminos de los campos componentes del
multiplete vectorial es
\be
 {\cal L}_{Maxwell}^{SUSY}=\frac 1 {2e^2} (-\frac 1 2
 F_{\mu\nu}F^{\mu\nu}+D^2)-\frac {i} {2 e^2}\left( \lambda
 /\!\!\!\partial {\bar{\lambda }} + {\bar{ \lambda }} \bar
 {/\!\!\!\partial} \lambda \right)
 \label{maxc}
\ee

Para poder construir las potencias superiores de $F^2$ y $F\tilde
F$, que aparecen en (\ref{8}), necesarias en la expresi\'on de la
acci\'on de BI, se mostr\'o en  \cite{DP}, que deben incluirse en
el desarrollo potencias de los supercampos reales $X$ y $Y$
definidos como
\be
 X = \frac{1}{8} (D^2 W^2 + \bar D^2 \bar W^2)
 \label{13}
\ee
\be
 Y = -\frac{i}{16} (D^2 W^2 - \bar D^2 \bar W^2)
 \label{14}
\ee
El motivo es que, como se muestra en el ap\'endice,
\be
 X |_{0} = ( \frac{1}{2} F^{\mu \nu} F_{\mu \nu}-D^2 ) +i
 \lambda /\!\!\!\partial \bar \lambda +i \bar \lambda
 \bar{/\!\!\!\partial} \lambda
 \label{uno}
\ee
\be
 Y|_{0} = \frac{1}{2} (\frac{1}{2 } F^{\mu \nu}\tilde F _{\mu \nu}
 +  \lambda /\!\!\!\partial \bar \lambda - \bar \lambda
 \bar{/\!\!\!\partial}  \lambda).
 \label{dos}
\ee
De manera que las componentes mas bajas ($\theta=\bar\theta=0$) de
los supercampos $X$ y $Y$ incluyen los invariantes $F^2$ y $F
\tilde F$. Potencias arbitrarias  $(F^2)^n$ y $(F\tilde F)^m$ se
obtienen en las componentes mas bajas de $X^n$ y $Y^m$
respectivamente.

Consideremos entonces el siguiente lagrangiano supersim\'etrico
que como veremos dar\'a origen en su parte bos\'onica a teor\' \i
a de BI
\begin{equation}
 {\cal L}_{{BI}}^{{SUSY}} = \frac{1}{4e^2}\left[
 \int\! d^2\theta ~W^2+\int\!
 d^2\bar \theta ~\bar W^2\right] + \sum_{s,t=0}^\infty
 a_{st}^{(\beta)}\int\! d^4\theta~ W^2\bar W^2~ X^s Y^t
 \label{15}
\end{equation}
donde $a_{st}^{(\beta)}$ son coeficientes a determinar. La
ec.(\ref{15}) corresponde al lagrangiano propuesto en
\cite{DP}-\cite{CF}. En lo que respecta al \'ultimo t\'ermino de
(\ref{15}), es importante se\~nalar que para la construcci\'on del
lagrangiano de BI, no solo es necesario considerar $W^2$ y $\bar
W^2$ sino que tambi\'en se deben introducir productos de estos dos
supercampos quirales. En efecto, dado que la componente mas alta
$\theta^2 \bar\theta^2$ de $W^2 \bar W^2$ toma la forma
\be
 W^2 \bar W^2\vert_{\theta^2 \bar \theta^2} = \theta^2 \bar
 \theta^2 \left( (D^2 - \frac{1}{2} F_{\mu \nu}F^{\mu \nu})^2 +
 (\frac{1}{2}\tilde F_{\mu \nu}F^{\mu \nu})^2 +{\rm fermiones}\right)
 \label{termino}
\ee
podemos ver multiplicando esta expresi\'on por los supercampos
(\ref{13}-\ref{14}) ser\'a posible reproducir las potencias
superiores de $F^2$ y $F\tilde F$ necesarias en la expansi\'on de
(\ref{8})\footnote{No existen en $W^2 \bar W^2$ otros t\'erminos
puramente bos\'onicos mas que los expresados en (\ref{termino}),
luego para obtener luego de la integraci\'on en el superespacio
componentes puramente bos\'onicas necesitamos supercampos que
contenga en sus componentes $\theta=\bar\theta=0$ t\'erminos
puramente bos\'onicos, y dichos supercampos son
(\ref{13})-(\ref{14})}.

Mediante la construcci\'on (\ref{15}) es posible supersimetrizar
lagrangianos $\cal L$ de la forma
\be
 {\cal L}=-\frac 1 {4e^2} F_{\mu \nu}F^{\mu \nu}+\sum_{s,t=0}^\infty
 c_{st} \left( (\frac{1}{2}F_{\mu \nu}F^{\mu
 \nu})^2 + (\frac{1}{2}\tilde F_{\mu \nu}F^{\mu \nu})^2 \right)
 \left(F_{\mu \nu}F^{\mu \nu}\right)^s \left(\tilde F_{\mu
 \nu}F^{\mu \nu}\right)^t
 \label{expr}
\ee
Una consecuencia inmediata es que todos los lagrangianos
supersimetrizables se reducen al t\'ermino topol\'ogico
\be
 {\cal L}^{SUSY}|_{F=\pm i \tilde F}
 =\mp\frac i {4e^2} \tilde F_{\mu \nu}F^{\mu \nu}
 \label{topo}
\ee
sobre soluciones autoduales, $F_{\mu\nu}=\pm i\tilde F_{\mu\nu}$,
o sea que saturan una cota de Bogomol'nyi \cite{inst},
\cite{Bogo},\cite{Bre},\cite{HT}. El lagrangiano de BI satisface
esta propiedad ya que en el caso abeliano es posible escribir
\footnote{En la expresi\'on (\ref{det}) se debe entender
$F^2=F_{\mu\nu}F^{\mu\nu}$ y $F\tilde F=F_{\mu\nu}\tilde
F^{\mu\nu}$. La f\'ormula (\ref{det}) es v\'alida en el caso no
abeliano si se emplea la prescripci\'on de traza sim\'etrica (ver
cap. \ref{nstr}).}
\ba
 \sqrt{-\det(g_{\mu\nu}+\frac 1 \beta
 F_{\mu\nu})}\!\!\!&=\!\!\!&\sqrt{(1+\frac 1 {4\beta^2}
 F^2)^2-\frac 1 {16\beta^4}(F^2+ i \tilde F F)
 (F^2- i \tilde F F ) }\nonumber\\
 \!\!\!&=\!\!\!&\sqrt{(1+\frac 1 {4\beta^2} F^2)^2-\frac 1
 {64\beta^4}
 (F+i\tilde F)^2 (F-i\tilde F)^2 }\nonumber \\
 \!\!\!&=\!\!\!&\sqrt{(1\pm\frac i {4\beta^2} F\tilde F)^2+\frac 1
 {4\beta^2} (F\mp i\tilde F)^2 }
 \label{det}
\ea
Los coeficientes $a_{st}^{(\beta)}$ en la ec.(\ref{15})  deben ser
elegidos entonces de manera de obtener co\-mo par\-te
bo\-s\'o\-ni\-ca del lagrangiano ${\cal L}^{SUSY}_{BI}$ el
lagrangiano de BI ec.(\ref{2}).

Concentr\'emonos en los t\'erminos puramente bos\'onicos de ${\cal
L}_{{BI}}^{{SUSY}}$. Esto se logra po\-nien\-do los fermiones a
cero:
\begin{eqnarray}
 \left. {\cal L}_{{BI}}^{{SUSY}}\right| _{%
 {BOS}} \!\!\!& \equiv \!\!\!& {\tilde {\cal L}}_{{BI}}
 \nonumber\\
 \!\!\!& =\!\!\!& \frac{\beta ^2}{e^2}\left(1-\sqrt{1+\frac
 1{2\beta^2}\left( F^{\mu \nu}F_{\mu \nu}-2D^2\right)
 -\frac 1{16\beta ^4}\left(F^{\mu\nu}\tilde F_{\mu \nu}\right)^2}
 \;\right)
 \label{quien}
\end{eqnarray}
Notemos que al reemplazar la ecuaci\'on de movimiento para el
campo $D$ (que en el presente caso es $D=0$), la parte bos\'onica
del lagrangiano supersim\'etrico coincide con el lagrangiano de BI
${\tilde {\cal L}}_{BI}|_{D=0} = {\cal L}_{BI}$.

Los coeficientes $a_{st}^{(\beta)}$ pueden ser calculados
imponiendo la identidad (\ref{quien}). A partir de las componentes
bos\'onicas de $W^2 \bar W^2$ (dadas en el ap\'endice) es posible
encontrar una relaci\'on de recurrencia que conecta los
coeficientes $a$ en el lagrangiano supersim\'{e}trico de BI con los
coeficientes $q$ de la expansi\'on del lagrangiano de BI (ver
ec.(\ref{5})),
\begin{eqnarray}
 & & a_{00}^{(\beta)}=\frac{1}{8e^2\beta^2} \nonumber\\
 & & a_{n - 2m ~2m}^{(\beta)}=\frac{(-1)^m}{e^2\beta^{2n + 2}}
 \sum_{j=0}^m 4^{m-j}{{n + 2 - j} \choose j}\,
 q_{n+1-j}\nonumber\\
 & & a_{n~2m+1}^{(\beta)}=0
 \label{21}
\end{eqnarray}
($a_{00}^{(\beta)}=\frac 1 8 ,~a_{10}^{(\beta)}=-\frac 1
{16},~a_{20}^{(\beta)}=\frac 5 {128},~a_{02}^{(\beta)}=\frac 1
{32},~a_{30}^{(\beta)}=-\frac 7 {256},~a_{12}^{(\beta)}=-\frac 3
{64},...$, donde $e^2=\beta^2=1$).

Podemos a esta altura escribir una f\'ormula compacta en
t\'erminos de supercampos para la extensi\'on supersim\'{e}trica del
lagrangiano de Born-Infeld, usando la siguiente identidad
\ba
 {\cal L}_{BI}\!\!\!&=\!\!\!&\frac {\beta^2}{e^2}\left(1-\sqrt{1+\frac 1
 {2\beta^2} F_{\mu\nu}F^{\mu\nu}-\frac 1
 {16\beta^4}(F_{\mu\nu}\tilde F^{\mu\nu})^2} \right)\nonumber \\
 \!\!\!&=\!\!\!&-\frac 1 {4e^2} F_{\mu\nu}F^{\mu\nu} +
 \frac 1 {4e^2\beta^2} (( F_{\mu\nu}F^{\mu\nu})^2+
 (F_{\mu\nu}\tilde F^{\mu\nu})^2) {\cal R} %
 \label{tata}%
\ea
donde $\cal R$ es la funci\'onal de $F^2$ y $F\tilde F$ definida
como
\be
 {\cal R}=\frac 1 {4}\left(1+\frac 1 {4\beta^2} {F_{\mu\nu}F^{\mu\nu}}
 +\sqrt{1+ \frac 1 {2\beta^2} {F_{\mu\nu}F^{\mu\nu}} -\frac 1
 {16\beta^4}(F_{\mu\nu}\tilde F^{\mu\nu})^2 }\;
 \right)^{-1} \nonumber
\ee
A partir de la ec.(\ref{tata}) encontramos entonces la expresi\'on
compacta en t\'erminos de supercampos\footnote{Entonces, los
coeficientes $a$ son los correspondientes al desarrollo
$1/4(1+X/2+\sqrt{1+X-Y^2})^{-1}=\sum a_{st}X^sY^t$.}
\be
 {\cal L}^{SUSY}_{BI}=\frac 1 {4e^2} \left( \int d^2\theta
 W^2(y,\theta) +~h.c. \right)+\frac 1 {4e^2\beta^2} \int d^2\theta
 d^2\bar\theta \frac {W^2 \bar W^2}{1+\frac 1 {2\beta^2}
 X+\sqrt{1+\frac 1 {\beta^2} X-\frac 1 {\beta^4} Y^2}}
\ee
Esta expresi\'on fue hallada en \cite{CF} y discutida como
realizaci\'on no lineal de una segunda supersimetr\' \i a en
\cite{BG},\cite{RT},\cite{T}(ver tambi\'en \cite{APS}).

Con el conocimiento de los coeficientes $a^{(\beta)}_{st}$, el
lagrangiano supersim\'etrico de Born-Infeld puede ser escrito
expl\' \i citamente en la forma
\be
 {\cal L}_{{BI}}^{{SUSY}} = \tilde {\cal L}_{BI} + {\cal L}_{fer} +
 {\cal L}_{fb}
 \label{22}
\ee
donde ${\cal L}_{fer}$ contiene t\'erminos puramente fermi\'onicos
de autointeraccion mientras que  ${\cal L}_{fb}$ incluye los
t\'erminos cin\'eticos de los fermiones y t\'erminos cruzados
bos\'on-fermi\'on. Ambos t\'erminos pueden ser calculados como
expansiones en potencias crecientes de campos fermi\'onicos y
bos\'onicos. Para la discusi\'on que nos proponemos hacer de las
relaciones de Bogomol'nyi a trav\'es del \'algebra
supersim\'etrica, veremos que ser\'a necesario conocer solo
ciertos t\'erminos cuadr\'aticos en los fermiones (ver secci\'on
[1.5]).

Para ser precisos, solo los t\'erminos cuadr\'aticos de la forma $
\lambda\partial_\mu \bar \lambda$ y $ \bar \lambda \partial_\mu
\lambda$ contribuyen al \'algebra de cargas\footnote{T\'erminos de
orden mayor a dos en fermiones se anulan al poner los fermiones a
cero en el \'algebra de cargas luego de calcular los conmutadores.
Los t\'erminos sin derivadas no contribuyen a las cargas
supersim\'{e}tricas.}. Por lo que escribiremos expl\' \i citamente,
solo tales t\'erminos en ${\cal L}_{fb}$.

Analizando las componentes de los supercampos $W^2\bar W^2$, $X$ e
$Y$ (ver ap\'endice \ref{a2}), vemos que los t\'erminos que nos
interesan en el producto $W^2\bar W^2 X^s Y^t$ son cuatro y
corresponden esquem\'aticamente a las siguientes componentes en
t\'erminos de fermiones
\ba
 \underbrace{W^2\bar W^2} &\cdot&
 \underbrace{X^sY^t}|_{\theta^2
 \bar \theta^2} \nonumber \\ %
 \underline{\theta^2\bar\theta}:\bar\lambda&\longleftrightarrow&
 \underline{\bar\theta}:\partial \lambda \nonumber \\ %
 \underline{\bar\theta^2\theta}:\lambda&\longleftrightarrow&
 \underline{\theta}:\partial \bar \lambda \nonumber \\ %
 \underline{\theta^2\bar\theta^2}:\lambda\partial\bar\lambda&
 \longleftrightarrow& \underline{\theta=\bar\theta=0}:0 \nonumber \\ %
 \underline{\theta^2\bar\theta^2}:0&\longleftrightarrow&
 \underline{\theta=\bar\theta=0}:\lambda\partial\bar\lambda\nonumber
\ea

Denotaremos como ${\cal L}_{fb}^{I}+{\cal L}_{fb}^{II}$ a la suma
de los t\'erminos relevantes (los t\'erminos restantes de orden
mayor o igual en campos fermi\'onicos pueden ser calculados
facilmente), tenemos entonces
\be
 {\cal L}_{fb} = {\cal L}_{fb}^I[ \lambda,  \partial \bar \lambda]
 + {\cal L}_{fb}^{II}[ \bar \lambda, \partial\lambda] +
 {\rm otros~t\acute{e}rminos}
 \label{40}
\ee
\begin{eqnarray}
 & &  {\cal L}_{fb}^I\left[ \lambda ,\partial \bar \lambda \right]
 = -\frac{i}{2e^2} \lambda /\!\!\!\partial \bar \lambda
 -i\sum_{s,t=0}^\infty a_{st}^{(\beta)}\lambda \sigma ^\nu
 \partial_\mu \bar \lambda (X_{{BOS}})^{s-1}
 (Y_{{BOS}})^{t-1} \nonumber \\
 & & \left( -2iX_{{BOS}}Y_{{BOS}} +A^{*}(isY_{{BOS}}+\frac
 t2X_{{BOS}}) \right) \left( A\delta _\nu ^\mu
 +\frac 12\Omega^{*\mu \rho }\Omega _{\rho \nu }\right)
 \label{222}
\end{eqnarray}
\be
 {\cal L}_{fb}^{II}[ \bar \lambda, \partial\lambda]
 = {\cal L}_{fb}^I\left[ \lambda ,\partial \bar \lambda \right]^\dagger
 \label{qqq}
\ee
donde
\ba
 X_{{BOS}} \!\!\!&=\!\!\!&\frac 12F_{\mu \nu }F^{\mu \nu }-D^2 \nonumber\\
 Y_{{BOS}} \!\!\!&=\!\!\!&\frac 14F_{\mu \nu }\tilde F^{\mu \nu }
 \label{xy}
\ea
Las expresiones de  A y $\Omega ^{*\mu \rho }\Omega _{\rho \nu }$,
calculadas en el ap\'endice, est\'an dadas por
\be
 A=D^2-\frac 12F^{\mu \nu }F_{\mu \nu }-\frac i2
 F^{\mu \nu }\tilde F_{\mu \nu}
\ee
\be
 \Omega ^{*\nu \rho }\Omega _{\rho \mu }=\left( D^2+\frac 12
 F_{\alpha \beta}F^{\alpha \beta }\right) \delta _\mu ^\nu
 -2D\eta ^{\nu \rho}\tilde F_{\rho \mu }+2F^{\nu \rho }
 F_{\rho \mu}
\ee

Reescribiendo los t\'erminos fermi\'onicos mediante espinores de
Majorana de cuatro componentes $\Lambda$ (ver ap\'endice
\ref{a1}), tenemos para ${\cal L}_{fb}$
\begin{eqnarray}
 {\cal L}_{fb} \!\!\!&=\!\!\!& -\frac{i}{2e^2} \bar \Lambda /\!\!\!\partial
 \Lambda  + \sum_{s,t=0}^\infty  a_{st}^{(\beta)} X_{{BOS}} ^ {s-1}
 Y_ {{BOS}} ^ {2t-1} \nonumber\\
 \!\!\!& \!\!\!& \left(i\bar \Lambda /\!\!\!\partial \Lambda Y_{{BOS}}
 \left[
 s(X_{{BOS}}^2+4Y_{{BOS}}^2)-X_{{BOS}}
 \left( Z_{{BOS}}- 2X_{{BOS}}\right)
 \right]
 \right.
 \nonumber \\
 \!\!\!& \!\!\!&
 \left.
 +2i\bar \Lambda \Gamma ^\mu \partial ^\nu \Lambda (D\tilde
 F_{\nu\mu }-F_{\nu \rho }F_{\;\mu }^\rho )
 \left [ X_{{BOS}}Y_{{BOS}}+ 2 \left ( 2sY_{{BOS}}^2+tX_{{BOS}}^2
 \right) \right]
 \right.\nonumber \\
 \!\!\!& \!\!\!& +\bar \Lambda \Gamma ^5 /\!\!\!\partial \Lambda
 X_{BOS} \left[ t(X_{BOS}^2+4Y_{BOS}^2)+ 4Y_{{BOS}}^2+
 Y_{{BOS}}Z_{{BOS}}(s-2t)\right]\nonumber \\
 \!\!\!& \!\!\!&\left. -2\bar \Lambda \Gamma ^5\Gamma ^\mu
 \partial^\nu \Lambda (D\tilde F_{\nu \mu }
 -F_{\nu \rho }F_{\;\mu }^\rho )X_{{BOS}}Y_{{BOS}
 }(s-2t)\right) +{\rm otros~t\acute{e}rm.}
 \label{ui}
\end{eqnarray}
donde $X_{BOS}$ y $Y_{BOS}$ fueron definidos en (\ref{xy}) y
\begin{eqnarray}
 Z_{{BOS}} \!\!\!&=\!\!\!&\frac 12F_{\mu \nu }F^{\mu \nu }+D^2
\end{eqnarray}
\noindent Terminamos esta secci\'on se\~nalando que el lagrangiano
supersim\'etrico de Born-Infeld (\ref{22})
\[
 {\cal L}_{{BI}}^{{SUSY}} = {\tilde {\cal L}}_{BI} + {\cal L}_{fer}
 + {\cal L}_{fb}
\]
es invariante bajo las siguientes transformaciones de
supersimetr\' \i a $N=1$
\begin{eqnarray}
 & & \delta_\Upsilon A_\mu  =  - i \bar \Upsilon \Gamma_\mu \Lambda
 \;\;\;\;\; \;\;\;\;\;~~~~~~~~\;\;\;\;~~~  \delta_\Upsilon D  = - \bar \Upsilon
 \Gamma^5 /\!\!\!\partial \Lambda \nonumber\\
 & & \delta_\Upsilon \Lambda = i (-F_{\mu \nu} \Sigma^{\mu
 \nu} +D \Gamma^5 )\Upsilon
 \label{SN1}
\end{eqnarray}
donde $\Upsilon$ es el par\'ametro de la transformaci\'on
supersim\'{e}trica (fermi\'on de Majorana) construido a partir de los
par\'ametros de Weyl $\epsilon$ y $\bar \epsilon$.

\section{El modelo supersim\'etrico de Higgs}

En la secci\'on anterior construimos  la teor\' \i a de
Born-Infeld supersim\'etrica $N=1$ en $d=4$. El lagrangiano
construido determina la din\'amica del campo de gauge. Dado que
estamos buscando relaciones de Bogomol'nyi para la teor\' \i a de
gauge espont\'aneamente rota, consideraremos, adem\'as del
lagrangiano ya obtenido, un lagrangiano supersim\'etrico de Higgs
invariante de gauge al que a\-gre\-ga\-re\-mos el t\'ermino de
Fayet-Iliopoulos para que la simetr\' \i a de gauge este
espont\'aneamente rota (realizada en el modo de Goldstone). Este
lagrangiano puede ser construido a partir del multiplete escalar
quiral $\Phi$ acoplado al supercampo vectorial real $V$. El modelo
es bien conocido y lo describiremos fundamentalmente para
establecer nuestras convenciones.

El bloque de materia necesario en la construcci\'on de la acci\'on
supersim\'etrica de Higgs es el supermultiplete chiral
$\Phi=(\phi,\psi,G)$ cuya expansi\'on es
\be
 \Phi(y,\theta)=\phi+\sqrt 2 \theta \psi+ \theta^2 G
 \label{fi}
\ee
Contiene un campo escalar complejo $\phi$ (campo de Higgs), un
espinor de Weyl $\psi$ (higgsino) y un campo auxiliar complejo
$G$. La acci\'on supersim\'etrica de Higgs invariante de gauge
requiere adem\'as introducir el multiplete vectorial real
$V_{WZ}=(A_\mu,\lambda,D)$ que describimos en la secci\'on
anterior.

La acci\'on de Higgs invariante de gauge con ruptura espont\'anea
de simetr\' \i a se escribe
\be
 {\cal L}^{SUSY}_{Higgs}=\int d^2\theta d^2\bar\theta
 \left(\Phi^\dagger e^{2V}\Phi-2\xi^2\, V\right).
 \label{higgs}
\ee
El primer t\'ermino da lugar a los t\'erminos cin\'eticos
invariantes de gauge para los campos pertenecientes al
supermultiplete $\Phi$, mientras que el segundo (t\'ermino de
Fayet-Iliopoulos) genera la ruptura de simetr\' \i a, cuando se
reemplaza en la acci\'on la ecuaci\'on de movimiento para el campo
auxiliar $D$ ($\xi$ es el par\'ametro dimensional de
Fayet-Iliopoulos). La acci\'on (\ref{higgs}) en t\'erminos de los
campos componentes de los supermultipletes queda, en el gauge de
WZ,
\ba
 {\cal L}^{SUSY}_{Higgs}\!\!\!&=\!\!\!&D_\mu \phi^\dagger D^\mu \phi-\frac
 i 2 \bar \Psi \Dsl^{(5)} \Psi+G^\dagger G+\frac i {\sqrt
 2}\left(\bar \Lambda \Gamma^5 \Psi (\phi+\phi^\dagger)- \bar
 \Lambda \Psi
 (\phi-\phi^\dagger)\right)\nonumber \\
 \!\!\!&\!\!\!&+D\phi^\dagger \phi-\xi^2 D
 \label{higgsc}
\ea
Hemos reagrupado los espinores de Weyl $\psi$ y $\bar \psi$ en un
espinor de Majorana $\Psi$. Las derivadas covariantes son
$D_\mu=\partial_\mu + i A_\mu$ si act\'uan sobre el campo escalar
$\phi$ y $D_\mu^{(5)}=\partial_\mu + i \Gamma^5 A_\mu$ si act\'uan
sobre el fermi\'on de Majorana $\Psi$\footnote{El fermi\'on de
Majorana solo puede aparecer acoplado quiralmente al campo de
gauge debido a que $\bar\Psi\Gamma^\mu\Psi=0$.}. La acci\'on
(\ref{higgsc}) es invariante frente a las siguientes
transformaciones de supersimetr\' \i a (en notaci\'on de Weyl)
\footnote{Las transformaciones (\ref{susyh}) corresponden a la
composici\'on de una transformaci\'on de supersimetr\'{\i}a y una
transformaci\'on de supergauge para permanecer en el gauge de WZ,
esta es la raz\'on por la que aparece la derivada covariante en la
transformaci\'on para el higgsino y el campo auxiliar (ver
discusi\'on en el cap. \ref{susy}).}
\ba
 &&\delta^{(WZ)}_\epsilon\phi=\sqrt 2 \epsilon\psi
 ~~~~~~~~~~~~~~~~~~~~~\;\;\;\;\;\;\;\;
 \delta^{(WZ)}_\epsilon G=i\sqrt 2 \bar\epsilon \bar{\Dsl}\psi+2i
 \bar\epsilon\bar\lambda~\phi \nonumber\\
 && \delta^{(WZ)}_\epsilon\psi=\sqrt 2 G\epsilon+i\sqrt 2
 \Dsl\phi~\bar\epsilon
 \label{susyh}
\ea
donde $\epsilon$ es el par\'ametro (Weyl) de la transformaci\'on
supersim\'{e}trica y $\bar{\Dsl}=\bar\sigma^\mu D_\mu$ con
$D_\mu=\partial_\mu + i A_\mu$.

Recordemos brevemente c\'omo se genera la ruptura espont\'anea de
simetr\' \i a en el modelo supersim\'{e}trico de Maxwell-Higgs. Al
agregar a la acci\'on (\ref{higgs}) la din\'amica de Maxwell
(\ref{12'}), los t\'erminos que contienen al campo auxiliar $D$
son (ver (\ref{maxc}),(\ref{higgsc}))
\be
 {\cal L}_D= \frac  1 {2e^2} D^2 +D(|\phi|^2-\xi^2) %
\ee
La ecuaci\'on de movimiento para $D$ resulta
\be
 \frac 1 {e^2} D + (|\phi|^2-\xi^2)=0 %
 \label{dmax}
\ee
obteniendose, al reinsertar la expresi\'on para $D$ en el
lagrangiano, un potencial efectivo para el campo de Higgs $\phi$
$$
 {\cal L}_{bos}=-\frac 1 {4e^2} F_{\mu\nu}F^{\mu\nu}+D_\mu \phi^\dagger
 D^\mu \phi-V[\phi] \nonumber
$$
\be
 V[\phi]=\frac 1{2e^2}D^2=\frac {e^2} 2 (|\phi|^2-\xi^2)^2
 \label{maxhg}
\ee
Vemos claramente que la invarianza frente a supersimetr\'{\i}a impone
una relaci\'on entre la constante de a\-co\-pla\-mien\-to de gauge
$e$ y el acoplamiento en el potencial de Higgs $\lambda$ que
coincide con la condici\'on de Bogomol'nyi\footnote{Esta misma
relaci\'on se deduce, en un an\'alisis puramente algebr\'aico,
completando cuadrados en la expresi\'on para la energ\' \i a de
las soluciones\cite{Bogo}.}
\be
 \lambda=\frac{e^2} 2
\ee
para la existencia de una cota para la energ\' \i a y ecuaciones
de primer orden cuyas soluciones satisfacen las ecuaciones de
Euler-Lagrange del modelo \cite{dVS},\cite{Bogo},\cite{ed}.

Estudiaremos ahora el caso en el que la din\'amica del campo de
gauge est\'a dada por  la acci\'on de Born-Infeld. Veremos c\'omo
el potencial ad-hoc para el campo de Higgs pro\-pues\-to en
\cite{NS1} para lograr una cota de Bogomol'nyi, est\'a determinado
un\' \i vocamente por la invarianza supersim\'{e}trica del modelo BIH.


\section{El modelo supersim\'{e}trico $N=2$ de Born-Infeld-Higgs en $d=3$}

Procederemos ahora a la reducci\'on dimensional a $d=3$
dimensiones espacio-temporales. Mediante este procedimiento
obtendremos una teor\' \i a supersim\'etrica $N=2$. Es en este
caso en el que las relaciones de Bogomol'nyi aparecen relacionadas
con la supersimetr\' \i a.

La reducci\'on dimensional se realiza de la siguiente forma: los
campos se toman independientes de $x_3$ (
$\partial_3=0$)\footnote{Desde el punto de vista de Kaluza-Klein,
la idea es pensar que $x^3$ es una coordenada compacta. Al
desarrollar los campos en Fourier para $x^3$ solo excitamos el
modo cero.}. Para el campo vectorial $A_{\mu}$ se toma $A_3 = N$
donde $N$ transforma como un escalar bajo $SO(2,1)$. Tenemos
entonces un campo escalar real extra con din\'amica al reducir a
un modelo $3$-dimensional. La reducci\'on dimensional de un
fermi\'on de Majorana en $d=4$ da origen a dos fermiones de
Majorana $d=3$ que se pueden acomodar como un fermi\'on de Dirac
en $d=3$~(ver ap\'endice \ref{a3})\footnote{La aparici\'on de $N$
y los dos fermiones de Majorana de dos componentes es natural
cuando se estudia la descomposici\'on de representaciones de
$SO(3,1) \supset SO(2,1)$. }.

Reduciendo dimensionalmente (\ref{22}), obtenemos la acci\'on de
Born-Infeld $N=2$
\be
 S^{(3)} = S_{bos}^{(3)} + S_{fb}^{(3)} + S_{fer}^{(3)}
 \label{c9}
\ee
Aqu\' \i~
\begin{eqnarray}
 S_{bos}^{(3)}   \!\!\!&=\!\!\!& \frac{\beta^2}{e^2} \int d^3x
 \nonumber\\
 \!\!\!&\!\!\!&\left(1- \sqrt{1 - \frac{1}{\beta^2}D^2   + \frac{1}{2\beta^2}
 F^{ij}F_{ij} -\frac{1}{\beta^2}\partial_iN\partial^iN  -
 \frac{1}{\beta^4}(\tilde F_k\partial^k N)^2}
 \right)\nonumber\\
 \label{c10}
\end{eqnarray}
En lo que respecta a $S_{fb}^{(3)}$, puede ser escrita como
\begin{eqnarray}
 S_{fb}^{(3)} \!\!\!&=\!\!\!&-\frac i {2e^2} \int d^3x\bar \Sigma
 /\!\!\!\partial\Sigma +i\sum_{s,t=0}^\infty {a_{st}^{(\beta)}} \left(
 X^{\left(3\right) } \right) ^{s-1} \left( Y^{ \left( 3\right) }
 \right) ^{2t-1} \nonumber \\
 \!\!\!&\!\!\!& \left\{ \bar \Sigma
 /\!\!\!\partial \Sigma \left[ s \left( \left( X^{\left(3\right)
 }\right) ^2+ \left( Y^{\left( 3\right) }\right) ^2 \right)
 Y^{\left(3\right) }+ \left(2 X^{\left( 3\right)} - Z^ { \left(
 3\right) } \right) X^{ \left( 3\right)} Y^{ \left( 3\right)}
 \right. \right.\nonumber \\
 \!\!\!&\!\!\!& +\left. \frac{1}{2} t X^{ \left( 3
 \right)} \left( \left( X^{\left(3\right) } \right) ^2+ \left(
 Y^{\left( 3\right) } \right)^2 \right)
 +\left( Y^{\left( 3\right) } \right)^2 X^{\left( 3\right) }
 \right] -\bar \Sigma \gamma ^i \partial ^j \Sigma \nonumber\\
 \!\!\!&\!\!\!&\left(X^{ \left( 3\right) } \eta _{ij}+2F_{ik}F_{\;j}^k \right)
 \left[ X^{\left( 3\right) } Y^{\left(3\right) } \right. (
 F_{ik}F_{\;j}^k  +  \frac{s}{2}-t)
 \nonumber\\
 \!\!\!& \!\!\!& \left. + \left( s\left( Y^{\left( 3\right)
 }\right) ^2
 + t\left( X^{\left( 3\right) }\right) ^2
 \right) \right] \nonumber\\
 \!\!\!&\!\!\!& + \left. \bar \Sigma \partial ^j\Sigma D\tilde F_j \left[
 \left( Y^{\left(3\right) }\right) ^2 + 2t\left( X^{\left( 3\right)
 }\right) ^2- \left( 1 + 2s - 4t \right) X^{\left( 3\right) }
 Y^{\left( 3\right) }\right] \right\} +{\cal O}(\Sigma^4)
 \label{ff}
\end{eqnarray}
donde
\begin{eqnarray}
X^{\left( 3\right) } = \frac 12 F_{ij}F^{ij}  - D^2  - \left(
\partial _iN\right)^2 \\ Y^{\left( 3\right) } = \tilde F^i\partial
_iN \\ Z^{\left( 3\right) } = D^2+\frac 12 F_{ij}F^{ij}
\label{ff1}
\end{eqnarray}
y $\Sigma$ es un fermi\'on de Dirac construido a partir de la
reducci\'on dimensional de $\Lambda$
\be
 \Sigma = \lambda_2 + i\lambda_1
 \label{c12}
\ee
Finalmente, $S_{fer}^{(3)}$ es la acci\'on puramente fermi\'onica
obtenida mediante reducci\'on dimensional cuya forma expl\' \i
cita es irrelevante para nuestros prop\'ositos, o sea el c\'alculo
del \'algebra supersim\'etrica.

Procedamos ahora a reducir dimensionalmente la acci\'on de Higgs
(\ref{higgs}). La misma toma la forma \cite{ed}
\begin{eqnarray}
 S_{Higgs}^{(3)} \!\!\!& = \!\!\!& \int d^3x \left(|D_i\phi|^2 -
 \frac{i}{2} \bar \Omega \Dsl \Omega
 + |G|^2 - \frac{1}{\sqrt 2}(\bar\Omega \Sigma \phi +  \bar \Sigma
 \Omega \phi^\dagger) +  \right. \nonumber\\
 \!\!\!& \!\!\!& D(|\phi|^2 -\xi^2) + \left. \frac 1 2 N
 \bar \Omega \Omega-N^2 |\phi|^2\right)~~~.
 \label{c3}
\end{eqnarray}
Aqu\' \i~ $\phi$ es un escalar complejo cargado, $\Omega$ es el
espinor de Dirac construido a partir de la reducci\'on dimensional
de $\Psi$ como
\be
 \Omega=\psi_1+i\psi_2
\ee
$N$ un escalar real y $G$ un campo auxiliar complejo. La derivada
covariante est\'a definida como
\be
 D_i = \partial_i + i A_i.
 \label{c4}
\ee
Usando la trivial ecuaci\'on de movimiento para el campo auxiliar
$G$, la acci\'on (\ref{c3}) se reduce a
\begin{eqnarray}
 S_{Higgs}^{(3)} \!\!\!& = \!\!\!& \int d^3x \left(|D_i\phi|^2 - \frac{i}{2}
 \bar \Omega \Dsl \Omega - \frac{1} {\sqrt 2} (\bar\Omega \Sigma
 \phi +  \bar \Sigma \Omega \phi^\dagger) + \right. \nonumber\\
 \!\!\!&\!\!\!&D(|\phi|^2 - \xi^2) + \left. \frac 1 2 N \bar \Omega \Omega-N^2
 |\phi|^2 \right).
 \label{F}
\end{eqnarray}

La acci\'on supersim\'etrica $N=2$ de Born-Infeld-Higgs en $d=3$
est\'a dada finalmente por
\be
S^{(3)}_{SUSY} = S^{(3)}_{bos}  + S^{(3)}_{fb} + S^{(3)}_{fer} +
S_{Higgs}^{(3)} \label{c6} \ee
donde las diferentes acciones fueron definidas en las
ecs.(\ref{c9})-(\ref{ff}) y (\ref{F}). Las dimensiones de los
par\'ametros y campos en unidades de masa son:
\be
 [\beta] = m^2~~~~~[e] = m^\frac 1 2~~~~~[\xi] = m^\frac 1 2
 \label{mas}
\ee
\ba
 [(A_\mu,N,\Sigma,D)]\!\!\!& = \!\!\!&(m,m,m^\frac32,m^2)\nonumber\\
 \left[(\phi,\Omega,G)\right]\!\!\!& = \!\!\!&(m^\frac 1 2, m,m^\frac 32)
 \label{masas}
\ea
La acci\'on  (\ref{c6}) es invariante bajo las siguientes
transformaciones de supersimetr\' \i a $N=2$
\be
 \begin{array}{lll}
 \delta_\Upsilon \phi =\frac i {\sqrt 2}  \bar \Upsilon \Omega
 &~~\delta_\Upsilon\Omega= i\sqrt 2(i \Dsl\phi + N \phi)\Upsilon
 &~~\delta_\Upsilon N=-\frac i2\bar\Upsilon\Sigma+{\rm h.c.}\nonumber\\
 \delta_\Upsilon A_i=-\frac i2\bar\Upsilon\gamma_i\Sigma+{\rm h.c.}
 &~~\delta_\Upsilon\Sigma =-i(\Delta_{ij}F^{ij}+ D-
 \frac i2 /\!\!\!\partial N)\Upsilon
 &~~\delta_\Upsilon D=-\frac{1}{2}\bar\Upsilon /\!\!\!\partial\Sigma+
 {\rm h.c.}
 \end{array}
 \label{susyy}
\ee
donde hemos reagrupado los 2 par\'ametros independientes en un
fermi\'on de Dirac $\Upsilon=\epsilon_2+i\epsilon_1$.

Recapitulemos el contenido de campos luego de la reducci\'on
dimensional y veamos su origen
$$~~~~~~~~~d=4~~~~~~~~~~~~~~~~d=3$$
$$
 {V}:\left\{  \begin{array}{ccc}
   A_{\mu} & \to & A_i\oplus N \\
   \Lambda~({\rm Majorana}) & \to & \Sigma~({\rm Dirac}) \\
   D & \to & D
 \end{array}  \right.
$$ $$
 {\Phi}:\left\{ \begin{array}{ccc}
   \phi & \to & \phi \\
   \Psi~({\rm Majorana}) & \to & \Omega~({\rm Dirac}) \\
   G & \to & G \\
 \end{array} \right.
$$

La reducci\'on dimensional genera a partir del supercampo real $V$
en $d=4$: dos multipletes $N=1$ en $d=3$ uno espinorial real
$(A_i,\lambda_1)$ y otro escalar real $(N,\lambda_2)$ que resultan
estar acoplados de manera de poder ser acomodados en un multiplete
supersim\'{e}trico $N=2$ en $d=3$. De manera an\'aloga tenemos para el
supercampo quiral $\Phi$ en $d=4$: dos multipletes escalares
reales $(\phi_1,\psi_1)$ y $(\phi_2,\psi_2)$, donde en estos
\'ultimos $\phi_1$ y $\phi_2$ son las componentes real e
imaginaria de $\phi$, y $\psi_1$ y $\psi_2$ los fermiones que
provienen de la reducci\'on dimensional de $\Psi$, ambos
multipletes acoplados de manera que es posible acomodarlos en un
supermultiplete $N=2$ en $d=3$.

La acci\'on (\ref{c6}) contiene a\'un el campo auxiliar $D$.
Partiendo de la acci\'on (\ref{c6}) con todos los fermiones
puestos a cero, tenemos para el campo auxiliar
\be
 {\cal L}_D=-\frac{\beta^2}{e^2} \sqrt{1 - \frac{1}{\beta^2}D^2
 + \frac{1}{2\beta^2}F^{ij}F_{ij} -\frac{1}{\beta^2}
 \partial_iN\partial^iN-\frac 1{\beta^4}(\varepsilon_{ijk}F^{ij}\partial^k N)^2}
 + D(|\phi|^2 -\xi^2)
 \label{ld}
\ee
La ecuaci\'on de movimiento para $D$ resulta,
\be
 D =  -  \frac{e^2(|\phi|^2 - \xi^2)}{\sqrt{1 +
 \frac{e^4}{\beta^2} (|\phi|^2 - \xi^2)^2}}
 \sqrt{1 + \frac{1}{2\beta^2} F^{ij}F_{ij} - \frac{1}{\beta^2}
 \partial_iN \partial^iN- \frac{1}{\beta^4}
 (\varepsilon_{ijk}F^{ij}\partial^k N)^2 }
 \label{lab}
\ee
Reinsertando en la acci\'on (\ref{ld}) y poniendo los fermiones a
cero obtenemos la acci\'on bos\'onica
\ba
 S^{(3)}_{SUSY}|_{fer=0}\!\!\!&=&\!\!\! \int d^3x
 \frac{\beta^2}{e^2}\left(1- \sqrt{1 + \frac{e^4}{\beta^2}
 (|\phi|^2 - \xi^2)^2}\right.\times\nonumber\\
 &&\!\!\!\!\!\!\left.\sqrt{ 1 +\frac{1}{2\beta^2} F^{ij}F_{ij}-
 \frac{1}{\beta^2} \partial_iN\partial^iN - \frac{1}{\beta^4}
 (\varepsilon_{ijk}F^{ij}\partial^k
 N)^2} \right)+|D_i\phi|^2 -N^2 |\phi|^2 \nonumber \\
 \label{ac}
\ea
que en el l\' \i mite $\beta\to \infty$ coincide como era de
esperar con la reducci\'on a $d=3$ del modelo de Maxwell-Higgs
(\ref{maxhg}).

Es interesante destacar que en nuestro tratamiento, el potencial
de ruptura de simetr\' \i a aparece como un factor multiplicativo
dentro de la ra\' \i z cuadrada de BI, como resultado de la
extensi\'on supersim\'etrica de la teor\' \i a bos\'onica
(\ref{1}) acoplada a un campo de Higgs . Esto sig\-ni\-fi\-ca que,
{\it la invarianza supersim\'{e}trica $N = 2$ fuerza esta particular
forma funcional para la acci\'on} (lo mismo sucede si uno
permanece en $d=4$ con una teor\' \i a $N=1$). En  \cite{NS1} esta
forma funcional fue seleccionada del conjunto infinito de
posibilidades de acoplar el campo de Higgs y su potencial de
ruptura de simetr\' \i a a la teor\' \i a de BI, pidiendo que el
modelo tuviera las ecuaciones de Bogomol'nyi usuales \cite{Bogo}.
Luego, la supersimetr\' \i a explica la raz\'on de tal elecci\'on
asociada al modelo de Born-Infeld-Higgs.

\section{Cota  $\grave{a}$ la Bogomol'nyi}

Veamos c\'omo, a partir de la acci\'on (\ref{ac}), es posible
obtener una cota {\it $\grave{a}$ la Bogomol'nyi} para la energ\'
\i a de las soluciones (cf.\cite{Bogo},\cite{NS1}). Partiendo de
la acci\'on bos\'onica (\ref{ac}) consideremos el caso est\'atico
con $A_0=N=0$. En este caso la densidad de energ\' \i a de una
configuraci\'on est\'a dada por
(cf.\cite{Bogo})\footnote{Analizamos el caso est\'atico por lo que
los \' \i ndices que emplearemos $a,b=1,2$ corren solo sobre las
componentes espaciales del plano bidimensional.

$\bullet$ El tensor $\varepsilon_{ab}$ lo definimos completamente
antisim\'etrico con $\varepsilon_{12}=1$.

$\bullet$ Descomponenemos la derivada covariante sobre el campo de
Higgs en sus partes real e imaginaria
\be
 D_a\phi=(\partial_a+iA_a)(\phi_1+i\phi_2)=(\partial_a\phi_1
 -A_a\phi_2)+i(\partial_a\phi_2+A_a\phi_2)
\ee
definiendo
\be
 D_a\phi_b=(\partial_a\phi_b-\varepsilon_{bc}A_a\phi_c)
\ee
donde como es usual los \' \i ndices internos (en el espacio
complejo) se mezclan con los de espacio tiempo. Con estas
definiciones vale la siguiente igualdad
\be
 \varepsilon_{ab}(D_a\phi)^\ast D_b\phi=i\varepsilon_{ab}
 \varepsilon_{cd}D_a\phi_cD_b\phi_d
\ee }
\ba
 {\cal E}\!\!\!&=\!\!\!&  \frac{\beta^2}{e^2}
 \sqrt{1 + \frac{e^4}{\beta^2}(|\phi|^2 - \xi^2)^2}
 \sqrt{ 1 +\frac{1}{2\beta^2} F^{ab}F_{ab}}
 -\frac{\beta^2}{e^2}+ |D_a\phi|^2
 \label{energy} \\
 \!\!\!&=\!\!\!&\frac{\beta^2}{e^2}\sqrt{\left(1 \mp \frac{e^2}{\beta^2}
 B(|\phi|^2 - \xi^2)\right)^2+\frac 1 {\beta^2}
 \left(\phantom{\frac{\beta^2}{e^2}}\!\!\!\!\!\!\! B\pm e^2
 (|\phi|^2 - \xi^2)\right)^2}-\frac{\beta^2}{e^2}\nonumber \\
 \!\!\!&+\!\!\!&\frac 12(\varepsilon_{ab} D_a \phi_c \mp
 \varepsilon_{cd} D_b\phi_d)^2\pm\varepsilon_{ab}\varepsilon_{cd}
 D_a\phi_cD_b\phi_d \label{pupu} \\
 \!\!\!&\geq\!\!\!& \mp B(|\phi|^2 - \xi^2)\pm
 \varepsilon_{ab}\varepsilon_{cd}D_a\phi_cD_b\phi_d\nonumber \\
 \!\!\!&\geq\!\!\!& \pm\left(\frac12 \varepsilon_{ab}F_{ab} (|\phi|^2-\xi^2)+
 \varepsilon_{ab}\varepsilon_{cd}D_a\phi_cD_b\phi_d\right)\nonumber \\
 \!\!\!&\geq\!\!\!&\pm\partial_aS_a \label{end}
\ea
\be
 S_a=\varepsilon_{ab}(\varepsilon_{cd}\phi_cD_b\phi_d-A_b\xi^2)
 \label{curr}
\ee
La igualdad se cumple cuando se satisfacen las ecuaciones de
Bogomol'nyi
\ba
 F_{ab}\!\!\!&=\!\!\!&\pm\varepsilon_{ab} e^2 (|\phi|^2 - \xi^2)
 \label{bogoh0}\\
 \varepsilon_{ab}D_a\phi_c\!\!\!&=\!\!\!&\pm\varepsilon_{cd}D_b\phi_d~~~,
 \label{bogoh}
\ea
que para el presente modelo coinciden con las del modelo de
Maxwell-Higgs cf.\cite{Bogo} (ver sec. \ref{unici}). La ecuaci\'on
(\ref{bogoh}) se suele escribir en forma compacta como
\be
 {\BB D} \phi=0~~~~~~~~o~~~~~~~~\bar{\BB D} \phi=0
 \label{conche}
\ee
definiendo ${\BB D}=D_1+iD_2$ y $\phi=\phi_1+i\phi_2$.

Para hallar la energ\' \i a $E$ de las soluci\'ones integramos
$\cal E$ en el plano. Cuando la configuraci\'on tiene energ\' \i a
finita, necesariamente $D_a\phi_c\to 0$ en el infinito. Luego,
solo contribuye el segundo t\'ermino de (\ref{curr}) quedando
\be
 E^{(vort)}\geq|\int d^2x~ {\rm div}\, \vec S ~|
 =\xi^2|\oint_{S^1_\infty} \!\! \vec A\cdot\vec dx|=\xi^2 \Theta
\ee
donde denotamos por $\Theta$ al flujo magn\'etico que, en las
soluciones solit\'onicas, resulta estar cuantizado
\cite{NO},\cite{tofpol},\cite{Bogo}
\be
 \Theta=2\pi |n|
\ee
La energ\' \i a del v\'ortice  BPS resulta ser independiente de la
constante de acoplamiento $e$\footnote{Expresadando la relaci\'on
(\ref{mvort}) en t\'erminos de la masa del vector de gauge
$M_v^2=2e^2\xi^2$ queda una expresi\'on comparable a la del
monopolo \cite{NO},\cite{tofpol}
\be
 E^{(vort)}_{BPS}=M_v^2 \frac {\pi n} {e^2}
 ~~~~~~~~~~~~~~~~~~~~E^{(mon)}_{BPS}=M_w \frac {4\pi m}{e^2}
 \label{mvort}
\ee
La diferencia a se\~nalar entre $d=2+1$ y $d=3+1$ es que en el
primer caso la constante de acoplamiento es dimensional
$[e]=m^{1/2}$ mientras que en el segundo es adimensional.}
\be
 E_{BPS}^{(vort)}=\xi^2\, 2\pi|n|
\ee

\section {Ecuaciones BPS en el modelo de Born-Infeld-Higgs}

Como vimos en el cap\' \i tulo anterior, al extender las
supersimetr\' \i as del modelo de $N=1$ a $N=2$, y en presencia de
cargas centrales,  es posible tener representaciones masivas
``cortas" del \'algebra\footnote{Estas representaciones masivas
llamadas BPS como se\~nalamos, son las \'unicas que pueden
volverse no-masivas mediante mecanismos
din\'amicos\cite{sw},\cite{kir}.}. Veremos que los estados
bos\'onicos correspondientes a estas representaciones satisfacen
ecuaciones diferenciales de primer orden o ecuaciones de
Bogomol'nyi-Prasad-Sommerfield \cite{LLW}-\cite{ed}. El \'algebra
supersim\'etrica que da origen a la cota de Bogomol'nyi, deducida
algebraicamente en las ecs.(\ref{energy})-(\ref{end}), ser\'a
discutida en la pr\'oxima secci\'on.

Desde el punto de vista supersim\'etrico, las ecuaciones de
Bogomol'nyi (\ref{bogoh0})-(\ref{bogoh}) se derivan del an\'alisis
de las variaciones supersim\'etricas del gaugino y el higgsino.
Considerando las variaciones supersim\'{e}tricas para estos fermiones
ecs.(\ref{susyy}) en el caso est\'atico con $A_0=N=0$ tenemos
\ba
 \delta_\Upsilon\Omega\!\!\!&=
 \!\!\!&-\sqrt 2\gamma^a D_a\phi\Upsilon \nonumber\\
 \delta_\Upsilon\Sigma\!\!\!&=\!\!\!&-i(\Delta_{ab}F^{ab}+ D) \Upsilon
 \label{susy3}
\ea
Eligiendo para las matrices $\gamma$ la siguiente representaci\'on
\be
 \gamma^0=\sigma^3\!=\left(
 \begin{array}{cc}
   1 & 0 \\
   0 & -1\! \
 \end{array}\right)~~~~~~~~~~
 \gamma^1=i\sigma^1\!=\left(
 \begin{array}{cc}
   0 & i \\
   i & 0 \
 \end{array}\!\right)~~~~~~~~~~
 \gamma^2=i\sigma^2\!=\left(
 \begin{array}{cc}
   0  & 1 \\
   -1 & 0 \
 \end{array}\!\!\right)
\ee
obtenemos
\ba
 \delta_\Upsilon\Omega\!\!\!&=\!\!\!&-i\sqrt 2\left( \begin{array}{cc}
   0                & {\bb D}\phi \\
   \bar{\bb D}\phi & 0 \
 \end{array}\right)\Upsilon
 \label{bogohc}\\
 \delta_\Upsilon\Sigma\!\!\!&=\!\!\!&-i\left(\begin{array}{cc}
   D+B & 0 \\
   0 & D-B \
 \end{array}\right) \Upsilon~~~~~.
 \label{susyd}
\ea
donde hemos usado (\ref{conche}).

Descomponiendo el par\'ametro supersim\'{e}trico $N=2$ como $\Upsilon=
\left(
 \begin{array}{c}
  \epsilon_1 \\
  \epsilon_2
 \end{array}
\right)$ vemos que las ecuaciones  BPS para un ``fondo"
solit\'onico (anti-solit\'onico) resultan de imponer la
anulaci\'on de las variaciones supersim\'etricas (\ref{bogohc}) y
(\ref{susyd}) generadas por $\epsilon_1$ ($\epsilon_2$). La otra
mitad de los generadores supersim\'etricos es no nula y se dice
que la mitad de la supersimetr\' \i a est\'a rota. El efecto de
estas cargas rotas actuando sobre el estado bos\'onico BPS es el
de generar los estados fermi\'onicos del multiplete ``corto''
$\frac12$-BPS.

Las ecuaciones BPS para un v\'ortice con flujo magn\'etico
\underline{positivo} y el modo cero fermi\'onico sobre el mismo
(compa\~nero SUSY) se obtiene mediante las condiciones
\begin{eqnarray}
 \delta_{\epsilon_2} \Sigma = 0 & \to &
 \frac{1}{2}\varepsilon_{0ij}F^{ij}=-D
 \label{bo1}\\
 \delta_{\epsilon_2} \Omega = 0 & \to &~ D_1\phi = -i D_2 \phi
 \label{bogo}
\end{eqnarray}
\begin{eqnarray}
 \delta_{\epsilon_1}\Sigma \!\!\!&\ne\!\!\!& 0 \nonumber\\
 \delta_{\epsilon_1}\Omega \!\!\!&\ne\!\!\!& 0
 \label{bogos}
\end{eqnarray}
Usando la f\'ormula expl\' \i cita para $D$ podemos reescribir
(\ref{bo1}) en la forma
\be
 \frac{1}{2}\varepsilon_{0ij}F^{ij}=
 \frac{e^2(|\phi|^2 - \xi^2)}{\sqrt{1 +
 \frac{e^4}{\beta^2} (|\phi|^2 - \xi^2)^2}}
 \sqrt{1 + \frac{1}{2\beta^2} F^{ij}F_{ij} }
 \label{epa}
\ee

De estas ecuaciones, obtenemos una expresi\'on simple para el
campo magn\'etico que de hecho coincide con la ecuaci\'on de
Bogomol'nyi usual obtenida cuando la din\'amica esta determinada
por la acci\'on de Maxwell (ver ec.(\ref{dmax}))
\be
 B \equiv -\frac 12\varepsilon_{0ij}F^{ij}=e^2( \xi^2-|\phi|^2)
 \label{B}
\ee
Esto se sigue de la expresi\'on expl\' \i cita dada por
(\ref{lab}); elevando al cuadrado y separando potencias en
$1/\beta^2$ queda
\be
 D^2-\left(e^2(|\phi|^2 - \xi^2)\right)^2+
 \frac{\left(e^2(|\phi|^2 - \xi^2)\right)^2}{\beta^2}(D^2-\frac
 12F^2)=0
\ee
Luego, $D=-e^2(|\phi|^2 - \xi^2)$ v\'alida a orden cero en
$1/\beta^2$, tambi\'en es soluci\'on a todo orden ya que de
(\ref{lab}) se sigue que $D^2=\frac 12 F^2$.

Las ecuaci\'ones (\ref{B}) y (\ref{bogo}) son las ecuaciones de
Bogomol'nyi para el sistema de Born-Infeld-Higgs. Coinciden con
las que aparecen en el sistema de Maxwell-Higgs (las ecuaciones
originales de Bogomol'nyi \cite{Bogo}-\cite{dVS}) y tienen
entonces las mismas soluciones exactas encontradas originalmente
en \cite{dVS}.

El hecho de que las ecuaciones de Bogomol'nyi coincidan con las
del modelo de Maxwell-Higgs no es casualidad y ser\'a analizado en
la \'ultima secci\'on. En la proxima secci\'on obtendremos la cota
en la energ\' \i a desde la perspectiva supersim\'{e}trica analizando
el \'algebra supersim\'etrica.

\section{Algebra supersim\'etrica $N=2$ y cota de Bogomol'nyi}

A partir del  modelo $3$-dimensional definido por la acci\'on
(\ref{c9}), se puede f\'acilmente construir la  corriente
conservada asociada a la supersimetr\' \i a, mediante el
procedimiento de Noether, y, a partir de ella, los conmutadores de
las cargas. Las correspondientes supercargas $\bar Q$ y $Q$ pueden
ser escritas de la siguiente forma
\be
 \bar Q \Upsilon \equiv \int d^2x \,( \frac{\partial
 \cal L}{\partial(\partial_0\Sigma)} \delta_\Upsilon\Sigma +
 \frac{\partial\cal L}{\partial(\partial_0\Omega)}\delta_\Upsilon\Omega)
 \label{def}
\ee
\be
 Q \equiv \gamma^0  \bar Q^\dagger
 \label{defi}
\ee
Luego de un poco de \'algebra se obtiene
\begin{eqnarray}
 \bar Q \!\!\! & = & \!\!\!-\frac{1}{2} \int d^2x \,\Sigma^\dagger
 \left( \frac 1{e^2} + 2\sum_{s=0}^\infty a_{s0}^{(\beta)}
 (B^2 - D^2)^s ((2s+ 3)D^2 - \right.\nonumber\\
 &&\left.\vphantom{\sum_0^\infty}B^2  - 2(s+1) \gamma^0 BD )
 \right)(\gamma^0 B+D)+\frac{i}{\sqrt 2}
 \int d^2x\, \Omega^\dagger {\Dsl \phi}
 \label{qra}
\end{eqnarray}
\begin{eqnarray}
 Q\!\!\! & = & \!\!\!-\frac 12 \int d^2x\, (B + \gamma^0 D) \left( \frac 1
 {e^2}+ 2\sum_{s=0}^\infty a_{s0}^{(\beta)}(B^2 - D^2)^s
 \right.\nonumber \\
 & &\left. \vphantom{\sum_0^\infty}((2s + 3)D^2 -
 B^2  - 2(s+1) \gamma^0 BD )
 \right) \Sigma -
 \frac{i}{\sqrt 2} \int d^2x\,
 \gamma^0({\Dsl \phi})^\dagger
 \Omega
 \label{qraf}
\end{eqnarray}
Consideramos aqu\' \i~ v\'ortices de Nielsen-Olesen, tomando
$A_0=N=0$ luego de la diferenciaci\'on, y nos restringimos al caso
est\'atico. M\'as importante, solo hemos incluido t\'erminos
lineales en los campos fermionicos, pues estamos interesados en
derivar, a partir del \'algebra supersim\'{e}trica, los t\'erminos
puramente bos\'onicos que dan origen a las ecuaciones de
Bogomol'nyi. T\'erminos no lineales en fermiones necesariamente
dan contribuciones fermi\'onicas que se anulan al poner los
fermiones a cero una vez calculada el \'algebra.

Del \'algebra de cargas supersim\'etricas (\ref{techito}) para
nuestro modelo podemos obtener ex\-pl\' \i \-ci\-ta\-men\-te la
cota de Bogomol'nyi en t\'erminos de la energ\' \i a y  la carga
central. Dado que la expansi\'on del lagrangiano de BI en
potencias de $1/\beta^2$ da origen a primer orden a la teor\' \i a
de Maxwell ser\'a instructivo mostrar c\'omo  obtenemos a este
orden las correspondientes contribuciones al \'algebra
supersim\'{e}trica y presentar luego los argumentos que conducen al
resultado completo. Tenemos entonces para las cargas
supersim\'{e}tricas a orden cero en $1/\beta^2$, que denotamos como
$\bar Q^{(0)}$ y $ Q^{(0)}$,
\be
 \bar Q^{(0)} =- \frac{1}{2e^2} \int d^2x\, \Sigma^\dagger (\gamma^0
 B + D)+ \frac{i}{\sqrt 2} \int d^2x\, \Omega^\dagger\! \Dsl \phi
 \label{mm}
\ee
Con esta expresi\'on y  $Q^{(0)}$ que puede ser calculada de la
ec.(\ref{defi}), podemos computar el \'algebra supersim\'{e}trica que
toma la forma (ver ap\'endice \ref{a3})
\be
 \{ Q^{(0)},\bar Q^{(0)}\} = -\left( /\!\!\!\!P^{(0)} +
 Z^{(0)}\right)
 \label{z}
\ee
con $P_i$ el $3$-momento y $Z$ la carga central. Usando las
expresiones expl\' \i citas para $\bar Q^{(0)}$ y $ Q^{(0)} $
obtenidas arriba podemos calcular el anticonmutador de las mismas
correspondiente al lado izquierdo en (\ref{z}). Comparando con el
lado derecho en la \'ultima ecuaci\'on podemos
identificar\footnote{Las reglas de anticonmutaci\'on para los
campos fermi\'onicos que se derivan del formalismo can\'onico son
\ba
 \left\{ \Sigma^\dagger_\alpha(\vec x),\Sigma_\beta(\vec y)\right\}
 \!\!\!&=\!\!\!&-2e^2
 \delta_{\alpha\beta}\delta^{(2)}(\vec x-\vec y)\nonumber\\
 \left\{ \Omega^\dagger_\alpha(\vec x),\Omega_\beta(\vec y)\right\}
 \!\!\!&=\!\!\!&-2
 \delta_{\alpha\beta}\delta^{(2)}(\vec x-\vec y)
\ea}
\ba
 P^{(0)}_0\!\!\!&=\!\!\!&M^{(0)}\nonumber\\
 \!\!\!&=\!\!\!&-\frac 12{\rm tr}
 (\gamma^0 \{ Q^{(0)}, \bar Q^{(0)} \}) \nonumber \\
 \!\!\!&=\!\!\!& \int d^2x \left(\frac{1}{2e^2} B^2 + |D_a\phi|^2
 +\frac{1}{2e^2} D^{2}\right)\nonumber \\
 \!\!\!&=\!\!\!&\int d^2x \left(\frac{1}{2e^2} B^2 + |D_a\phi|^2
 +\frac{e^2}{2}(|\phi|^2 - \xi^2)^2\right)
 \label{qqs}\\
 &&\nonumber\\
 Z^{(0)}\!\!\!&=\!\!\!&-\frac 12 {\rm tr} (\{ Q^{(0)},\bar
 Q^{(0)} \}) \nonumber \\
 \!\!\!&=\!\!\!&\int d^2x \left( \frac{1}{e^2}DB -i \varepsilon_{ab}
 (D_a\phi)^\ast D_b\phi\right)\nonumber \\
 \!\!\!&=\!\!\!&\int d^2x \left( -B(|\phi|^2 - \xi^2) -i \varepsilon_{ab}
 (D_a\phi)^\ast D_b\phi\right)= -\xi^2 \oint_{S^1_\infty}\!
 \vec A \cdot \vec dx
 \label{uu}
\ea
Nos restringimos al caso est\'atico, $a,b=1,2$ toman valores sobre
\' \i ndices espaciales, y hemos usado la soluci\'on de la
ecuaci\'on de movimiento para $D$ (\ref{lab}), a orden cero en
$1/\beta^2$ (\ref{dmax}). El resultado (\ref{uu}) muestra que la
carga central del \'algebra supersim\'{e}trica, a orden Maxwell, es
igual a la carga topol\'ogica de la soluci\'on. De hecho $Z^{(0)}$
es igual al flujo del vector de Bogomol'nyi (\ref{curr}).

Consideremos ahora el orden siguiente en la expansi\'on en
$1/\beta^2$. En este caso, en lugar de (\ref{mm}) tenemos
\be
 \bar Q^{(1)} = \bar Q^{(0)} - \frac{1}{4e^2\beta^2}
 \int d^2x\, \Sigma^\dagger \left( \frac{1}{2}(3D^2 - B^2)
 -\gamma^0 B D\right)(\gamma^0 B + D)
 \label{eee}
\ee
y la soluci\'on (\ref{lab}) para el campo auxiliar a primer orden
en $1/\beta^2$ es
\be
 D^{(1)}=-e^2(|\phi|^2-\xi^2)\left( 1+\frac 1 {2\beta^2}
 \left( B^2-e^4(|\phi|^2-\xi^2)^2\right) \right)
\ee
Del anticonmutador de las cargas supersim\'{e}tricas en este caso
resulta
\begin{eqnarray}
 P^{(1)}_0\!\!\!&=\!\!\!&
 \int d^2x \left(\frac{1}{2e^2} B^2 +|D_a\phi|^2 +
 \frac{1}{2e^2}D^{2} +
 \frac{1}{4\beta^2e^2}(D^2-B^2)(B^2+3D^2)
 \right)\nonumber\\
 \!\!\!&=\!\!\!&\int d^2x \left(\frac{1}{2e^2} B^2 + |D_a\phi|^2
 +\frac{e^2}{2}(|\phi|^2 - \xi^2)^2\right.\nonumber\\
 \!\!\!&\!\!\!&\left. -\frac 1 {4e^2\beta^2}
 \left( B^2-e^4(|\phi|^2-\xi^2)^2\right)
 \left( B^2+e^4(|\phi|^2-\xi^2)^2\right) \right)
 \label{qqse}\\
 & & \nonumber \\
 Z^{(1)}\!\!\!&=\!\!\!&\int d^2x \left( \frac{1}{e^2}DB -i \varepsilon_{ab}
 (D_a\phi)^\ast D_b\phi+\frac 1 {e^2\beta^2}DB(D^2-B^2)
 \right)\nonumber\\
 \!\!\!&=\!\!\!&\int d^2x \left( -B(\phi^2 - \xi^2) -i \varepsilon_{ab}
 (D_a\phi)^\ast D_b\phi\right)\nonumber \\
 \!\!\!&\!\!\!&+\frac 1 {2\beta^2}B(|\phi|^2-\xi^2)
 \left( B^2-e^4(|\phi|^2-\xi^2)^2\right)
\ea
Es posible resumar la serie en la expresi\'on de la carga
supersim\'{e}tricas ec.(\ref{qra}) obteni\'endose como resultado
\be
 \bar Q = -\frac{1}{2e^2}\int d^2x \,\Sigma^\dagger (1 + f +
 \gamma^0g) (\gamma^0 B + D) + \frac{i}{\sqrt 2}\int d^2x\,
 \Omega^\dagger\!\Dsl\phi
 \label{silvaesmuycatolico}
\ee
con
\be
f = - 1 + \frac{B^2+D^2}{B^2-D^2} M - \frac{2BD}{(B^2-D^2)^2} N
\label{fs} \ee
\be
 g = \frac{B^2+D^2}{B^2-D^2}\left(
 N - \frac{2BD}{(B^2-D^2)^2} M
 \right)
 \label{gs}
\ee
donde
\be
 M = \frac 2 {\beta^2}\left(\frac{\beta^2 + B^2}{R} - \beta^2 \right)
 \label{ms}
\ee
\be
 N = \frac{2BD}{R}
 \label{ns}
\ee
\be
 R = \sqrt{1 + \frac{1}{\beta^2}(B^2 + D^2)}
 \label{rs}
\ee
Con esto ya es posible calcular el \'algebra supersim\'{e}trica
\be
 \{ Q,\bar Q\} =-\left( /\!\!\!\!P + Z\right)
 \label{zzz}
\ee
e identificar la energ\' \i  a y la carga central en t\'erminos de
$f$ y $g$,
\be
 E = \frac{1}{2e^2} \int d^2x \left( (B^2 + D^2)(f +1) +  2g BD
 \right) + \int d^2x |D_a\phi|^2
 \label{ene}
\ee
\be
 Z = \frac{1}{2e^2} \int d^2x \left( (B^2 + D^2) g +  2(f+1) BD
 \right) -i \int d^2x \varepsilon_{ab} (D_a\phi)^\ast D_b\phi
 \label{zene}
\ee
Ahora, usando la ecuaci\'on de movimiento para el campo auxiliar
$D$ (ec.(\ref{lab})) podemos ver que el lado derecho en las
ecs.(\ref{ene})-(\ref{zene}) toma la forma (\ref{energy})
\be
 E= \frac {\beta^2}{e^2} \sqrt{ 1 + \frac{e^4}{\beta^2}
 (|\phi|^2 -\xi^2)^2} \sqrt{1 + \frac{1}{\beta^2}B^2}
 -\frac{\beta^2}{e^2} + |D_a\phi|^2
 \label{w}
\ee
\be
 Z = \int d^2x \left(- B(|\phi|^2 - \xi^2)-i
 \varepsilon_{ab}(D_a\phi)^\ast D_b\phi \right)
 =-\xi^2 \oint_{S^1_\infty} \vec A \cdot \vec dx =- \xi^2 2 \pi n
 \label{u}
\ee

Vemos que, finalmente, $Z$ est\'a dada por la misma expresi\'on
que en el caso de Maxwell. De hecho, esta coincidencia no es
accidental y puede entenderse de la siguiente manera. Al obtener
la cota en la energ\' \i a $\grave a$ la Bogomol'nyi, mirando la
teor\' \i a puramente bos\'onica  de Born-Infeld-Higgs, escribimos
la energ\' \i a como una suma de cuadrados (\ref{pupu}). Al
efectuar esta manipulaci\'on se obtiene en adici\'on a los
t\'erminos cuadr\'aticos un t\'ermino de superficie. Este
t\'ermino de superficie es el responsable de la aparici\'on de la
carga topol\'ogica como cota de la energ\' \i a. Ahora bien, este
t\'ermino de superficie no es modificado por el hecho de trabajar
con el lagrangiano de BI o de Maxwell. Es m\'as, las ecuaciones de
Bogomol'nyi, que provienen de pedir que valga la igualdad en
(\ref{end}) coinciden en ambas teor\' \i as (veremos en la
\'ultima secci\'on que esto no es casualidad). Visto desde el
punto de vista supersim\'etrico, la cota en la energ\' \i a se
debe a la presencia de la carga central, la cual no depende de la
forma del t\'ermino cin\'etico para el campo de gauge. Hagamos
hincapi\'e en que los resultados obtenidos en las
ecs.(\ref{silvaesmuycatolico})-(\ref{u}) corresponden al modelo
supersim\'etrico de Born-Infeld exacto donde no hay aproximaciones
en potencias de $1/\beta^2$.

La cota para la energ\' \i a en el modelo supersim\'{e}trico de BIH
resulta de una propiedad general de las representaciones
supersim\'{e}tricas con $N\geq 2$. Al estudiar las representaciones
del \'algebra supersim\'{e}trica $N=2$ para el caso masivo (ver cap\'
\i tulo anterior y \cite{susy}) notamos que en presencia de cargas
centarles las representaciones satisfacen
\be
 M\geq|Z|
 \label{bps}
\ee
Esta es la interpretaci\'on supersim\'{e}trica de la cota de
Bogomol'nyi-Prasad-Sommerfield. En particular, la igualdad es
v\'alida cuando existe un estado f\' \i sico que es aniquilado por
alguna de las cargas supersim\'{e}tricas $Q|{BPS}\rangle=0$, a este
estado particular se lo llama estado BPS\footnote{Lo que sucede en
un backgroud BPS es que el mismo forma un multiplete
supersim\'etrico con menos estados que los que uno esperar\' \i a
para un multiplete masivo. En particular un multiplete $1/2$-BPS
tiene el mismo n\'umero de estados que un multiplete no masivo y
es el candidato natural para representar bosones de gauge que
sufren ruptura espont\'anea de sim\'etria\cite{OW}.}. Esta es la
raz\'on de haber impuesto la anulaci\'on de las variaciones
supersim\'{e}tricas de los fermiones de la teor\' \i a (ver ec.
(\ref{def}),(\ref{bo1})-(\ref{bogo}))\footnote{ La propiedad
(\ref{bps}) para el estado $\psi_{BPS}$ es exacta para la teor\'
\i a que consideremos, mientras la supersimetr\'{\i}a no est\'e rota,
no solo cl\'asica sino cu\'anticamente, ya que fue derivada
utilizando teor\' \i a de representaciones y \'algebra. Este es el
fundamento del estudio de dualidades en teor\' \i as
supersim\'{e}tricas de campos y de cuerdas.}. La cota Bogomol'nyi en
el presente caso coincide, como se se\~nal\'o anteriormente, con
la de la teor\' \i a de Maxwell-Higgs,
\be
 P^0 = E \ge |Z|
 \label{bm}
\ee
o
\be
 E \ge \xi^2 2\pi |n|
 \label{etu}
\ee
donde $n$ es  el n\'umero de lineas de flujo de campo magn\'etico
medidas por la carga central $Z$ que coincide con la carga
topol\'ogica \cite{ed}.

\section{Unicidad de las ecuaciones de Bogomol'nyi \label{unici}}

En esta secci\'on discutiremos las ecuaciones de Bogomol'nyi en
teor\' \i as de gauge generales (que dependen de los dos unicos
invariantes fundamentales $ F^{\mu \nu} F_{\mu \nu}$ y $ \tilde
F^{\mu \nu} F_{\mu \nu}$) acopladas a campos escalares de Higgs de
forma m\' \i nima \cite{CS}. Analizando su extensi\'on
supersim\'etrica, mostraremos expl\' \i citamente porqu\'e la
estructura BPS resultante es insensible a la forma particular del
lagrangiano de gauge. El lagrangiano de Maxwell, el de Born-Infeld
o lagrangianos no polin\'omicos mas complicados, Dan todos lugar a
las mismas ecuaciones y cotas de Bogomol'nyi, las cuales est\'an
determinadas por el \'algebra supersim\'etrica subyacente.

Con el prop\'osito de poder hallar una cota \`a la Bogomol'nyi
para la teor\' \i a de Born-Infeld-Higgs los campos escalares de
Higgs fueron acoplados en \cite{NS1}-\cite{NS2} de manera de
reproducir las relaciones BPS ordinarias (p.ej. Maxwell o
Yang-Mills). En las secci\'ones anteriores mostramos que el
lagrangiano sugerido en \cite{NS1} asi como las ecuaciones BPS
propuestas resultaron univocamente determinadas en el punto de
vista supersim\'etrico. Como consecuencia de estos resultados, es
v\'alido preguntarse por la sensibilidad de las relaciones BPS al
tipo de din\'amica para el campo de gauge. El prop\'osito de la
presente secci\'on es responder a tal pregunta.

Una respuesta r\'apida surge de observar que, en el contexto
supersim\'{e}trico, como se se\~nal\'o anteriormente, las ecuaciones
de Bogomol'nyi se obtienen imponinedo que se anulen las
variaciones supersim\'{e}tricas del gaugino y del higgsino, y como
estas variaciones son las mismas independientemenete del tipo de
lagrangiano que se est\'e analizando es de esperar que las
relaciones resultantes no dependan de la forma del lagrangiano.
Sin embargo, la din\'amica asociada con el lagrangiano interviene
a trav\'es de la ecuaci\'on de movimiento para el campo auxiliar
$D$ (perteneciente al supermultiplete vectorial) el cual aparece
precisamente en la transformaci\'on supersim\'etrica del gaugino.
Es entonces a trav\'es del campo $D$ que la forma del lagrangiano
puede, en principio, alterar la forma general de las relaciones
BPS.

Lo que mostraremos en esta secci\'on es que supersimetr\'{\i}a junto
con la ecuaci\'on de movimiento (algebr\'aica) para el campo
auxuliar $D$ hace que las realciones BPS permanezcan inalteradas
independientemente de la elecci\'on del lagrangiano.

Presentaremos el an\'alisis considerando un modelo abeliano
invariante de gauge general en $d=3$. Nuestros argumentos deber\'
\i an ser v\'alidos, sin embargo, para otros modelos como por
ejemplo generalizaciones del modelo de Georgi-Glashow $SO(3)$ y
tambi\'en en otras dimensiones de espacio-tiempo.

La construcci\'on del lagrangiano general invariante de gauge
acoplado a un campo de Higgs la realizaremos nuevamente en $d=4$.
Luego, reduciendo a $d=3$, obtendremos el modelo supersim\'{e}trico
$N=2$.

Como se\~nalamos en secciones anteriores, el supercampo b\'asico
para la construcci\'on de una acci\'on invariante de gauge
supersim\'etrica en $d=4$ es el supercampo (quiral) de curvatura
$W_\alpha$ definido a partir del supercampo real $V$ (\ref{9}).
Las extensiones supersim\'{e}tricas de las teor\' \i as de Maxwell y
Yang-Mills se construyen precisamente a partir de $W_\alpha$
considerando $W^2$ y su conjugado herm\' \i tico $\bar W^2$
(\ref{12'}). Ahora bien, dado que deseamos construir lagrangianos
\underline{mas generales} vimos que  es necesario considerar los
supercampos $X$ e $Y$ definidos en
(\ref{13})-(\ref{14})\footnote{Mediante los supercampos $W,X,Y$
podemos construir los lagrangianos generales supersimetrizables
que dependen de $F^2$ y $F\tilde F$ pero no de sus derivadas.}.
Recordemos las componentes puramente bos\'onicas de los mismos
\begin{eqnarray}
 X |^{(BOS)}_{0} \!\!\!&=\!\!\!& - (
 D^2 - \frac{1}{2} F^{\mu \nu} F_{\mu \nu}) \nonumber\\
 X |^{(BOS)}_{\theta \bar \theta}
 \!\!\!&=\!\!\!& \frac 12 \theta \sigma^\rho \bar \theta \,
 \partial_\rho  (\tilde F^{\mu \nu} F_{\mu \nu}) \nonumber\\
 X |^{(BOS)}_{\theta^2 \bar \theta^2} \!\!\!&=\!\!\!& \frac 14 \theta^2\bar\theta^2
 \,\Box \, (D^2 - \frac{1}{2} F^{\mu \nu} F_{\mu\nu})
 \label{3pp}
\end{eqnarray}
y
\ba
 Y|_{0}^{(BOS)}\!\!\!&=\!\!\!& \frac 14\tilde F^{\mu \nu}
 F _{\mu \nu}\nonumber\\
 Y|_{\theta\bar \theta}^{(BOS)}
 \!\!\!&=\!\!\!& \frac 12 \theta \sigma^\rho \bar \theta\, \partial_\rho
 (D^2 - \frac{1}{2} F^{\mu \nu} F_{\mu\nu}) \nonumber\\
 Y|_{\theta^2 \bar \theta^2}^{(BOS)} \!\!\!&=\!\!\!&
 -\frac 1 {16} \theta^2 \bar \theta^2
 \, \Box \, (\tilde F^{\mu \nu} F _{\mu \nu})
 \label{4pp}
\ea
El tercer supercampo necesario en la construcci\'on del
lagrangiano invariante de gauge mas general, es la combinaci\'on
$W^2 \bar W^2$ cuya \'unica componente puramente bos\'onica se
encuentra en la componente mas alta del supercampo
\be
 W^2 \bar W^2\vert_{\theta^2 \bar \theta^2}^{(BOS)} = \theta^2 \bar
 \theta^2 \left( (D^2 - \frac{1}{2} F_{\mu \nu}F^{\mu \nu})^2 +
 (\frac{1}{2}\tilde F_{\mu \nu}F^{\mu \nu})^2 \right)
 \label{5pp}
\ee
Remarquemos entonces, que toda la dependencia de
(\ref{3pp})-(\ref{5pp}) en la curvatura $F_{\mu \nu}$ y el campo
auxiliar $D$ es en la combinaci\'on
\be
 {\cal F} =\frac 1 {\beta^2} (D^2 - \frac{1}{2} F^{\mu \nu}
 F_{\mu \nu})
 \label{6pp}
\ee
y este hecho tendr\'a consecuencias importantes en nuestra
discusi\'on. Aqu\' \i , para poder definir la variable
adimensional $\cal F$ hemos introducido el par\'ametro dimensional
$\beta$ con las mismas dimensiones que $F_{\mu\nu}$ (ver
ec.(\ref{mas})). Su interpretaci\'on es,  como vimos en el cap\'
\i tulo \ref{herra}, el m\'aximo valor de  campo que admite la
teor\' \i a de Born-Infeld \cite{B}-\cite{BI}.

Estamos entonces en condiciones de escribir el lagrangiano
supersim\'{e}trico $N=1$ e invariante de gauge  mas general en
t\'erminos de $X$, $Y$ y $W^2\bar W^2$ (ver ec.(\ref{15}))
\begin{equation}
 {\cal L}^{SUSY}_{d=4} =
 \frac{1}{4e^2} \int\left( W^2 d^2\theta  + \bar W^2  d^2\bar
 \theta \,\right) + \frac{1}{e^2} \sum_{s,t=0}^\infty b_{st}^{(\beta)}\int
 d^4\theta ~ W^2\bar W^2~ X^s Y^t
 \label{7}
\end{equation}
donde hemos factorizado la constante de acoplamiento $e$.

El primer t\'ermino del lado de derecho de la ec.(\ref{7}) depende
tambi\'en de $F_{\mu \nu}$ y $D$ a trav\'es de la combinaci\'on
(\ref{6pp}), propiedad que ya observamos al construir la
extensi\'on supersim\'{e}trica de la teor\' \i a de Maxwell (ver
ecs.(\ref{12'})-(\ref{maxc}))\footnote{Por completitud recordemos
que la \'ultima componente de $W^2$ ($\bar W^2$) contiene el
t\'ermino $D^2 - \frac{1}{2} F^{\mu \nu} F_{\mu \nu} + i F_{\mu
\nu}\tilde F^{\mu \nu}$ ($D^2 - \frac{1}{2} F^{\mu \nu} F_{\mu
\nu} - i F_{\mu \nu} \tilde F^{\mu \nu}$) de manera que la suma de
integrales en $\theta$ y $\bar \theta$ da como resultado la
conocida extensi\'on supersim\'{e}trica de la teor\' \i a de
Maxwell.}. El segundo t\'ermino genera una teor\' \i a no
polin\'omica. En el caso abeliano que estamos analizando, el
lagrangiano bos\'onico invariante de gauge mas general que se
puede construir para el campo de gauge $A_\mu$ es una funci\'on
arbitraria $f(A,B)$ donde $A=F_{\mu \nu} F^{\mu \nu}$ y $B=F_{\mu
\nu}\tilde F^{\mu \nu}$ (ver ap\'endice \ref{a1}). Como fue
discutido en las secciones anteriores y en \cite{DP}, la
expresi\'on supersim\'etrica (\ref{7}) permite construir solo
extensiones supersim\'etricas de lagrangianos bos\'onicos que
puedan ser expresados en serie de potencias como
\be
 \sum_{s,t=0}^\infty b_{st}^{(\beta)} \left(A^2+B^2 \right) A^s B^t%
\ee
Por supuesto, como vimos en la ec.(\ref{tata}), la teor\' \i a de
Born-Infeld puede ser escrita con un lagrangiano de esta forma.

Del lagrangiano (\ref{7}), para una adecuada elecci\'on de los
coeficientes $b^{(\beta)}_{st}$ (ecs.(\ref{21})) y por reducci\'on
dimensional, obtuvimos la teor\' \i a supersim\'{e}trica $N=2$ de
Born-Infeld en $d=3$ (ec.(\ref{c9})). El procedimiento est\'andar
de reducci\'on dimensional (en la coordenada espacial $x_3$)
implica identificar $A_3$ con un campo escalar $N$. Al buscar
soluciones auto-duales asociadas con v\'ortices magn\'eticos
est\'aticos, el campo $A_0$ (asi como el campo $N$) pueden ser
puestos a cero, entonces podemos identificar la intensidad de
campo $F$ con el campo magn\'etico  $B$ como
\be
 \frac{1}{2} F_{ij} F^{ij} = B^2~~~~~~i,j=0,1,2
 \label{8pp}
\ee
donde
\be
 B =- \frac{1}{2} \varepsilon_{ab}F_{ab} ~~~ ~ ~ ~  a,b=1,2%
 \label{9pp}
\ee
Una vez realizada la reducci\'on dimensional, obtenemos la
versi\'on $d=3$ del lagrangiano supersim\'{e}trico dado en la
ec.(\ref{7}). Como ya se\~nalamos anteriormente, mediante este
procedimiento la supersimetr\' \i a se extiende de $N=1$ a $N=2$.

De lo discutido resulta que los t\'erminos dependientes solo del
campo de gauge en la parte bos\'onica del lagrangiano
supersim\'{e}trico $N=2$ pueden ser escritos en forma compacta como
\be
 {\cal L}^{(BOS)}_{d=3} = \frac{\beta^2}{e^2}
 \sum_{n=1}^{\infty} c_n {\cal F}^n
 \label{10pp}
\ee
donde los coeficientes $c_n$ son adimensionales y $\cal F$
(definido en (\ref{6pp})) toma ahora la forma
\be
 {\cal F} = \frac{D^2 - B^2}{\beta^2}
 \label{11pp}
\ee
Notemos que la elecci\'on $c_1 = \frac 12$ y $c_n = 0$ para $n \ne
1$ corresponde al lagrangiano de Maxwell mientras que la
elecci\'on $c_1 = \frac 12$, $c_2 = \frac 18$, $c_3 =-\frac
1{16}$, $\ldots~$, da el lagrangiano de Born-Infeld para el campo
de gauge.

En lo que concierne al sector del campo de Higgs, en $d=4$, el
acoplamiento entre el campo escalar de Higgs $\phi$ y el campo de
gauge $A_\mu$, y la realizaci\'on de la simetr\' \i a en el modo
de goldstone (simetr\' \i a espont\'aneamente rota), surge del
acoplamiento de los supercampos dado por la ec.(\ref{higgs}). La
reducci\'on dimensional no altera los t\'erminos que contienen al
campo auxiliar $D$ perteneciente al multiplete vectorial real $V$,
de manera que podemos escribir los t\'erminos dependientes de $D$
en la parte bos\'onica del lagrangiano supersim\'{e}trico $d=3$ como
\be
 {\cal L}_D= \frac{\beta^2}{e^2} \sum_{n=1}^{\infty} c_n
 \left(\frac {D^2 - B^2}{\beta^2}\right)^n  + D(|\phi|^2 -
 \xi^2)
 \label{15pp}
\ee
Podemos ahora obtener la ecuaci\'on de movimiento para $D$ de
manera de eliminarlo. Escribiendo tal ecuaci\'on
\be
 \frac {1}{e^2} \sum_{n=1}^{\infty}  n c_n
 \left(\frac{D^2 - B^2}{\beta^2}\right)^{n - 1}\! 2D +
 (|\phi|^2 - \xi^2) = 0
 \label{16pp}
\ee
podemos ver, comparando potencias en $1/\beta^2$ que
\begin{eqnarray}
 D \!\!\!& = \!\!\!& - \frac{e^2}{2c_1}(|\phi|^2 - \xi^2)\nonumber\\
 B \!\!\!&= \!\!\!& \pm D%
 \label{17v}
\end{eqnarray}
es la \'unica soluci\'on no trivial de la ec.(\ref{16pp}). Estas
dos ecuaciones pueden ser combinadas en una sola que no es mas que
la famosa ecuaci\'on de  Bogomol'nyi para el campo magn\'etico de
los v\'ortices de Nielsen-Olesen
\be
 B = \pm \frac{e^2}{2c_1}(|\phi|^2 - \xi^2)
 \label{18}
\ee
Esto muestra que la ecuaci\'on de Bogomol'nyi para el campo de
gauge en configuraciones de v\'ortice (magn\'eticos) es
\underline{independiente} de la forma particular del lagrangiano
elegido para el campo de gauge dado que hemos probado la f\'ormula
(\ref{18}) para el lagrangiano supersim\'etrico general
(\ref{7})+(\ref{higgs}).

Analicemos ahora las transformaciones $N=2$ que dejan invariante
la teor\' \i a supersim\'{e}trica $d=3$ de Born-Infeld-Higgs.
Escribiremos solo las transformaciones que son relevantes para la
discusi\'on de las ecuaciones de Bogomol'nyi, esto es las del
higgsino y el gaugino (que denotamos como $\Omega$ y $\Sigma$), en
el caso est\'atico y con $A_0=N=0$ ecs.(\ref{susy3})-(\ref{susyd})
\be
 \delta_\Upsilon\Omega=-\sqrt 2\gamma^a D_a\phi\Upsilon=
 -i\sqrt 2\left( \begin{array}{cc}
   0                & (D_1+iD_2)\phi \\
   (D_1-iD_2)\phi & 0 \
 \end{array}\right)\left(\begin{array}{c}
   \epsilon_1 \\
   \epsilon_2 \
 \end{array}
 \right)
 \label{19}
\ee
\be
 \delta_\Upsilon\Sigma=-i(-\frac12\varepsilon_{abc}F^{ab}\gamma^c+ D)
 \Upsilon=-i\left(\begin{array}{cc}
   D-\frac 12\varepsilon_{ab}F_{ab} & 0 \\
   0 & D+\frac 12\varepsilon_{ab}F_{ab} \
 \end{array}\right) \left(\begin{array}{c}
   \epsilon_1 \\
   \epsilon_2 \
 \end{array}
 \right)
 \label{20}
\ee
donde el par\'ametro de la transformaci\'on (Dirac) $\Upsilon$ lo
hemos descompuesto en componentes.

Como discutimos, pidiendo que se anulen simult\'aneamente las
variaciones supersim\'{e}tricas asociadas con el higgsino y el
gaugino, se obtienen las ecuaciones de Bogomol'nyi. De hecho solo
es efectivamente posible anular las variaciones generadas por
$\epsilon_1$ o $\epsilon_2$ obteniendo consecuentemente las
ecuaciones para un solit\'on o un antisolit\'on. Por ejemplo,
imponiendo que las generadas por $\epsilon_2$ sean cero, obtenemos
de la variaci\'on del higgsino la siguiente ecuaci\'on de
autodualidad
\be
 \delta_{\epsilon_2} \Omega = 0 \to D_1 \phi = -i D_2 \phi
 \label{21pp}
\ee
Notemos que esta ley de transformaci\'on depende de la manera en
que el transporte paralelo es definido en t\'erminos del campo de
gauge y no de la forma expl\' \i cita de la acci\'on para el campo
de gauge. Es posible entender entonces porqu\'e la ec.(\ref{21pp})
es completamente independiente de la forma particular de la
acci\'on elegida para el campo de gauge, al menos para teor\' \i
as con acoplamiento de gauge m\' \i nimo\footnote{Para un
an\'alisis de las ecuaciones de Bogomoln'yi en teor\' \i as de
gauge con acoplamiento no m\' \i nimo ver ref.\cite{cbpf}.}. En lo
que respecta a la ecuaci\'on que se deriva de la transformaci\'on
correspondiente al gaugino tenemos,
\be
 \delta_{\epsilon_2} \Sigma =0 \to -\frac{1}{2}
 \varepsilon_{ab}F_{ab} = D
 \label{22pp}
\ee
esta ecuacui\'on podr\' \i a en principio depender del lagrangiano
elegido para el campo de gauge a trav\' es del t\'ermino $D$. Sin
embargo, como hemos visto (ec.(\ref{17v})), la soluci\'on de la
ecuaci\'on de movimiento para $D$ toma la misma forma
independientemente del lagrangiano elegido para la din\'amica del
campo de gauge, dado que $D$ siempre aparece como $D^2 - B^2$.

Esta propiedad puede ser verificada tambi\'en analizando las dos
supercargas que pueden ser obtenidas mediante el procedimiento de
Noether. Como mostramos anteriormente para el caso de Born-Infeld,
las supercargas $Q$ y $\bar Q$ siempre pueden ser escritas en la
forma
\ba
 \bar Q \!\!\!&=\!\!\!& -\frac 1 {2e^2} \int d^2x\, \Sigma^\dagger
 {\cal H}[B,D](\gamma^0 B +D) + \frac{i}{\sqrt 2} \int d^2x\,
 \Omega^\dagger\!\Dsl\phi\nonumber \\
 Q \!\!\!&=\!\!\!& -\frac 1{2e^2} \int d^2x\,  (B +\gamma^0 D){\cal H}[B,D]
 \Sigma - \frac{i}{\sqrt2} \int d^2x\, \gamma^0 ({\Dsl \phi})^\dagger
 \Omega
 \label{24}
\ea
donde  ${\cal H}$ es una funcional real de los campos $D$ y $B$
que puede ser calculada orden a orden en $1/\beta^2$. Mas a\'un,
las ecs.(\ref{24}) son v\'alidas no solo al considerar la
extensi\'on supersim\'{e}trica de la teor\' \i a de Born-Infeld
ec.(\ref{silvaesmuycatolico}) sino tambi\'en el lagrangiano
gen\'erico (\ref{7}). La forma de ${\cal H}$ depender\'a del
conjunto de coeficientes $b_{st}^{(\beta)}$. Lo que podemos ver
facilmente es que vale la siguiente f\'ormula
\be
 {\cal H} = {\cal H}_{Maxwell}+ \sum_{n=1}^{\infty}{\cal H}_n[B,D]
 \label{gene}
\ee
donde
\be
{\cal H}_{Maxwell} = 1 \label{mu} \ee
\be
\left.{\cal H}_n [B,D]\right\vert_{B^2 = D^2} = 0 \label{que} \ee
Luego, la condici\'on $\bar Q |BPS\rangle = 0$ se satisface cuando
$(B  +\gamma^0 D)\Upsilon =0$ y $ {\Dsl\phi} \Upsilon = 0$ para
alg\'un $\Upsilon\ne0$, independendientemente de la forma que tome
la funcional ${\cal H}$. Si eligimos que se anulen las
transformaciones supersim\'{e}tricas generadas por el par\'ametro
$\epsilon_2$, obtenemos las dos ecuaciones de Bogomol'nyi
(\ref{21pp}),(\ref{22pp}). Por supuesto que esto era de esperarse
ya que las expresiones $(B + \gamma^0 D)\Upsilon$ y $ \Dsl \phi
\Upsilon$, que aparecen en (\ref{24}), generan las leyes de
transformaci\'on para el  gaugino y el higgsino respectivamente.

En lo que concierne al \'algebra de supercargas, al usar la
condici\'on de Bogomol'nyi $B = \pm D$, solo la contribuci\'on
Maxwell de ${\cal H}$ sobrevive. Esto demuestra nuevamente que la
estructura BPS no es sensible a la forma particular del
lagrangiano para el campo de gauge.

\section{Resumen y Discusi\'on}

Estudiando la extensi\'on supersim\'{e}trica $N=2$ del modelo de
Born-Infeld-Higgs en $d=3$ dimensiones, hemos hallado las
relaciones de Bogomol'nyi para el sector bos\'onico. El inter\'es
en $3$ dimensiones se debe a que en este espacio-tiempo se conocen
soluciones de v\'ortice para las ecuaciones de Bogomol'nyi. Hemos
encontrado un hecho notable: las mismas ecuaciones (y luego el
mismo conjunto de soluciones) siguen valiendo cuando la din\'amica
del campo de gauge est\'a determinada por el lagrangiano de
Born-Infeld.

Originalmente, las extensiones supersim\'etricas a la teor\' \i a
de Born-Infeld fueron cons\-tru\-\' \i\-das usando el formalismo
de supercampos \cite{DP}-\cite{CF} y solo el sector bos\'onico
hab\' \i a sido expl\' \i citamente escrito en campos componentes.
Dado que una de nuestras metas era derivar las relaciones de
Bogomol'nyi partiendo del \'algebra supersim\'{e}trica $N=2$, fue
necesario explicitar el lagrangiano fermi\'onico, al menos, a
orden cuadr\'atico en campos fermi\'onicos los cuales dieron
ordenes lineales en las corrientes de Noether siendo estas las
\'unicas contribuciones no nulas al \'algebra en el sector de
background puramente bos\'onico.

Nuestro an\'alisis muestra que la supersimetr\' \i a fuerza una
particular forma funcional para la acci\'on bos\'onica en donde el
potencial de Higgs entra dentro de la raiz cuadrada de Born-Infeld
(ver ec.(\ref{ac}) de tal manera que las relaciones de Bogomol'nyi
resultantes son las mismas tanto para la teor\' \i a de  Maxwell
como para la de Born-Infeld.

Como era de esperar, la carga central del \'algebra supersim\'{e}trica
$N=2$ coincide con la carga topol\'ogica (el n\'umero de unidades
de flujo magn\'etico) del modelo mostrando que la cota de
Bogomol'nyi no se modifica cuando se analiza la teor\' \i a de
Born-Infeld. Esto se mostr\'o construyendo el \'algebra
supersim\'{e}trica y derivando la desigualdad de Bogomol'nyi en la
forma usual.

Analizamos finalmente el lagrangiano supersim\'etrico mas general
que es posible construir para el campo de gauge, tal que su parte
bos\'onica dependa de los invariantes fundamentales $ F^{\mu \nu}
F_{\mu \nu}$ y $ \tilde F^{\mu \nu} F_{\mu \nu}$. Este lagrangiano
general incluye, para una particular elecci\'on de coeficientes,
el lagrangiano supersim\'etrico de Born-Infeld, y tambi\'en una
clase infinita de lagrangianos que tienen propagaci\'on causal
\cite{DP}. Hemos mostrado porqu\'e las relaciones de Bogomol'nyi
asociadas con el sector bos\'onico permanecen, para esta dada
familia de lagrangianos, inalteradas a pesar del lagrangiano
elegido como din\'amica para el campo de gauge: los lagrangianos
de Maxwell, de Born-Infeld y lagrangianos no polin\'omicos mas
complicadas tienen la misma estructura BPS.

Finalmente se\~nalemos que se podr\' \i a realizar un an\'alisis
similar para el caso de teor\' \i as de gauge no-abelianas. A este
respecto existe una ambig\"uedad en la acci\'on relacionada con la
forma de definir la estructura de traza. Para el caso de
Born-Infeld fue se\~nalado  \cite{Bre} que la traza sim\'etrica
definida en \cite{Tse2} parece quedar singularizada por
consideraciones BPS a diferencia de otras definiciones. Esto
sugerir\' \i a la existencia de una extensi\'on supersim\'{e}trica
para dicha acci\'on. Este problema ser\'a analizado en el
pr\'oximo cap\' \i tulo.



\chapter{Teor\' \i a de Born-Infeld no Abeliana\label{nstr}}

\begin{center}
 {\begin{minipage}{6truein}
 { \sl Usando los invariantes de curvatura naturales como bloques de
construcci\'on en el formalismo de supercampos, mostramos que el
uso de la traza sim\'etrica en la definici\'on no-abeliana de la
acci\'on de Born-Infeld es compatible con supersimetr\' \i a.
Analizamos su relaci\'on con la estructura de ra\' \i z cuadrada
del lagrangiano no abeliano de Born-Infeld en el sector bos\'onico
y discutimos tambi\'en relaciones BPS en conexi\'on con la
construcci\'on supersim\'etrica.

Estos desarrollos son parte de los resultados originales de esta
tesis \cite{gui2}. }
 \end{minipage}}
\end{center}

\section{Introducci\'on}

Acciones del tipo Dirac-Born-Infeld (DBI) surgen en el estudio de
la din\'amica de bajas energ\' \i as de  Dp-branas
\cite{Tse}-\cite{AN},\cite{Lei}-\cite{Pol},
\cite{Tse2}-\cite{Bre},\cite{T},\cite{Tay}-\cite{G1}. En el caso
de la teor\' \i a de supercuerdas, se debe tratar con la
extensi\'on supersim\'etrica de la acci\'on DBI. Dado que cuando
un n\'umero $\cal N$ de D-branas coinciden, existe un aumento de
simetr\' \i a, la acci\'on abeliana de DBI debe ser generalizada
al caso no abeliano con grupo de gauge $U(\cal N)$. La
explicaci\'on de este hecho  debida a E. Witten \cite{Wi} es la
siguiente: agregando cargas en los extremos de las cuerdas
abiertas (factores de Chan-Paton) generamos un teor\' \i a de
gauge en el espacio-tiempo. Restringiendo las cuerdas a un
hiperplano $(p+1)$-dimensional (Dp-brana) obtenemos una teor\' \i
a de gauge en el volumen de mundo de la brana. Si estamos en
presencia de $\cal N$ Dp-branas paralelas y separadas, al calcular
los posibles estados de la teor\' \i a observamos que existir\'an
$\cal N$ estados no masivos correspondientes a cuerdas abiertas
terminando en la misma brana, que dar\'an origen a una teor\' \i a
de gauge $U(1)^{\cal N}$, y ${\cal N}^2-{\cal N}$ estados masivos
para las cuerdas (orientadas) terminando en distintas branas cuya
masa ser\'a proporcional a la separaci\'on entre branas. Si
superponemos las branas en un mismo punto todos estos estados
masivos se vuelven no masivos y la teor\' \i a sufre un aumento de
simetr\' \i a $U(1)\to U({\cal N})\sim U(1)\times SU({\cal N})$,
obteniendose los ${\cal N}^2$ vectores necesarios para reproducir
la adjunta del grupo de gauge $U({\cal N})$.

En la literatura han sido discutidas varias posibilidades para la
extensi\'on no abeliana de la acci\'on de DBI
\cite{H}-\cite{Bre},\cite{T}. B\'asicamente, difieren en la manera
en que se define la operaci\'on de traza sobre el grupo. En el
contexto de cuerdas, la operaci\'on de traza sim\'etrica sugerida
por Tseytlin \cite{Tse2} parece ser la apropiada. Entre sus
ventajas podemos se\~nalar que:

(i) Elimina potencias impares de la curvatura no deseadas,
implicando esto que la fuerza de campo $F$ (pudiendo ser grande)
debe variar lentamente dado que $ F^3 \sim [D,D]F^2$. Con este
tipo de aproximaci\'on abeliana (implica $F's$ conmutantes) es
posible hacer contacto con la acci\'on efectiva  a orden \'{a}rbol de
la teor\' \i a de cuerdas abiertas \cite{Tse2}\footnote{Es
importante se\~nalar que la acci\'on efectiva no es \'unica.
Precisamente los t\'erminos que dependen de derivadas de
$F_{\mu\nu}$, en la expansi\'on perturbativa en $\acute\alpha$,
son ambiguos dado que no est\'an determinados por las amplitudes
de scattering on-shell de cuerdas \cite{tse}. Este ambig\"uedad
est\'a relacionada con el hecho de que distintos lagrangianos
pueden dar origen a la misma matriz S (teorema de equivalencia).
Desde el punto de vista de integral funcional tiene que ver con
cambios de variable en la medida de integraci\'on que no afectan
las condiciones de capa de masa.}.

(ii) Es la \'unica que da origen a una acci\'on que es linealizada
por las condiciones BPS y a ecuaciones de movimiento que coinciden
con aquellas que resultan de imponer la anulaci\'on de las
funciones $\beta$ para la teor\' \i a de supercuerdas abiertas en
presencia de campos de fondo \cite{BP}-\cite{Bre}.

Debemos mencionar, sin embargo, que existen algunos problemas no
resueltos en lo que concierne al uso de la traza sim\'etrica en la
acci\'on no abeliana de Born-Infeld. En particular, en \cite{HT}
se se\~nalan discrepancias entre los resultados obtenidos con la
teor\' \i a no abeliana de Born-Infeld sim\'etrica y el espectro
esperado en teor\' \i as de branas.

Como fue mencionado en \cite{Bre}, el hecho de que la
prescripci\'on de traza sim\'etrica fuera singularizada por ser la
que da origen a relaciones BPS deber\' \i a estar conectado con la
posibilidad de supersimetrizar la teor\' \i a de BI. Es esta la
motivaci\'on de este cap\' \i tulo, donde construimos la versi\'on
supersim\'etrica de la acci\'on no abeliana de BI, discutiendo el
tema de la traza en los \' \i ndices internos del grupo de gauge y
las relaciones de Bogomol'nyi que resultan en el sector
bos\'onico.

Nuestro an\'alisis extiende al caso no abeliano los resultados
obtenidos en \cite{DP} y en el cap\' \i tulo anterior.

\section{Construcci\'on de la acci\'on}

Como vimos en el cap\' \i tulo anterior, el objeto b\'asico para
la construcci\'on de una teor\' \i a invariante de gauge
supersim\'etrica es el supercampo vectorial (real) de gauge $V$
que escribimos en la forma
\begin{eqnarray}
 &&V(x,\theta,\bar\theta)  =  C + i \theta \chi - i \bar \theta
 \bar \chi + \frac{i}{2} \theta^2 (M + iN) - \frac{i}{2} \bar
 \theta^2 (M - iN)
 \nonumber\\
 &&-\theta \sigma^\mu \bar \theta A_\mu + i \theta^2 \bar \theta
 (\bar \lambda +\frac{i}{2} \bar{/\!\!\!\partial} \chi)
 -   i \bar \theta^2 \theta (\lambda + \frac{i}{2} /\!\!\!\partial \bar
 \chi)
 + \frac{1}{2} \theta^2  \bar \theta^2 (D + \frac{1}{2} \Box C)
 \label{1prima}
\end{eqnarray}
donde $C,M,N,D$ son campos escalares reales, $\lambda,\chi$
fermiones de Weyl de dos componentes y $A_\mu$ el campo de gauge.
En el caso no abeliano que nos ocupa, el supercampo $V$ toma
valores en el \'algebra de Lie del grupo de gauge, luego
\be
 V=V^a t^a
\ee
de manera que todos los campos componentes llevan un \' \i ndice
interno de grupo\footnote{En el presente cap\' \i tulo usaremos
letras griegas $\mu,\nu,..$ para los \' \i ndices
espacio-temporales y letras latinas $a,b,..$ para referirnos a los
\' \i ndices internos de grupo.}
\ba
 V\!\!\!&=\!\!\!&(C^a,\xi^a,M^a,N^a,A^a_\mu,\lambda^a,D^a)\nonumber\\
 A_\mu\!\!\!&=\!\!\!&A_\mu^a t^a \nonumber\\
 \lambda\!\!\!&=\!\!\!&\lambda ^a t^a \nonumber\\
 D\!\!\!&=\!\!\!&D^a t^a
 \label{10prima}
\ea
donde $t^a$ son los generadores hermiticos
\be
 [t^a,t^b] = i f^{abc} t^c
 \label{11prima}
\ee
\be
 {\rm tr} \, t^a t^b =  {\cal C} \delta^{ab}
 \label{12prima}
\ee
La invarianza de gauge se generaliza en el caso supersim\'{e}trico,
implement\'an\-do\-se en t\'er\-mi\-nos de su\-per\-cam\-pos
seg\'un la ley de transformaci\'on
\be
 e^{2V} \to e^{-2i\Phi^\dagger} e^{2V} e^{2i\Phi}
 \label{12b}
\ee
donde $\Phi(\Phi^\dagger)$ es un supercampo quiral (antiquiral)
que toma valores en el \'algebra de Lie del grupo de gauge.
\be
 \bar D_{\dot \alpha}\Phi= 0
 \label{condi}
\ee
Expl\' \i citamente,
\be
 \Phi^a(y,\theta)= \phi^a + \theta \psi^a +\theta^2 F^a
 \label{ex}
\ee
aqu\' \i~ $\phi^a=\phi_1^a+i\phi^a_2$ y $F^a$ son campos escalares
complejos y $\psi^a$ es un espinor de Weyl. El supercampo $\Phi$
representa la generalizaci\'on supersim\'{e}trica de los par\'ametros
ordinarios de la transformaci\'on de gauge.

La invarianza de supergauge (gauge generalizada) nos permite
elegir un gauge conveniente llamado de Wess-Zumino, en el cual
$C,\chi,M$ y $N$ son puestos a cero, quedando un multiplete con el
campo de gauge $A_\mu$, el campo fermi\'onico de Majorana
$\Lambda$ y el campo auxiliar real $D$ (tomando todos valores en
el \'algebra de Lie del grupo de gauge)\footnote{El fijado del
gauge de WZ es mas complicado en el caso no abeliano quedando una
relaci\'on no lineal entre las componentes $(\phi_2,\chi,F)\subset
\Phi$ y los campos redundantes $(C,\chi,M,N)\subset V$. Sin
embargo el resultado es an\'alogo al del caso abeliano: fijado el
gauge de Wess-Zumino tenemos a\'{u}n la libertad de realizar
transformaciones de gauge en $A_\mu$.}.La transformaci\'on
(\ref{12b}) a primer orden en $\Phi$ implica, en el gauge de WZ,
para la componente $\theta\bar\theta$ correspondiente al campo de
gauge $A_\mu$,
\be
 \delta^{(WZ)} A_\mu= \nabla_\mu \phi_1
 =(\partial_\mu+i\left[A_\mu,~~\right])\,\phi_1
\ee
donde $\phi_1$ es la componente real de $\phi$ en (\ref{ex}). La
invarianza de gauge usual para $A_\mu$, entonces,  est\'a
contenida en (\ref{12b}), que generaliza la transformaci\'on de
simetr\' \i a a todo el supermultiplete.

A partir de  $V$, se construye el supercampo de curvatura (quiral)
$W_\alpha$\footnote{An\'alogamente a lo que sucede en el caso
abeliano, en el gauge de WZ la expansi\'on en serie de $e^{2V}$
contiene solo un n\'umero fininto de t\'erminos.}
\begin{equation}
 W_\alpha \left( y,\theta \right) = \frac{1}{8} \bar D^2 (e^{-2 V}
 D_\alpha  e^{2 V})
 \label{2p}
\end{equation}
A diferencia de  (\ref{12b}), bajo una transformaci\'on de
supergauge  $W_\alpha$ transforma covariantemente,
\be
 W_\alpha \to e^{-i2\Phi} W_\alpha e^{i2\Phi}~~.
 \label{12c}
\ee
En lo que concierne al conjugado herm\' \i tico,  $\bar W_\alpha$,
el mismo transforma como
\be
 \bar W_{\dot \alpha} \to e^{-i2\Phi^\dagger}\bar W_{\dot\alpha}
 e^{i2\Phi^\dagger}
 \label{12d}
\ee
Escrito en componentes, $W_\alpha$ toma la forma
\begin{equation}
 W_\alpha \left( y,\theta \right) =i\lambda _\alpha
 -\theta_\alpha D
 -\frac{i}{2}\left( \theta\sigma ^\mu\bar \sigma ^\nu \right) _\alpha
 F_{\mu\nu}
 -\theta^2  (\not\!\nabla \bar\lambda)_\alpha
 \label{6}
\end{equation}
donde
\be
 F_{\mu \nu} = \partial_\mu A_\nu - \partial_\nu A_\mu +i[A_\mu,A_\nu]
 \label{8p}
\ee
y
\be
 ( \not\!\nabla \bar\lambda)_\alpha =
 \sigma^\mu_{~\alpha \dot\alpha} \nabla_\mu\bar
 \lambda^{\dot\alpha}
 \label{nuevi}
\ee
La conocida extensi\'on supersim\'{e}trica  $N=1$ de la teor\' \i a de
Yang-Mills se construye a partir de $W$ considerando $W^2$ y su
conjugado herm\' \i tico $\bar W^2$. Conociendo que  $W^2$ toma la
forma
\begin{eqnarray}
 W^2 \!\!\!&=\!\!\!& -\lambda^2 - i (\theta \lambda\, D+D\,
 \theta\lambda)+\frac 12 ( \theta \sigma^\mu \bar \sigma^\nu
 \lambda\, F_{\mu \nu}+F_{\mu \nu}\,\theta\sigma^\mu\bar \sigma^\nu
 \lambda ) + \nonumber\\ & & \theta^2 \left(- i\lambda\!\! \not\!
 \nabla \bar \lambda -i(\not\! \nabla \bar \lambda)\lambda + D^2 -
 \frac{1}{2}(F_{\mu\nu}F^{\mu \nu} +
 i \tilde{F}_{\mu\nu}F^{\mu \nu}) \right)
 \label{9p}
\end{eqnarray}
donde
\be
 \tilde F_{\mu \nu} = \frac{1}{2} \varepsilon_{\mu \nu \alpha
 \beta} F^{\alpha \beta}
 \label{dual}
\ee
Escribiendo expl\'\i citamente los generadores del grupo de gauge
se tiene
\begin{eqnarray}
 W^2 \!\!\!&=\!\!\!&\frac12 \{t^a,t^b\}\!\left(\!-
 \lambda^a \lambda^b - 2i \,\theta \lambda^a\, D^b
 +\theta\sigma^\mu\bar\sigma^\nu\!\lambda^a\, F_{\mu \nu}^b
 \right.\nonumber\\
 \!\!\!&\!\!\!&\left. -2i\, \theta^2\,
 \lambda^a (\delta^{bc}{/\!\!\!\partial}
 + f^{bcd} /\!\!\!\!A^d) \bar \lambda^c
 + \theta^2
 \left(D^aD^b - \frac{1}{2}(F_{\mu\nu}^aF^{b\,\mu \nu } + i
 \tilde{F}_{\mu\nu}^aF^{b\,\mu \nu})\! \right)\right)
 \label{13p}
\end{eqnarray}
donde
\be
 \{t^a,t^b\}= t^at^b + t^b t^a
\ee
\ba
 {/\!\!\!\partial}\!\!\!&=\!\!\!& \sigma^\mu \partial_\mu \\
 /\!\!\!\!A\!\!\!&=\!\!\!& \sigma^\mu A_\mu
\ea
De la  ec.(\ref{9p}) y su an\'aloga para $\bar W_{\dot\alpha} \bar
W^{\dot\alpha}$ vemos que la extensi\'on supersim\'{e}trica de la
teor\' \i a de Yang-Mills puede ser escrita en la forma
\be
 {\cal L}_{Y\!M}^{SU\!SY} =  \frac{1}{4g^2 {\cal C}}\; {\rm tr}\!\int
 d^2\theta\, W^2  +  h.c.
 \label{15rere}
\ee
cuya parte puramente bos\'onica est\'a dada por (cf.(\ref{maxc}))
\be
 \left. {\cal L}_{Y\!M}^{SU\!SY}\right|_{bos}= -\frac{1}{4g^2} F_{\mu \nu}^a
 F^{a \mu \nu }+\frac 1{2g^2}D^aD^a
 \label{re15}
\ee

Fijadas las convenciones estamos listos para extender el
tratamiento del cap\' \i tulo anterior y encontrar el lagrangiano
supersim\'etrico $N=1$ de Born-Infeld en el caso no abeliano NBI.
Este lagrangiano estar\'a b\'asicamente construido en t\'erminos
de $W$, $\bar W$ y $e^{\pm 2V}$. Es importante mencionar que la
operaci\'on de traza sobre los \' \i ndices del grupo de gauge que
es necesario realizar para obtener un lagrangiano escalar, podr\'
\i a diferir en principio, de la traza usual ``${\rm tr}$''
definida en (\ref{12prima}) y usada en (\ref{15rere}). Como
veremos a continuaci\'on este hecho esta relacionado con la
prescripci\'on para el ordenamiento de los generadores tanto al
definir el determinante en espacio tiempo como al expandir la ra\'
\i z cuadrada de BI.

En principio se busca definir el lagrangiano de NBI como un
determinante en espacio-tiempo. De manera naive, al calcular el
determinante para el caso no abeliano en $d=4$ tenemos
\cite{T}\footnote{Dado que $\det (g+F)=\det(g+F^T)=\det (g-F)$ las
expansi\'ones tanto en los casos abeliano como no abeliano
contienen solo potencias pares de $F$.}$^,$\footnote{En el
presente cap\' \i tulo ponemos por simplicidad el par\'ametro
dimensional de BI $\beta=1$.},
\be
 -\det\left(g_{\mu \nu} I + F_{\mu \nu}\right)= I +
 \frac{1}{2} F^2 -
 \frac{1}{4}(``F^4"-\frac{1}{2}(F^2)^2)
 \label{ddd}
\ee
En el caso abeliano, el t\'ermino $F^4$ en el lado derecho de la
ec.(\ref{bi4}) fue expresado en t\'erminos de $F^2$ y $F\tilde F$
(ver ec.(\ref{f4})). Con $``F^4"$ en el caso no abeliano
(ec.(\ref{ddd})) queremos hacer hincapie en que existe una
ambig\"{u}edad en el ordenamiento de los factores del t\'ermino de
orden $F^4$ (ver ecs.(\ref{f41})-(\ref{f4e}))
\footnote{Aparecer\'an tambi\'en ambig\"{u}edades al definir
$\sqrt{-\det(g+F)}$ relacionadas con el ordenamiento de los
factores en todo desarrollo de Taylor de una funci\'on de
matrices. As\'\i~ p.ej. al definir una funci\'on $f(X,Y)$ mediante
su desarrollo de Taylor la misma no queda un\' \i vocamente
definida si $X$ e $Y$ no conmutan, es necesario dar una
prescripci\'on para el ordenamiento de las potencias $X^m Y^n$.},
ambig\"{u}edad que se elimina al determinar una elecci\'on para la
traza.

Comencemos entonces la b\'usqueda de la extensi\'on supersim\'{e}trica
del modelo NBI. Con el objeto de conseguir potencias (pares) de
orden mayor a dos de $F_{\mu \nu} F^{\mu \nu}$ y $\tilde F_{\mu
\nu} F^{\mu \nu}$ que necesariamente aparecen en un lagrangiano
tipo BI, deberemos incluir, como en el cap\' \i tulo anterior,
potencias superiores de  $W$ y $\bar W$ combinadas de manera de
respetar la invarianza de gauge\footnote{An\'alogamente al caso
abeliano, existen relaciones algebr\'aicas que permiten reducir
expresiones del tipo
$F^n=F_{\mu_1}^{~\mu_2}F_{\mu_2}^{~\mu_3}...F_{\mu_n}^{~\mu_1}$ a
expresiones en terminos de $F^2$ y $F\tilde F$ (ver
ec.(\ref{f41})-(\ref{f4e})).}. En el caso abeliano, conseguimos
esto combinando $W^2 \bar W^2$ con potencias adecuadas de $D^2W^2$
y $\bar D^2 \bar W^2$ ecs.(\ref{13})-(\ref{15}) y
refs.\cite{DP}-\cite{CF},\cite{GNSS}. Para el caso no abeliano, en
vista de las leyes de transformacion
(\ref{12b}),(\ref{12c}),(\ref{12d}), la situaci\'on es mas sutil.
Consideremos entonces los supercampos invariantes de gauge que
pueden dar origen a t\'erminos de orden cu\'artico. Hay dos
candidatos naturales,
\be
 F_1 \equiv\!\int\! d^2\theta d^2 \bar \theta\, W^2 e^{-2V} \bar W^2
 e^{2V}
 \label{1q1}
\ee
\be
 F_2\equiv\int\!d^2 \theta d^2\bar\theta\,W^\alpha e^{-2V}\bar
 W^{\dot\beta} e^{2V} W_\alpha e^{-2V}\bar W_{\dot \beta} e^{2V}
 \label{2q2}
\ee
cuyas componentes puramente bos\'onicas son (poniendo el campo
auxiliar $D$ a cero)
\be
 {\rm tr}\left. F_1\right\vert_{bos}=
 \frac{1}{4}\theta^2\bar\theta^2 \, {\rm tr}\left(( F^2)^2
 +(F\tilde F)^2 \right)= \theta^2\bar\theta^2{\cal F}_1
 \label{qq1}
\ee
\be
 {\rm tr}\left. F_2\right\vert_{bos} =
 \frac{1}{4}\theta^2\bar\theta^2\,  {\rm tr}\left( F_{\mu\nu}
 F_{\rho\sigma} F^{\mu\nu} F^{\rho\sigma} +
 F_{\mu\nu} F_{\rho\sigma} \tilde F^{\mu\nu}
 \tilde F^{\rho\sigma} \right)= \theta^2\bar\theta^2{\cal F}_2
 \label{qq2}
\ee
Podemos ver que una particular combinaci\'on de ${\cal F}_1$ y
${\cal F}_2$ genera los t\'erminos de orden cu\'artico que
resultan de la expansi\'on de la raiz cuadrada del lagrangiano
NBI, siempre que este sea definido usando la traza sim\'etrica:
\be
 {\rm Str} \left(t_1,t_2,\ldots, t_N \right) \equiv
 \frac{1}{N!}\sum_{\pi} {\rm tr} \left( t_{\pi(1)}
 t_{\pi(2)} \ldots t_{\pi(N)}\right) \, .
 \label{tra}
\ee
En efecto, tenemos que
\be
 {\rm Str} \left. \left(1 - \sqrt{1 +
 \frac{1}{2} F_{\mu \nu} F^{\mu \nu} -\frac{1}{16} ( F_{\mu \nu}
 \tilde F^{\mu\nu})^2}\; \right) \right|^{4to~ord.}\!\!= \frac 1 {32}
 {\rm Str}\left( (F^2)^2+(F\tilde F)^2\right)
 \label{za}
\ee
donde en la expresi\'on anterior debemos tener en cuenta que cada
$F$ lleva un generador $F=F^at^a$. Calculando el lado derecho de
(\ref{za}) obtenemos
\ba
 {\rm Str}\left( (F^2)^2+(F\tilde F)^2\right)\!\!\!&=\!\!\!&
 \frac 13 \left\{ 2\, {\rm tr}\left( (F^2)^2+(F\tilde F)^2\right)
 +\, {\rm tr}\left(  F_{\mu\nu}
 F_{\rho\sigma} F^{\mu\nu} F^{\rho\sigma} +
 F_{\mu\nu} F_{\rho\sigma} \tilde F^{\mu\nu}
 \tilde F^{\rho\sigma} \right)  \right\}\nonumber\\
 \!\!\!&=\!\!\!&\frac 43\left\{ 2{\cal F}_1+{\cal F}_2  \right\}
 \label{tapos}
\ea
lo que nos muestra que
\be
 \left. \frac{1}{24}{\rm tr}\int d^2\theta d^2\bar\theta
 \left( 2 F_1+ F_2\right)
 \right \vert_{bos} =  {\rm Str} \left. \left(1 - \sqrt{1 +
 \frac{1}{2} F_{\mu \nu} F^{\mu \nu} -\frac{1}{16} ( F_{\mu \nu}
 \tilde F^{\mu\nu})^2}\; \right) \right|^{4to~ord.}
 \label{sr}
\ee

Otra raz\'on en favor del uso de la traza sim\'etrica es que
resuelve de manera natural las ambig\"uedades en la definici\'on
del lagrangiano de BI como determinante, obteni\'endose
\cite{tse}, \cite{Bre},
\be
 {\rm Str} \left(1 - \sqrt{1 + \frac{1}{2} F_{\mu \nu} F^{\mu \nu}
 -\frac{1}{16} (F_{\mu \nu}\tilde F^{\mu \nu})^2}\; \right) = {\rm
 Str}\left(1 -  \sqrt{\det(g_{\mu \nu} + F_{\mu \nu})}\; \right)
 \label{chu}
\ee
donde ambos lados est\'an un\' \i vocamente definidos por la
prescripci\'on Str.

Es natural entonces tomar como definici\'on de la extensi\'on no
abeliana del lagrangiano de Born-Infeld la propuesta de Tseytlin
\cite{Tse2} usando ``Str"
\ba
 {\cal L}_{NBI}\!\!\!&=\!\!\!&{\rm Str}\left(1-
 \sqrt{\det(g_{\mu \nu} + F_{\mu \nu})}\, \right)\\
 \!\!\!&=\!\!\!& -\frac 14 {\rm tr}\, F^2+\frac 1 {48}
 \left[\phantom{\frac 12}\!\!\!{\rm tr}(F^2)^2\right.\nonumber\\
 \!\!\!&\!\!\!&\left. +{\rm tr}(F\tilde F)^2+
 \frac 12 {\rm tr} ( F_{\mu\nu}
 F_{\rho\sigma} F^{\mu\nu} F^{\rho\sigma} +
 F_{\mu\nu} F_{\rho\sigma} \tilde F^{\mu\nu}
 \tilde F^{\rho\sigma})\right]+{\cal O}(F^6)
 \label{nbi}
\ea
En particular obtenemos de la definici\'on de traza sim\'etrica
(\ref{tra}) que
\be
 {\rm Str} \left( ``F^4"-\frac 12 (F^2)^2\right)=\frac 14 {\rm
 Str}\left( F\tilde F\right)^2
 \label{f44}
\ee
o sea que la prescripci\'on Str es una suerte de aproximaci\'on
abeliana (cf.(\ref{f4})) pues bajo la operaci\'on ``Str" los
factores $F$ pueden ser tratados como conmutantes. Esta
definici\'on para el lagrangiano NBI es en cierto sentido la m\'
\i nima extensi\'on no abeliana de la teor\' \i a abeliana
(\ref{1}), que adem\'as es consistente con el requerimiento
b\'asico para el orden \'arbol de la teor\' \i a de cuerdas, esto
es, una sola operaci\'on de traza para el producto de los tensores
de campo $F$ entendidos como matrices. Sorprendentemente la
prescripci\'on ``Str" reproduce tambi\'en exactamente los
t\'erminos $F^2+\alpha '^2F^4$ para la acci\'on efectiva no
abeliana de cuerdas abiertas
\cite{tse},\cite{growit}\footnote{Mientras que supersimetr\'{\i}a $N=1$
no fija un\' \i vocamente la componente bos\'onica de la acci\'on
no lineal abeliana a la forma de BI \cite{DP}, actualmente se
sospecha fuertemente que la imposici\'on de supersimetr\'{\i}a $N=4$ en
$d=4$ (o $N=1$ en $d=10$) deber\'ia forzarla; ya que en
particular, en \cite{berg} se muestra que la estructura mas
general de los t\'erminos $F^4$ pidiendo supersimetr\'{\i}a global en
$d=10$ coincide con (\ref{f44}) y mas sugestivamente en la
ref.\cite{t3} para los t\'erminos $F^6$.}.

En resumen, la ec.(\ref{sr}) es uno de los pasos im\-por\-tan\-tes
en nues\-tra de\-ri\-va\-ci\'on de la extensi\'on supersim\'{e}trica
del lagrangiano NBI, muestra que: al analizar los posibles
t\'erminos cu\'articos de $F$ en la expansi\'on de la raiz
cuadrada del determinante de BI, en principio no cualquier
combinaci\'on de los mismos es supersimetrizable, debemos elegir
com\-bi\-na\-ci\'ones contenidas en los candidatos naturales $F_1$
y $F_2$. La prescripci\'on de traza sim\'etrica Str corresponde a
una particular combinaci\'on de trazas normales de $F_1$ y $F_2$,
ec.(\ref{tapos}), luego  {\it la definici\'on de la extensi\'on no
abeliana del lagrangiano de Born-Infeld usando Str es
su\-per\-si\-me\-tri\-za\-ble}. Esta combinaci\'on fue
originalmente propuesta por Tseytlin \cite{tse} para la acci\'on
de la teor\' \i a de BI de manera de hacer contacto con la teor\'
\i a de bajas energ\' \i as que resulta de la teor\' \i a de
supercuerdas. Vale la pena mencionar que el lado derecho de
(\ref{sr}) se anula para $F = \pm i\tilde F$. Esto garantiza, al
menos para el orden cu\'artico que hemos estamos discutiendo hasta
aqu\' \i~, que el lagrangiano supersim\'{e}trico se reducir\'a al de
Yang-Mills cuando la cota de Bogomol'nyi (en versi\'on eucl\' \i
dea) sea saturada \cite{inst},\cite{Hashi},\cite{HT},\cite{CS}.

El an\'alisis previo fue efectuado en el sector puramente
bos\'onico. Es natural extenderlo considerando la combinaci\'on
completa de los supercampos  $2F_1 + F_2$ ya que la traza de los
mismos nuevamente se acomoda en la forma de una traza sim\'etrica
\begin{eqnarray}
 \frac{1}{3}
 (2\!\!\!\!\!\!\!\!\! & & \!{\rm tr}  F_1 +  {\rm tr}  F_2)  = \nonumber\\
 & & {\rm Str}  \left(
 W^\alpha,W_\alpha, e^{-2V} \bar W_{\dot \beta}
 e^{2V}, \right.
 \left. e^{-2V} \bar W^{\dot \beta}
 e^{2V}
 \right)
 \label{sr2}
\end{eqnarray}
Para poder construir las  potencias superiores de $F^2$ y $F\tilde
F$, necesarias para obtener el lagrangiano de BI, definimos,
extendiendo el tratamiento del cap\' \i tulo anterior \cite{GNSS},
los supercampos $X$ y $Y$ como,
\begin{eqnarray}
 X \!\!\!&=\!\!\!&\frac 1{16}\left(\bar D^2
  \bar {\cal W}_{\dot\beta}\;\bar{\cal W}^{\dot\beta}-2\bar
 D_{\dot\alpha}  \bar {\cal W}_{\dot\beta}\; \bar
 D^{\dot\alpha}  \bar {\cal W}^{\dot\beta}+ \bar{\cal W}_{\dot\beta}\; D^2
 \bar {\cal W}^{\dot\beta}\right.\nonumber\\
 \!\!\!&\!\!\!&\quad\left.
 + e^{-2V}\left( D^2 {\cal W}^\beta\; {\cal W}_\beta-
 2D^\alpha {\cal W}^\beta\;D_\alpha {\cal W}_\beta+{\cal W}^\beta\;
 D^2 {\cal W}_\beta \right)e^{2V}\right)
 \label{x}
\end{eqnarray}
\begin{eqnarray}
 Y \!\!\!&=\!\!\!& \frac{i}{32}\left(\bar D^2
  \bar {\cal W}_{\dot\beta}\;\bar{\cal W}^{\dot\beta}-2\bar
 D_{\dot\alpha}  \bar {\cal W}_{\dot\beta}\; \bar
 D^{\dot\alpha}  \bar {\cal W}^{\dot\beta}+ \bar{\cal W}_{\dot\beta}\; D^2
 \bar {\cal W}^{\dot\beta}\right.\nonumber\\
 \!\!\!&\!\!\!&\quad\left.
 - e^{-2V}\left( D^2 {\cal W}^\beta\; {\cal W}_\beta-
 2D^\alpha {\cal W}^\beta\;D_\alpha {\cal W}_\beta+{\cal W}^\beta\;
 D^2 {\cal W}_\beta \right)e^{2V}\right)
 \label{y}
\end{eqnarray}
donde hemos definido
\ba
 {\cal W}_\beta&\equiv&e^{2V}W_\beta e^{-2V}\nonumber\\
 \bar{\cal W}_{\dot\beta}&\equiv&e^{-2V}
 \bar W_{\dot\beta}e^{2V}\nonumber
\ea
Los supercampos (\ref{x})-(\ref{y}) transforman de manera
covariante bajo transformaciones de supergauge
\be
 X \to e^{-2i\Phi} X e^{2i\Phi} ~, ~~~  Y \to
 e^{-2i\Phi} Y e^{2i\Phi}
 \label{t}
\ee
y sus componentes para  $\theta = 0$ son, como en el caso
abeliano, los dos invariantes b\'asicos
\be
 X\vert_{\theta =0} = \frac{1}{4} F_{\mu \nu}  F^{\mu \nu} ~, ~ ~ ~
 Y\vert_{\theta =0} = \frac{1}{8} \tilde F_{\mu \nu}  F^{\mu \nu}
 \label{bto}
\ee
Inspirados en el resultado (\ref{sr2}) obtenido de manera de
reproducir adecuadamente las potencias cu\'articas en $F$ y
$\tilde F$, proponemos el siguiente lagrangiano supersim\'{e}trico no
abeliano como candidato para reproducir la din\'amica de BI en su
sector bos\'onico,
\begin{eqnarray}
 {\cal L}_{NBI}^{(SUSY)}={\cal L}_{YM}^{(SUSY)}+
 \sum_{n,m} C_{st} \int d^4\theta\; {\rm Str}
 \left(W^\alpha ,
 W_\alpha , \bar{\cal W}_{\dot\beta}
 ,  \bar{\cal W}^{\dot\beta},X^s,Y^t
 \right)
 \label{L}
\end{eqnarray}
Debemos se\~nalar en este punto que la expresi\'on (\ref{L}) da
una familia de lagrangianos correspondiente a la extensi\'on
supersim\'{e}trica de lagrangianos bos\'onicos invariante de gauge
dependiendo de la fuerza de campo $F$ a trav\'es de los
invariantes algebr\'aicos $F_{\mu\nu} F^{\mu \nu}$ y $\tilde
F_{\mu\nu} F^{\mu \nu}$. Las combinaciones de $F^2$ y $F\tilde F$
no son arbitrarias sino que est\'an restringidas por
supersimetr\'{\i}a. La versi\'on abeliana de (\ref{L}) fue ajustada en
el cap\' \i tulo anterior (ver \cite{DP}-\cite{CF},\cite{GNSS})
mediante una elecci\'on apropiada de los coeficientes $C_{st}$ de
manera de reproducir el lagrangiano de BI.  Lo mismo sucede en el
caso no abeliano: para una elecci\'on particular de los
coeficientes $C_{st}$ obtenemos la teor\' \i a no abeliana de
Born-Infeld,
\begin{eqnarray}
 C_{st}=a_{st}^{(\beta)}\vert_{\beta=1}
 \label{coe}
\end{eqnarray}
donde los $a_{st}^{(\beta)}$ fueron definidos en el cap\' \i tulo
anterior por las ecs.(\ref{21}). Con esta elecci\'on tenemos como
parte bos\'onica del lagrangiano (\ref{L}),
\be
 \left. {\cal L}_{NBI}^{(SUSY)}\right\vert_{bos}
 = {\rm Str} \left( 1 - \sqrt{-\det(g_{\mu \nu} + F_{\mu \nu})}
 \right)
\label{LL} \ee
Este es el lagrangiano NBI en la forma propuesta en \cite{tse}.

Como en el caso no abeliano, existen otras posibles elecciones de
coeficientes $C_{st}$ que tambi\'en dan lugar a una teor\' \i a
invariante de gauge supersim\'etrica y causal con din\'amica para
el campo de gauge no polin\'omica. En particular, la alternativa
de una acci\'on de BI para $SO(N)$, propuesta recientemente en
\cite{RT}, deber\' \i a corresponder a alguna de estas elecciones.

Hemos construido el lagrangiano supersim\'{e}trico $N=1$
(ec.(\ref{L})) empleando el formalismo de supercampos, cuya parte
bos\'onica se encuentra expresada en t\'erminos de la raiz
cuadrada de $\det (g_{\mu \nu} + F_{\mu \nu})$. Empleamos los
invariantes de curvatura naturales como bloques de construcci\'on
obteniendo un lagrangiano que, en su sector bos\'onico, depende
solo de los invariantes $F_{\mu\nu}F^{\mu \nu}$ y
$F_{\mu\nu}\tilde F^{\mu \nu}$ y puede ser expresado en t\'erminos
de la traza sim\'etrica de un determinante.

Como mencionamos arriba, la extructura de traza de la teor\' \i a
de la teor\' \i a de BI fue fijada en \cite{Bre} pidiendo que la
acci\'on quedase linealizada en las configuraciones BPS
(instantones, monopolos, v\'ortices). En el presente an\'alisis
hemos visto que la traza sim\'etrica surge naturalmente en el
formalismo de supercampos cuando se construye la raiz cuadrada del
lagrangiano de BI. Esta confluencia de resultados no es mas que
una manifestaci\'on de la conocida conexi\'on entre supersimetr\'{\i}a
y relaciones BPS. Para terminar el cap\' \i tulo discutiremos
entonces los aspectos BPS del modelo.

\section{Aspectos BPS}

Para fijar ideas nos concentraremos en configuraciones de
instantones en $d=4$.  En el gauge de  Wess-Zumino, el
supermultiplete vectorial $N=1$ esta formado por
$(A_\mu,\Lambda,D)$, donde $\Lambda$ es un fermi\'on de Majorana.
Con el objeto de analizar relaciones BPS, consideraremos un
mo\-de\-lo supersim\'{e}trico $N=2$ que incluye, aparte de estos
campos, un multiplete escalar quiral. En analog\' \i a con los
procedimientos realizados para la construcci\'on del lagrangiano
supersim\'{e}trico $N=1$ (\ref{L}), es posible construir el
lagrangiano supersim\'{e}trico $N=2$ agregando al multiplete vectorial
un multiplete escalar quiral \cite{ketov}. No detallaremos la
construcci\'on aqu\' \i~ sino que consideraremos las
transformaciones supersim\'{e}tricas $N=2$ relevantes para derivar las
relaciones BPS.

El supermultiplete quiral $\Psi$ de supersimetr\'{\i}a $N=2$ puede ser
expresado en t\'erminos de supermultipletes $N=1$ como
$\Psi=(V,\Phi)$ donde $V$ es un supermultiplete vectorial y $\Phi$
un supermultiplete escalar quiral, ambos de $N=1$. En t\'erminos
de campos tenemos $(A_\mu,\Sigma,\phi,D,F)$ donde $\Sigma$ es
ahora un fermi\'on de Dirac, $\Sigma=(\lambda_1,\lambda_2)$,
$\phi$ un escalar complejo, $\phi = \phi_1 + i\phi_2$ y, $D$ y
$F$, campos auxiliares. Las transformaciones supersim\'{e}tricas del
gaugino toman la forma (llamando $\epsilon =
(\epsilon_1,\epsilon_2)$ al par\'ametro de la transformaci\'on
$N=2$)\cite{susy}
\be
 \delta \lambda_i = (\Gamma^{\mu \nu}F_{\mu\nu} + \gamma_5D)\,
 \epsilon_i +i \varepsilon_{ij}
 (F + \gamma^\mu\nabla_\mu (\phi_1 + \gamma_5 \phi_2))\,\epsilon_j
 \label{sus}
\ee
donde
\be
 \Gamma^{\mu \nu} = \frac{i}{4} [\gamma^\mu,\gamma^\nu]
 \label{G}
\ee
Configuraciones de instantones corresponden a  $D=F=\phi=0$ de
manera que (\ref{sus}) se simplifica a
\be
 \delta \lambda_i = \Gamma^{\mu \nu}
 F_{\mu\nu}\,\epsilon_i\quad\quad (i=1,2)
 \label{be}
\ee
Construyendo un fermi\'on de Dirac
$\Psi\equiv\lambda_1+i\lambda_2$, a partir de los fermiones de
Majorana $\lambda$ y usando la propiedad (\ref{dudu}) obtenemos
\be
 \delta \Psi = \frac{1}{2}\Gamma^{\mu \nu}(F_{\mu\nu} + i
 \gamma_5 \tilde{F}_{\mu\nu} )\,\Upsilon
 \label{be2}
\ee
donde $\Upsilon$ es el par\'ametro de supersimetr\' \i a $N=2$
(fermi\'on de Dirac). Como vimos en el cap\' \i tulo anterior, una
forma simple de encontrar relaciones BPS es imponer $\delta \Psi =
0$. Denotando $\Upsilon=\left(\begin{array}{c}
  \chi_\alpha \\
  \bar\xi^{\dot\alpha}
\end{array}\right)$, obtenemos
\begin{eqnarray}
 (F_{\mu\nu} + i \tilde{F}_{\mu\nu})\, \chi & = & 0 \nonumber\\
 (F_{\mu\nu} - i \tilde{F}_{\mu\nu})\,\bar\xi & = & 0
 \label{be3}
\end{eqnarray}
En espacio eucl\' \i deo, las ecs.(\ref{be3}) toman la forma
\begin{eqnarray}
 (F_{\mu\nu} + \tilde{F}_{\mu\nu})\, \chi& = & 0
 \nonumber\\
 (F_{\mu\nu} - \tilde{F}_{\mu\nu})\,\bar\xi & = & 0
 \label{be4}
\end{eqnarray}
donde $\chi$ y $\xi$ son los par\'ametros fermi\'onicos de Weyl
correspondientes a las dos supersimetr\' \i as. Esta condiciones
dan origen a las ecuaciones de instant\'on y anti-instant\'on
\be
 F_{\mu\nu} = \pm  \tilde{F}_{\mu\nu}
 \label{inst}
\ee
Claramente se ve que cada una de estas soluciones es invariante
frente a la mitad de las supersimetr\' \i as.

El hecho de que las ecuaciones de autodualidad de Yang-Mills
surjan tambi\'en cuando la din\'amica del campo de gauge est\'a
gobernada por el lagrangiano no abeliano de BI fue observado con
anterioridad en \cite{Hashi},\cite{HT},\cite{G1}. En el contexto
supersim\'etrico, esto puede ser entendido siguiendo las
conclusiones del cap\' \i tulo anterior (ver tambi\'en \cite{CS})
donde vimos que las transformaciones supersim\'etricas del
gaugino, junto con las ecuaciones (algebraicas) de movimiento para
los campos auxiliares, hacen que las relaciones BPS permanezcan
inalteradas independientemenete de la elecci\'on del lagrangiano
para el campo de gauge. Mas a\'un es posible ver que las cargas
supersim\'{e}tricas $N=2$ para una teor\' \i a no polin\'omica,
obtenidas mediante la construcci\'on de Noether, coinciden {\it on
shell}\,, con las que resultan en las teor\' \i as de Maxwell o
Yang-Mills.

\section{Discusi\'on}

En s\' \i ntesis, usando el formalismo de supercampos, derivamos
el lagrangiano supersim\'{e}trico no abeliano de Born-Infeld que
presenta la estructura BPS esperada, o sea  la de la teor\' \i a
normal de Yang-Mills. En nuestra construcci\'on hemos visto com\'o
los supercampos naturales a partir de los cuales se obtienen la
teor\' \i as no abelianas usuales constrinien severamente los
posibles lagrangianos no abelianos supersim\'etricos. La
derivaci\'on mostr\'o la posiblidad de supersimetrizar la
prescripci\'on de traza sim\'etrica Str para la extensi\'on
no-abeliana del lagrangiano de Born-Infeld reescribi\'endola en
t\'erminos de los bloques naturales en la construcci\'on
supersim\'{e}trica en t\'erminos de supercampos. Debe ser remarcado
que el hecho de que el lagrangiano puramente bos\'onico dependa
\'unicamente de los invariantes b\'asicos $F^2$ y $F\tilde F$ no
se debe a la elecci\'on de la traza sim\'etrica sino a haber
elegido a $W^2$ y $D^2W^2$ como bloques de construcci\'on de la
teor\' \i a supersim\'etrica. Finalmente, mencionemos que no solo
el lagrangiano supersim\'{e}trico no abeliano de BI sino toda una
familia de lagrangiano no polin\'omicos se encuentra inclu\' \i da
en nuestro resultado ec.(\ref{L}). Todos ellos son linealizados
por configuraciones BPS, que coinciden con las usuales de la
teor\' \i a de Yang-Mills.



\chapter{ Diones no-BPS y branas en la teor\' \i a de
Dirac-Born-Infeld\label{nobps}}

\begin{center}
 {\begin{minipage}{6truein}
 { \sl Construiremos soluciones di\'onicas no-BPS para la acci\'on
 de bajas energ\' \i as de una D3-brana excitando los campos
 escalares que describen las oscilaciones transversales de la
 brana. Analizaremos la imagen que emerge de tales configuraciones
 y, en particular, la respuesta de la soluci\'on D3-brana+cuerda a
 peque\~nas perurbaciones.

 Estas soluciones y su an\'{a}lisis corresponden a parte de los resultados
 originales de la tesis \cite{gui3}.}
 \end{minipage}}
\end{center}

\section{Introducci\'on}
Recientemente han sido discutidas en la literatura ciertas
soluciones a la teor\' \i a de Dirac-Born-Infeld (DBI) en
conexi\'on con la din\'amica de Dp-branas
\cite{Pol2}-\cite{Pol},\cite{CM}-\cite{BLM},\cite{Bre},\cite{G1},
\cite{t3}-\cite{GK}. Mas precisamente, la acci\'on de DBI para
campos de gauge $(p+1)$-dimensionales y un cierto n\'umero de
campos escalares, que describen las fluctuaciones transversales de
la brana, admite como soluciones configuraciones est\'aticas tipo
BPS y no-BPS, que pueden ser interpretadas en t\'erminos de
intersecciones de branas de distinta dimensionalidad, en
particular, branas y cuerdas. A pesar de que muchas propiedades
est\'aticas de estas intersecciones provienen de la invarianza
frente a supersimetr\'{\i}a y de argumentos BPS, las propiedades
din\'amicas de las soluciones dependen fuertemente de la no
linealidad de la acci\'on de DBI. En particular, las condiciones
de borde efectivas impuestas a las cuerdas sujetas a las branas
deben ser investigadas usando la acci\'on de DBI completa. Mas
a\'un, ciertas configuraciones no-BPS ser\' \i an relevantes en el
estudio de aspectos no perturbativos de las teor\' \i as de campos
que describen la din\'amica de bajas energ\' \i a de las Dp-branas
\cite{BGa}.

En \cite{CM}-\cite{G} se construyeron soluciones puramente
el\'ectricas BPS y no-BPS. Asimismo, la propagaci\'on de una
perturbaci\'on normal tanto a la cuerda como a la D3-brana fue
investigada en \cite{CM} para una soluci\'on de fondo de tipo BPS.
Los resultados obtenidos mostraron que la imagen de una cuerda
sujeta a la brana con condiciones de contorno de Dirichlet emerge
naturalmente de la din\'amica de DBI. En \cite{Svv2}-\cite{savi}
se estudiaron perturbaciones polarizadas a lo largo de la brana
para un fondo de tipo BPS y se mostr\'o que en este caso se
realizaban condiciones de borde de Neumann. De esta manera se
reafirm\'o la imagen de Polchinski de las Dp-branas como
hiperplanos en el espacion tiempo donde pueden terminar cuerdas
abiertas. Nuevas soluciones no-BPS para la acci\'on de DBI fueron
construidas en \cite{Hashi} donde tambi\'en fueron discutidas
soluciones BPS magn\'eticas. Un estudio detallado de las
soluciones BPS di\'onicas fue presentado en \cite{BLM}.

En este cap\' \i tulo nos concentraremos en el caso de D3-branas y
construiremos ex\-pl\' \i\-ci\-ta\-men\-te soluciones di\'onicas
no-BPS cuando el campo de gauge U(1) se acopla a un campo escalar.
Analizaremos luego las soluciones en conexi\'on con la geometr\'
\i a de la deformaci\'on de la brana por efecto de la tensi\'on de
la cuerda-$(n,m)$ que soporta cargas el\'ectricas y magn\'eticas
\cite{Wi},\cite{Sch}. Estudiando la energ\' \i a de estas
configuraciones no-BPS, compararemos los resultados con los
obtenidos en los casos BPS y no-BPS puramente el\'ectricos
\cite{CM}-\cite{BLM}. Estudiaremos tambi\'en peque\~nas
excitaciones, transversales tanto a la brana como a la cuerda, de
manera de examinar si la respuesta de las soluciones no-BPS es
consistente con la interpretaci\'on en la cual el sistema
D3-brana+cuerda descripto corresponde a las condiciones de borde
adecuadas (Dirichlet).

El plan del cap\' \i tulo es el siguiente: en la secci\'on
\ref{due} construiremos las soluciones no-BPS, con cargas
el\'ectricas y magn\'eticas, para el modelo DBI de un campo de
gauge abeliano acoplado a un campo escalar. Discutiremos las
propiedades de las soluciones y las compararemos con las
soluciones ya existentes en la literatura. Calcularemos luego la
energ\' \i a renormalizada e interpretaremos las soluciones
di\'onicas no-BPS en t\'erminos de cuerdas tirando de las
D3-branas. En la secci\'on \ref{tri} consideraremos peque\~nas
perturbaciones al fondo no-BPS, normales a la brana y a la cuerda,
de manera de examinar las condiciones de borde resultantes.
Finalmente discutiremos los resultados en la secci\'on
\ref{trocua}.

\section{Soluciones a la acci\'on de Dirac-Born-Infeld \label{due}}

\subsection{Acci\'on de Dirac-Born-Infeld}

Una Dp-brana (o membrana de Dirichlet p-dimensional) movi\'endose
en ${\BB M}_{10}$ describe una superficie $p+1$-dimensional. La
din\'amica que se deduce de la teor\' \i a de cuerdas para el
regimen de bajas energ\' \i as de las Dp-branas se expresa en
t\'erminos de funciones $z^M(x^{\mu})$ ($M=0,..,10$ y
$x^\mu=0,..,p$ ), que representan la posici\'on del volumen de
mundo (worldvolume) de la Dp-brana en ${\BB M}_{10}$, y un campo
de gauge $A_\nu(x^\mu)$ que vive en el volumen de mundo de la
Dp-brana. Las ecuaciones de movimiento para estos campos se
obtienen de la acci\'on de Dirac-Born-Infeld
\cite{BI},\cite{dirac},\cite{lei2}
\be
  S_{DBI}^{(p)} = -T_p \int d^{p+1}x \sqrt{-{\rm det} (G_{\mu \nu} +
  T^{-1} F_{\mu \nu})}
  \label{dbi}
\ee
donde
\be
 G_{\mu\nu}=\eta_{MN}\partial_\mu z^M \partial_\nu z^N
\ee
es la m\'etrica inducida en el volumen de mundo de la brana por la
m\'etrica plana $\eta_{MN}$ 10-dimensional  y
\be
 F_{\mu\nu}=\partial_\mu A_\nu-\partial_\nu A_\mu
\ee
es el tensor de campo electromagn\'etico. La propiedad BPS de la
Dp-brana permite deducir de la teor\' \i a de cuerdas que
\cite{Pol2}
\be
 T_p=\frac 1 {g_s}(2\pi)^{(1-p)/2}T^{(p+1)/2}
\ee
donde $T=(2\pi\alpha')^{-1}$ es la tensi\'on de la cuer\-da
fun\-da\-men\-tal y  $g_s$ es la cons\-tan\-te de
a\-co\-pla\-mien\-to para la interacci\'on de cuerdas.

La acci\'on (\ref{dbi}) es invariante frente a difeomorfismos en
el volumen de mundo de la brana. La forma habitual de fijar esta
libertad es imponer el gauge est\'atico que consiste en
identificar las coordenadas del volumen de mundo con las primeras
$p+1$ coordenadas del espacio-tiempo 10-dimensional
\be
 z^M=x^\mu,~~~~~~~~~~~~~~~~~M=0,1,..,p
\ee
Llamando a las restantes coordenadas transversales $X^a$
\be
 z^M=X^a,~~~~~~~~~~~~M=p+1,..,9,~~~~~a=p+1,..,9
\ee
tenemos luego que la acci\'on (\ref{dbi}) en el gauge est\'atico,
para el caso $p=3$ toma la forma\footnote{Para la discusi\'on de
la energ\' \i a de las soluci\'ones es conveniente normalizar la
acci\'on de manera que sea nula para campos nulos.}
\be
 S_{DBI}^{(3)} = T_3 \int d^4x \left(1-\sqrt{-{\rm det}
 (\eta_{\mu \nu} + T^{-1} F_{\mu \nu} +
 \partial_\mu X^a \partial_\nu X^a)}\right)
 \label{D3}
\ee
donde $\eta_{\mu \nu}$ es la m\'etrica de Minkowski en $3+1$
dimensiones, $\eta_{\mu \nu} = {\rm diag}(-,+,+,+)$\footnote{En
este cap\' \i tulo cambiamos la convenci\'on de la m\'etrica para
coincidir con la convenci\'on mayormente usada en el contexto de
Dp-branas.}, $F_{\mu\nu}$ es el tensor de campo electromagn\'etico
en el volumen de mundo (world volume) de la D3-brana, $X^a$ ($a =
4,5,\ldots,9$) son los campos escalares que describen las
fluctuaciones transversales de la brana y
\be
 T_3 = \frac{1}{2\pi g_s} T^2  \, , \;\;\; T = \frac{1}{2\pi \alpha'}
 \label{T3}
\ee

La acci\'on (\ref{D3}) puede ser obtenida tambi\'en por
reducci\'on dimensional de la acci\'on de Born-Infeld en espacio
plano 10-dim ($x^M$, $M=0,1,2, \ldots, 9$), si suponemos que los
campos solo dependen de las primeras $1+3$ coordenadas $x^\mu$ y
que las componentes extra $A_4, A_5 \ldots A_9$ del campo de gauge
representan los campos escalares transversales a la brana
\cite{CM},\cite{G}.

Consideraremos el caso en que excitamos solo un campo escalar,
$X^a = \delta^{a9} X$. En este caso la ec.(\ref{D3}) toma la forma
\begin{eqnarray}
 S_{DBI}^{(3)} \!\!\!&=\!\!\!& T_3 \int d^4x
 \left(\phantom{\sum^2} \!\!\! \!\!\!1-\left( \left(1 +
 \partial_\mu X\partial^\mu X \right) \left(1
 + \frac{1}{2T^2} F_{\mu \nu} F^{\mu \nu}\right)
 - \frac{1}{16 T^4}
 \left(\tilde F_{\mu \nu} F^{\mu \nu}
 \right)^2 \right.  \right.  \nonumber\\
 \!\!\!&\!\!\!&
 \left. \left. +
 \frac{1}{T^2}\partial^\mu XF_{\mu \nu} F^{\nu \rho}\partial_\rho X
 \right)^{1/2}\,\right)
 \label{tutix}
\end{eqnarray}
Reescribi\'endola en t\'erminos de los cam\-pos $\vec E$ y $\vec
B$ que\-da (ver ap\'en\-dice para las convenciones) \cite{CM},
\cite{G},\cite{savi}
\ba
 S_{DBI}^{(3)}\!\!\!&=\!\!\!& T_3 \int d^4x
 \left(\phantom{\sum^2} \!\!\! \!\!\! 1-
 \left(1+\frac 1 {T^2}(\vec B^2-\vec E^2)
 -\frac 1 {T^4}(\vec B\cdot\vec E)^2
 +(\vec \nabla X)^2-\dot X^2 \right.\right. \nonumber\\
 \!\!\!&\!\!\!&\!\!\!\!\!\left.\left.-\frac 1{T^2}\dot X^2\vec B^2
 +\frac 1{T^2}(\vec B\cdot\vec\nabla X)^2
 -\frac 1{T^2}(\vec E\times\vec\nabla X)^2-\frac 2{T^2}\dot X^2
 (\vec E\times\vec B\cdot\vec\nabla X)\right)^{1/2}\,\right)
 \label{dbic}
\ea

\subsection{Di\'on en la teor\' \i a de Born-Infeld}

Buscaremos soluciones a la acci\'on de Born-Infeld por dos
razones: corresponden a un contexto simplificado y son relevantes
de por si \footnote{De la acci\'on (\ref{tutix}) poniendo el campo
escalar $X=0$ recuperamos la acci\'on original de Born-Infeld
(\ref{tuti}).}. Las ecuaciones de movimiento que se derivan de la
acci\'on de Born-Infeld (\ref{tuti}) son
\be
 \partial_\mu \frac{F^{\mu\nu}-\frac 1{4T^2}(F\tilde F)
 \tilde F^{\mu\nu}}{\sqrt{1+\frac
 1{2T^2}F^2 -\frac 1{16T^4}(F\tilde F)^2}}=0
 \label{bi1}
\ee
Estas ecuaciones deben suplementarse con las identidades de
Bianchi
\be
 \partial_\mu \tilde F^{\mu\nu}=0
 \label{bi2}
\ee
El conjunto de ecuaciones (\ref{bi1})-(\ref{bi2}) puede ser
reescrito generalizando las ecuaciones de Maxwell
como\footnote{Las formulaciones de dualidad de la presente teor\'
\i a corresponde a escribir las ecuaciones (\ref{e1})-(\ref{e3})
en la forma
\be
 \partial_\mu (D^{\mu\nu}+i\tilde F^{\mu\nu})=0
\ee
y considerar rotaciones $(D^{\mu\nu}+i\tilde F^{\mu\nu})\to
e^{i\phi}(D^{\mu\nu}+i\tilde F^{\mu\nu})$
\cite{bialy}-\cite{zumi}. }
\ba
 \partial_\mu D^{\mu\nu}=0
 \label{e1}\\
 \partial_\mu \tilde F^{\mu\nu}=0
 \label{e3}
\ea
si definimos \cite{G},\cite{bialy}
\ba
 D^{\mu\nu}\!\!\!&=\!\!\!&-\frac{\delta S_{BI}}
 {\delta F_{\mu\nu}}\nonumber\\
 \!\!\!&=\!\!\!&\frac{F^{\mu\nu}-\frac 1{4T^2}(F\tilde F)
 \tilde F^{\mu\nu}}{\sqrt{1+\frac
 1{2T^2}F^2 -\frac 1{16T^4}(F\tilde F)^2}}
\ea
El l\' \i mite Maxwell corresponde a tomar $T\to\infty$ en la
ecuaci\'on (\ref{bi1}). Asociamos con $D^{\mu\nu}$ los vectores de
desplazamiento $D^{\mu\nu}=(\vec D, \vec H)$. La propiedad
fundamental de la teor\' \i a de Born-Infeld es la diferencia
entre los vectores ponderomotrices $(\vec E,\vec B)$ y los
vectores de desplazamiento $(\vec D,\vec H)$.

Veamos c\'omo la particular dependencia funcional de $F_{\mu\nu}$
en $D^{\mu\nu}$ genera soluciones de energ\' \i a finita. Dado que
estamos interesados en soluciones est\'aticas, las ecuaciones que
debemos resolver son
\ba
 {\rm div}~ \frac{\vec E+\frac 1 {T^2}(\vec E \cdot \vec B)\vec B}
 {\sqrt{1+\frac 1{T^2}(\vec B^2-\vec E^2)-\frac 1{T^4}
 (\vec E \cdot \vec B)^2}}\!\!\!&=\!\!\!&0
 \label{e}\\
 {\rm rot}~ \frac{\vec B-\frac 1 {T^2}(\vec E \cdot \vec B)\vec E}
 {\sqrt{1+\frac 1{T^2}(\vec B^2-\vec E^2)-\frac 1{T^4}
 (\vec E \cdot \vec B)^2}}\!\!\!&=\!\!\!&0
 \label{e2}
\ea
que provienen de (\ref{bi1}) y
\ba
 {\rm div}~\vec B\!\!\!&=\!\!\!&0
 \label{b}\\
 {\rm rot}~\vec E\!\!\!&=\!\!\!&0
 \label{b2}
\ea que provienen de las identidades de Bianchi (\ref{bi2}).
Proponiendo un ansatz radial para el di\'on y agregando fuentes en
las ecuaciones (\ref{e}) y (\ref{b}) obtenemos\footnote{Las
ecuaciones (\ref{e2}) y (\ref{b2}) se satisfacen trivialmente para
soluciones esf\'ericas. La fuente en la ecuaci\'on para $\vec B$
es ficticia y se interpreta  \`a la Dirac \cite{mono}
(cuantizaci\'on de la carga magn\'etica).}
\begin{eqnarray}
 \vec E \!\!\!&=\!\!\!& \frac{q_e}{4\pi\sqrt{ a_0^4 + r^4}~}~
 \check r
 \label{no}\\
 \vec B \!\!\!&=\!\!\!&  \frac{q_m}{4\pi r^2}  \check r
 \label{BE2}
\end{eqnarray}
donde $a_0^4=(q_e^2+q_m^2)/(4\pi T)^2$. Los campos de
desplazamiento del di\'on contenidos en $D^{\mu\nu}$ son
\ba
 \vec D\!\!\!&=\!\!\!&\frac{\vec E+\frac 1 {T^2}(\vec E \cdot \vec B)\vec B}
 {\sqrt{1+\frac 1{T^2}(\vec B^2-\vec E^2)-\frac 1{T^4}
 (\vec E \cdot \vec B)^2}}=\frac {q_e}{4\pi r^2}\check r\\
 \vec H \!\!\!&=\!\!\!& \frac{\vec B-\frac 1 {T^2}(\vec E \cdot \vec B)\vec E}
 {\sqrt{1+\frac 1{T^2}(\vec B^2-\vec E^2)-\frac 1{T^4}
 (\vec E \cdot \vec B)^2}}=\frac{q_m}{4\pi\sqrt{ a_0^4 + r^4}~}~
 \check r
\ea

El potencial vector $A_\mu=(-A^0,A_r,A_\theta,A_\phi)$ que origina
los campos (\ref{no})-(\ref{BE2}) es
\ba
 A^0\!\!\!&=\!\!\!&\int_r^\infty dr \frac {q_e} {4\pi\sqrt{ a_0^4 +
 r^4}}
 \label{ab}\\
 A_\phi\!\!\!&=\!\!\!&\frac {q_m (1-\cos \theta)}{4\pi r \sin
 \theta}\\
 A_r\!\!\!&=\!\!\!&A_\theta=0
 \label{ba}
\ea
Esta soluci\'on puede ser obtenida a partir de la soluci\'on
original de Born-Infeld \cite{BI} con fuentes puramente
el\'ectricas mediante una rotaci\'on de dualidad \cite{GR}. La
presente soluci\'on (\ref{ab})-(\ref{ba}) corresponde a la
generalizaci\'on di\'onica de dicha soluci\'on. La soluci\'on
(\ref{no}) muestra que el campo el\'ectrico (ponderomotriz) no
diverge en $r=0$. Computando la energ\' \i a tenemos \cite{bialy}
\ba
 {\cal E}\!\!\!&=\!\!\!&T^{00}=\vec D \cdot \vec E-{\cal
 L}_{BI} \nonumber\\
 \!\!\!&=\!\!\!&\frac {T^2}{2\pi g_s}\left(\sqrt{1+\frac 1 {T^2}
 (\vec D^2+\vec B^2)+\frac 1 {T^4}(\vec D \times \vec B)^2}-1\right)
\ea
que para la soluci\'on (\ref{no})-(\ref{BE2}) implica
\ba
 E_{di\acute{o}n}\!\!\!&=\!\!\!&4\pi \int_0^\infty dr\,r^2{\cal E}
 \nonumber\\
 \!\!\!&=\!\!\!&\frac {4 a_0^3T^2} {3g_s}\int_0^\infty dx \frac 1
 {\sqrt{1+x^4}~}\\
 \!\!\!&=\!\!\!&\frac{\Gamma^2(1/4)} {48\pi^{5/2}g_s} \,
 (q_e^2+q_m^2) \frac 1 {a_0}
 \label{pelea}
\ea
donde hemos usado que
\be
 \int_0^\infty dx \frac 1
 {\sqrt{1+x^4}~}=\frac {\Gamma^2(1/4)}{4\sqrt\pi}
\ee
De la expresi\'on (\ref{pelea}) vemos que en la teor\' \i a de
Born-Infeld los diones tienen energ\' \i a finita y que en el l\'
\i mite Maxwell ($a_0\to0$) la energ\' \i a de los diones
puntuales diverge. En la secci\'on siguiente reinterpretaremos
esta divergencia en el contexto de D-branas y cuerdas.

\subsection{Di\'on en la teor\' \i a de Dirac-Born-Infeld}

La existencia de soluciones BPS para la teor\' \i a DBI requiere
que excitemos al menos un escalar \cite{CM},\cite{T} con el objeto
de hallar estas soluciones y soluciones mas generales que
contemplen las que encontramos en la secci\'on anterior
escribiremos las ecuaciones de movimiento para la teor\' \i a de
DBI con un escalar excitado. Partiendo entonces de la acci\'on
(\ref{dbic}), las ecuaciones para soluciones independientes del
tiempo son
\begin{eqnarray}
 \!\!\!&\!\!\!&\vec \nabla \cdot \left( \frac{1}{R} \left( \vec \nabla X + \frac{1}{T^2}
 (\vec B \cdot  \vec \nabla X) \vec B +\frac{1}{T^2} \vec E \times (\vec E \times
 \vec \nabla X)
 \right)\right) = 0\nonumber\\
 \!\!\!&\!\!\!&\vec \nabla \cdot \left( \frac{1}{R} \left( \vec E
 +\vec \nabla X \times (\vec E \times
 \vec \nabla X)
 + \frac{1}{T^2} (\vec E \cdot \vec B) \vec B
 \right)\right) = 0\nonumber\\
 \!\!\!&\!\!\!&\vec \nabla \times \left( \frac{1}{R} \left( \vec B
 +  (\vec B \cdot
 \vec \nabla X) \vec \nabla X
 -\frac{1}{T^2} (\vec E \cdot \vec B) \vec E
 \right)\right)= 0
 \label{tch}
\end{eqnarray}
En lo que respecta a $R$, se la define como
\begin{eqnarray}
 R^2 \!\!\!&=\!\!\!&1+(\vec \nabla X)^2+\frac{1}{T^2}\left(\vec B^2-
 \vec E^2+(\vec B \cdot \vec \nabla X)^2-
 (\vec E \times \vec \nabla X)^2\right)
 \nonumber\\
 \!\!\!&\!\!\!& -\frac{1}{T^4}(\vec E \cdot \vec B)^2
 \label{R}
\end{eqnarray}

Dado que estamos interesados en soluciones de bi\'on
\cite{CM}-\cite{G} con cargas el\'ectrica $q_e$ y magn\'etica
$q_m$, necesariamente $\vec E$ y $\vec B$ tienen fuentes
puntuales. Las soluci\'ones para el caso $X = 0$ fueron halladas
en la secci\'on anterior (ecs.(\ref{ab})-(\ref{ba})).

Siguiendo \cite{G}-\cite{Hashi}, construimos la soluci\'on
empleando la simetr\' \i a residual del espacio 10-dimensional.
Efectuando un boost a la soluci\'on (\ref{ab})-(\ref{ba}) en la
direcci\'on $x^9$ obtenemos
\begin{eqnarray}
 X \!\!\!&=\!\!\!& -\frac{q_e\sqrt a}{4\pi T} \int_r^\infty dr \frac{1}{\sqrt{r^4_0 + r^4}}
 \nonumber\\
 A_0 \!\!\!&=\!\!\!&  -\frac{q_e}{4\pi} \int_r^\infty dr \frac{1}{\sqrt{r_0^4+ r^4}}
 \nonumber\\
 A_\varphi \!\!\!&=\!\!\!& \frac{q_m}{4\pi r} \frac{(1 - \cos \theta)}{\sin \theta}
 \; , \;\;\; A_\theta = A_r = 0
 \label{xab}
\end{eqnarray}
donde
\be
 r_0^4 = (4\pi T)^{-2}((1-a)q_e^2 + q_m^2)
 \label{cerraria}
\ee
y $a$ est\'a relacionada con el cuadrado de la velocidad del
boost. Los campos el\'ectrico y magn\'etico asociados con
(\ref{xab}) toman la forma
\begin{eqnarray}
 \vec E \!\!\!&=\!\!\!& \frac{q_e}{4\pi\sqrt{ r_0^4 + r^4}} \check r \nonumber\\
 \vec B \!\!\!&=\!\!\!&  \frac{q_m}{4\pi r^2}  \check r
 \label{BE}
\end{eqnarray}
Notemos que dado que el boost es la direcci\'on  $x^9$, no afecta
las direcciones transversales $x^i \, , i=1,2, \ldots,8$. M\'as
a\'un, dado que estamos considerando soluciones est\'aticas,
$A_\varphi$ en (\ref{xab}) no es afectado y el boost deja intacto
campo magn\'etico.

Estas soluciones, otra contribuci\'on original de esta tesis,
generalizan todas las soluciones conocidas (de una fuente) en el
contexto DBI-branas discutidas en la literatura, tanto las BPS
como las no-BPS. Poninedo $q_m=0$ recuperamos, para $a<1$, los
BIones el\'ectricos, para $a>1$, las soluciones el\'ectricas de
garganta (throat/catenoid) y, para $a=1$, las soluciones BPS
el\'ectricas \cite{CM},\cite{G},\cite{G1},\cite{Hashi}. Las nuevas
soluciones que hemos encontrado generalizan las soluciones de
BIones y gargantas el\'ectricas al caso de diones. En lo que
respecta al campo magn\'etico $\vec B$, es importante destacar que
su valor es independiente de $a$ siendo esto consistente con la
afirmaci\'on de la cuantizaci\'on de su carga. Es m\'as, la
inducc\'on magn\'etica $\vec H$ est\'a dada por
\be
 \vec H \equiv \frac{1}{R}\left( \vec B + \left(
 \vec B \cdot \vec \nabla X
 \right) \vec \nabla X - \frac{1}{T^2}\left(\vec E \cdot \vec B
 \right) \vec E
 \right) = \frac{q_m}{4\pi\sqrt{ r_0^4 + r^4}} \check r
 \label{HH}
\ee
Las cargas el\'ectricas y magn\'eticas de las soluciones fueron
ajustadas de manera que
\be
 \int_{S_\infty} dS_i E^i = q_e  \, , \;\;\;\;   \;\;\;\;  \int
 _{S_\infty} dS_i B^i = q_m
 \label{cargas}
\ee
Es conveniente definir la carga del campo escalar $q_s$ como
\be
 q_s \equiv T \int_{S_\infty} dS_i \partial^i X = \sqrt a \, q_e
 \label{lades}
\ee
En t\'erminos de estas cargas, $r_0$ en (\ref{cerraria}) toma la
forma
\be
 r_0^4 = (4\pi T)^{-2} \left(q_e^2 +q_m^2 - q_s^2\right)
 \label{qq}
\ee
%
\begin{figure}
 \vspace{-3.7cm}
 \centerline{ \psfig{figure=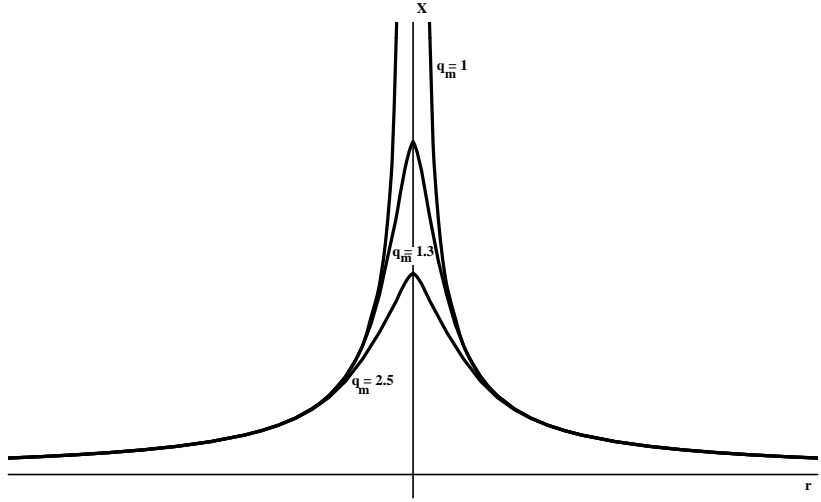,height=14cm,angle=0} }
 \vspace {-3.5cm}
 \caption{ Campo escalar $X$ como funci\'on de $r$ para un valor de carga
 el\'ectrica fija, $q_e=1$ ,y para distintos valores de la carga magn\'etica $q_m$.
 El l\' \i mite BPS se obtiene para $q_m^{BPS}= 1$ cuando $q_e=1$ y $q_s= \sqrt 2$.
 \label{fig-1} }
\end{figure}
De (\ref{xab}) podemos ver que, para el regimen $q_s^2 \leq q_m^2
+ q_e^2$, la soluci\'on para el campo escalar toma esencialmente
la forma descripta en la Fig. \ref{fig-1}. Cualitativamente, su
comportamiento es similar a la soluci\'on puramente el\'ectrica en
la teor\' \i a pura BI y tambi\'en al encontrado en \cite{Hashi}
excepto por la existencia de la carga magn\'etica no nula que
disminuye la altura del pico.

Para $q_s^2> q_e^2 +q_m^2$ el campo escalar toma la forma que se
muestra en la Fig. \ref{fig-2} que puede ser interpretada como dos
branas asint\'oticamente planas (de hecho un par brana-antibrana)
unidas por una garganta (throat) de radio $r_t$. Estas branas se
encuentran separadas por una distancia $\Delta = 2\; |\, X(r_t)|$
que corresponde a la diferencia entre los dos valores
asint\'oticos de $X(r)$. El radio de la garganta est\'a dado por
\be
 r_t^4 = -r_0^4 = (4\pi T)^{-2}\left(q_s^2 - q_e^2 - q_m^2\right)
 \label{th}
\ee
Notemos que el aumento de la carga magn\'etica hace que la
garganta adelgace y que $\Delta$ aumente. Las soluciones BPS
corresponden al caso $r_0 = 0$. Esto es, cuando la carga escalar
satisface
\be
 q_s^{BPS} = \pm \sqrt{q_e^2 + q_m^2}
 \label{has}
\ee
%
\begin{figure}
 \vspace{-3.3cm}
 \centerline{ \psfig{figure=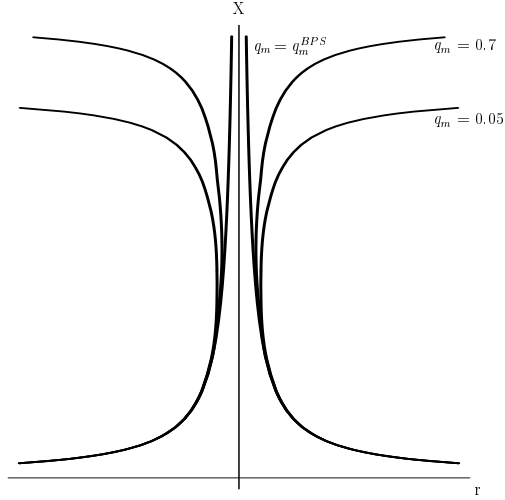,height=16cm,angle=0}}
 \vspace{-6cm}
 \caption{Campo escalar $X$ como funci\'on de $r$ para un valor de carga
 el\'ectrica fija, $q_e=1$, y para distintos valores de la carga magn\'etica $q_m$.
 El l\' \i mite BPS se obtiene para $q_m^{BPS}= 1$ cuando $q_e=1$ y $q_s= \sqrt 2$.
 \label{fig-2} }
\end{figure}

Cuando se satisface (\ref{has}), las soluciones de las ecuaciones
de Bogomol'nyi
\begin{eqnarray}
 \vec E &=& T \cos \xi \vec \nabla X\label{quese}
 \\
 \vec B &=& T\sin \xi \vec \nabla X
 \label{quesecalle}
\end{eqnarray}
son
\begin{eqnarray}
 X&=& - \frac{\sqrt{q_e^2 + q_m^2}} {4\pi T r} \nonumber\\
 \vec E &=& \frac{q_e}{4\pi r^2}\check r\nonumber\\
 \vec B &=& \frac{q_m}{4\pi r^2}\check r
 \label{grandiestadormido}
\end{eqnarray}
con\footnote{En el l\'\i mite BPS se tiene
\ba
 \vec D\!\!\!&=\!\!\!& \vec E \\
 \vec H\!\!\!&=\!\!\!& \vec B \\
  e^{i\xi}q_s \!\!\!&=\!\!\!& q_e+iq_m~~~~.
\ea}
\begin{eqnarray}
 \cos \xi = \frac{q_e}{\sqrt{q_e^2 + q_m^2}}
 \label{cadacosa}
\end{eqnarray}
Se\~nalemos que $r_0 = 0$ implica que el par\'ametro de boost
$a=1+ q_m^2/q_e^2 >1$. En particular, para $q_m=0$ el boost es al
cono de luz. De hecho, $q_m = 0$ corresponde a $\xi = 0$ y
entonces (\ref{xab}) se reducen a las soluciones BPS el\'ectricas
discutidas en \cite{CM},\cite{G}. La elecci\'on $\xi = \pi/2$
($q_e=0$) corresponde a las soluciones BPS magn\'eticas discutidas
en \cite{G}. Para $\xi$ arbitrario nuestras soluciones BPS
coinciden con las analizadas en \cite{BLM}. Computaremos ahora la
energ\' \i a para las configuraci\'ones no-BPS descriptas
anteriormente y la interpretaremos en t\'erminos de cuerdas unidas
a branas.

\begin{figure}
 \vspace{-3.5cm}
 \centerline{ \psfig{figure=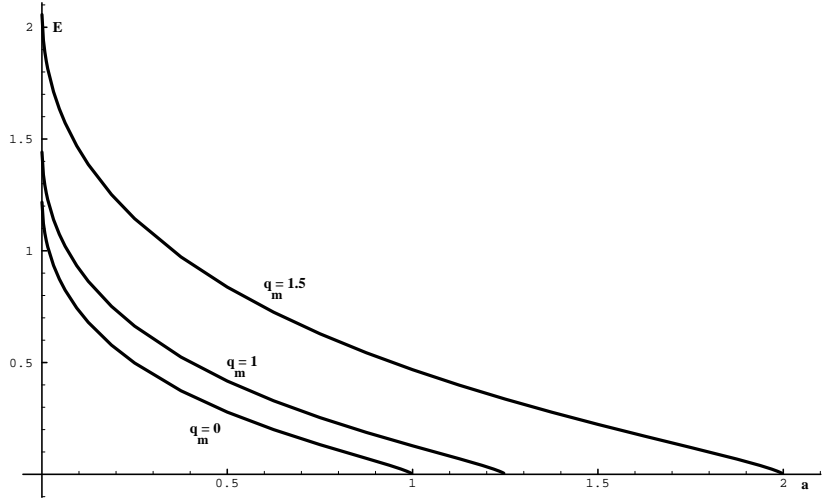,height=14cm,angle=0}}
 \vspace{-3.5cm}
 \caption{Energ\' \i a de la configuraci\'on no-BPS (una D3-brana
 deformada por una cuerda adherida a ella) como funci\'on de
 $a = q_s^2/q_e^2$ para distintos valores de carga magn\'etica $q_m$.
 \label{fig-3} }
\end{figure}

Consideremos el caso de una sola D3-brana y calculemos la energ\'
\i a almacenada en el volumen de mundo de la brana para la
configuraci\'on  (\ref{xab})-(\ref{BE}) cuando $r_0^4 \geq 0$,
\begin{eqnarray}
 E_{wv} &=& \int d^3x \, T_{00} \nonumber\\
 &=& \frac{2}{g_s} \int r^2 dr \left(
 \frac{q_e^2 + q_m^2 }{(4\pi r)^2\sqrt{r^4 + r_0^4}} + T^2 \left(
 \frac{r^2}{\sqrt{r^4 + r_0^4}} - 1\right) \right)\nonumber\\
 &=& \frac{\Gamma^2(1/4)}{96 \pi^{5/2} g_s}
 \left(  {2}  (q_e^2 + q_m^2) + q_s^2 \right) \frac{1}{r_0}
 \label{divergetodo}
\end{eqnarray}
Dado que el l\' \i mite BPS se alcanza cuando $r_0 = 0$, vemos que
la energ\' \i a $E_{wv}$ diverge precisamente en el punto que
deber\' \i a corresponder a la cota m\' \i nima de la energ\' \i
a. La manera de evitar este problema es normalizar la energ\' \i a
respecto del valor de Bogomol'nyi. Con este fin definimos
\begin{eqnarray}
 E &\equiv&  E_{wv} - E_{sub} = E_{wv} - \frac{\Gamma^2(1/4)}{ 32 \pi^{5/2} g_s }
 |\, q_s| \sqrt{q_e^2 + q_m^2} \frac{1}{r_0} \nonumber\\
 &=&
 \frac{\Gamma^2(1/4)}{6 \sqrt \pi  g_s}T^2
 \left( 2 - \frac{3|\, q_s|}{|\, q_s| + \sqrt{q_e^2 + q_m^2}   }
 \right) r_0^3
 \label{sub}
\end{eqnarray}
Claramente tenemos que $E=0$ para el caso BPS ($r_0 = 0$). En
general, $0 \leq E < \infty$ de manera que la configuraci\'on BPS
da una cota m\' \i nima para la energ\' \i a. En la Fig.
\ref{fig-3} se muestra la energ\' \i a dada por (\ref{sub}) como
funci\'on de la carga escalar. A carga el\'ectrica fija, podemos
ver que, al crecer la carga magn\'etica, la cota de Bogomol'nyi se
alcanza para valores mayores de la carga escalar.

Podemos interpretar la substracci\'on realizada en (\ref{sub}) de
la siguiente forma: usando la ec.(\ref{xab}), $E_{wv}$, dada por
la  ec.(\ref{divergetodo}), puede ser reescrita como
\be
 E_{wv} =    \frac{T}{6\pi g_s}  \frac{1}{|\, q_s|} \left(2(q_e^2 + q_m^2) + q_s^2 \right)
 \, |X(0)|
 \label{lar}
\ee
En lo que concierne al t\'ermino substraido, el mismo toma la
forma
\be
 E_{sub}
 =  \frac{T}{2\pi g_s} \sqrt{q_e^2 + q_m^2} \, |X(0)|
 \label{latex}
\ee
La conexi\'on entre el campo el\'ectrico del di\'on y las cuerdas
fundamentales (cargadas frente al campo $B_{\mu\nu}$) da origen a
la cuantizaci\'on del flujo del campo el\'ectrico
\cite{stro}-\cite{G} de manera que $q_e =2\pi g_s n $. Para la
carga magn\'etica, escribimos $q_m = 2\pi m$. Luego, $E_{sub}$
puede ser reescrita en la forma
\be
 E_{sub}
 =T\sqrt{n^2 + \frac{1}{g_s^2}m^2} |X(0)|
 \label{nm}
\ee
Es posible finalmente reescribir la energ\' \i a renormalizada $E$
definida en (\ref{sub}) como
\begin{eqnarray}
 E&=&E_{wv} - T_{(n,m)}\int_0^{|X(0)|} dX
 \label{joder}\\
 &=& E_{wv} + T_{(n,m)} \int_{|X(0)|}^\infty dX  -
 T_{(n,m)} \int_0^\infty dX
 \label{tbdral}
\end{eqnarray}
donde
\be
 T_{(n,m)} =  T \sqrt{n^2 + \frac{1}{g_s^2}m^2}
 \label{sss}
\ee
La f\'ormula (\ref{tbdral}) pone de manifiesto la raz\'on de la
substracci\'on: el segundo t\'ermino del lado derecho de
(\ref{tbdral}) representa la energ\' \i a de una cuerda
semi-infinita (de tensi\'on $T_{(n,m)}$) extendi\'endose desde la
punta del pico hasta el infinito. El tercer t\'ermino substrae la
energ\' \i a (infinita) de una cuerda que se extiende desde $0$
(donde se encuentra la brana plana) hasta el infinito. Estamos
computando entonces, la energ\' \i a de una brana deformada por
una cuerda con respecto a la energ\' \i a de una configuraci\'on
no-interactuante brana-cuerda (que de hecho resulta ser, como es
habitual una configuraci\'on BPS)\footnote{En una configuraci\'on
BPS no existe energ\' \i a de interacci\'on entre  los
constituyentes del sistema.}.
%
\begin{figure}
 \vspace{-3.5cm}
 \centerline{ \psfig{figure=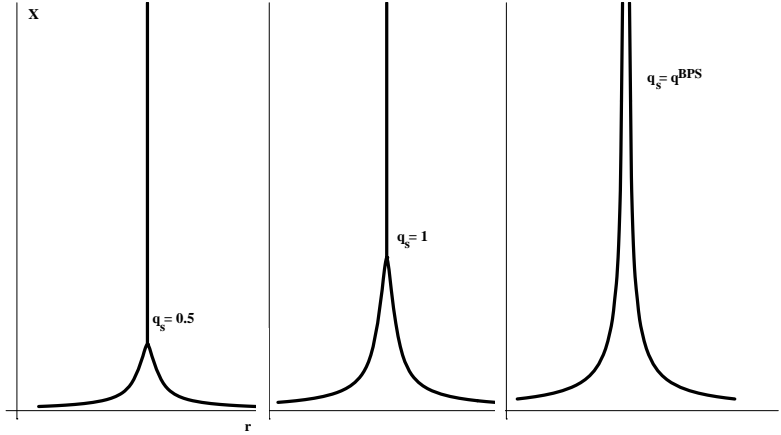,height=14cm,angle=0}}
 \vspace{-4cm}
 \caption{ Deformaci\'on de la brana, debida a la cuerda, como funci\'on
 de $r$ para valores de la carga escalar (en unidades apropiadas)
 variando entre $q_s = 0.5$ (izq.) y el valor BPS
 $q_s^{BPS}= \sqrt{q_e^2 + q_m^2} = \sqrt2$ (der.).
 \label{fig-4} }
\end{figure}

En la Fig. \ref{fig-4} se representa una secuencia de
deformaciones cuando se aumenta la carga escalar hasta alcanzar el
valor BPS $q_s^{BPS}$. Es posible tambi\'en computar la energ\' \i
a  est\'atica alamcenada en el volumen de mundo de la soluci\'on
de tipo garganta ($r_t^2 =-r_0^2>0$). Obtenemos
\be
 E_{wv} = \frac{\Gamma^2(1/4)}{48\sqrt 2 \pi^{5/2}g_s}
 \left(  {2}  (q_e^2 + q_m^2) + q_s^2 \right) \frac{1}{r_t} +
 \frac{4}{3}\frac{T^2}{g_s} r_t^3
 \label{eth}
\ee
que tambi\'en diverge en el l\' \i mite BPS ($r_t \to 0$). La
substracci\'on adecuada de manera de obtener un resultado finito
es
\be
 E_{sub}= \frac{\Gamma^2(1/4)}{ 16\sqrt 2 \pi^{5/2} g_s }
 |q_s| \sqrt{q_e^2 + q_m^2} \frac{1}{r_t}
 \label{susm}
\ee
Tenemos entonces para la garganta
\be
 E =  E_{wv} - E_{sub} = \frac{4}{3} \frac{T^2}{g_s}
 \left(1 +
 \frac{\Gamma^2(1/4)}{4 \sqrt{2\pi}}
 \frac{|q_s| -
  2\sqrt{q_e^2 + q_m^2}}
  {|q_s| +\sqrt{q_e^2 + q_m^2}}
 \right)r_t^3
 \label{beg}
\ee
La energ\' \i a substraida $E_{sub}$ definida por la
ec.(\ref{susm}) puede ser reescrita como
\be
 E_{sub} =  \frac{T}{2\pi g_s}\sqrt{q_e^2 + q_m^2} 2\vert X(r_t) | =
 T_{(n,m)} \Delta
 \label{bis}
\ee
donde $T_{(n,m)}$ est\'a definida por la ec.(\ref{sss}). Luego, la
energ\' \i a finita $E$ en (\ref{beg}) corresponde a la diferencia
entrela soluci\'on de garganta y la configuraci\'on
no-interactuante brana-cuerda-antibrana (separaci\'on infinita).
La interpretaci\'on de esta \'ultima soluci\'on en el contexto de
cuerdas permite computar el decaimiento por efecto tunel de un par
brana-antibrana \cite{CM},\cite{Svv},\cite{savi}.

Resumiendo, en esta secci\'on hemos construido soluciones a la
acci\'on  DBI, las hemos reinterpretado en terminos de
intersecciones de cuerdas y branas y le hemos dado sentido a las
expresiones para la energ\' \i a de dichas soluciones. En la
pr\'oxima secci\'on analizaremos peque\~nas perturbaciones
alrededor de las soluciones halladas.

\section{Din\'amica y condiciones de borde efectivas\label{tri}}

Analizaremos en la presente secci\'on, en el esp\' \i ritu de
\cite{CM}, la respuesta de la teor\' \i a a peque\~nas
fluctuaciones alrededor de las soluciones est\'aticas no-BPS
di\'onicas que hemos encontrado anteriormente. Tomamos como fondo
la soluci\'on (\ref{xab}) y estudiemos la propagaci\'on de una
perturbaci\'on $\eta$ del tipo onda-$s$, polarizada en una
direcci\'on perpendicular a la brana y a $\check X^9$, digamos
$\check X^8$. Partiendo de la acci\'on (\ref{D3}) y escribiendo la
perturbaci\'on $\eta$ alrededor de la soluci\'on est\'atica
(\ref{xab}), obtenemos desarrollando a segundo orden en la
fluctuaci\'on
\begin{equation}
 -\left(r^4 + \frac{q_e^2 + q_m^2}{(4\pi T)^2}
 \right) \ddot \eta(r,t)+ 2r^3 \eta'(r,t)
 +\left(r^4 +  r_0^4\right)\eta''(r,t) = 0
 \label{sipero}
\ee
Escrbiendo $\eta(r,t) = \eta(r) \exp(i \omega t)$ y definiendo $x
= \omega r$ la ecuaci\'on estacionaria est\'a dada por
\be
 \frac{1}{x^2}f(x) (x^2 f(x) \eta'(x))'+
 \frac{\kappa^2 + x^4}{x^4}\eta(x)  = 0
 \label{lr}
\ee
donde
\begin{equation}
 \kappa = \frac{\sqrt{q_e^2 + q_m^2}}{4\pi T}\omega^2
\end{equation}
y
\be
 f(x) = \frac{\sqrt{x^4 + \omega^4 r_0^4}}{x^2}
 \label{seju}
\ee
En el l\' \i mite BPS ($r_0 \to 0\Rightarrow f(x) \to 1$)
recuperamos los casos estudiados originalmente en
\cite{CM},\cite{Svv}. Para estudiar la ec.(\ref{lr}) hacemos un
cambio de variables de $x\to\xi$, donde entonces $\xi$ mide la
longitud de arco a lo largo de la superficie subtendida por $X$
\footnote{La elecci\'on del l\' \i mite inferior es para que el
pico del potencial est\'e situado en $\xi=0\longleftrightarrow
x=\sqrt k$}
\be
 \xi(r)=\omega\int_{\sqrt{\kappa}/\omega}^r d\tilde r
 \sqrt{1 +  {X'}^2(\tilde r)}
\ee
Usando la forma expl\' \i cita para $X$ dada por (\ref{xab}),
$\xi$ puede ser escrita en la forma
\be
 \xi(x) = \int_{\sqrt \kappa}^x dy
 \sqrt{\frac{y^4 + \kappa^2}{y^4 +  r_0^4 \omega^4}}
 \label{zz}
\ee
Definiendo
\be
 \tilde \eta (x) = \left( x^4 +{\kappa^2}\right)^\frac{1}{4}
 \eta(x)
 \label{zzz1}
\ee
la ec.(\ref{lr}) se transforma en una ecuaci\'on de Schr\"{o}dinger
unidimensional
\be
 \left(- \frac{d^2}{d \xi^2}   + V(\xi)
 \right) \tilde \eta(\xi) = \tilde \eta(\xi)
 \label{cho}
\ee
con potencial
\be
 V(\xi) = \frac{5\kappa^2 x^6}{\left(x^4 + \kappa^2\right)^3}
 +\frac{(4 \pi T)^2 }{q_e^2 + q_m^2}r_0^4\kappa^2 x^2
 \frac{3\kappa^2 - 2 x^4}{\left(\kappa^2 + x^4\right)^3}
 \label{malv}
\ee
El primer t\'ermino en (\ref{malv}) es formalmente id\'entico al
potencial en el l\' \i mite BPS \cite{CM} excepto que la
relaci\'on entre $\xi$ y $x$, dada por (\ref{zz}) depende de $r_0$
y, luego, coincide con el caso BPS solo para $r_0=0$. Otra
diferencia importante con el l\' \i mite BPS concierne al dominio
unidimensional donde el potencial (\ref{malv}) est\'a definido:
siendo nuestra soluci\'on no-BPS, $\xi$ se extiende desde un valor
finito (negativo) $\xi(0)$ hasta $+\infty$, dado que la c\'uspide
de la soluci\'on tiene una altura finita $X(0)$. Ahora, desde
$X^9= X(0)$ a infinito (esto es, en el intervalo
$\xi\in(-\infty,\xi(0))$) la perturbaci\'on actua directamente
sobre la acci\'on escalar libre de la cuerda semi-infinita unida a
la brana. Luego, en esta regi\'on tenemos en lugar de (\ref{cho}),
\be
 -\frac{d^2 \tilde \eta(\xi)}{d \xi^2}
  = \tilde \eta(\xi)
 \label{chos}
\ee
Podemos considerar entonces la ec.(\ref{cho}) en el dominio
unidimensional completo definiendo
\be
 V_{eff}(\xi) = \left \{
 \begin{array}{ll}
 0 & \mbox{if $-\infty <\xi<\xi(0)$}
 \\
 V(\xi) & \mbox{if $ \xi(0) <\xi <\infty$}
 \end{array}
 \right.
 \label{compv}
\ee

El potencial (\ref{compv}), correspondiente a una configuraci\'on
no-BPS, es m\'as complicado que el que se obtiene en el l\' \i
mite BPS, que fue estudiado originalmente en el l\' \i mite
$\kappa \to 0$ usando  aproximaciones de potencial tipo delta
\cite{CM} y barrera cuadrada \cite{Svv}. Utilizaremos este segundo
m\'etodo y aproximaremos el potencial por una barrera de
potencial, ajustando su altura y ancho de manera que la integral
de $V$ y $\sqrt V$ coincidan con la del potencial (\ref{compv}).
Definiendo
\be
 S = \int \sqrt{V(\xi)} d\xi \, , \;\;\;\;
 U = \sqrt \kappa\int V(\xi) d\xi
 \label{ss}
\ee
Podemos ver que, mediante un apropiado cambio de variables, tanto
$S$ como $U$ no dependen de $\kappa$. En t\'erminos de estas
cantidades, encontramos para las amplitudes de reflecci\'on y
transmisi\'on
\begin{eqnarray}
 R & = &  \frac{\exp \left(- i\sqrt \kappa S^2/U
 \right)}{-1 + ({2i \sqrt \kappa S}/{U}) {\rm coth} S} \nonumber\\
 T & = &  i\left(\frac{2\sqrt \kappa S}{U} {\rm cosech} S \right) R
 \label{mex}
\end{eqnarray}
La ec.(\ref{mex}) muestra que obtenemos una reflecci\'on total con
cambio de fase que se aproxima a $\pi$ en el l\' \i mite de bajas
energ\' \i as ($\kappa \to 0$). Calculando num\'ericamente $S$ y
$U$ es posible ver tambi\'en que el coeficiente de reflecci\'on
no-BPS $|R(r_0)|$ es ligeramente mayor que el BPS,
$|R(r_0)|>|R(0)|$.

Concluimos del an\'alisis anterior, que una perturbaci\'on
transversal a la cuerda unida a la brana, se refleja en
concordancia con los resultados esperados para condiciones de
borde de tipo Dirichlet: la amplitud de reflecci\'on $R$ va a $-1$
en el l\' \i mite de bajas energ\' \i as ($\kappa \to 0$). La
imagen que emerge est\'a de acuerdo con la idea de la D3-brana
actuando como frontera de las cuerdas abiertas
\cite{Pol2},\cite{Pol}.

\section{Resumen y discusi\'on \label{trocua}}

En resumen hemos construido soluciones di\'onicas no-BPS
correspondientes a la acci\'on de Dirac-Born-Infeld para un campo
de gauge U(1) en el volumen de mundo acoplado a un campo escalar y
las hemos analizado en el contexto de din\'amica de branas.  Si
bien nuestras soluciones incluyen las BPS ya discutidas en la
literatura, nos hemos concentrado en el sector no-BPS de manera de
examinar si esta caracter\' \i stica afecta la imagen debida a
Polchinski de cuerdas terminando en las branas. Una cantidad
importante en el an\'alisis de las soluciones no-BPS es el valor
de la carga escalar $q_s$ que puede ser expresado en t\'erminos de
las cargas el\'ectrica y magn\'etica como
\be
 q_s^2 = q_e^2 + q_m^2 - (4\pi T)^2 r_0^4
 \label{ul}
\ee
Para $r_0^4 > 0$ nuestras soluciones corresponden a una brana con
una punta (spike), para $r_0^4 < 0$ tenemos una soluci\'on de
brana-antibrana unidas por una garganta (throat). La energ\' \i a
substraida (renormalizada) de estas soluciones di\'onicas no-BPS
puede ser acomodada de manera que resulta natural adoptar la
imagen de una brana deformada por una cuerda de tensi\'on
$T_{(n,m)} = T\sqrt{n^2 + m^2/g_s^2}$ ($m$ y $n$ son el n\'umero
de flujos unitarios el\'ectricos y magn\'eticos de las soluci\'on)
unida a ella. Como se mostr\'o gr\'aficamente en la Fig.
\ref{fig-4}, al aumentar la carga escalar hasta el valor BPS
$q_s^{BPS}$, la punta en la brana crece y cuando el valor
$q_s^{BPS}$ es superado, la soluci\'on se convierte en un par
brana-antibrana unidas por una garganta. Finalmente hemos
estudiado el efecto de peque\~nas perturbaciones transversales
tanto a la cuerda como a la brana sobre la soluci\'on no-BPS,
mostrando mediante un an\'alisis de dispersi\'on que los
resultados corresponden a las condiciones de contorno de Dirichlet
esperadas. En particular, la amplitud de reflecci\'on para el
fondo no-BPS es ligeramente mayor que la hallada en el caso BPS y
tiende a  $-1$ en el l\' \i mite de bajas energ\' \i as.



\chapter{Conclusiones\label{conc}}

En esta tesis hemos estudiado distintos aspectos de modelos de
Dirac-Born-Infeld, cuyo inter\'{e}s ha crecido al aparecer como teor\'
\i as efectivas, a bajas energ\' \i as, de los modelos de cuerdas
que unifican todas las interacciones. Resumiremos en estas
conclusiones los principales resultados originales que hemos
obtenido en el trabajo de tesis.

\vspace{.5cm}

En primer lugar, en  el cap\'{\i}tulo \ref{susya} estudiamos la
extensi\'on supersim\'{e}trica, $N=2$, del modelo de Born-Infeld-Higgs
en 3 dimensiones de espacio-tiempo y sus relaciones de
Bogomol'nyi. En las extensiones supersim\'etricas de la teor\' \i
a de Born-Infeld que hab\'{\i}an sido discutidas previamente
\cite{DP}-\cite{CF}, solo el sector bos\'{o}nico hab\'{\i}a sido construido
expl\'{\i}citamente. Para derivar las relaciones de Bogomol'nyi a
partir del \'{a}lgebra de supersimetr\'{\i}a $N=2$, es necesario conocer
tambi\'{e}n el lagrangiano puramente fermi\'onico y de interacci\'{o}n (al
menos a orden cuadr\'atico) pues origina t\'{e}rminos en las
corrientes de Noether que de hecho constituyen las \'unicas
contribuciones no nulas al \'algebra de supersimetr\'{\i}a cuando \'{e}sta
se escribe en el sector puramente bos\'onico.

Llevamos a cabo entonces tal construcci\'{o}n y a partir del
an\'alisis del \'{a}lgebra, pudimos mostrar que la supersimetr\' \i a
fuerza una forma funcional particular para la acci\'on bos\'onica,
en la que el potencial de Higgs aparece dentro de la ra\'{\i}z cuadrada
de Born-Infeld (ver ec.(\ref{ac})). De esta manera la forma
funcional del potencial de ruptura de simetr\'{\i}a de gauge, que hab\'{\i}a
sido seleccionado en \cite{NS1} con el fin de tener las ecuaciones
de Bogomol'nyi usuales, result\'{o} un\' \i vocamente determinada por
la extensi\'{o}n supersim\'{e}trica del modelo. Como era de esperar,
encontramos que la carga central del \'algebra de supersimetr\'{\i}a
$N=2$ coincide con la carga topol\'ogica (el n\'umero de unidades
de flujo magn\'etico) del modelo, lo que muestra que la cota de
Bogomol'nyi no se modifica cuando la acci\'{o}n de Born-Infeld
reemplaza a la de Maxwell.

Trabajamos en 3 dimensiones de espacio-tiempo porque en este caso
se conocen soluciones de v\'ortice del modelo de Maxwell-Higgs que
satisfacen ecuaciones de Bogomol'nyi. Como hecho notable,
encontramos que las mismas ecuaciones (y luego, el mismo conjunto
de soluciones) son v\'{a}lidas cuando la din\'amica del campo de gauge
est\'a determinada por el lagrangiano de Born-Infeld. Esto nos
indujo a analizar el conjunto de lagrangianos supersim\'etricos
con un sector bos\'{o}nico dependiente de los invariantes
fundamentales $ F^{\mu \nu} F_{\mu \nu}$ y $ \tilde F^{\mu \nu}
F_{\mu \nu}$. Este conjunto incluye, para una particular
elecci\'on de coeficientes, el lagrangiano supersim\'etrico de
Born-Infeld, y tambi\'en una clase infinita de lagrangianos que
tienen propagaci\'on causal \cite{DP}. Mostramos en este contexto
el porqu\'e las relaciones de Bogomol'nyi son insensibles a la
elecci\'{o}n del lagrangiano para el campo de gauge: lagrangianos de
Maxwell, de Born-Infeld y no polin\'omicos mas complicados tienen
la misma estructura BPS, siempre que el acoplamiento con el campo
de Higgs sea m\' \i nimo.

En la definici\'on de la acci\'{o}n de Born-Infeld no abeliana existe
una ambig\"uedad relacionada con la forma de definir la estructura
de traza en los \' \i ndices de grupo. Motivados por un trabajo
previo \cite{Bre} donde se se\~nal\'{o} que  una cierta elecci\'{o}n de
traza (sim\'etrica), propuesta en \cite{Tse2}, permite la
existencia de una cota de Bogomol'nyi, y conocida la relaci\'{o}n
entre cotas y supersimetr\'{\i}a, decidimos estudiar la extensi\'{o}n
supersim\'{e}trica de dicha acci\'{o}n no abeliana. Presentamos nuestros
resultados en el cap\'{\i}tulo \ref{nstr}.

En particular, mostramos que la acci\'{o}n definida en base a la
prescripci\'{o}n de traza sim\'{e}trica es com\-pa\-ti\-ble con
supersimetr\'{\i}a. Para ello construimos el lagrangiano supersim\'{e}trico
$N=1$ en $d=4$ (ec.(\ref{L})), empleando los supercampos de
curvatura usuales como bloques b\'{a}sicos. Obtuvimos un lagrangiano
que, en su sector bos\'onico, depende solo de los invariantes
$F_{\mu\nu}F^{\mu \nu}$ y $F_{\mu\nu}\tilde F^{\mu \nu}$. La
prescripci\'{o}n ``Str" permiti\'{o} definir la extensi\'{o}n no abeliana de
la teor\' \i a de Born-Infeld como una ra\'{\i}z cuadrada del
determinante de $g_{\mu \nu} + F_{\mu \nu}$, eliminando las
ambig\"uedades en el ordenamiento del determinante, como asi
tambi\'{e}n del desarrollo en serie de la raiz cuadrada. El m\'{e}todo de
construcci\'{o}n di\'{o} origen a una familia de lagrangianos que,
similarmente al caso abeliano, resultan estar acotados por una
cantidad topol\'{o}gica, saturando la cota (en espacio plano de
Minkowski) cuando $F_{\mu\nu}=\pm i\tilde F_{\mu\nu}$. Mostramos
tambi\'{e}n c\'{o}mo estas ecuaciones BPS resultan del an\'{a}lisis de las
variaciones supersim\'{e}tricas.

Como mencionamos arriba, la extructura de traza de la teor\' \i a
de BI fue fijada en \cite{Bre} requiriendo que la acci\'on quedase
linealizada en las configuraciones BPS (instantones, monopolos,
v\'ortices). En nuestro an\'alisis, pudimos ver que la traza
sim\'etrica surge naturalmente en el formalismo de supercampos,
cuando se construye la ra\'{\i}z cuadrada del lagrangiano de BI. Esta
confluencia de resultados no es mas que una manifestaci\'on de la
conocida conexi\'on entre supersimetr\'{\i}a y relaciones de
Bogomol'nyi.

\vspace{.5in}

En el cap\'{\i}tulo \ref{nobps} estudiamos la acci\'{o}n de
Dirac-Born-Infeld en $d=4$, en relaci\'{o}n a la din\'{a}mica de bajas
energ\' \i as de una D3-brana, en la teor\' \i a de supercuerdas
tipo IIB. Extendimos las soluciones de las ecuaciones de
movimiento conocidas, el\'{e}ctricas, BPS y no-BPS, al caso magn\'{e}tico.
As\'{\i}, obtuvimos soluciones di\'onicas no-BPS de las ecuaciones de
movimiento de Dirac-Born-Infeld para un campo de gauge U(1) en el
volumen de mundo, acoplado a un campo escalar. Analizando luego
estas soluciones en el contexto de din\'amica de branas. Nos
concentramos en el sector no-BPS de las soluciones,de manera de
examinar si esta caracter\' \i stica afecta la imagen de
Polchinski \cite{Pol2} en la que las cuerdas terminan en branas.
La carga del campo  escalar $q_s$ se expresa, para las soluciones
de branas, en t\'erminos de las cargas el\'ectrica y magn\'etica
como
\be
 q_s^2 = q_e^2 + q_m^2 - (4\pi T)^2 r_0^4
\ee
Para $r_0^4 > 0$ nuestras soluciones corresponden a una brana con
una punta, para $r_0^4 < 0$ se tiene una soluci\'on de
brana-antibrana unidas por una garganta. La energ\' \i a
substra\'{\i}da (renormalizada) de estas soluciones di\'onicas no-BPS
puede ser acomodada de manera que resulta natural interpretar a la
soluci\'{o}n como una brana deformada por una cuerda de tensi\'on
$T_{(n,m)} = T\sqrt{n^2 + m^2/g_s^2}$ ($m$ y $n$ son el n\'umero
de flujos unitarios el\'ectricos y magn\'eticos de las soluci\'on)
unida a ella. Este resultado es esperable en base  a la simetr\'
\i a S de la teor\' \i a de supercuerdas IIB. Como se mostr\'o
gr\'aficamente en la Fig. \ref{fig-4}, al aumentar la carga
escalar hasta el valor $q_s^{BPS}$, la punta en la brana crece y
cuando este valor es superado, la soluci\'on se convierte en un
par brana-antibrana unidas por una garganta. De esta manera,
muestras soluciones realizan los dos conjuntos de soluciones
discutidos en \cite{G}. Finalmente estudiado el efecto de
peque\~nas perturbaciones transversales a la cuerda y a la brana.
Mostramos,  mediante un an\'alisis de dispersi\'on, que los
resultados son consistentes con las condiciones de contorno de
Dirichlet esperadas. En particular, vimos que la amplitud de
reflecci\'on para el fondo no-BPS es ligeramente mayor que la
hallada en el caso BPS y tiende a $-1$ en el l\' \i mite de bajas
energ\' \i as.



\appendix

\chapter{Convenciones\label{a1}}

\underline{M\'etrica e Indices}
\begin{itemize}
\item Usamos \' \i ndices griegos $\mu ,\nu , ..=0,1,2,3$, para las
coordenadas del espacio-tiempo $x^{\mu}$.
\item Los \' \i ndices latinos $i,j,k$ denotan coordenadas
espaciales $x^i=(x^1,x^2,x^3)=(x,y,z)$. Alternativamente usamos
$\vec x=(x^1,x^2,x^3)$ para las componentes contravariantes.
\item La signatura de la m\'etrica  es $g_{\mu\nu}={\rm
diag}(+,-,-,-)$ (cap\' \i tulos 3 y 4) y $g_{\mu\nu}={\rm
diag}(-,+,+,+)$ (cap\' \i tulo 5).
\item  El tensor de Levi-Civita $\varepsilon_{\mu\nu\rho\sigma}$ es totalmente antisim\'etrico con
$\varepsilon_{0123}=1~(\varepsilon^{0123}=-1)$.
\item El tensor totalmente antisim\'etrico en dimensiones
espaciales se define con $\varepsilon^{123}=1$ sin distinguir
entre \' \i ndices covariantes o contravariantes.
\end{itemize}

\underline{Campo electromagn\'etico}

\begin{itemize}

\item i) M\'etrica $g_{\mu\nu}={\rm diag}(+,-,-,-)$

A partir de la 1-forma $A_{\mu}=(\phi,-\vec A)$ se define
\ba
 \vec E&=&-\vec \nabla \phi-\partial_t \vec A  \\ %
 \vec B&=&\vec \nabla \wedge \vec A  %
\ea
que se expresa en notaci\'on covariante como
\be
F_{\mu\nu}=\partial_\mu A_\nu-\partial_\nu A_\mu \ee
\be
F_{\mu\nu}=\left(
\begin{array}{cccc}
  0             & E_x           & E_y        & E_z  \\
  -E_x          & 0             & -B_z       & B_y  \\
  -E_y          & B_z           &  0         & -B_x \\
  -E_z          & -B_y          & B_x        & 0
\end{array}
\right) \ee
dando las siguientes relaciones
\ba
 &E^i=E_i=&F_{0i}  \nonumber \\
 &B^i=B_i=&-\frac 1 2 {\varepsilon_{ijk}}  F_{jk}
\ea
\ba
 &F_{0i}=&E_i \nonumber \\ %
 &F_{ij}=&-\varepsilon_{ijk} B_k
\ea

\item ii) M\'etrica $g_{\mu\nu}={\rm diag}(-,+,+,+)$

\noindent A partir de la 1-forma $A_{\mu}=(-\phi,\vec A)$ se
define
\ba
 \vec E&=&-\vec \nabla \phi-\partial_t \vec A  \\ %
 \vec B&=&\vec \nabla \wedge \vec A  %
\ea
que se expresa en notaci\'on covariante como
\be
F_{\mu\nu}=\partial_\mu A_\nu-\partial_\nu A_\mu \ee
\be
F_{\mu\nu}=\left(
\begin{array}{cccc}
  0             & -E_x           & -E_y        & -E_z  \\
  E_x           & 0              & B_z         & -B_y  \\
  E_y           & -B_z           &  0          & B_x \\
  E_z           & B_y            & -B_x        & 0
\end{array}
\right) \ee
dando las siguientes relaciones
\ba
 &E^i=E_i=&F_{i0}  \nonumber \\
 &B^i=B_i=&\frac 1 2 {\varepsilon_{ijk}}  F_{jk}
\ea
\ba
 &F_{i0}=&E_i \nonumber \\ %
 &F_{ij}=&\varepsilon_{ijk} B_k
\ea

\item El dual de Hodge se define como
\be
 \tilde F^{\mu\nu}=\frac 12 \varepsilon^{\mu\nu\rho\sigma}
 F_{\rho\sigma}
 \label{hodge}
\ee
que corresponde a hacer
\be
 \begin{array}{ccc}
  \vec E & \to & -\vec B \\
  \vec B & \to & \vec E
 \end{array}
\ee
La expresi\'on inversa a (\ref{hodge}) es
\be
 F^{\mu\nu}=-\frac 12 \varepsilon^{\mu\nu\rho\sigma} \tilde
 F_{\rho\sigma}
\ee

\item Identidades
\be
 F^2=F_{\mu\nu}F^{\mu\nu}=2\left( \vec B^2-\vec E^2  \right)
 \label{fff}
\ee
\be
 F\tilde F=F_{\mu\nu}\tilde F^{\mu \nu}=4 \vec E \cdot \vec B
\ee
\be
 {\tilde {\tilde {F}}}=-F
 \label{po}
\ee
\be
 \tilde F^2=\tilde F_{\mu\nu} \tilde F^{\mu\nu}=-F^2
\ee
\be
 \tilde F_{\mu}^{~\rho} \tilde F_{\rho}^{~\nu}=F_{\mu}^{~\rho}
 F_{\rho}^{~\nu}+ \frac 1 2 \delta_\mu^{\nu} F^2
 \label{fo}
\ee
\be
 F_{\mu}^{~\rho}\tilde F_\rho^{~\nu}=-\frac 1 4 \delta_\mu^{\nu} F \tilde
 F
\label{2f} \ee
Las identidades anteriores se dedujeron usando las siguientes
propiedades del tensor de Levi-Civita
\ba
 \varepsilon^{\mu\nu\rho\sigma}\varepsilon_{\mu\omega\tau\upsilon}
 &=&-(\delta^\nu_\omega\delta^\rho_\tau\delta^\sigma_\upsilon-
     \delta^\nu_\omega\delta^\rho_\upsilon\delta^\sigma_\tau+
     \delta^\nu_\tau\delta^\rho_\upsilon\delta^\sigma_\omega-
     \delta^\nu_\tau\delta^\rho_\omega\delta^\sigma_\upsilon+
     \delta^\nu_\upsilon\delta^\rho_\omega\delta^\sigma_\tau-
     \delta^\nu_\upsilon\delta^\rho_\tau\delta^\sigma_\omega)\nonumber\\
 &=&-\delta^{\nu\rho\sigma}_{\omega\tau\upsilon} \\
 \varepsilon^{\mu\nu\rho\sigma}\varepsilon_{\mu\nu\omega\tau}
 &=&-2(\delta^\rho_\omega\delta^\sigma_\tau-
     \delta^\rho_\tau\delta^\sigma_\omega)
 =-2\delta^{\rho\sigma}_{\omega\tau}\\
 \varepsilon^{\mu\nu\rho\sigma}\varepsilon_{\mu\nu\rho\omega}
 &=&-3!\delta^\sigma_\omega \\
 \varepsilon^{\mu\nu\rho\sigma}\varepsilon_{\mu\nu\rho\sigma}&=&-4!
\ea Usando las identidades (\ref{fo}) y (\ref{2f}) tenemos que
\ba
 F^4&=&F_\mu^{~\nu}F_\nu^{~\rho} F_\rho^{~\sigma}
 F_\sigma^{~\mu} \nonumber \\
 &=&\tilde F_\mu^{~\nu} \tilde F_\nu^{~\rho}F_\rho^{~\sigma}
 F_\sigma^{~\mu}-\frac 12 (F^2)
 \delta^\rho_\mu F_\rho^{~\sigma}F_\sigma^{~\mu} \nonumber \\
 &=&-\frac 14  \tilde F_\mu^{~\nu} \delta^\sigma_\nu (F\tilde F)
 F_\sigma^{~\mu}+\frac 12 \left(F^2\right)^2 \nonumber \\
 &=&\frac 14 (F\tilde
 F)^2+\frac 12 \left(F^2\right)^2
 \label{f4}
\ea
que permite reescribir el lagrangiano de Born-Infeld (\ref{1}) en
t\'erminos de $F^2$ y $F\tilde F$. Usando la antisimetr\' \i a de
$F_{\mu\nu}$ podemos mostrar que toda contracci\'on de un n\'umero
impar de $F's$ es cero
\be
 F^{2n+1}=F_{\mu_1}^{~\mu_2} F_{\mu_2}^{~\mu_3} \ldots
 F_{\mu_{2n+1}}^{~\mu_1}=0.
\ee
El resultado (\ref{f4}) se generaliza para un n\'umero par
arbitrario de $F's$ usando la siguiente relaci\'on
\ba
 F^{2n}&=&F_{\mu_1}^{~\mu_2} F_{\mu_2}^{~\mu_3} \ldots
 F_{\mu_{2n}}^{~\mu_1} \nonumber \\
 &=&-\frac 12 F^2 F^{2n-2}+\frac 1 {16} (F\tilde F)^2
 F^{2n-4}~~~~~~~~~~(n\geq 3)
\ea
Luego cualquier contracci\'on de $F's$ se puede expresar en
t\'erminos de los invariantes fundamentales $F^2$ y $F\tilde F$.
Para las primeras potencias tenemos
\ba
 F^6&=&-\frac 1 {2^2} (F^2)^3-\frac 3 {4^2} F^2 (F\tilde F)^2 \\
 F^8&=&\frac 1 {2^3} (F^2)^4+\frac 4{2~4^2} (F^2)^2 (F\tilde F)^2+
 \frac 1{4^3}(F\tilde F)^4
\ea

\item \underline{Caso no abeliano:} En este caso $F=F^at^a$ y las
identidades (\ref{fff})-(\ref{2f}) son reemplazadas por
\be
 \tilde F \tilde F=F\tilde{\tilde F}=-F_{\mu\nu} F^{\mu\nu}
\ee
\be
 \tilde F_{\mu\nu}\tilde F^{\nu\rho}=F^{\rho\gamma} F_{\gamma\mu}+
 \frac 12 \delta^\rho_\mu F^2
\ee
\be
 F_{\mu\nu}\tilde F^{\nu\rho}=-\tilde F^{\rho\gamma}
 F_{\gamma\mu}-\frac 12 \delta^\rho_\mu \tilde F F
\ee
obteni\'endose en particular
\ba
 F^4=F_\mu^{~\nu}F_\nu^{~\rho} F_\rho^{~\sigma}
 F_\sigma^{~\mu}
 \!\!\!&=\!\!\!&\frac 14 \left( (F^2)^2+ F_{\mu\nu}\, F^2 F^{\mu\nu}+
 F_{\mu\nu}F_{\rho\sigma} \tilde F^{\mu\nu}\tilde F^{\rho\sigma}
 \right)
 \label{f41}\\
 &&\nonumber\\
 F_{\mu\nu}F^{\nu\rho} F^{\mu\sigma}F_{\sigma\rho}\!\!\!&=\!\!\!&
 \frac 14 \left( (F^2)^2+F_{\mu\nu}\, F\tilde F \tilde F^{\mu\nu}+
 F_{\mu\nu}F_{\rho\sigma} F^{\mu\nu} F^{\rho\sigma}  \right)\\
 &&\nonumber\\
 F_{\mu\nu}F_{\rho\sigma} F^{\mu\sigma}F^{\rho\nu}\!\!\!&=\!\!\!&
 \frac 14 \left( (F\tilde F)^2+F_{\mu\nu}\, F^2 F^{\mu\nu}+
 F_{\mu\nu}F_{\rho\sigma} F^{\mu\nu} F^{\rho\sigma}  \right)
 \label{f4e}
\ea
que muestra que las potencias de orden $F^4$ en el desarrollo del
determinante de BI se factorizan en t\'erminos de $F^2$ y $F\tilde
F$, independientemente de la elecci\'on para la traza en los \' \i
ndices internos de grupo. Es de esperar que est\'a propiedad valga
tambi\'en para las potencias de orden superior.

\end{itemize}

\underline{Espinores}

\begin{itemize}

\item Seguimos en casi su totalidad la notaci\'on de Lykken
\cite{lykken} salvo ligeras modificaciones. Para un estudio
detallado de las representaciones espinoriales del grupo de
Lorentz ver \cite{corson}, tambien son discutidas en
 \cite{wess}, \cite{bailin},\cite{ramond}.

\item Usamos \' \i ndices griegos sin punto $\alpha,\beta,..=1,2$
para denotar las componentes de la representaci\'on $(\frac 12,0)$
del grupo de Lorentz $\psi^{(L)}=\psi_\alpha$.
\item Usamos \' \i ndices griegos con punto $\dot\alpha,\dot\beta,..=1,2$
para denotar las componentes de la representaci\'on $(0,\frac 12)$
del grupo de Lorentz $\psi^{(R)} =\bar\psi^{\dot\alpha}$
\footnote{La notaci\'on es redundante pues un espinor barrado es
sin\'onimo de que transforma con \' \i ndices punteados, pero
mantenemos la convenci\'on usual ya que ayuda a evitar
confusiones.}.
\item Las letras may\'usculas griegas $\Lambda,\Upsilon,\Sigma,..$
las usamos para denotar espinores de cuatro componentes, en
general fermiones de Majorana en $d=4$.
\item Las min\'usculas griegas $\lambda,\epsilon,\psi,..$ las usamos
en $d=4$ para denotar fermiones de Weyl de dos componentes.
\item La m\'etrica en el espacio de espinores de Weyl es \footnote{La
notaci\'on de Van der Waerden de \' \i ndices arriba y abajo,
punteados y sin puntear es para tener en cuenta la propiedad
$\psi^{(L)}\in(\frac 12,0)\Rightarrow i\sigma^2 \psi^\ast \in
(0,\frac 1 2)$ y que $\chi^{(L)}i\sigma^2\psi^{(L)}$ es un escalar
bajo transformaciones de Lorentz ver \cite{bailin},\cite{ramond}.}
\ba
 \epsilon_{\alpha\beta}=\epsilon_{\dot\alpha\dot\beta}=&\left(
 \begin{array}{cc}
   0 & -1 \\
   1 & 0 \
 \end{array}\right)& \\
 \epsilon^{\alpha\beta}=\epsilon^{\dot\alpha\dot\beta}=&\left(
 \begin{array}{cc}
   0 & 1 \\
   -1 & 0 \
 \end{array}\right)&=i\sigma^2
\ea
donde
\ba
 \epsilon_{\alpha\beta}\epsilon^{\beta\gamma}&=&\delta^\gamma_\alpha\\
 \epsilon_{\dot\alpha\dot\beta}\epsilon^{\dot\beta\dot\gamma}
 &=&\delta^{\dot\gamma}_{\dot\alpha}
\ea
\item La convenci\'on para subir y bajar \' \i ndices es
\ba
 \psi^\alpha=\epsilon^{\alpha\beta}\psi_\beta~~~~~~~~~~~~~~~~~~\psi_\alpha=
 \epsilon_{\alpha\beta} \psi^\beta\\
 \bar\psi^{\dot\alpha}=\epsilon^{\dot\alpha\dot\beta} \bar\psi_{\dot\beta}
 ~~~~~~~~~~~~~~~~~~\bar\psi_{\dot\alpha}=
 \epsilon_{\dot\alpha\dot\beta} \bar\psi^{\dot\beta}
\ea
\item Producto escalar para espinores de Weyl
\ba
 \psi\chi=\psi^\alpha\chi_\alpha=-\psi_\alpha\chi^\alpha= \chi^\alpha
 \psi_\alpha=\chi\psi\\
\label{pi1}
 \bar\psi\bar\chi=\bar\psi_{\dot\alpha}\bar\chi^{\dot\alpha}=
 -\bar\psi^{\dot\alpha}\bar\chi_{\dot\alpha}=
 \bar\chi_{\dot\alpha}\bar\psi^{\dot\alpha}=\bar\chi\bar\psi
 \label{pi2}
\ea
En particular denotaremos
\ba
 \psi^2&=&\psi^\alpha\psi_\alpha\\
 \bar\psi^2&=&\bar\psi_{\dot\alpha}\bar\psi^{\dot\alpha}
\ea
\item El adjunto de un espinor de Weyl se define como
\be
 (\psi_\alpha)^\dagger\equiv\bar\psi_{\dot\alpha}~~~~~~~~~~~~~~~~~
 ~~~~~~~(\bar\psi^{\dot\alpha})^\dagger\equiv\psi^{\alpha}
\ee
La operaci\'on de conjugaci\'on $\dagger$ intercambia el orden de
los fermiones en un producto escalar sin tener en cuenta el
caracter grassmann de los mismos y las definiciones
(\ref{pi1})-(\ref{pi2}) fueron hechas para que
\be
 (\chi\psi)^\dagger=(\chi^\alpha\psi_\alpha)^\dagger=
 \bar\psi_{\dot\alpha}\bar\chi^{\dot\alpha}=\bar\chi\bar\psi
\ee
Dado que las $\sigma^\mu$ son herm\' \i ticas tenemos
\be
 (\chi\sigma^\mu\bar\psi)^\dagger=\psi\sigma^\mu\bar\chi
\ee
\item Las matrices de Pauli
\be
 \sigma^1=\left(\begin{array}{cc}
   0 & 1 \\
   1 & 0 \
 \end{array}\right)
 ~~~~~~~~~~\sigma^2=\left(\begin{array}{cc}
   0 & -i \\
   i & 0 \
 \end{array}\right)
 ~~~~~~~~~~\sigma^3=\left(\begin{array}{cc}
   1 & 0 \\
   0 & -1 \
 \end{array}\right)
\ee
satisfacen
\ba
 \sigma^i\sigma^j&=&\delta^{ij}+i\varepsilon^{ijk}\sigma^k \\
 \sigma^i\sigma^j\sigma^k&=&\delta^{ij}\sigma^k-\delta^{ik}\sigma^j+
 \delta^{jk}\sigma^i+i\varepsilon^{ijk} \\
 {\rm tr}~\sigma^i\sigma^j&=&2\delta^{ij}\\
 \left[ \sigma^i , \sigma^j \right] &=&2 i \varepsilon^{ijk} \sigma^k \\
 \{\sigma^i , \sigma^j \} &=&2\delta^{ij}I_2
\ea
donde $I_2$ denota la matriz identidad $2\times2$.
\item A partir de ellas definimos las matrices herm\' \i ticas
\ba
 \sigma^\mu&=&(I_2,\vec\sigma) \\
 \bar\sigma^\mu&=&(I_2,-\vec\sigma)
\ea
tambi\'en usaremos $\sigma^0\equiv\bar\sigma^0\equiv I_2$.
\ba
 \sigma^\mu~{\rm tiene~\acute{\i}ndices~sin~punto-punto}:&~
 \sigma^\mu_{~\alpha\dot\beta} \nonumber \\
 \bar\sigma^\mu~{\rm tiene~\acute{\i}ndices~punto-sin~punto}:&~
 \bar\sigma^{\mu~\dot\alpha\beta} \nonumber
\ea
Podemos construir vectores de Lorentz haciendo
\ba
 \chi\sigma^\mu\bar\psi&=&\chi^\alpha\sigma^\mu_{~\alpha\dot\alpha}
 \bar\psi^{\dot\alpha}\\
 \bar\chi\bar\sigma^\mu\psi&=&\bar\chi_{\dot\alpha}
 \sigma^{\mu~\dot\alpha\alpha} \psi_\alpha
\ea Tenemos las siguientes relaciones de completitud
\ba
 {\rm tr}~\sigma^\mu\bar\sigma^\nu&=&2g^{\mu\nu}\\
 \sigma^\mu_{~\alpha\dot\beta}\sigma_\mu^{~\dot\gamma\delta}&=&
 2\delta^\delta_\alpha\delta^{\dot\gamma}_{\dot\beta}
\ea
Las matrices $\sigma^\mu$ y $\bar\sigma^\mu$ se encuentran
relacionadas
\ba
 \bar\sigma^{\mu~\dot\alpha\beta}&=
 \epsilon^{\dot\alpha\dot\beta} \epsilon^{\beta\alpha}
 \sigma^\mu_{~\alpha\dot\beta}&=\sigma^{\mu~\beta\dot\alpha} \\
 \sigma^\mu_{~\alpha\dot\beta}&=
 \epsilon_{\alpha\beta} \epsilon_{\dot\beta\dot\alpha}
 \bar\sigma^{\mu~\dot\alpha\beta}&=\bar\sigma^\mu_{~\dot\beta\alpha}
\ea
y satisfacen las identidades
\ba
 (\sigma^\mu\bar\sigma^\nu+\sigma^\nu\bar\sigma^\mu)^\beta_\alpha&=&
 2g^{\mu\nu}\delta^\beta_\alpha \\
 (\bar\sigma^\mu\sigma^\nu+\bar\sigma^\nu\sigma^\mu)^{\dot\beta}_{\dot
 \alpha}&=&2g^{\mu\nu}\delta^{\dot\beta}_{\dot\alpha}
\ea
\ba
 \sigma^\mu\bar\sigma^\nu\sigma^\rho=(g^{\mu\nu}g^{\rho\omega}-g^{\mu\rho}g^{\nu\omega}+
 g^{\nu\rho}g^{\mu\omega})\sigma_\omega-i\varepsilon^{\mu\nu\rho\omega}\sigma_\omega\\
 \bar\sigma^\mu\sigma^\nu\bar\sigma^\rho=(g^{\mu\nu}g^{\rho\omega}-g^{\mu\rho}g^{\nu\omega}+
 g^{\nu\rho}g^{\mu\omega})\bar\sigma_\omega+i\varepsilon^{\mu\nu\rho\omega}\bar\sigma_\omega
\ea

\item Para las matrices de Dirac tomamos la representaci\'on
quiral
\be
 \Gamma ^\mu =\left(
 \begin{array}{cc}
 0 & \sigma ^\mu  \\
 \bar \sigma ^\mu  & 0
 \end{array}
 \right)
 \label{rep}
\ee
\item La convenci\'on para la matriz de quiralidad es
\be
 \Gamma^5 =i \Gamma ^1 \Gamma ^2 \Gamma ^3 \Gamma^0=
 \left(
 \begin{array}{cc}
 I_2 & 0  \\
 0  & -I_2
 \end{array}
 \right)
 \label {g5}
\ee
\item La convenci\'on para los generadores de Lorentz es
\footnote{El generador de rotaciones en el espacio-tiempo
$M^{\mu\nu}$ del grupo de Lorentz es $M^{\mu\nu}=x^\mu P^\nu
-x^\nu P^\mu +\Sigma^{\mu\nu}$.}
\be
 \Sigma^{\mu\nu}\equiv\frac i4\left[\Gamma^\mu,\Gamma^\nu\right]=
 \left( \begin{array}{cc}
    \sigma^{\mu\nu} & 0 \\
   0 & \bar\sigma^{\mu\nu} \
 \end{array}\right)
 \label{lor}
\ee
donde
\ba
 \sigma^{\mu\nu}&=&\frac i4\left[\sigma^\mu\bar\sigma^\nu-
 \sigma^\nu\bar\sigma^\mu\right]\\
 \bar\sigma^{\mu\nu}&=&\frac i4\left[\bar\sigma^\mu\sigma^\nu-
 \bar\sigma^\nu\sigma^\mu\right]
\ea
que satisfacen la relaci\'on de trazas
\be
 {\rm tr} \sigma^{\mu\nu}\sigma^{\rho\sigma}=\frac 12 (g^{\mu\rho}g^{\nu\sigma}-
 g^{\mu\sigma}g^{\nu\rho})+\frac i2 \varepsilon^{\mu\nu\rho\sigma}
\ee
y las relaciones de dualidad (cf.(\ref{dudu}))
\ba
 \sigma^{\mu\nu}&=&\frac
 i2\epsilon^{\mu\nu\rho\sigma}\sigma_{\rho\sigma}
 \label{dudu1}\\
 \bar\sigma^{\mu\nu}&=&-\frac
 i2\epsilon^{\mu\nu\rho\sigma}\bar\sigma_{\rho\sigma}
\ea
La representaci\'on (\ref{rep}) para las matrices $\Gamma$ muestra
que aparecen contenidos, naturalmente, en un espinor de Dirac dos
espinores de Weyl de quiralidades opuestas. Dado que frente a
transformaciones de Lorentz un espinor de Dirac $\Psi_D$
transforma como
\be
 \Psi\longrightarrow {\cal S}\Psi
\ee
donde
\be
 {\cal S}=e^{-\frac i2\omega_{\mu\nu}\Sigma^{\mu\nu}}
\ee
vemos que si denotamos
\be
 \Psi=\left(  \begin{array}{c}
  \chi_{\alpha} \\ 
  \bar\xi^{ \dot\alpha} \ 
 \end{array} \right)
 \label{dirac}
\ee
debido a (\ref{lor}) $\chi$ y $\bar\xi$ transforman
independientemente frente a transformaciones de Lorentz
\be
 \psi_\alpha\longrightarrow (S)_\alpha^{~\beta}\psi_\beta
 \equiv\left(e^{-\frac i2\omega_{\mu\nu}\sigma^{\mu\nu}}
 \right)_\alpha^{~\beta}\psi_\beta
\ee
\be
 \bar\xi^{\dot\alpha}\longrightarrow
 ([S^\dagger]^{-1})^{\dot\alpha}_{~\dot\beta}\bar\xi^{\dot\beta}
 \equiv\left(e^{-\frac i2\omega_{\mu\nu}\bar\sigma^{\mu\nu}}
 \right)^{\dot\alpha}_{~\dot\beta}\bar\xi^{\dot\beta}
\ee
\item El adjunto de Dirac se define como
\be
 \bar\Psi\equiv\Psi^\dagger \Gamma^0=\left( \begin{array}{cc}
   \xi^{\alpha} &  \bar\chi_{\dot\alpha} \
 \end{array} \right)
 \label{adj}
\ee
\item El conjugado de carga se define a partir de este \'ultimo
espinor como\footnote{Partiendo de $\Psi$ que satisface
$(i/\!\!\!\partial+e~\!/\!\!\!\!A-m)\Psi=0$ se busca definir
$\Psi^c$ en t\'erminos de $\bar\Psi$ de manera que  satisfaga
$(i/\!\!\!\partial-e~\!/\!\!\!\!A-m)\Psi^c=0$.}
\be
 \Psi^c\equiv C\bar\Psi^T~~~~~~~~{\rm con}~~~~~~~~
 C\Gamma^{\mu\,T}=-\Gamma^\mu C
 \label{carga}
\ee
Para nuestra convenci\'on de matrices $\Gamma$ ec.(\ref{rep})
tenemos
\be
 C=-i\,\Gamma^2\Gamma^0=\left( \begin{array}{cc}
   \epsilon_{\alpha\beta} & 0 \\
   0 & \epsilon^{\dot\alpha\dot\beta} \
 \end{array}\right)~~~~\Longrightarrow~~~~C^T=-C
 \label{ce}
\ee
luego
\be
 \Psi^c=C\bar\Psi^T=\left( \begin{array}{cc}
 \epsilon_{\alpha\beta} & 0 \\
 0 & \epsilon^{\dot\alpha\dot\beta} \
 \end{array}\right)\left(  \begin{array}{c}
  \xi^\beta \\
  \bar\chi_{\dot\beta} \
 \end{array} \right)=\left(  \begin{array}{c}
  \xi_\alpha \\
  \bar\chi^{\dot\alpha} \
 \end{array} \right)
\ee
El n\'umero de grados de libertad puede ser reducido mediante
distintos procedimientos. Los fermiones $\chi$ y $\bar\xi$ en
(\ref{dirac}) son las componentes de Weyl de
$\Psi_D=\Psi_L+\Psi_R$ donde
\ba
 \Psi_L\equiv\frac 12 (1+\Gamma^5)\Psi&=&\left(  \begin{array}{c}
  \chi_{\alpha} \\
  0 \
 \end{array} \right)
  \label{psil}\\
 \Psi_R\equiv\frac 12 (1-\Gamma^5)\Psi&=&\left(  \begin{array}{c}
  0 \\
  \bar\xi^{\dot\alpha} \
 \end{array} \right)
 \label{psir}
\ea
\item El espinor de Majorana $\Psi_M$ se define como aquel que
satisface la propiedad de ser autoconjugado de carga
\be
 {\underline{\rm Espinor~de~Majorana}}:\Psi_M~~~\Leftrightarrow~~~\Psi_M^c=\Psi_M
 \label{mayorana}
\ee
entonces
\be
 C\bar\Psi_M^T=\Psi_M~~~~\Longrightarrow~~~~\left(  \begin{array}{c}
  \xi_\alpha \\
  \bar\chi^{\dot\alpha} \
 \end{array} \right)=\left(  \begin{array}{c}
  \chi_{\alpha} \\
  \bar\xi^{ \dot\alpha} \
 \end{array} \right)~~~~\Longrightarrow~~~~\xi=\chi
\ee
Luego los fermiones de Majorana en nuestra notaci\'on se escriben
\ba
 \Psi_M &=&\left(
 \begin{array}{c}
 \psi _\alpha  \\
 \bar \psi ^{\dot \alpha }
 \end{array}
 \right)
 \label{c88}\\
 \bar\Psi_M&=&\left(
 \begin{array}{cc}
   \psi^\alpha & \bar\psi_{\dot\alpha} \
 \end{array}
 \right)
 \label{c89}
\ea
De donde deducimos que no pueden existir espinores de
Majorana-Weyl en $d=4$ por ser incompatible (\ref{c88}) con las
ecs.(\ref{psil})-(\ref{psir}).
\end{itemize}
\underline{Identidades} \ba
 \psi^\alpha\psi^\beta&=&-\frac 12 \epsilon^{\alpha\beta}\psi^2\\
 \psi_\alpha\psi_\beta&=&\frac 12 \epsilon_{\alpha\beta}\psi^2\\
 \bar\psi^{\dot\alpha}\bar\psi^{\dot\beta}&=&\frac 12
 \epsilon^{\dot\alpha\dot\beta}\bar\psi^2\\
 \bar\psi_{\dot\alpha}\bar\psi_{\dot\beta}&=&-\frac 12
 \epsilon_{\dot\alpha\dot\beta}\bar\psi^2
\ea
\ba
 ((\phi\sigma^\mu)_{\dot\alpha})^\dagger&=&(\sigma^\mu\bar\phi)_\alpha\\
 ((\sigma^\mu\bar\sigma^\nu\theta)_\alpha)^\dagger&=&
 (\bar\sigma^\mu\sigma^\nu\bar\theta)_{\dot\alpha}\\
 \sigma^{\mu\nu\dagger}&=&\bar\sigma^{\mu\nu}
\ea
\ba
 (\theta\chi)(\theta\psi)&=&-\frac 12 \theta^2\chi\psi\\
 (\theta\sigma^\mu\bar\sigma^\nu\theta)&=& \theta^2g^{\mu\nu}\\
 (\theta\sigma^\mu\bar\theta)(\theta\sigma^\nu\bar\theta)&=&
 \frac 12 \theta^2\bar\theta^2g^{\mu\nu}\\
 \theta \sigma^\mu\bar\theta(\sigma^\nu\bar\theta)_\alpha &=&
 \frac 12 \bar\theta^2 (\theta\sigma^\mu\bar\sigma^\nu)_\alpha\\
 (\lambda\sigma^\mu\bar\theta)(\theta\sigma^\nu\bar\theta)&=&
 \frac 12\bar\theta^2\lambda\sigma^\mu\bar\sigma^\nu\theta
\ea
\be
 \theta\sigma^\mu\bar\chi=\bar\chi\theta\sigma^\mu=\sigma^\mu\bar\chi\theta=
 -\bar\chi\bar\sigma^\mu\theta
\ee
\be
 \sigma^\rho\bar\theta\sigma^\mu\bar\sigma^\nu\theta=
 -\bar\theta\bar\sigma^\rho\sigma^\mu\bar\sigma^\nu\theta
\ee
\ba
 (\theta\sigma^\omega\bar\theta)(\bar\theta\sigma^\rho\sigma^\mu
 \bar\sigma^\nu\theta)&=&-\frac 12 \theta^2\bar\theta^2
 (g^{\rho\mu}g^{\nu\omega}-g^{\rho\nu}g^{\mu\omega}+g^{\mu\nu}
 g^{\rho\omega}+i\varepsilon^{\mu\nu\rho\omega})\\
 \theta\sigma^\rho\bar\sigma^\omega\sigma^\mu\bar\sigma^\nu\theta&=&
 \theta^2 (g^{\rho\omega}g^{\mu\nu}-g^{\rho\mu}g^{\omega\nu}+g^{\rho\nu}
 g^{\omega\mu}-i\varepsilon^{\rho\omega\mu\nu})
\ea
\be
 \epsilon^{\alpha\beta}\frac\partial{\partial\theta^\beta}
 =-\frac\partial{\partial\theta_\alpha}
\ee
\underline{Propiedades de las matrices $\Gamma$}
\begin{itemize}
\item Para nuestra convenci\'on de $\Gamma^5$ ec.(\ref{g5}) tenemos
\be
 (\Gamma^5)^2=1
\ee
\item Toda representaci\'on para las matrices $\Gamma$ satisface
\ba
 \Gamma^{0\dagger}&=&\Gamma^{0}\\
 \Gamma^{i\dagger}&=&-\Gamma^{i}\\
 \Gamma^{\mu\dagger}&=&\Gamma^{0}\Gamma^\mu\Gamma^0\\
 \Gamma^{\mu\,T}&=&-C^{-1}\Gamma^\mu C
\ea
\item Para la representaci\'on quiral (\ref{rep}) vale
\ba
 \Gamma^{0\,T}&=&\Gamma^{0}\\
 \Gamma^{1\,T}&=&-\Gamma^{1}\\
 \Gamma^{2\,T}&=&\Gamma^{2}\\
 \Gamma^{3\,T}&=&-\Gamma^{3}
\ea
\be
 C=-i\,\Gamma^2\Gamma^0
\ee
\item Para el producto de varias $\Gamma$'s tenemos
\ba
 \Gamma^\mu\Gamma^\nu&=&2g^{\mu\nu}-\Gamma^\nu\Gamma^\mu\\
 \Gamma^\mu\Gamma^\nu\Gamma^\rho&=&g^{\mu\nu}\Gamma^\rho-
 g^{\mu\rho}\Gamma^\nu+g^{\nu\rho}\Gamma^\mu-i
 \varepsilon^{\mu\nu\rho\sigma}\Gamma^5\Gamma_\sigma\\
 \Sigma^{\mu\nu}\Gamma^\rho&=&\frac i2(\Gamma^\mu g^{\nu\rho}-
 \Gamma^\nu g^{\mu\rho})+\frac 12 \varepsilon^{\mu\nu\rho\sigma}
 \Gamma^5\Gamma_\sigma\\
 \Sigma^{\mu\nu}\Gamma^\rho&=&-\Gamma^\rho\Sigma^{\mu\nu}+
 \varepsilon^{\mu\nu\rho\sigma}\Gamma^5\Gamma_\sigma\\
 \Sigma^{\mu\nu}&=&\frac i2\varepsilon^{\mu\nu\rho\sigma}\Gamma^5
 \Sigma_{\rho\sigma}
 \label{dudu}\\
 \Sigma^{\mu\nu\dagger}&=&\Gamma^0\Sigma^{\mu\nu}\Gamma^0
\ea
\item Contracciones y trazas
\ba
 \Gamma^\mu\Gamma_\mu&=&4\\
 {\rm tr}\Gamma^\mu&=&0\\
 {\rm tr}\Gamma^5&=&0\\
 {\rm tr}\Gamma^5\Gamma^\mu&=&0\\
 {\rm tr}\Gamma^\mu\Gamma^\nu&=&4g^{\mu\nu}\\
 {\rm tr}\Gamma^5\Gamma^\mu\Gamma^\nu&=&0\\
 {\rm tr}\Gamma^{\mu_1}\ldots\Gamma^{\mu_{2n+1}}&=&0\\
 {\rm tr}\Gamma^5\Gamma^{\mu_1}\ldots\Gamma^{\mu_{2n+1}}&=&0
\ea
\end{itemize}
\underline{Relaciones entre espinores de Weyl y de Majorana}

Ecuaciones escritas en t\'erminos de espinores de Weyl pueden ser
reexpresadas, teniendo en cuenta las expresiones (\ref{c88}) y
(\ref{c89}, en t\'erminos de espinores de Majorana. Los espinores
de Majorana tienen las siguientes propiedades
\ba
 \bar\Psi\Gamma^\mu\Psi&=&0
 \label{maj1}\\
 \bar\Lambda\Psi&=&\bar\Psi\Lambda\\
 \bar\Lambda\Gamma^5\Psi&=&\bar\Psi\Gamma^5\Lambda\\
 \bar\Lambda\Gamma^\mu\Psi&=&-\bar\Psi\Gamma^\mu\Lambda\\
 \bar\Lambda\Gamma^5\Gamma^\mu\Psi&=&\bar\Psi\Gamma^5\Gamma^\mu\Lambda\\
 \bar\Lambda\Sigma^{\mu\nu}\Gamma^\rho\Psi&=&
 \bar\Psi\Gamma^\rho\Sigma^{\mu\nu}\Lambda
 \label{maj2}
\ea
\begin{itemize}
\item T\'erminos cin\'eticos:
\ba
 \psi/\!\!\!\partial\bar\psi&=&\frac
 12\bar\Psi(1+\Gamma^5)/\!\!\!\partial\Psi\\
 \bar\psi\bar{/\!\!\!\partial}\psi&=&\frac
 12\bar\Psi(1-\Gamma^5)/\!\!\!\partial\Psi
\ea
donde se debe entender
\ba
 /\!\!\!\partial&=&\sigma^\mu\partial_\mu\\
 \bar{/\!\!\!\partial}&=&\bar\sigma^\mu\partial_\mu
\ea

\item T\'erminos de masa
\ba
 \psi\psi&=&\frac 12 \bar\Psi(1+\Gamma^5)\Psi\\
 \bar\psi\bar\psi&=&\frac 12 \bar\Psi(1-\Gamma^5)\Psi\\
 \psi\lambda&=&\frac 12 \bar\Psi(1+\Gamma^5)\Lambda\\
 \bar\psi\bar\lambda&=&\frac 12 \bar\Psi(1-\Gamma^5)\Lambda
\ea

\item T\'erminos de corriente
\ba
 \psi\sigma^\mu\bar\psi&=&\frac 12 \bar\Psi\Gamma^5\Gamma^\mu\Psi\\
 \bar\psi\bar\sigma^\mu\psi&=&-\frac 12 \bar\Psi\Gamma^5\Gamma^\mu\Psi\\
 \lambda\sigma^\mu\bar\psi&=&\frac 12
 \bar\Lambda(1+\Gamma^5)\Gamma^\mu\Psi\\
 \bar\lambda\bar\sigma^\mu\psi&=&
 \frac 12 \bar\Lambda(1-\Gamma^5)\Gamma^\mu\Psi
\ea
\end{itemize}
\underline{Acci\'on de Born-Infeld ('34)}

Usando la siguiente identidad
\ba
 {\rm det}(g_{\mu\nu}+F_{\mu\nu})&=&{\rm det}\left(g_{\mu\rho}
 (\delta^\rho_\nu+F_{~\nu}^\rho)\right)\\
 &=&g\cdot {\rm det}(\delta_{\nu}^\rho+F_{~\nu}^\rho)
\ea
donde $g=\det g_{\mu\nu}$ y $F_{~\nu}^\rho=g^{\rho\mu}F_{\mu\nu}$,
y recordando que el tensor de Levi-Civita en espacio curvo se
define como\footnote{La presencia de $\sqrt{-g}$ en el denominador
de $ j^{\mu\nu\rho\sigma}$ hace que la generalizaci\'on a espacio
curvo de $F\tilde F$ definida por la ec.(\ref{hochi}) sea
efectivamente un invariante topol\'ogico.}
\be
 j^{\mu\nu\rho\sigma}=\frac {\varepsilon^{\mu\nu\rho\sigma}}
 {\sqrt{-g}}~~~~~~~\Longrightarrow ~~~~~~~
 j_{\mu\nu\rho\sigma}={\sqrt{-g}}~{\varepsilon_{\mu\nu\rho\sigma}}
\ee
(donde hemos usado que $\varepsilon_{0123}=-\varepsilon^{0123}$).
La definici\'on del lagrangiano de Born-Infeld en espacio curvo
queda
\ba
 {\cal L}_{BI}&=&\left( \sqrt{-\det (
 g_{\mu \nu })} - \sqrt{-\det \left(
 g_{\mu \nu }+F_{\mu \nu }\right) }\, \right)\\
 &=&\sqrt{-g} \left(1 + {\cal F}- {\cal G}^2 \right)\label{pipu}\\
 &=&\sqrt{-g}~ L\label{pipu1}
\ea
donde
\be
 {\cal F}=\frac 1 2F_{\mu\nu}F^{\mu\nu}=\frac 12
 g^{\mu\rho}g^{\nu\sigma}F_{\mu\nu}F_{\rho\sigma}~~~~~~~~~
 {\cal G}=\frac 14 F_{\mu\nu}\tilde F^{\mu\nu}=
 \frac {\varepsilon^{\mu\nu\rho\sigma}}
 {8\sqrt{-g}\,} F_{\mu\nu}F_{\rho\sigma}
 \label{hochi}
\ee
En espacio curvo, la ec.(\ref{po}) sigue valiendo si el dual se
define como $\tilde F^{\mu\nu}= \frac12 j^{\mu\nu\rho\sigma}
F_{\rho\sigma}$.

\noindent \underline{Acci\'on, ecuaciones de movimiento y leyes de
conservaci\'on}
\begin {itemize}
\item Partiendo de una acci\'on general
\be
 {\cal S}=\int d^4x\sqrt{-g}~L(g_{\mu\nu},{\cal F,G})
\ee
donde $F$ y $G$ fueron definidos por las ecs.(\ref{hochi}),
deducimos las siguientes ecuaciones de movimiento (teniendo en
cuenta que $F=dA$)
\be
 \nabla_\mu D^{\mu\nu}=0~~~\Longrightarrow~~~
 \frac 1 {\sqrt{-g}} \partial_\mu(\sqrt{-g}\, D^{\mu\nu})
 \label{ecu}
\ee
\be
 dF=0~~~\Longrightarrow~~~
 \frac 1 {\sqrt{-g}} \partial_\mu(\sqrt{-g}\, \tilde F^{\mu\nu})=0
 \label{bichi}
\ee
donde $\nabla=\partial+\Gamma$ y hemos definido $D^{\mu\nu}$
como\cite{BI},\cite{GR}\footnote{La ausencia de un 2 en la
ec.(3.3) del paper original de Born-Infeld se debe a que ellos
derivan el lagrangiano considerando la antisimetr\' \i a de
$F_{\mu\nu}$.}
\be
 D^{\mu\nu}=-\frac 2 {\sqrt{-g}}\frac {\delta {\cal S}}
 {\delta F_{\mu\nu}}=-2\frac {\delta { L}}
 {\delta F_{\mu\nu}}
 \label{dmu}
\ee
En t\'erminos de vectores 3-dimensionales tenemos que
$F_{\mu\nu}=(\vec E, \vec B)$ contiene los campos ponderomotrices
y que $D^{\mu\nu}=(\vec D,\vec H)$ contiene los vectores de
desplazamiento el\'ectrico e inducci\'on magn\'etica. Para el caso
de BI en el que $L$ esta dado por (\ref{pipu})-(\ref{pipu1}) se
obtiene
\be
 D^{\mu\nu}=\frac {F^{\mu\nu}-{\cal G}\,\tilde F^{\mu\nu}}{\sqrt{1+{\cal F-G}^2}}
\ee
\item Las leyes de conservaci\'on, debidas a la invariancia frente a
difeomorfismos, pueden ser deducidas de las identidades de Bianchi
(\ref{bichi}) y las ecuaciones de movimiento (\ref{ecu})
obteni\'endose\footnote{Obviamente tambi\'en podemos deducirlas
usando el teorema de Noether.}
\be
 \nabla_\mu T^{\mu}_{~\rho}=0
\ee
donde el tensor de energ\' \i a impulso $T^{\mu}_{~\rho}$ est\'a
definido como
\be
 T_{\mu\nu}=\frac 2 {\sqrt{-g}}\frac {\delta {\cal S}}{\delta
 g^{\mu\nu}}=
  D_{\mu\rho}F^{\rho}_{~\nu}-g_{\mu\nu}\, L
\ee
\item Con el objeto de definir el hamiltoniano y mostrar una
formulaci\'on equivalente (dual) de la teor\' \i a de BI,
definimos
\be
 {\cal P}=\frac 12 \tilde D^{\mu\nu} D_{\mu\nu}=
 \frac {-{\cal F+G}^2{\cal F}+4{\cal G}^2}
 {1+{\cal F-G}^2}
\ee
\be
 {\cal Q}=\frac 14 D^{\mu\nu}\tilde D_{\mu\nu}=\cal G
\ee
Con estas definiciones es posible mostrar que es posible expresar,
en la teor\' \i a de BI, $F_{\mu\nu}$ en t\'erminos de
$D_{\mu\nu}$ como
\be
 F_{\mu\nu}=\frac{D_{\mu\nu}+{\cal Q}\,\tilde D_{\mu\nu}}{\sqrt{1+{\cal P-Q}^2}}
\ee
Definiendo entonces \footnote{El factor $1/2$ en la definici\'on
del hamiltoniano viene de la definici\'on de $D^{\mu\nu}$ dada por
la ecuaci\'on (\ref{dmu}).}
\ba
 H({\cal P,Q})&=&-\frac 12 D^{\mu\nu}F_{\mu\nu}-L\\
 &=&{\sqrt{1+{\cal P-Q}^2}}-1~~,
\ea
tenemos que
\be
 \frac {\delta H}{\delta \tilde D_{\mu\nu}}=
 \frac{\tilde D^{\mu\nu}-{\cal Q}\,D^{\mu\nu}}{2\sqrt{1+{\cal
 P-Q}^2}}=
 \frac 12\tilde F^{\mu\nu}
 \label{pero}
\ee
La formulaci\'on equivalente de la teor\' \i a se obtiene
definiendo el hamiltoniano como
\be
 {\cal H}=\sqrt{-g}\,H=\sqrt{-\det(g_{\mu\nu}+\tilde
 D_{\mu\nu})}-\sqrt{-\det(g_{\mu\nu})}
 \label{hami}
\ee
y tomando que
\be
 \tilde D_{\mu\nu}=\partial_\mu B_\nu-\partial_\nu B_\mu
 \label{duli}
\ee
Luego, no solo tenemos las ecuaciones de movimiento que se derivan
de (\ref{hami})
\be
 \frac 1 {\sqrt{-g}}\partial_\mu(\sqrt{-g}
 \frac {\delta H}{\delta \tilde
 D_{\mu\nu}})=0\Longleftrightarrow \frac 1 {\sqrt{-g}}
 \partial_\mu(\sqrt{-g}\, \tilde F^{\mu\nu})=0
 \label{lopa}
\ee
(donde hemos usado (\ref{pero})) sino tambi\'en las identidades de
Bianchi correspondientes a (\ref{duli})
\be
 d\tilde D=0~~~\Longrightarrow~~~
 \frac 1 {\sqrt{-g}} \partial_\mu(\sqrt{-g}\, D^{\mu\nu})=0
 \label{rio}
\ee
Tenemos entonces dos formulaciones equivalentes para la teor\' \i
a sin fuentes en t\'erminos de $\cal L$ y en t\'erminos de ${\cal
H }$ cf. (\ref{ecu})-(\ref{bichi}) con (\ref{lopa})-(\ref{rio})
(ver tambi\'en \cite{bialy}). Estas formulaciones son la base para
el estudio de la simetr\' \i a de dualidad en la teor\' \i a de BI
\cite{schro},\cite{zumi}.

\end{itemize}



\chapter{Componentes de los supercampos\label{a2}}

\begin{itemize}

\item Derivadas en el superespacio:
\be
 \begin{array}{cccccccc}
   \partial_\alpha&=&\partial/{\partial \theta^\alpha}
   && \partial^\alpha&=&\partial/{\partial \theta_\alpha}
   &=-\epsilon^{\alpha\beta}\partial_\beta \\
   \bar\partial^{\dot\alpha}&=&\partial/{\partial
   \bar\theta_{\dot\alpha}}
   && \bar\partial_{\dot\alpha}
   &=&\partial/{\partial
   \bar\theta^{\dot\alpha}}&=-\epsilon_{\dot\alpha\dot\beta}\bar\partial^{\dot\beta} \\
   \partial_\alpha\theta^\beta&=&\delta^\beta_\alpha
   && \bar\partial^{\dot\alpha}\theta_{\dot\beta}
   &=&\delta_{\dot\beta}^{\dot\alpha} \\
   \partial^\alpha\theta^\beta&=&-\epsilon^{\alpha\beta}
   && \partial_\alpha\theta_\beta&=&-\epsilon_{\alpha\beta}\\
   \bar\partial^{\dot\alpha}\bar\theta^{\dot\beta}
   &=&-\epsilon^{\dot\alpha\dot\beta}
   && \bar\partial_{\dot\alpha}\bar\theta_{\dot\beta}
   &=&-\epsilon_{\dot\alpha\dot\beta}\\
   \partial_\alpha \theta^2&=&2\theta_\alpha
   & & \bar\partial_{\dot\alpha} \bar\theta^2
   &=&-2\bar\theta_{\dot\alpha}\\
   \partial^2\theta^2&=&4
   && \bar\partial^2\bar\theta^2&=&4 \\
   \partial^\alpha\theta^2&=&-2\theta^\alpha
   && \bar\partial^{\dot\alpha}\bar\theta^2&=&2\bar\theta^{\dot\alpha}\\
   \partial^2&=&2\partial_1\partial_2
   &~~~~~~~~~~~~~~~~~&\partial_\alpha\partial_\beta
   &=&-\frac12\epsilon_{\alpha\beta}\partial^2 \
  \end{array}
\ee
Notar que las derivadas espinoriales no suben y bajan los \' \i
ndices con $\epsilon$ sino con $-\epsilon$.

\item Integraci\'on en el superespacio:

Las integrales en las variables $\theta$ (grassmanns) es la usual
siguiendo la definici\'on de Berezin
\ba
 \int d\theta~\theta&=&1\\
 \int d\theta&=&0\\
 \int d\theta~f(\theta)&=&f_1
\ea
donde usamos que una funci\'on arbitraria de una \'unica variable
de Grassmann $\theta$ tiene una expansi\'on de Taylor
$f(\theta)=f_0+\theta f_1$.

En el caso de superespacio $N=1$ la convenci\'on es
\ba
 d^2\theta&=&-\frac 14d\theta^\alpha d\theta^\beta
 \epsilon_{\alpha\beta}\\
 d^2\bar\theta&=&-\frac 14d\bar\theta_{\dot\alpha}
 d\bar\theta_{\dot\beta} \epsilon^{\dot\alpha\dot\beta}\\
 d^4\theta&=&d^2\theta d^2\bar\theta
\ea
Obteni\'endose las identidades
\ba
 \int d^2\theta~\theta^2&=&1\\
 \int d^2\bar\theta~\bar\theta^2&=&1
\ea
\item Para obtener lagrangianos supersim\'{e}tricos debemos tomar la
componente mas alta del supercampo. En el formalismo esto se logra
integrando en $d^4\theta$ para un supercampo real y en $d^2\theta$
para un supercampo quiral. Las convenciones son:

{Supercampo quiral}:
\be
 \Phi=\phi+\sqrt2\theta\psi+\theta^2F
\ee

{Supercampo vectorial}:
\be
 V_{WZ} = - \theta \sigma^\mu \bar \theta A_\mu + i \theta^2 \bar
 \theta \bar \lambda - i \bar \theta^2 \theta \lambda
 + \frac{1}{2} \theta^2  \bar \theta^2 D%
 \label{V}%
\ee

Combinando convenientemente supercampos se obtienen distintos
modelos su\-per\-si\-m\'{e}\-tri\-cos \footnote{A menos de t\'erminos
de superficie.}:

\noindent\underline{Modelo~de~Wess-Zumino} \ba
 {Cin\acute e ticos}:~~~~~~~~~~~~~~\int d^4\theta\, \Phi^\dagger\Phi
 &=&\partial_\mu\phi^\dagger\partial^\mu\phi
 +F^\dagger F -i\bar{\psi} \bar{\sigma}^\mu \partial_\mu \psi\\
 {Superpotencial}:~~~\int  d^2\theta\,{\cal W}(\Phi)+h.c.&=&
 \left (\frac{\partial {\cal W}}{\partial\phi}F
 -\frac{1}{2}\,\frac{\partial^{2}{\cal W}}{\partial^2 \phi}\psi^2
 \right)+h.c.
\ea \underline{Modelo~de~Higgs~supersim\'{e}trico} \ba
 {Cin\acute e ticos}:\int d^4\theta\, \Phi^\dagger e^{2V}\Phi
 &=&(D_\mu \phi)^\dagger(D^\mu \phi)-i\,\bar{\psi}\bar{\sigma}^\mu
 D_\mu\psi+ F^\dagger F\nonumber\\
 &&+ D \phi^\dagger\phi+i\sqrt{2}\,\phi^\dagger \lambda\psi
 -i\sqrt{2}\,\phi\, \bar{\psi}\bar{\lambda}\\
 {Fayet-Iliopoulos}:~~~\int d^4\theta\,2\xi^2\,V&=&
 \xi^2 D
\ea
donde la derivada covariante $D_\mu=\partial_\mu+iA_\mu$
\footnote{El acoplamiento con el campo de gauge $A_\mu$ en la
derivada covariante actuando sobre los fermiones de Weyl se
expresa en t\'{e}rminos de fermiones de Majorana como
$\bar\Psi\Gamma^\mu {\cal D}^{(5)}_\mu\Psi$ donde ${\cal
D}^{(5)}_\mu=\partial_\mu+i\Gamma^5A_\mu$. No puede ser de otra
manera ya que para majoranas $\bar\Psi\Gamma^\mu\Psi=0$.}. La
invarianza de supergauge restringe los posibles superpotenciales
que ahora deben ser invariantes frente a (\ref{agi}).

\noindent\underline{Super Yang-Mills}
\be
 \frac{1}{4\pi}{\rm Im}\left(\,\tau\,{\rm Tr}\int d^2\theta\,
 W^\alpha W_\alpha \right)=-\frac{1}{4g^2} F^a_{\mu\nu}F^{a\mu\nu}- \frac{\theta}{32\pi^2}
 F^a_{\mu\nu}\tilde F^{a\mu\nu}+ \frac{1}{2g^2}D^a D^a-\frac i{g^2}\lambda^a\sigma^\mu
 (D_\mu\bar{\lambda})^a
\ee
donde la normalizaci\'{o}n de los generadores es ${\rm Tr }\,t^a
t^b=\frac12\delta^{ab}$ y la derivada covariante $D_\mu$ actua
sobre el fermi\'{o}n en la adjunta $\bar\lambda=\bar\lambda^a t^a$
como $D_\mu\bar\lambda=\partial_\mu+i{[}A_\mu, \bar\lambda{]}$.
$F_{\mu\nu}$ tambi\'{e}n en la adjunta se escribe como
$F_{\mu\nu}=\partial_\mu A_\nu-\partial_\nu A_\mu+i[A_\mu,A_\nu]$.

\item Componentes de los distintos supercampos necesarios para la
construcci\'{o}n de la extensi\'{o}n supersim\'{e}trica de la acci\'{o}n de
Born-Infeld abeliana.

\underline{Supermultipletes de curvatura $W_\alpha$ y $\bar
W_{\dot\alpha}$}

\ba
 W_\alpha(y,\theta) &=&\frac 14\bar D_{\dot\beta}\bar D^{\dot\beta}
 D_\alpha V\\
 &=&i \lambda_\alpha-\theta_\alpha D+\frac i 2
         (\sigma^\mu\bar\sigma^\nu\theta)_\alpha F_{\mu\nu}
         -\theta^2 (/\!\!\!\partial\bar\lambda)_\alpha
\ea

\ba
 \bar W_{\dot\alpha}(y^\dagger,\bar\theta)&=&\frac 14 D^{\beta}
 D_{\beta} \bar D_{\dot\alpha} V\\
 &=&-i \bar\lambda_{\dot\alpha}-\bar\theta_{\dot\alpha} D-\frac i 2
 (\bar\sigma^\mu\sigma^\nu\bar\theta)_{\dot\alpha} F_{\mu\nu}
 +\bar\theta^2 (/\!\!\!\bar\partial\lambda)_{\dot\alpha}
\ea
donde
\ba
 (/\!\!\!\partial\bar\lambda)_\alpha&=&\sigma^\mu_{~\alpha\dot\alpha}
 \partial_\mu\bar\lambda^{\dot\alpha}\\
 (/\!\!\!\bar\partial\lambda)_{\dot\alpha}&=&\bar\sigma^{\mu~\beta}_{~\dot\alpha}
 \partial_\mu\lambda_{\beta}
\ea
Notemos que
\be
 \bar W_{\dot\alpha}=(W_\alpha)^\dagger
\ee
Solo hay componentes puramente bos\'{o}nicas en
$\theta^2(\bar\theta^2)$ de $W_\alpha(\bar W_{\dot\alpha})$.

\underline{Supercampo $W^2$}: En variables $(x,\theta,\bar\theta)$

\be
 \left. W^2\left( x\right) \right| _0=-\lambda^2
\ee

\be
 \left. W^2\left( x\right) \right| _\theta =-2iD\,\theta \lambda
 +F_{\mu \nu } \,\theta \sigma ^\mu \bar \sigma ^\nu \lambda
\ee

\be
 \left. W^2\left( x\right) \right| _{\theta^2 }=\theta^2 \left(
 -2i\lambda /\!\!\!\partial \bar \lambda +A\right)
\ee

\be
 \left. W^2\left( x\right) \right| _{\theta \bar \theta }=-i\theta
 \sigma ^\mu \bar \theta\, \partial _\mu \lambda^2
\ee

\be
 \left. W^2\left( x\right) \right| _{\theta^2 \bar \theta }=
 -\theta^2 \partial _\mu \left( \Omega ^{\mu \nu }\eta _{\nu \rho
 }\,\lambda \sigma ^\rho \bar \theta \right)
\ee

\be
 \left. W^2\left( x\right) \right| _{\theta^2 \bar \theta^2 }=\frac
 14\theta^2 \bar \theta^2 \Box  \lambda^2
\ee
donde
\begin{eqnarray*}
 A &=&D^2-\frac 12F^{\mu \nu }F_{\mu \nu }-\frac i2F^{\mu \nu
 }\tilde F_{\mu \nu } \\ A^{*} &=&D^2-\frac 12F^{\mu \nu }F_{\mu
 \nu }+\frac i2F^{\mu \nu }\tilde F_{\mu \nu }
\end{eqnarray*}
\be
\Omega ^{\mu\nu}=D\eta ^{\mu\nu}+iF^{\mu\nu}-\tilde F^{\mu\nu} \ee
En todos lados, salvo que se diga lo contrario, $D,F_{\mu \nu },$
$\tilde F_{\mu \nu }$ y $\lambda $ son funciones de $x$.

Las componentes de $\bar W^2$ pueden ser obtenidas de estas
\'ultimas expresiones calculando el adjunto.

\underline{Supercampo $W^2\bar W^2(x)$}: En variables x

\be
 \left. W^2\bar W^2\left( x\right) \right| _0=\lambda^2 \bar
 \lambda^2 \
\ee

\be
 \left. W^2\bar W^2\left( x\right) \right| _\theta =2i\bar
 \lambda^2 \left( D\, \theta \lambda -\frac i2\theta \sigma ^\mu \bar
 \sigma ^\nu \lambda F_{\mu \nu }\right)
\ee

\be
 \left. W^2\bar W^2\left( x\right) \right| _{\theta^2 }=\theta^2
 \bar\lambda^2 \left(2i\,\lambda /\!\!\!\partial \bar \lambda
 -A\right)
\ee

\begin{eqnarray}
 \left. W^2\bar W^2\left( x\right) \right| _{\theta \bar \theta }
 &=& -i\theta\sigma^\mu\bar \theta~ \lambda^2
 \stackrel{\longleftrightarrow }{\partial_\mu} \bar\lambda^2
 +4\left(D\,\theta\lambda+\frac i2F_{\mu \nu }\,
 \theta\sigma^\mu\bar\sigma^\nu\lambda
 \right) \nonumber\\
 &&\cdot\left(D\,\bar\theta\bar\lambda-\frac i2 F_{\rho\sigma}\,
 \bar\theta\bar\sigma^\rho\sigma^\sigma\bar\lambda\right)
\end{eqnarray}
\begin{eqnarray}
 \left. W^2\bar W^2\left( x\right) \right| _{\theta^2 \bar \theta }
 &=&\theta^2 \left\{ \left( 2iD\,\bar\theta\bar\lambda-\bar
 \lambda \bar \sigma ^\mu \sigma ^\nu \bar \theta F_{\mu \nu
 }\right) \left( -2i\lambda /\!\!\!\partial \bar \lambda +A\right)
 +\right.
 \nonumber\\
 & & \left.\left(\Omega^{\tau\rho}\eta_{\tau\mu}
 \bar\theta\bar\sigma^\mu\lambda\right)
 \stackrel{\longleftrightarrow }{\partial_\rho}\left(\bar\lambda^2\right)
 \right \}
\end{eqnarray}
\begin{eqnarray*}
 \left. W^2\bar W^2 \right|_{\theta^2 \bar \theta^2 } &=&\theta^2
 \bar \theta^2 \left\{ -\frac 14 \left( \lambda^2 \Box \bar
 \lambda^2 +\bar \lambda^2 \Box \lambda^2 \right) +\frac 12
 \partial _\mu
 \left( \lambda^2 \right)
 \partial ^\mu
 \left( \bar \lambda^2 \right) \right.  \\ && -4 \left( \lambda
 /\!\!\!\partial \bar \lambda \right) \left( \bar \lambda \bar
 {/\!\!\!\partial}\lambda \right) \left. -2iA^{*}\lambda
 /\!\!\!\partial \bar \lambda- 2iA\,\bar\lambda/\!\!\!\bar\partial\lambda
 +AA^{*}\right.\nonumber \\
 && \left.+2i\,{\rm Im}U \right \}
\end{eqnarray*}
donde
\[
A\!\stackrel{\leftrightarrow }{\partial }\!B=A\partial B-\left(
\partial A\right) B
\]
\be
 U=-i\lambda\,\partial_\rho\left( (\Omega^{\rho\mu})^*\sigma^\nu\bar\lambda\right)
 \Omega_{\mu\nu}
\ee

\be
 \Omega ^{*\nu \rho }\Omega _{\rho \mu }= \left( D^2+\frac
 12F_{\alpha \beta }F^{\alpha \beta } \right) \delta _\mu ^\nu
 -2D\eta ^{\nu \rho }\tilde F_{\rho \mu }+2F^{\nu \rho }F_{\rho \mu}
\ee
y
\be
 {Im} \left(
 \partial _\nu
 \left( \Omega ^{*\nu \rho } \right) \Omega _{\rho \mu } \right)
 =-D\partial _\nu F^{\nu \mu }+\partial _\nu \left( F^{\nu \alpha }
 \right) \tilde F_{\alpha \mu }+\tilde F^{\nu \alpha }\partial _\nu
 F_{\alpha \mu }+\frac 12\partial _\mu (F_{\alpha \beta })\tilde
 F^{\alpha \beta }
 \label{sonia}
\ee

\vspace{0.4 cm}

\underline{Supercampo $X\left( x\right) \equiv \frac 18\left(
D^2W^2(x)+\bar D^2\bar W^2(x)\right) $}

\be
 \left. X\right| _0=\left( i\left( \lambda /\!\!\!\partial \bar
 \lambda +\bar \lambda \bar {/\!\!\!\partial }\lambda \right)
 -D^2+\frac 12F^2\right)
\ee

\be
 \left. X\right| _\theta =-\partial_\mu(D\,\bar\lambda\bar\sigma^\mu
 \theta)-\frac i2 \partial_\rho(F_{\mu\nu}\,\bar\lambda
 \bar\sigma^\mu\sigma^\nu\bar\sigma^\rho\theta)
\ee

\be
 \left. X\right| _{\bar\theta }=\partial_\mu(D\,\lambda\sigma^\mu
 \bar\theta)-\frac i2 \partial_\rho(F_{\mu\nu}\,\lambda
 \sigma^\mu\bar\sigma^\nu\sigma^\rho\bar\theta)
\ee

\be
 \left. X\right| _{\theta \bar \theta }=i\theta\sigma^\mu
 \bar\theta\,\partial_\mu\left(\lambda /\!\!\!\partial\bar\lambda
 -\bar\lambda\bar {/\!\!\!\partial }\lambda+\frac 12F\tilde F\right)
\ee

\be
 \left. X\right| _{\theta^2 }=-\frac 12\theta^2 \Box \bar \lambda^2
\ee

\be
 \left. X\right|_{\bar\theta^2}=-\frac 12\bar\theta^2\Box\lambda^2
\ee

\be
 \left. X\right| _{\theta^2 \bar \theta }=\theta^2 \Box\left[\frac i2
 \left( D\, \bar\lambda\bar\theta\right)
 -\frac14\left(F_{\mu\nu}\bar\lambda\bar\sigma^\mu\sigma^\nu\bar\theta
 \right) \right]
\ee

\be
 \left. X\right| _{\theta \bar \theta^2 }=-\bar\theta^2\Box \left[\frac i2
 \left( D\, \lambda\theta\right)
 +\frac14\left(F_{\mu\nu}\lambda\sigma^\mu\bar\sigma^\nu\theta
 \right) \right]
\ee

\be
 \left. X\right| _{\theta^2 \bar \theta^2 }=-\frac 14\theta^2 \bar
 \theta^2 \Box \left( i\left( \lambda /\!\!\!\partial \bar \lambda
 +\bar \lambda \bar { /\!\!\!\partial}\lambda \right) -D^2+\frac
 12F^2\right)
\ee

\vspace{0.4 cm}

\underline{Supercampo $Y\left( x\right) \equiv -\frac i{16}\left(
D^2W^2(x)-\bar D^2\bar W^2(x)\right) $}

\be
 \left. Y\right| _0=\frac 12 \left( \lambda /\!\!\!\partial
 \bar \lambda -\bar \lambda \bar {/\!\!\!\partial }\lambda \right)
 +\frac 14 F\tilde F
\ee

\be
 \left. Y\right| _\theta =-\frac i2\partial_\mu(D\,\bar\lambda\bar\sigma^\mu
 \theta)+\frac 14 \partial_\rho(F_{\mu\nu}\,\bar\lambda
 \bar\sigma^\mu\sigma^\nu\bar\sigma^\rho\theta)
\ee

\be
 \left. Y\right| _{\bar \theta }=-\frac i2 \partial_\mu(D\,\lambda\sigma^\mu
 \bar\theta)-\frac 14 \partial_\rho(F_{\mu\nu}\,\lambda
 \sigma^\mu\bar\sigma^\nu\sigma^\rho\bar\theta)
\ee

\be
 \left. Y\right| _{\theta \bar \theta }=\frac 12\theta\sigma^\mu
 \bar\theta\,\partial_\mu\left(-i(\lambda /\!\!\!\partial\bar\lambda
 +\bar\lambda\bar {/\!\!\!\partial }\lambda)+D^2-\frac 12 F^2\right)
\ee

\be
 \left. Y\right| _{\theta^2 }=-\frac i4\theta^2 \Box\bar\lambda^2
\ee

\be
 \left. Y\right| _{\bar \theta^2 }=\frac i4\bar \theta^2 \Box\lambda^2
\ee

\be
 \left. Y\right| _{\theta^2 \bar \theta }=-\theta^2\Box\left[\frac 14
 (D\,\bar\lambda\bar\theta)+\frac i8(F_{\mu\nu}\, \bar\lambda\bar\sigma^\mu
 \sigma^\nu\bar\theta) \right]
\ee

\be
 \left. Y\right| _{\bar\theta^2\theta }=-\bar\theta^2\Box\left[\frac 14
 (D\,\lambda\theta)-\frac i8(F_{\mu\nu}\, \lambda\sigma^\mu
 \bar\sigma^\nu\theta) \right]
\ee

\be
 \left. Y\right| _{\theta^2 \bar \theta^2 }=-\frac 18 \theta^2 \bar
 \theta^2 \Box \left( \left( \lambda /\!\!\!\partial \bar \lambda
 -\bar \lambda \bar {/\!\!\!\partial }\lambda \right) +\frac
 12F\tilde F\right)
\ee

\end{itemize}



\chapter{Reducci\'on dimensional $d=4 \to d=3$\label{a3}}

\begin{itemize}

\item La reducci\'on dimensional en la coordenada $x^3$ consiste en tomar
\ba .~~\partial_3&=&0 \nonumber \\ .~~A_3&\to& {\rm campo~escalar}
\nonumber \ea
\item Para los \' \i ndices en $d=4$ usamos letras griegas $\mu ,\nu
,..=0,1,2,3$.

\item Para los \' \i ndices en $d=3$ letras latinas $i, j,..=0,1,2$.

\item La convenci\'on para la m\'etrica es $g={\rm diag}(+,-,-,...)$.

\item Los tensores de Levi-Civita valen
\ba {\underline
{d=4}}:&~\varepsilon_{0123}=1,~&\varepsilon^{0123}=-1 \nonumber \\
{\underline{d=3}}:&~\varepsilon_{012}=1,~&~\varepsilon^{012}=1
\nonumber \ea
\underline{Tensor de campo electromagn\'etico}

\item  La 1-forma $A_\mu$ se reduce
$$ A_\mu=(A_0,A_1,A_2,A_3)\longrightarrow
\left[A_i=(A_0,A_1,A_2)\right]\oplus N$$

\item La 2-forma $F_{\mu\nu}=\partial_\mu
A_\nu-\partial_\nu A_\mu$ en $d=4$ queda
\be
 F_{\mu\nu}=\left(
 \begin{array}{cccc}
  0             & F_{01}        & F_{02}        & \partial_0 N \\
  F_{10}        & 0             & F_{12}        & \partial_1 N \\
  F_{20}        & F_{21}        & 0             & \partial_2 N \\
  -\partial_0 N & -\partial_1 N & -\partial_2 N & 0
 \end{array}
 \right)
\ee
Luego resulta que
\be
 F_{\mu\nu}F^{\mu\nu}|_{d=4} \longrightarrow F_{ij}F^{ij}-2
 \partial_i N \partial^i N|_{d=3}
\ee
\item El dual de Hodge definido en $d=4$ como $\tilde F^{\mu\nu}=\frac 1 2
\varepsilon^{\mu\nu\rho\sigma}F_{\rho\sigma}$ queda
\be
 \tilde F^{\mu\nu}=\left(
 \begin{array}{cccc}
  0              & -\partial_2 N   & \partial_1 N   & -F_{12} \\
  \partial_2 N   & 0               & -\partial_0 N  & F_{02} \\
  -\partial_1 N  & \partial_0 N    & 0              & -F_{01}\\
  F_{12}         & -F_{02}         & F_{01}         & 0
 \end{array}
 \right)
\ee
que da como resultado
\be
 F_{\mu\nu}\tilde F^{\mu\nu}|_{d=4} \longrightarrow
 -4\partial_i N \tilde F^i|_{d=3}
\ee
donde se define $\tilde F^i=\frac {\varepsilon^{ijk}} 2
F_{jk}=(F_{12},-F_{02},F_{01})=(-B,-E_y,E_x)$.

\underline{Contracciones del tensor de Levi-Civita}
\item En $d=3$ tenemos
\ba
 \varepsilon^{ijk}\varepsilon_{iab}&=&
 \delta^i_a\delta^j_b-\delta^i_b\delta^j_a=\delta^{jk}_{ab}\\
 \varepsilon^{ijk}\varepsilon_{ija}&=&2\delta^k_a\\
 \varepsilon^{ijk}\varepsilon_{ijk}&=&3!
\ea

\underline{Espinores}

\item Denotamos las matrices de Dirac como $\Gamma$ en
$d=4$ y como $\gamma$ en $d=3$.

\item Los espinores (Majorana) en $d=4$ los denotamos con letras
may\'usculas griegas $\Psi,\Lambda,\Upsilon$. Los espinores de
Dirac en $d=3$ por may\'usculas griegas $\Sigma,\Omega$ y los
espinores de Majorana en $d=3$ mediante min\'usculas griegas
$\psi,\lambda,\epsilon$.

\item Las matrices de Dirac en $d=4$ las escribimos
\footnote{La matriz $\Gamma^5$ fue definida en (\ref{g5}).}
\ba
 \Gamma^i=& \gamma^i \otimes \sigma^3&=\left(\begin{array}{cc}
   \gamma^i & 0 \\
   0        & -\gamma^i \
 \end{array}\!\!\right)
 \label{tu}\\
 \Gamma^3=& 1 \otimes i\sigma^1&=\left(\begin{array}{cc}
   0  & i \\
   i & 0 \
 \end{array}\!\right)\\
 \Gamma^5=& 1 \otimes -\sigma^2&=\left(\begin{array}{cc}
   0 & i \\
   -i & 0 \
 \end{array}\right)
 \label{ti}
\ea
donde las matrices de Dirac $ (2\times2)$ en $d=3$ son
\ba
 \gamma^0=&\sigma^2=&\left(
 \begin{array}{cc}
   0 & -i \\
   i & 0\! \
 \end{array}\right)\\ 
 \gamma^1=&i\sigma^1=&\left(
 \begin{array}{cc}
   0 & i \\
   i & 0 \
 \end{array}\!\right)\\ 
 \gamma^2=&i\sigma^3=&\left(
 \begin{array}{cc}
   i & 0 \\
   0 & -i \
 \end{array}\!\!\right)
\ea
de manera que tenemos una representaci\'on de Majorana (imaginaria
pura) tanto en $d=3$ como en $d=4$. Para poder hallar la matriz de
conjugaci\'on de carga $C$ definida por la ec.(\ref{carga})
necesitamos conocer $\Gamma^T$
\be
 {\underline {d=4}}: \left\{
 \begin{array}{c}
   \Gamma^{0\,T}=-\Gamma^0~~~~~~~~~~~~~~~~~~~ \\
    \Gamma^{i\,T}=\Gamma^i~~~~~(i=1,2,3) \
 \end{array}\right.
\ee
\be
 {\underline {d=3}}:\left\{
 \begin{array}{c}
   \gamma^{0\,T}=-\gamma^0~~~~~~~~~~~~~~~~~~~ \\
   \gamma^{a\,T}=\gamma^a~~~~~(a=1,2) \
 \end{array}\right.
\ee
Si definimos la matriz de conjugaci\'on de carga $C$ como (cf.
ec.(\ref{ce}))
\be
 \underline {d=4}:~~C=-\Gamma^0
\ee
\be
 \underline {d=3}:~~C=-\gamma^0
\ee
con esta definici\'on $C^2=1$ y los espinores autoconjugados de
carga $\Psi_M$ tienen componentes reales )\footnote{Confrontar con
la representaci\'on quiral (\ref{rep}) donde el espinor de
Majorana se expresa en t\'erminos de espinores de Weyl como
\be
Q_a^A=\left(  \begin{array}{c}
  Q_{\alpha}^A \\
  \bar Q^{\dot\alpha}_A \
 \end{array} \right)
\ee
y satisface (ver ec.(\ref{mayorana}))
\be
 C_{ab}\bar Q_b^A=Q_a^A~~~~\Longrightarrow~~~~\bar
 Q_a^A=-C_{ab}Q^A_b=Q^A_b C_{ba}
\ee
Para $C$ dada por la ec.(\ref{ce}) se tiene $C^2=-1$.}
\be
 {\underline{\rm Espinor~de~Majorana}}:\Psi_M~~~
 \Leftrightarrow~~~\Psi_M^c=\Psi_M~~~\Longrightarrow
 \Psi^\ast=\Psi
 \label{lalala}
\ee
Al efectuar la reducci\'on dimensional de un fermi\'on de Majorana
en $d=4$ obtenemos
\ba
 \Psi_M&=&\left(\begin{array}{c}
   \psi_1 \\
   \psi_2 \
 \end{array}\right)\\
 \bar\Psi_M&=&\left( \begin{array}{cc}
   \bar\psi_1 & -\bar\psi_2 \
 \end{array}\right)
 \label{relac}
\ea
donde $\psi_1,~\psi_2$ son dos fermiones de Majorana en $d=3$ (o
sea fermiones de 2 componentes reales)\footnote{La definici\'on
del adjunto de Dirac en $d=3$ es (\ref{adj}):
$\bar\psi=\psi^\dagger\gamma^0$.}. Por conveniencia de notaci\'on
acomodaremos estos dos fermiones en un fermi\'on de Dirac
$\Sigma$.

\item  Denotamos los generadores de Lorentz como $\Sigma^{\mu\nu}$ en
$d=4$ y como $\Delta^{ij}$ en $d=3$.
\ba
 \Sigma^{\mu\nu}&=&\frac i4 \left[ \Gamma^\mu,\Gamma^\nu\right]\\
 \Delta^{ij}&=&\frac i4 \left[ \gamma^i,\gamma^j\right]=-\frac 12
 \varepsilon^{ijk}\gamma_k
\ea
Entonces para la representaci\'on de Majorana
(\ref{tu})-(\ref{ti}) tenemos
\ba
 \Sigma^{ij}=&\Delta^{ij}\otimes I_2&=~\left(
 \begin{array}{cc}
   \Delta^{ij} & 0 \\
   0   & \Delta^{ij} \
 \end{array}\!\!\right)\\
 \Sigma^{3i}=&\gamma^{i}\otimes \frac i2\sigma^2&=\frac12\left(\!
 \begin{array}{cc}
   0          & \gamma^i \\
   -\gamma^i  & 0 \
 \end{array}\right)
\ea

\underline{Propiedades de espinores de Majorana en $d=3$}

\item En $d=3$ tenemos propiedades an\'alogas a
(\ref{maj1})-(\ref{maj2}).
\ba
 \bar\psi\gamma^i\psi&=&0\\
 \bar\lambda\psi&=&\bar\psi\lambda\\
 \bar\lambda\gamma^i\psi&=&-\bar\psi\gamma^i\lambda
\ea
\underline{Propiedades de las matrices $\gamma$}
\ba
 \gamma^{0\dagger}&=&\gamma^0\\
 \gamma^{j\dagger}&=&-\gamma^j~~~~~~~(j=1,2)\\
 \gamma^{i\dagger}&=&\gamma^0\gamma^i\gamma^0\\
 \gamma^0\gamma^1\gamma^2&=&iI_2\\
 \gamma^i\gamma^j&=& g^{ij}+i\varepsilon^{ijk}\gamma_k\\
 \left[\gamma^i,\gamma^j\right]&=&2i\varepsilon^{ijk}\gamma_k
\ea

\vskip 2cm
\underline{Reducci\'on del \'algebra supersim\'{e}trica $N=1$ ($d=4$)
a $N=2$ ($d=3$)}

\item Teniendo en cuenta las relaciones (\ref{relac}) denotamos
las supercargas Majorana en $d=4$ como
\be
 Q=\left(
 \begin{array}{c}
   Q^{1} \\
   Q^{2} \
 \end{array}
 \right)
\ee
donde $Q^{1,2}$ son las supercargas Majorana en $d=3$. Partiendo
del \'algebra supersim\'{e}trica $N=1$ en $d=4$ ec.(\ref{techito})
tenemos
\be
 \{Q_a,\bar Q_b\}=\left(
 \begin{array}{cc}
   \{Q^{1},\bar Q^{1}\} & -\{Q^{1},\bar Q^{2}\} \\
   \{Q^{2},\bar Q^{1}\} & -\{Q^{2},\bar Q^{2}\} \
 \end{array}
 \right)=-2\left(
 \begin{array}{cc}
    /\!\!\!\!P & iP_3 \\
   iP_3 &  -/\!\!\!\!P \
 \end{array}
 \right)
\ee
donde $ /\!\!\!\!P=\gamma^iP_i$. Obteni\'endose
\ba
 \{Q^{1},\bar Q^{1}\}&=&-2 /\!\!\!\!P
 \label{no1}\\
 \{Q^{2},\bar Q^{2}\}&=&-2 /\!\!\!\!P
 \label{no2}\\
 \{Q^{1},\bar Q^{2}\}&=&2i P_3
 \label{no3}\\
 \{Q^{2},\bar Q^{1}\}&=&-2 iP_3
 \label{no4}
\ea
Multiplicando por $C$ a derecha en las ecuaciones anteriores y
utilizando la propiedad de que $C\bar Q^{1,2\,T}=Q^{1,2}$
obtenemos (comparar con ec.(\ref{amsa}))
\be
 \{Q^{A}_a, Q^{B}_b\}=2 (\gamma^iC)_{ab}P_i\delta^{AB}-
 2i C_{ab}P_3\epsilon^{AB}
\ee
donde $A,B=1,2$. Definiendo un espinor de Dirac $Q=\frac 12
(Q^1+iQ^2)$ tenemos
\be
 \{Q,\bar Q\}=-( /\!\!\!\!P-Z)
 \label{potato}
\ee
donde $Z=P_3$.
\end{itemize}



\newpage



\begin{thebibliography}{99}


\bibitem{B} M.~Born, Nature {\bf 132} (1933) 282, 1004.

M.~Born, {\sl On the Quantum Theory of Electromagnetic Field}
Proc. R. Soc. (London) {\bf A143} (1934) 410.

\bibitem{BI} M.~Born and L.~Infeld, {\sl Foundations of the New
Field Theory}, Proc. Roy. Soc. (London) {\bf A144} (1934) 425.

\bibitem {diri} P.A.M. Dirac, {\sl Classical Theory of Radiating
Electrons}, Proc. Roy. Soc. (London) {\bf A167} (1938) 148.

M.~Born, {\sl Th\'eorie non-lin\'eare du champ
\'electromagn\'etique}, Ann. Inst. Poincar\'e {\bf 7} (1939) 155.

\bibitem{gl} Y.A.~Gol'fand and E.S.~Likhtman, {\sl Extension
of the Algebra of Poincar\'e Group Ge\-ne\-ra\-tors and Violation
of P-Invariance}, JETP Lett. {\bf13} (1971) 323.

\bibitem{wezu} J. Wess y B. Zumino, {\sl Supergauge Transformations
in Four Dimensions}, Nucl. Phys. {\bf B70} (1974) 39.

\bibitem{wess} J.~Wess y J.~Bagger, {\sl Supersymmetry and
Supergravity}, Princeton, Princeton University Press, (1983).

\bibitem{soh} M.~Sohnius, {\sl Introducing Supersymmetry},
Phys. Rep. {\bf 128} (1985) 39.

\bibitem{freund} P.~Freund, {\sl Introduction to Supersymmetry},
Cambridge University Press, 1986.

\bibitem{susy} C.G.~Bollini, {\sl Supersimetr\' \i a
(Notas de Clase)}, Centro At\'{o}mico Bariloche, 1985.

P.C.~West, {\sl Introduction to Supersymmetry and Supergravity},
World Scientific, 1990.

\bibitem{bailin} D.~Bailin y A.~Love, {\sl Supersymmetric
gauge field theory and string theory}, Institute of Physics
Publishing, Bristol, (1994).

\bibitem{lykken}  J.~Lykken, {\sl Introduction to Supersymmetry},
TASI lectures 1996, {\tt hep-th/9612114}.

\bibitem{haag} R.~Haag, J.T.~{\L}opusza\'nski y M.~Sohnius,
{\sl All Possible Generators of Supersymmetries of the S-Matrix},
Nucl. Phys. {\bf B88} (1975) 257.

\bibitem{sw} N.~Seiberg y E.~Witten, {\sl Electric-Magnetic Duality,
Monopole Condensation, and Confinement in N=2 Supersymmetric
Yang-Mills Theory}, Nucl. Phys. {\bf B426} (1994) 19, Erratum-{\it
ibid.} {\bf B430} (1994) 485,  {\tt hep-th/9407087}.

{\sl Monopoles, Duality and Chiral Symmetry Breaking in N=2
Supersymmetric QCD}, Nucl. Phys. {\bf B431} (1994) 484, {\tt
hep-th/9408099 }.

\bibitem{inst} A.A. Belavin, A.M. Polyakov, A.S. Schwartz y Y.S.
Tyupkin, {\sl Pseudoparticle Solutions of the Yang-Mills
Equations}, Phys. Lett {\bf B59} (1975) 85.

\bibitem{dVS} H.~de Vega and F.A.~Schaposnik, {\sl A classical vortex
solution of the abelian Higgs model}, Phys. Rev. {\bf D14} (1976)
1100.

\bibitem{ps} M.K.~Prasad y C.M.~Sommerfield, {\sl An Exact
Classical Solution for the 't Hooft Monopole and the Julia-Zee
dyon}, Phys. Rev. Lett. {\bf 35} (1975) 760.

\bibitem{Bogo} E.B. Bogomol'nyi, {\sl The stability of classical
solutions}, Sov. J. Nucl. Phys. {\bf 24} (1976) 449.

\bibitem{NS1} K.~Shiraishi and S.~Hirenzaki, {\sl Bogomol'nyi equations
for vortices in Born-Infeld-Higgs systems}, Int. Jour. of Mod.
Phys. {\bf A6} (1991) 2635.

\bibitem{NS2} A.~Nakamura and K.~Shiraishi, {\sl Born-Infeld Monopoles
and Instantons}, Hadronic Journal {\bf 14} (1991) 369.

\bibitem{raja} R. Rajaraman, {\sl Solitons and Instantons in
Quantum Field Theory}, North-holland, 1982.


\bibitem{OW} E. Witten and D. Olive, {\sl Supersymmetry Algebras
that include Topological Charges}, Phys. Lett. {\bf B78} (1978)
97.

\bibitem{LLW} C.~Lee, K.~Lee and E.J.~Weinberg, {\sl Supersymmetry
and Selfdual Chern-Simons systems}, Phys. Lett. {\bf B243} (1990)
105.

C.~Lee, K.~Lee and H.~Min, {\sl Selfdual Maxwell Chern-Simons
Solitons}, Phys. Lett.  {\bf B252} (1990) 79.

B. Damski, {\sl Supersymmetry and Bogomol'nyi equations in the
Maxwell-Chern-Simons systems}, {\tt  hep-th/0001022 }

\bibitem{ed} J.D.~Edelstein, C.~N\'u\~nez and F.A.~Schaposnik, {\sl
Supersymmetry and Bogomol'nyi equations in the abelian Higgs
model}, Phys. Lett. {\bf B329} (1994) 39, {\tt hep-th/9311055}.

\bibitem{cbpf} P.~Navratil, {\sl N=2 supersymmetry in a Chern-Simons
system with the magnetic moment interaction}, Phys. Lett.  {\bf
B365} (1996) 119.

H.R. Christiansen, M.S. Cunha, J.A. Helay\"{e}l-Neto, L.R.U. Manssur
and A.L.M. Nogueira {\sl N = 2 Maxwell-Chern-Simons model with
anomalous magnetic moment  coupling via dimensional reduction},
Int. J. Mod. Phys. {\bf A14} (1999) 147,  {\tt hep-th/9802096}.

\bibitem{tring} M.B. Green, J.H. Schwarz y E.Witten, {\sl Superstring
Theory: Vol I \& II}, Cambridge University Press, 1987.

J. Polchinski, {\sl String Theory}, Cambridge University Press,
1998.

\bibitem{ss} J. Scherk y J.H. Schwarz, {\sl Dual Models of Non-Hadrons},
Nucl. Phys. {\bf B81} (1974) 118

J. Scherk y J.H. Schwarz, {\sl Dual Models and the Geometry of
Space-Time}, Phys. Lett. {\bf B52} (1974) 347.

\bibitem{poly} A.M.~Polyakov, {\sl Quantum geometry of bosonic
strings}, Phys. Lett. {\bf B103} (1981) 207.

A.M.~Polyakov, {\sl Quantum geometry of fermionic strings}, Phys.
Lett. {\bf B103} (1981) 211.

\bibitem{Tse} E.S.~Fradkin and A.A.~Tsey\-tlin,
{\sl Non-linear Electrodynamics From Quantized Strings}, Phys.
Lett. {\bf B163} (1985) 123.

\bibitem{tse} A.A. Tseytlin, {\sl Vector Field
Effective Action in the Open Superstring Theory}, Nucl. Phys. {\bf
B276} (1986) 391, Erratum-{\it ibid.} {\bf B291} (1987) 876.

\bibitem{AN} A.~Abouelsaood, C.G.~Callan, C.R.~Nappi and
S.A.~Yost, {\sl Open Strings in Background Fields}, Nucl. Phys.
{\bf B280} (1987) 599.

\bibitem{Lei} J.~Dai, R.G.~Leigh and J.~Polchinski, {\sl New
Connections Between String Theories}, Mod. Phys. Lett. {\bf A4}
(1989) 2073.

\bibitem{lei2} R.G.~Leigh, {\sl Dirac-Born-Infeld Action
from the Dirichlet Sigma Model}, Mod. Phys. Lett. {\bf A4} (1989)
2767.

\bibitem{Pol2} J.~Polchinski, {\sl Dirichlet branes and
Ramond-Ramond charges}, Phys. Rev. Lett. {\bf 75} (1995) 4724,
{\tt hep-th/9510017}.

\bibitem{Pol} J.Polchinski, {\sl TASI Lectures on D-branes}, TASI 96,
World Scientific (1997), {\tt hep-th/9611050}.

\bibitem{ads} J. Maldacena, {\sl The Large N Limit of Superconformal
Field Theories and Supergravity}, Adv. Theor. Math. Phys. {\bf 2}
(1998) 231, {\tt hep-th/9711200}.

\bibitem{Wi} E.~Witten, {\sl Bound States of Strings and
p-Branes}, Nucl. Phys. {\bf B460} (1996) 335, {\tt
hep-th/9510135}.

\bibitem{stro} A.~Strominger, {\sl Open p-branes}, Phys. Lett. {\bf
B383} (1996) 44, {\tt hep-th/9512059}.

\bibitem{CM} C.G.~Callan and J.M.~Maldacena, {\sl Brane dynamics from
the Born-Infeld action}, Nucl. Phys. {\bf B513} (1998) 198, {\tt
hep-th/9708147}.

\bibitem{G} G.~Gibbons, {\sl Born-Infeld particles and Dirichlet p-branes},
Nucl. Phys. {\bf B514} (1998) 603, {\tt hep-th/9709027}.

\bibitem{Hashi} A.~Hashimoto, {\sl The Shape of Branes
Pulled by Strings}, Phys. Rev. {\bf D57} (1998) 6441, {\tt
hep-th/9711097}.

\bibitem{BLM} D.~Bak, J.~Lee and H.~Min, {\sl Dynamics of BPS States
in the Dirac-Born-Infeld Theory}, Phys. Rev. {\bf D59} (1999)
045011, {\tt hep-th/9806149}.

\bibitem{H} T.~Hagiwara, {\sl An Effective Lagrangian for
Multi-Photon Processes and a Nonlinear Born-Infeld Lagrangian},
Nucl. Phys. {\bf B189} (1981) 135.

T.~Hagiwara, {\sl A Nonabelian Born-Infeld Lagrangian}, J. Phys.
{\bf A14} (1981) 3059.

\bibitem{arg} P.C. Argyres y C.R. Nappi, {\sl Spin 1 Effective
Actions From Open Strings}, Nucl. Phys. {\bf B330} (1990) 151.

\bibitem{Tse2} A.A.~Tseytlin, {\sl On Non-Abelian generalization of
Born-Infeld action in string theory}, Nucl. Phys. {\bf B501}
(1997) 41, {\tt hep-th/9701125}.

\bibitem{BP} D.~Brecher and M.J.~Perry, {\sl Bound States of D-Branes
and the Non-Abelian Born-Infeld Action}, Nucl. Phys. {\bf B527}
(1998) 121, {\tt hep-th/9801127}.

\bibitem{Bre}  D.~Brecher, {\sl BPS States of the Non-Abelian
Born-Infeld Action}, Phys. Lett. {\bf B442} (1998) 117, {\tt
hep-th/9804180}.

\bibitem{Sch} J.H.~Schwarz, {\sl An SL(2,Z) Multiplet of Type IIB
Superstrings}, Phys. Lett. {\bf B360} (1995) 13, Erratum-{\it
ibid.} {\bf B364} (1995) 252, {\tt hep-th/9508143}.

\bibitem{dirac} P.A.M.~Dirac, {\sl An extensible model of the
electron}, Proc. Roy. Soc. {\bf A268} (1962) 57.

\bibitem{tow} J.A.~de Azcarraga, J.P. Gauntlett, J.M. Izquierdo y
P.K. Townsend, {\sl Topologocal Extensions of the Supersymmetry
Algebra for Extended Objects}, Phys. Rev. Lett. {\bf 63} (1989)
2443.

P.K. Townsend, {\sl M-Theory from its Superalgebra}, Lectures at
Cargese 1997, {\tt hep-th/9712004}.

\bibitem{cole} S.~Coleman y J.~Mandula, {\sl All Possible
Symmetries of the S-Matrix}, Phys. Rev. {\bf 159} (1967) 1251.

\bibitem{mack} G.~Mack y A.~Salam, {\sl Finite-Component
Field Representations of the Conformal Group}, Ann. Phys.(NY) {\bf
53} (1969) 174.

\bibitem{wein} S.~Weinberg, {\sl The Quantum Theory of Fields},
Cambridge University Press, 1996.

\bibitem{dual} P. Di Vecchia, {\sl Duality in N=2, N=4 Supersymmetric
Gauge Theories}, Lectures given at Les Houches Summer School in
Theoretical Physics, Session 68: Probing the Standard Model of
Particle Interactions, Les Houches, France, {\tt hep-th/9803026}.

L. Alvarez-Gaume y S.F. Hassan, {\sl Introduction to S duality in
$N=2$ supersymmetric gauge theories: A pedagogical review of the
work of Seiberg and Witten}, Fortsch. Phys. {\bf 45} (1997) 159,
{\tt hep-th/9701069}.

A. Harvey, {\sl Magnetic Monopoles, Duality and Supersymmetry},
Published in Trieste HEP Cosmology 1995, {\tt hep-th/9603086}.

\bibitem{kir} E. Kiritsis, {\sl Supersymmetry and Duality in Field
and String Theory}, Lectures given at NATO Advanced Study
Institute: TMR Summer School on Progress in String Theory and
M-Theory (Cargese 99), Cargese,  {\tt hep-ph/9911525}.

\bibitem{DP}  S.~Deser and R.~Puzalowski, {\sl Supersymmetric
Non-Polynomial Vector Multiplets and Causal Propagation}, J. Phys.
{\bf A13} (1980) 2501.

\bibitem{CF}  S.~Ceccotti and S.~Ferrara, {\sl Supersymmetric
Born-Infeld Lagrangians}, Phys. Lett. {\bf B187} (1987) 335.

\bibitem{BG} J.~Bagger and A.~Galperin, {\sl New Goldstone Multiplet
for Partially Broken Supersymmetry}, Phys. Rev. {\bf D55} (1997)
1091, {\tt hep-th/9608177}.

J.~Bagger, {\sl Partial Breaking of Extended Supersymmetry}, Nucl.
Phys. Proc. Suppl. {\bf A52} (1997) 362,  {\tt hep-th/9610022}.

\bibitem{APS} M.~Aganagic, C.~Popescu and J.H.~Schwarz, {\sl D-brane
actions with local kappa symmetry}, Phys. Lett. {\bf B393} (1997)
311, {\tt hep-th/9610249}.

{\sl Gauge invariant and gauge fixed D-brane actions}, Nucl. Phys.
{\bf B 495} (1997) 99, {\tt hep-th/9612080}.

\bibitem{ketov} S. Ketov, {\sl A Manifestly  N=2 Supersymmetric
Born-Infeld Action}, Mod. Phys. Lett. {\bf A14} (1999) 501, {\tt
hep-th/9809121}.

{\sl Born-Infeld-Goldstone Superfield Actions for Gauge Fixed D-5
Branes and D-3 Branes in 6-D}, Nucl. Phys. {\bf B553} (1999) 250,
{\tt hep-th/9812051}.

\bibitem{RT} M.~Rocek and A.A.~Tseytlin, {\sl Partial
breaking of global D=4 supresymmetry, cons\-trained superfields,
and 3-brane actions}, Phys. Rev. {\bf D59} (1999) 106001, {\tt
hep-th/9811232}.

\bibitem{T} A.A.~Tseytlin, {\sl Born-Infeld action, supersymmetry
and string theory}, en {\it Yuri Golfand memorial volume}, ed.
M.~Shifman, World. Sci. 2000 {\tt hep-th/9908105}.

\bibitem{HT} A.~Hashimoto and W.I.~Taylor IV, {\sl Fluctuation spectra
of tilted and intersecting D-branes from the Born-Infeld action},
Nucl. Phys. {\bf 503} (1997) 193, {\tt hep-th/9703217}.

\bibitem{NO} H. B. Nielsen and P. Olesen, {\sl Vortex line models for
dual strings}, Nucl. Phys. {\bf B61} (1973) 45.

\bibitem{tofpol} G. 't Hooft, {\sl Magnetic Monopole in Unified
Gauge Theories}, Nucl. Phys. {\bf B79} (1974) 276.

A.M. Polyakov, {Particle Spectrum in the Quantum Filed Theory},
JETP Lett. {\bf 20} (1974) 194.

\bibitem{GNSS} S.~Gonorazky, C.~N\'u\~nez, F.A.~Schaposnik
and G.~Silva, {\sl Bogomol'nyi Bounds and the Supersymmetric
Born-Infeld Theory}, Nucl.Phys. {\bf B531} (1998) 168, {\tt
hep-th/9805054}.

\bibitem{CS} H.~Christiansen, C.~N\'u\~nez and F.A.~Schaposnik,
{\sl Uni\-que\-ness of Bo\-go\-mol'nyi e\-qua\-tions and
Born-Infeld li\-ke Su\-per\-sy\-mme\-tric theories},  Phys. Lett
{\bf B441} (1998) 185, {\tt hep-th/9807197}.

\bibitem{Tay} W.~Taylor IV, {\sl Lectures on D-Branes, Gauge Theory and
M(atrices)}, Lectures at 2nd Trieste Conference on Duality in
String Theory, {\tt hep-th/9801182}.

\bibitem{G1} J.P.~Gauntlett, J.~Gomis and P.K.~Townsend, {\sl Bounds
for Worldvolume Branes}, JHEP {\bf 01} (1998) 003,  {\tt hep-th/
9711205}

\bibitem{growit} D.J.~Gross y E.~Witten, {\sl Superstring
modifications of Einstein equations}, Nucl.Phys. {\bf B277} (1986)
1.

\bibitem{berg} E.~Bergshoeff, M.~Rakowski y E.~Sezgin, {\sl Higher
derivative super Yang-Mills theories}, Phys. Lett. {\bf B185}
(1987) 371.

\bibitem{t3} R.R.~Metsaev, M.A.Rakhmanov y A.A.~Tseytlin,
{\sl The Born-Infeld action as the effective action in the open
superstring theory}, Phys. Lett. {\bf B193} (1987) 207.

E.~Bergshoeff, E.~Sezgin, C.N. Pope y P.K. Townsend, {\sl The
Born-Infeld Action from Conformal Invariance of the Open
Superstring}, Phys. Lett. {\bf B188} (1987) 70.

\bibitem{Svv} R.~Empar\'an, {\sl Born-Infeld Strings tunneling to
D-branes}, Phys. Lett. {\bf B423} (1998) 71, {\tt hep-th/9711106}.

K.G.~Savvidy, {\sl Brane Death via Born-Infeld String}, {\tt
hep-th/9810163}.

\bibitem{Svv2} C.G.~Savvidy and K.G. Savvidy, {\sl Neumann Boundary
Conditions from Born-Infeld Dynamics}, Nucl. Phys. {\bf B561}
(1999) 117, {\tt hep-th/9902023}.

\bibitem{savi} K.G.~Savvidy, {\sl Born-Infeld action in String
Theory}, (1999), PhD Thesis at Princeton University, {\tt
hep-th/9906075}.

\bibitem{KH} K.~Hashimoto, {\sl Born-Infeld Dynamics in
Uniform Electric Field}, JHEP {\bf 9907} (1999) 016.

\bibitem{GKMTZ} J.P.~Gauntlett, C.~Koehl, D.~Mateos, P.K.~Townsend and
M.~Zamaklar, {\sl Finite energy Dirac-Born-Infeld monopoles and
string junctions}, Phys. Rev. {\bf D60} (1999) 045004, {\tt
hep-th/9903156}.

\bibitem{g2} G.W.~Gibbons, {\sl Branes as BIons}, Class. Quant. Grav.
{\bf 16} (1999) 1471, {\tt hep-th/9803203}.

\bibitem{g3} G.W.~Gibbons, {\sl Wormholes on the World Volume:
Born-Infeld particles and Dirichlet p-branes}, Lecture at
``Quantum Mechanics of Fundamental Systems VI", Santiago, Chile,
{\tt hep-th/9801106}.

\bibitem{GK} K.~Ghoroku and K.~Kaneko, {\sl  Born-Infeld strings
between D-branes}, Phys. Rev. {\bf D61} (2000) 066004, {\tt
hep-th/9908154}.

\bibitem{BGa} O.~Bergman and M.R. Gaberdiel, {\sl Non-BPS Dirichlet
branes}, Class. Quant. Grav. {\bf17} (2000) 961, {\tt
hep-th/9908126}.

\bibitem{bialy} I.~Bialynicki-Birula, {\sl Non-Linear
electrodynamics: Variations on a Theme of Born-Infeld}, en {\it
Quantum Theory of Particles and Fields}, eds. B.~Jancewicz y
J.~Lukierski, World Scientific, (1983).

\bibitem{schro} E. Schr\"odinger, Proc. R. Soc. (London) {\bf A150}
(1935) 465.

\bibitem{GR} G.~Gibbons and D.A.~Rasheed, {\sl Electric-Magnetic
duality rotations in non-linear electrodynamics}, Nucl. Phys. {\bf
B454} (1995) 185, {\tt hep-th/9506035}.

\bibitem{zumi} D. Brace, B. Morariu y B. Zumino, {\sl Duality Invariant
Born-Infeld Theory}, {\tt hep-th/9905218}.

M. Gaillard y B. Zumino, {\sl Nonlinear Electromagnetic
Selfduality and Legendre Transformations}, Contributed to Newton
Institute Euroconference on Duality and Supersymmetric Theories,
Cambridge, {\tt hep-th/9712103}.

A.A. Tseytlin, {\sl Self-duality of Born-Infeld action and
Dirichlet 3-brane of type IIB superstring theory}, Nucl. Phys.
{\bf B469} (1996) 51, {\tt hep-th/9602064}.

\bibitem{mono} P.A.M. Dirac, {\sl Quantized singularities in the
electromagnetic field}, Proc. Roy. Soc. (London) {\bf A133} (1931)
60.

P.A.M. Dirac, {\sl The theory of magnetic poles}, Phys. Rev. {\bf
74} (1948) 817.

L. Alvarez-Gaume y S.F. Hassan, {\sl Introduction to S duality in
$N=2$ supersymmetric gauge theories: A pedagogical review of the
work of Seiberg and Witten}, Fortsch. Phys. {\bf 45} (1997) 159,
{\tt hep-th/9701069}.

\bibitem{corson} E.M.~Corson, {\sl Introduction to Tensors,
Spinors and Relativistic Wave-Equations}, Blackie \& Son Limited,
London, (1953).

\bibitem{ramond} P.~Ramond, {\sl Field Theory: A Modern Primer},
Addison-Wesley, 1989.

\bibitem{gui2} S. Gonorazky, F.A. Schaposnik y G. Silva, {\sl Supersymmetric
Nonabelian Born-Infeld Theory}, Phys. Lett. {\bf B449} (1999) 187,
{\tt hep-th/9812094}.

\bibitem{gui3} H.R. Christiansen, N. Grandi, F.A.
Schaposnik y G. Silva, {\sl  Non-BPS Dyons and branes in the
Dirac-Born-Infeld theory}, Phys. Rev. {\bf D61} (2000) 105016,
{\tt hep-th/9911119}.

\end{thebibliography}
\end{document}